\documentclass[11pt,a4paper,oneside]{article}
\usepackage[utf8]{inputenc}
\usepackage{amsmath}
\usepackage{amsfonts}
\usepackage{amssymb}
\usepackage{enumerate}
\usepackage{siunitx}
\usepackage{makeidx}
\usepackage[toc,page]{appendix} % multiple appendices
\usepackage{pdflscape} % landscape page
\usepackage{booktabs,threeparttable,adjustbox,multirow}
\usepackage{makecell} % for \makecell in tables
\usepackage[left=1.85cm,right=1.85cm,top=1.85cm,bottom=1.85cm]{geometry} % margin
\usepackage{float}
\usepackage[table]{xcolor}

\definecolor{prp}{RGB}{108,43,145}

\usepackage{makecell}
\usepackage[bottom]{footmisc}
\usepackage{siunitx}
\usepackage{titling}

\usepackage{verbatim}
\usepackage{listing}
\usepackage{parskip} % inline spacing

\usepackage{setspace} % paragraph spacing

\usepackage{longtable}
\usepackage{booktabs,tablefootnote,url}
\usepackage{adjustbox}
\usepackage{float}
\usepackage{chngcntr} % allows counter linking

\usepackage{graphicx}

\usepackage{longtable}
\usepackage{multirow}
\usepackage{tabu}
\usepackage{etoolbox}
\usepackage{setspace}
\usepackage{array,ragged2e}

\newcolumntype{L}[1]{%
    >{\raggedright\arraybackslash}p{#1}%
}

\usepackage[normalem]{ulem}

\usepackage{threeparttablex}
\usepackage{adjustbox}
\usepackage{mathtools}

\usepackage[super]{nth}
\usepackage{colortbl}

\usepackage{subcaption}
\usepackage{caption}

\usepackage{xurl}
\usepackage{authblk}

\usepackage{natbib}
\usepackage{bibunits}

\defaultbibliographystyle{chicago}
\defaultbibliography{bib_list}

\usepackage{afterpage}
\usepackage{capt-of}

\usepackage{authblk}
\usepackage{etoolbox}

\DeclareMathSymbol{\mh}{\mathord}{operators}{`\-}

\usepackage{tikz}
\usepackage{graphicx}

\usetikzlibrary{positioning}

\makeatletter

\newcommand*{\rom}[1]{%
    \expandafter\@slowromancap\romannumeral #1@%
}

\makeatother

\newcounter{appendixCounter}

\usepackage{siunitx}

\definecolor{AcademicBlue}{RGB}{25,105,200}

\usepackage[
    unicode,
    breaklinks,
    colorlinks
]{hyperref}

\hypersetup{
    colorlinks=true,
    linkcolor=AcademicBlue,
    citecolor=AcademicBlue,
    urlcolor=AcademicBlue,
    filecolor=AcademicBlue,
    pdfborder={0 0 0}
}

\usepackage{xr-hyper}

\usepackage[nameinlink]{cleveref}

\crefname{section}{section}{sections}
\Crefname{section}{Section}{Sections}

\crefname{appendix}{section}{sections}
\Crefname{appendix}{Section}{Sections}

\crefname{subsection}{section}{sections}
\Crefname{subsection}{Section}{Sections}

\crefname{subsubsection}{section}{sections}
\Crefname{subsubsection}{Section}{Sections}

\newcommand{\appendixref}[1]{%
    \hyperref[#1]{Appendix~\ref{#1}}%
}

\newcommand{\RsqOOS}{R^2_{\mathrm{OOS}}}

\usepackage{fancyhdr}
\usepackage{pdflscape}
\usepackage{eso-pic}

\fancypagestyle{landscape}{
    \fancyhf{}

    \fancyfoot{}

    \AddToShipoutPictureFG*{%
        \AtPageLowerLeft{%
            \put(
                \LenToUnit{\dimexpr.5\paperheight\relax},
                \LenToUnit{14.8cm}
            ){%
                \makebox(10.4cm,0){%
                    \rotatebox{90}{\thepage}%
                }%
            }%
        }%
    }%
}

\begin{document}
\begin{bibunit}
% First page (title)
\pagenumbering{gobble}   % temporarily disable page numbers

\onehalfspacing

\linespread{0.01}\selectfont % slightly tighter spacing for title page

% --- Title setup ---
\setlength{\droptitle}{-4.4em} % moves the title upward
\pretitle{\begin{center}\Large\bfseries \end{center}}
\posttitle{\begin{center}\par\vspace{0.5em}\end{center}} % space between title and authors

\title{\LARGE\bfseries Re(Visiting) Time Series Foundation Models in Finance}

\author{
  \vspace{-1.9em}
  \large \textbf{Eghbal Rahimikia}\textsuperscript{*}\\
  {\vspace{-1em}}
  \normalsize Alliance Manchester Business School, University of Manchester\\
  {\vspace{-0.35em}}
  \normalsize \href{mailto:eghbal.rahimikia@manchester.ac.uk}{eghbal.rahimikia@manchester.ac.uk}
  \vspace{-1.75em}\and % ← added consistent vertical space
  \large \textbf{Hao Ni}\\
  {\vspace{-1em}}
  \normalsize Department of Mathematics, University College London (UCL)\\
  {\vspace{-0.35em}}
  \normalsize \href{mailto:h.ni@ucl.ac.uk}{h.ni@ucl.ac.uk}
  \vspace{-1.1em}\and % ← same space before next block
  \large \textbf{Weiguan Wang}\\
  {\vspace{-1em}}
  \normalsize School of Economics, Shanghai University\\
  {\vspace{-0.35em}}
  \normalsize \href{mailto:weiguanwang@shu.edu.cn}{weiguanwang@shu.edu.cn}}

\captionsetup{font=footnotesize}

% --- Date ---
\date{\vspace{-0.3em}{\footnotesize\textit{First Version (Nov. 2025) \hspace{1.5em} This Version (Nov. 2026)}}}
\maketitle
\begingroup
\renewcommand\thefootnote{}
\footnotetext{* Corresponding author.}
\addtocounter{footnote}{0}
\endgroup
\vspace{-4.7em}

% --- Acknowledgment footnote ---

\begingroup
\renewcommand\thefootnote{}
\footnotetext{
The authors acknowledge the use of resources provided by the Isambard-AI National AI Research Resource (AIRR). Isambard-AI is operated by the University of Bristol and funded by the UK Government's Department for Science, Innovation and Technology (DSIT) via UK Research and Innovation and the Science and Technology Facilities Council [ST/AIRR/I-A-I/1023]. The authors would also like to acknowledge the assistance provided by Research IT and the use of the Computational Shared Facility at The University of Manchester. This work also made use of the facilities of the N8 Centre of Excellence in Computationally Intensive Research (N8 CIR), funded by the N8 Research Partnership and the EPSRC (Grant No.~EP/T022167/1). Hao Ni is supported by the EPSRC under Programme Grant EP/S026347/1 and the Alan Turing Institute under EPSRC Grant EP/N510129/1. Weiguan Wang acknowledges financial support from the National Natural Science Foundation of China (No.~72201158) and the Shanghai Technical Service Center of Science and Engineering Computing, Shanghai University. We are grateful for the valuable feedback and insightful comments received from participants at the AllianceBernstein Q-Group Research Seminar, the PIMCO Research Seminar, the London Quant Intelligence Forum, S\&P Global (London, UK), the \nth{5} Frontiers of Factor Investing Conference at Lancaster University (Lancaster, UK), the New Methods Research Seminar Series at the Bank of England (London, UK), the Research Seminar at the School of Economics, Shanghai University (Shanghai, China), the Silicon Valley Certification Hub Seminar (Silicon Valley, USA), Future Finance Fest (3F) (Amsterdam, Netherlands), the NeurIPS 2025 Workshop on Generative AI in Finance (San Diego, USA), the STAC Summit (London, UK), the \nth{13} World Congress of the Bachelier Finance Society (Bologna, Italy), the \nth{3} AI in Finance 2026 Conference at Concordia University (Montreal, Canada), the Dataism Laboratory for Quantitative Finance (DLQF) at Virginia Tech (Blacksburg, USA), the \nth{18} Annual Society for Financial Econometrics (SoFiE) Conference at the University of Macau (Macau, China), the EcoMod2026 International Conference on Economic Modeling and Data Science at the University of Luxembourg (Luxembourg City, Luxembourg), the P2P Financial Systems Workshop 2026 at the National Bank of Kazakhstan (Astana, Kazakhstan), the Future of AI in Financial Services (London, UK), the \nth{4} International Conference on Empirical Economics,
Penn State Altoona (Pittsburgh, USA), the \nth{3} Conference on AI in Finance at TU Dresden (Dresden, Germany), and the \nth{6} Finance and Accounting 2026 Annual Symposium at the University of Westminster (London, UK). We are also grateful to Hang Lou and Callum Lau for proofreading the manuscript and providing valuable feedback. All models are available at \href{https://FinText.ai}{\texttt{FinText.ai}} and \href{https://huggingface.co/FinText}{\texttt{Hugging Face}}.
}
\addtocounter{footnote}{0}
\endgroup

\vspace{0.01cm}
% --- Abstract ---
\renewcommand{\abstractname}{\hspace{-1.4mm}Abstract} % moves left
\vspace{2.9em}\begin{abstract}
\normalsize
\onehalfspacing
%\noindent Financial time series forecasting is central to trading, portfolio optimization, and risk management, yet it remains challenging due to noisy, non-stationary, and heterogeneous data. Recent advances in time series foundation models (TSFMs), inspired by large language models, offer a new paradigm for learning generalizable temporal representations from large and diverse datasets. This paper presents the first comprehensive empirical study of TSFMs in global financial markets. Using a large-scale dataset of daily excess returns across diverse markets, we evaluate zero-shot inference, fine-tuning, and pre-training from scratch against strong benchmark models. We find that off-the-shelf pre-trained TSFMs perform poorly in zero-shot and fine-tuning settings, whereas models pre-trained from scratch on financial data achieve substantial forecasting and economic improvements, underscoring the value of domain-specific adaptation. Increasing the dataset size, incorporating synthetic data augmentation, and applying hyperparameter tuning further enhance performance.
%\end{abstract}
\vspace{-0.9em}
\noindent Financial time series forecasting is vital for trading, portfolio optimization, and risk management but is challenged by noisy, nonstationary, and heterogeneous data. Time series foundation models (TSFMs), inspired by large language models, aim to learn transferable temporal representations. Whether pre-training on massive, diverse datasets enables similar generalization to finance remains an open question. This study provides a comprehensive empirical evaluation of TSFMs in global financial markets using large-scale daily excess-return data. We assess zero-shot inference, fine-tuning, and pre-training from scratch against strong benchmarks. Off-the-shelf TSFMs generally perform poorly, while finance-native pre-training improves long-window gross portfolio performance. Larger pre-training datasets and synthetic augmentation improve selected specifications, particularly at longer input windows. Configuration changes have mixed effects across forecasting and portfolio metrics, and TSFMs do not consistently outperform strong benchmarks. These findings caution against uncritical AI adoption in finance, where insufficient validation can lead to unreliable forecasts and substantial financial losses.
\end{abstract}

\vspace{-0.3em}

% --- Keywords and JEL ---
{\small

\vspace{-0.05em}
}
\vspace{-0.05em}
%\endgroup
\clearpage

% Start numbering from page 1 on the next page
\clearpage
\pagenumbering{arabic}   % use Arabic numbers
\setcounter{page}{1}
\pagestyle{fancy}        % restore your fancy page style
\section{Introduction}
\doublespacing
Financial time series forecasting is a central problem in quantitative finance, underpinning trading, portfolio construction, and risk management. Decades of research have produced a wide spectrum of models, from classical econometric approaches (e.g., ARIMA, GARCH) to modern machine learning (ML) models such as tree-based ensembles and deep neural networks. The common empirical lesson is that forecasting gains are possible but difficult to obtain. Financial data are noisy, nonstationary, and heterogeneous across assets and horizons, with low signal-to-noise ratios, regime shifts, and limited effective sample sizes for many instruments. These properties complicate generalization and make out-of-sample evaluation, rather than in-sample fit, the central test of a forecasting model.

The success of large language models (LLMs) has popularized a pre-training and fine-tuning paradigm for learning general-purpose representations that transfer across tasks and domains. This paradigm has inspired the emergence of time series foundation models (TSFMs)\footnote{In \citet{das2024DecoderonlyFoundationModel}, TimesFM is used as the abbreviation for time series foundation model. However, most of the recent literature adopts TSFM as the standard abbreviation for time series foundation model. Accordingly, in this study, we use ‘TSFM’ as the generic term for time series foundation models, and reserve ‘TimesFM’ to refer to the specific model of \citet{das2024DecoderonlyFoundationModel}.}, large pre-trained architectures designed to learn reusable temporal representations from vast and diverse time series corpora. Research on TSFMs is expanding rapidly, with frequent releases of new models and successive generations extending their capabilities.
Development spans major technology companies, specialist AI firms, and academic research institutions, reflecting broad interest in general-purpose time series forecasting. This expansion strengthens the need to assess whether progress on general forecasting tasks translates into reliable performance in financial markets. The appeal of this approach for finance is clear. If TSFMs have learned generic temporal structure, they may reduce the need to estimate a separate model for every asset, market, or horizon, and they may transfer useful representations to financial datasets with little task-specific training.

An important issue, however, is whether this promise survives in the setting that matters for investors: real-time forecasting of noisy daily excess returns. A model trained on broad non-financial time series may learn seasonality, trend, or long-memory patterns, but those patterns need not translate into useful cross-sectional rankings of stock returns. Moreover, standard foundation-model workflows often treat fine-tuning as the natural remedy when zero-shot performance is weak. In financial applications, this assumption is not innocuous because the information set is chronological, the predictive signal is small, and economic performance depends on portfolio rankings rather than point-forecast calibration alone. The relevant question is therefore not simply whether TSFMs are powerful forecasting models in general, but which training regime, if any, makes them useful for financial return prediction.

\begin{figure}
    \caption{Forecasting Performance across TSFM Training Regimes}
    \label{fig:intro_training_regimes_forecasting}
    \centering
    \includegraphics[width=0.77\textwidth]{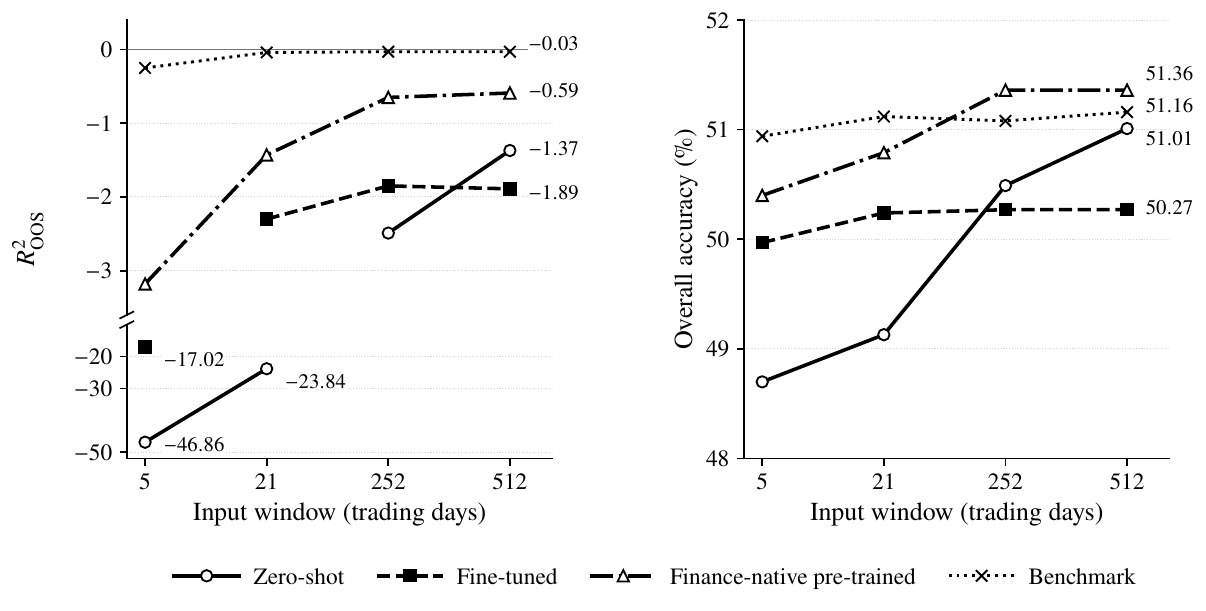}

    \begin{minipage}{0.82\textwidth}
        \scriptsize
        \textbf{Note:} The figure reports representative U.S. forecasting performance across input windows of 5, 21, 252, and 512 trading days. The left panel uses a broken vertical axis, with separate linear scales for the ranges $[-52,-10]$ and $[-3.6,0.4]$. Diagonal marks on the vertical axis indicate the omitted interval. The Chronos configurations are large for zero-shot inference and fine-tuning, and small for finance-native pre-training. The figure compares representative specifications, with differences reflecting both the training regime and model configuration.
    \end{minipage}
\end{figure}

This study examines this question using a large-scale panel of daily excess returns spanning 34 years across 94 countries and regions. We evaluate three TSFM regimes at the univariate level: (i) zero-shot inference with public pre-trained weights, (ii) fine-tuning public checkpoints on financial data, and (iii) pre-training from scratch on financial time series. We compare these approaches with a broad set of benchmarks, including linear models, ensemble models\footnote{Throughout this study, ensemble models refer specifically to tree-based ensemble models.}, and neural networks. Our largest pre-training corpus contains approximately 1.70 billion scalar observations across return and characteristic series. The analysis enables model development at the international level and testing across major markets, including the United States, Hong Kong, Taiwan, South Korea, Germany, the United Kingdom, India, and Australia. We assess model performance from both statistical and economic perspectives, using out-of-sample forecasting metrics under an expanding-window design and portfolio returns constructed from model forecasts. From an asset-pricing perspective, this design examines whether alternative pre-training regimes improve out-of-sample return prediction and cross-sectional portfolio rankings under a common chronological evaluation protocol.

\begin{figure}
    \caption{Portfolio Performance across TSFM Training Regimes}
    \label{fig:intro_training_regimes_portfolio}
    \centering
    \includegraphics[width=0.77\textwidth]{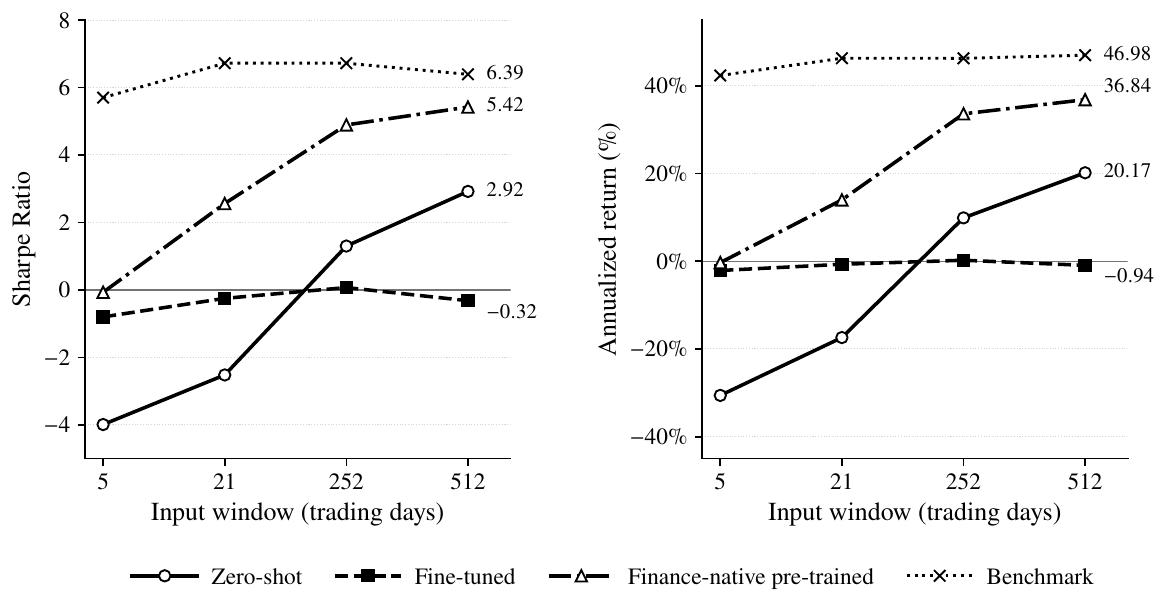}

    \begin{minipage}{0.82\textwidth}
        \scriptsize
        \textbf{Note:} The figure reports representative U.S. equal-weighted (EW) long--short portfolio performance across input windows of 5, 21, 252, and 512 trading days. The Chronos configurations are large for zero-shot inference and fine-tuning, and small for finance-native pre-training. The figure compares representative specifications, with differences reflecting both the training regime and model configuration.
    \end{minipage}
\end{figure}

\Cref{fig:intro_training_regimes_forecasting} and \Cref{fig:intro_training_regimes_portfolio} summarize the forecasting and portfolio performance evidence, respectively. The figures deliberately focus on a narrow comparison: CatBoost, which achieves the highest gross equal-weighted (EW) portfolio Sharpe ratio (SR) among the benchmarks at the 21-, 252-, and 512-day input windows, and representative Chronos specifications under zero-shot inference, fine-tuning, and finance-native pre-training from scratch. They therefore do not replace the full empirical analysis presented in this study; instead, they highlight its central comparison: when TSFMs are applied to daily equity returns, how much of their performance comes from generic transfer, from adapting public checkpoints, and from learning the return distribution directly.

Three takeaways are evident from the figures. First, the benchmark hurdle is high. Historical return forecasting is difficult, but conventional ML models are not weak baselines in this setting. CatBoost produces near-zero annual-average out-of-sample $R^2$ ($\RsqOOS$) values for the full stock sample across windows, from $-0.25\%$ at the 5-day window to $-0.03\%$ at the 252- and 512-day windows, while keeping overall directional accuracy around 51\%. Economically, the benchmark generates annualized returns of 42.35\%--46.98\% and SRs of 5.70--6.72. The portfolio quantities reported here refer to EW long--short portfolios that are rebalanced daily and reported before transaction costs.

%Thus, TSFMs are evaluated not only against simple linear specifications, but also against models that already extract economically useful rankings from noisy daily excess returns.

% as the $R^2$s achieved by CatBoost are $0.39\%$, $0.67\%$, $0.64\%$ and $0.60\%$, compared to $0.31\%$, $0.30\%$, $0.26\%$, and $0.19\%$ for the linear model.
% We also apply the Deobold-Mariano test to verify the significance of the outperformance of the CatBoost model over other benchmark models.

Second, generic transfer provides unreliable economic value in this setting. Our empirical analysis focuses on two widely adopted TSFMs, Chronos \citep{ansari2024chronos} and TimesFM \citep{das2024DecoderonlyFoundationModel}, and extends to additional architectures in the full results. In the representative forecasting comparison, zero-shot Chronos (Large) has $\RsqOOS$ values of $-46.86\%$, $-23.84\%$, $-2.49\%$, and $-1.37\%$ across the four input windows. Longer histories reduce the most extreme errors, and the zero-shot Chronos (Large) specification generates positive portfolio returns at the 252- and 512-day windows, reaching an annualized return of 20.17\% and an SR of 2.92 at the latter. These values indicate positive gross portfolio performance, but they remain far below the benchmark's 46.98\% annualized return and 6.39 SR at the same window. Results for the additional architectures are heterogeneous: a few zero-shot models produce considerably stronger long-window ranking signals, in one case approaching the benchmark, but none does so consistently across input windows and evaluation metrics. Fine-tuning public checkpoints also does not close the gap. Fine-tuned Chronos (Large) has $\RsqOOS$ values of $-17.02\%$, $-2.30\%$, $-1.85\%$, and $-1.89\%$ at the 5-, 21-, 252-, and 512-day input windows, respectively. Fine-tuning improves its squared-error performance relative to zero-shot inference at the first three windows but worsens it at 512 days, and all four values remain below CatBoost's corresponding results. Its gross EW long--short portfolio also remains weak, with a best SR of only 0.07 and annualized returns close to zero. Overall, these findings suggest that the evaluated TSFM specifications often struggle to translate improvements in point-forecast accuracy into economically valuable cross-sectional return forecasts, highlighting that strong performance on general forecasting benchmarks, including point and probabilistic metrics, does not necessarily translate into economic value.

% \notewg{@Eghbal, refer to the comment in conclusion for the "point estimate".}

Third, finance-native pre-training changes the interpretation of TSFMs. We pre-train TSFMs from scratch for each year using financial data dated through the training cutoff, avoiding look-ahead concerns inherited from external checkpoints. In \Cref{fig:intro_training_regimes_forecasting}, the finance-native Chronos model raises overall accuracy to 51.36\% at both the 252- and 512-day windows and avoids the severe short-window deterioration observed under zero-shot transfer. The remaining $\RsqOOS$ gap to the benchmark is still visible, but \Cref{fig:intro_training_regimes_portfolio} shows why the regime matters economically: finance-native pre-training raises Chronos long--short annualized returns to 33.61\% and 36.84\% at the 252- and 512-day windows, with SRs of 4.89 and 5.42. These values do not dominate the benchmark, but they substantially narrow the economic gap. The evidence therefore suggests that the weak zero-shot and fine-tuning results are not simply failures of the TSFM architecture; rather, they underscore the importance of learning representations directly from financial data instead of adapting generically pre-trained ones.

%  which achieves positive $R^2\approx 0.47$

The first-stage evidence motivates a second stage of the analysis, which examines the role of scale in finance-native pre-training. A central premise of the TSFM literature is that larger pre-training datasets enable models to learn more generalizable temporal representations and thereby improve downstream forecasting performance. We find that the benefit of scale depends on the model class. Expanding the training universe from U.S. to global markets does not uniformly strengthen conventional benchmarks, and in some cases weakens their portfolio performance. For TSFMs, however, larger and more heterogeneous finance-native samples become a central source of improvement. Relative to U.S.-only pre-training, TimesFM (20M) pre-trained on global returns augmented with Jensen--Kelly--Pedersen (JKP) characteristics \citep{jensen2023there} gains 11.95 and 13.47 percentage points in $\RsqOOS$ at the 252- and 512-day windows and raises SRs by 3.19 and 4.00; at the 512-day window, its SR reaches 7.62, above the U.S. CatBoost benchmark of 6.39. We further examine whether the benefits of scale extend to synthetic data, which has become an important component of many TSFMs and is frequently used to expand the effective size of pre-training corpora. Augmenting the finance-native pre-training dataset with synthetic time series yields comparable performance gains, suggesting that the improvement is not driven solely by the financial content of the added characteristic series. Instead, the evidence indicates that the scale and heterogeneity of the finance-native pre-training corpus are themselves important determinants of downstream performance. These scaling results are examined through broader checks over time, under transaction costs, across hyperparameter choices, and in seven major non-U.S. markets. The findings highlight both the relative benefits of scaling and the sensitivity of economic performance to implementation costs and market conditions. Under the position-based cost scenarios considered, net SRs fall sharply. Selected specifications retain positive net SRs under the 20 basis points (bps) or mixed-cost scenarios, whereas all specifications reported in the cost tables have negative net SRs under the 40 bps scenario. The portfolio evidence therefore measures the comparative strength of the models' cross-sectional ranking signals, while highlighting the implementation challenge posed by daily rebalancing.

Taken together, our findings reveal limits to transferring representations learned from large, general-purpose time-series corpora into equity-return prediction. Generic pre-training and subsequent financial fine-tuning do not reliably produce competitive cross-sectional return signals, while training the same model families from scratch on financial data substantially narrows the performance gap with strong benchmarks. TSFMs provide a useful setting for this comparison because their distinct pre-training and adaptation stages allow us to examine whether representations learned outside finance offer an advantage over those learned directly from financial histories. Scaling improves selected specifications, particularly at longer input windows, with gains depending on model family and pre-training composition. The contribution of this study is therefore not simply to apply foundation models to empirical asset pricing, but to identify the conditions under which they improve forecasting and gross portfolio performance, while documenting the limits imposed by trading costs. Our evidence highlights the importance of domain-aligned pre-training, sufficient pre-training scale, and economic rather than purely statistical evaluation, factors that have been largely overlooked in mainstream TSFM research.

\subsection{Related Work and Contributions} \label{related_work_section}
This study relates first to the literature on return predictability and machine learning in empirical asset pricing. Classic evidence on momentum, reversal, technical trading rules, and time-series momentum shows that historical prices contain weak but economically meaningful information about future returns \citep{jegadeesh1993ReturnsBuyingWinners,jegadeesh1990EvidencePredictableBehavior,brock1992simple,neely2014ForecastingEquityRisk,moskowitz2012time,asness2013value}. Automated technical-pattern recognition \citep{lo2000foundations} and the learning of predictive representations from price charts \citep{jiang2023re} extend this literature beyond predefined return signals. A related ML literature demonstrates gains from combining large predictor sets and capturing nonlinear interactions \citep{gu2020empirical,leippold2022MachineLearningChinese,chen2024deep,kelly2024virtue,kelly2025artificial}. The ability to learn weak signals also depends on the form of regularization \citep{shen2025can}, reinforcing the importance of comparing alternative learning methods in financial forecasting. In portfolio choice, \citet{simon2026deep} show that investor risk aversion acts as an economic regularizer in deep parametric portfolio policies, with the benefits of model complexity diminishing as risk aversion increases. These studies motivate our use of historical returns and discipline our benchmark design. Closely related work uses convolutional transformers to extract signals from factor-based residual portfolios for statistical arbitrage \citep{guijarro2026deep}. We complement these approaches by evaluating how the source and adaptation of pre-trained representations affect daily excess-return forecasting. Our comparison of generic pre-training, financial fine-tuning, and chronological finance-native pre-training, together with global data and augmentation, provides evidence on the conditions under which TSFMs become competitive with strong linear and nonlinear benchmarks.

This study also contributes to the literature on foundation models and domain adaptation. Large-scale pre-training has produced transferable representations in language models and, more recently, in time series forecasting, where TSFMs are designed to support zero-shot or few-shot transfer across datasets \citep{ouyang2022training,touvron2023llama,liu2024deepseek,ansari2024chronos,das2024DecoderonlyFoundationModel}. Work on financial language models shows, however, that finance often requires domain-specific adaptation and careful chronological design \citep{huang2023finbert,rahimikia2024r,he2025chronologically}. While \citet{hellum2023global} study transfer learning across countries, we also examine transfer from general-purpose time-series corpora
into equity-return prediction. Our contribution is threefold. First, we provide one of the largest empirical forecasting evaluations in asset pricing, comparing a broad range of linear, nonlinear, neural, and foundation models using pre-training corpora containing up to approximately 1.70 billion scalar observations, with return data spanning 34 years across global markets under an annual expanding-window evaluation design. Second, we combine statistical metrics with portfolio-level evaluation, which allows us to separate improvements in point-forecast fit from economically useful cross-sectional rankings. Third, we identify the conditions under which TSFMs become useful in this domain: finance-native pre-training, sufficient data scale and heterogeneity, and comparison against strong ML benchmarks. This evidence complements recent generative approaches to financial forecasting \citep{wiese2020quant,Vuletić01022024,liao2024sig,huang2024generative}, while providing a systematic benchmark of TSFMs at global scale.

The remainder of this study is organized as follows. \Cref{sec: ts_foundation_model} introduces the core concepts underlying TSFMs, detailing their architectures and training paradigms. \Cref{sec: methodology} outlines the proposed methodology, including the modeling framework, evaluation metrics, and experimental setup. \Cref{sec:data} describes the data sources, cleaning procedures, and computational resources used to conduct our experiments. \Cref{sec: numerical_results} presents the main numerical findings, comparing benchmark models with TSFMs across various configurations and data volumes. Finally, \Cref{sec:conclusion} summarizes the key insights, discusses implications for future research, and concludes the study.

\section{Time Series Foundation Models}\label{sec: ts_foundation_model}
This section outlines the core principles underlying TSFMs. We first provide a compact overview of the LLM concepts that are most relevant to time series forecasting. We then describe a unified TSFM framework and use it to compare two representative models, Chronos \citep{ansari2024chronos} and TimesFM \citep{das2024DecoderonlyFoundationModel}. Additional TSFMs used in our numerical experiments are summarized in \Cref{appendix: tsf_model} of the Internet Appendix. For a broader taxonomy of TSFMs by embedding design, architecture, pre-training objective, and adaptation strategy, see \citet{liang2024FoundationModelsTime}.

\subsection{LLMs as a Bridge to TSFMs} \label{llm_review}
Autoregressive LLMs are high-capacity sequence models trained to predict the next token from preceding tokens. Raw text is first mapped into a sequence of discrete indices \(z_{1:T}\), where each token belongs to a finite vocabulary \(\mathcal{V}\). These tokens are embedded into continuous vectors, augmented with positional information, and passed through a Transformer backbone \citep{vaswani2017attention}. The key operation in the Transformer is self-attention, which maps an input matrix \(X \in \mathbb{R}^{T \times d}\) into contextual representations by comparing all positions in the sequence:
\begin{equation}
\operatorname{Attention}(Q,K,V)
=
\operatorname{Softmax}\left(
\frac{QK^\top}{\sqrt{d_k}}+M
\right)V,
\end{equation}
where \(Q=XW^Q\), \(K=XW^K\), and \(V=XW^V\) are learned projections of the input, and \(d_k\) is the dimension of each key vector. For autoregressive decoder self-attention, the causal mask satisfies \(M_{ij}=0\) for \(j\leq i\) and \(M_{ij}=-\infty\) for \(j>i\), preventing each position from attending to future positions. The standard autoregressive objective is
\begin{equation}
\max_\theta \sum_{t=1}^T \log p_\theta \left( z_t \mid z_{<t}\right),
\end{equation}
where \(z_{<t}=(z_1,\ldots,z_{t-1})\). At inference time, the model recursively samples or selects the next token, appends it to the context, and continues until the desired output length is reached. This pre-training and inference paradigm is the conceptual basis for many TSFMs: a time series is converted into a sequence of model-readable units, processed by a scalable sequence model, and decoded into forecasts.

\subsection{A Unified Framework for TSFMs}
Consider a time series \(\mathbf{x}_{1:C+H}=(x_1,\ldots,x_{C+H})\), where \(\mathbf{x}_{1:C}\) is the historical context and \(\mathbf{x}_{C+1:C+H}\) is the forecasting target. A TSFM can be viewed as the composition of the following modules:
\begin{equation}
\mathbf{x}_{1:C}
\xrightarrow{\text{representation}}
\mathbf{u}_{1:m}
\xrightarrow{\text{sequence model}}
\mathcal{H}
\xrightarrow{\text{output and decoding}}
\begin{cases}
\widehat{P}_{\theta}(\cdot\mid\mathbf{x}_{1:C}),\\
\widehat{\mathbf{x}}_{C+1:C+H}.
\end{cases}
\end{equation}
Here, \(m\) is the number of input units, \(\mathcal{H}\) denotes the hidden representations used during forecasting, and \(\widehat{P}_{\theta}(\cdot\mid\mathbf{x}_{1:C})\) denotes a predictive distribution over the forecast horizon. The two outputs represent distributional and point forecasting, respectively. The diagram is schematic: autoregressive models repeatedly apply the sequence model and output head, conditioning subsequent predictions on previously generated outputs. The representation layer converts raw numeric observations into units suitable for sequence modeling. These units may be discrete value tokens, as in Chronos, or continuous patch embeddings, as in TimesFM. The sequence model processes the observed context and produces representations for forecasting. The Chronos-T5 variants used in our experiments employ an encoder--decoder Transformer with a bidirectional context encoder and an autoregressive decoder, whereas TimesFM uses a decoder-only Transformer with causal self-attention over input patches. The output head maps hidden states either to a probability distribution over future values or directly to continuous forecasts. The training objective and inference procedure depend on the output parameterization. Discrete token models can use categorical likelihood objectives, while continuous models can use point, quantile, or distributional objectives.

This unified view makes clear that Chronos and TimesFM share the same high-level pipeline but differ in how they represent time series values and how they train and decode forecasts. \Cref{tab:chronos_timesfm_framework} summarizes the main similarities and differences between the two models. Chronos follows an LLM-style discrete-token formulation, whereas TimesFM uses continuous patch embeddings and a regression-style forecasting objective. Representation and decoding strategy are distinct design choices. In our empirical application, \(H=1\): we predict next-day excess returns using the mean of sampled forecasts from Chronos and the point-forecast output from TimesFM.\footnote{The supplementary distributional illustration in \Cref{chronos_excess_return_30_days_forecast} considers horizons of 1--30 trading days. All main forecasting and portfolio results use a one-day forecasting horizon.}

\begin{table}
\centering
\begin{threeparttable}
\begin{adjustbox}{width=0.80\textwidth,center}
\scriptsize
\captionsetup{width=\linewidth}
\caption{Chronos and TimesFM under a Unified TSFM Framework}
\label{tab:chronos_timesfm_framework}
\begin{minipage}{\linewidth}
\begin{tabular}{lll}
\toprule
\textbf{Component} & \textbf{Chronos} & \textbf{TimesFM} \\
\midrule
Representation & Scaled and quantized value tokens & Continuous patches \\
Backbone & T5 encoder--decoder & Decoder-only Transformer \\
Output & Categorical distribution over tokens & Continuous forecast patch \\
Training loss & Token cross-entropy & MSE for the point-forecast formulation \\
Inference & Sampling and dequantization & Block prediction and autoregressive rollout \\
Evaluated forecast & Mean of sampled forecasts & Point-forecast output \\
\bottomrule
\end{tabular}
\vspace{0.1cm}
\begin{tablenotes}[para,flushleft]
\scriptsize
\textbf{Note:}  Comparison of Chronos and TimesFM under the unified TSFM framework. The two models share the same high-level pipeline but differ in representation, training objective, inference strategy, and forecasting output. The TimesFM loss entry describes the point-forecast formulation;
released checkpoints also expose experimental quantile outputs.
\end{tablenotes}
\end{minipage}
\end{adjustbox}
\end{threeparttable}
\end{table}

\subsection{Chronos}\label{subsec: chronos}
Chronos \citep{ansari2024chronos}, developed by researchers at Amazon, adapts the LLM pipeline to real-valued time series by replacing text tokenization with value quantization. For a fully observed historical context of length \(C\) containing at least one nonzero value, normalization by the mean absolute value is given by
\begin{equation}
s=\frac{1}{C}\sum_{i=1}^{C}|x_i|, \qquad \tilde{x}_i=\frac{x_i}{s}.
\end{equation}
The scaled observations are then mapped into a finite vocabulary through a quantization function \(q:\mathbb{R}\to\{1,\ldots,B\}\):
\begin{equation}
q(\tilde{x}) =
\begin{cases}
1 & \text{if } -\infty \leq \tilde{x} < b_1, \\
2 & \text{if } b_1 \leq \tilde{x} < b_2, \\
\vdots \\
B & \text{if } b_{B-1} \leq \tilde{x} < \infty,
\end{cases}
\end{equation}
where \((b_i)_i\) are bin edges. The resulting token sequence can be processed by a standard Transformer architecture. Following the main experimental setup of \citet{ansari2024chronos}, we use Chronos variants based on the T5 encoder--decoder architecture \citep{raffel2020exploring}.

Chronos is trained with a cross-entropy loss over discrete tokens and produces a categorical distribution for the next token. At inference time, future values are generated autoregressively by sampling from this distribution. Each predicted token \(j\) is mapped back to a bin center \(d(j)=c_j\), then rescaled by \(s\) to obtain the forecast on the original scale. Under the default Chronos configuration, reconstructed values lie in approximately \([-15s,15s]\), with adjacent bin-center spacing \(30s/(B-1)\). Values outside this range are assigned to the unbounded outer bins and reconstructed at the corresponding endpoint centers. Thus, a larger \(s\) coarsens the quantization resolution on the original scale, while a smaller \(s\) narrows the representable range. This design enables probabilistic forecasting through repeated sampling, but it also introduces discretization error and a finite representable range.

\subsection{TimesFM} \label{subsec: TimesFM}
TimesFM \citep{das2024DecoderonlyFoundationModel}, developed by researchers at Google, keeps the time series in a continuous representation rather than converting each scalar value into a discrete token. The following description summarizes the shared patch-based forecasting architecture. Our main experiments use TimesFM 1 and TimesFM 2 checkpoints, together with smaller architectures derived from TimesFM 2. The extended zero-shot results also include TimesFM 2.5. For a univariate historical context \(\mathbf{x}_{1:C}=(x_1,\ldots,x_C)\), with \(x_t\in\mathbb{R}\), the input is partitioned into non-overlapping patches of length \(L\). For simplicity, first suppose that \(C\) is divisible by \(L\). The input patches are \begin{equation}\label{TimesFM_eq_1}
\mathbf{p}_i
=
(x_{(i-1)L+1},\ldots,x_{iL})^\top
\in\mathbb{R}^{L},
\qquad i=1,\ldots,P,\quad P=C/L.
\end{equation}
When \(C\) is not divisible by \(L\), the context is padded to length \(L\lceil C/L\rceil\), and padding positions are masked. Each patch is embedded by a residual network, typically implemented as a multilayer perceptron:
\begin{equation}
\mathbf{e}_i=f_\phi(\mathbf{p}_i), \qquad \mathbf{e}_i\in\mathbb{R}^h,
\end{equation}
where \(h\) is the embedding dimension. TimesFM also uses masking during training so that the model is exposed to different effective context lengths. The embedded patches are passed through a decoder-style Transformer with causal self-attention.

The output head maps each hidden state back into the time domain through another residual network. Let \(L_{\mathrm{out}}\) denote the output patch length, which need not equal the input patch length \(L\). For an unpadded input prefix ending at time \(iL\),
\begin{equation}
\widehat{\mathbf{x}}_{iL+1:iL+L_{\mathrm{out}}}
=g_\psi(\mathbf{h}_i)
\in\mathbb{R}^{L_{\mathrm{out}}},
\end{equation}
where \(\mathbf{h}_i\) is the Transformer hidden state at input patch \(i\). Unlike Chronos's token cross-entropy objective, the point-forecast formulation of TimesFM described by
\citet{das2024DecoderonlyFoundationModel} uses mean squared error between predicted and realized values. Forecasts within an output patch are produced jointly; longer horizons are generated autoregressively by feeding predicted output blocks back into the model. This continuous formulation avoids quantization and dequantization. Although released TimesFM checkpoints also expose experimental quantile outputs, our empirical analysis uses the point-forecast output.

\section{Methodology}\label{sec: methodology}

\subsection{Problem Setup} \label{problem_steup}

Consider a financial market with \(M\) assets. Let \(S_t^{(i)}\) denote the price of asset \(i\) on trading day \(t\), and let \(D_t^{(i)}\) denote the cash dividend or other cash distribution paid between \(t-1\) and \(t\). The one-day total return is
\begin{equation}
r_t^{(i)}=\frac{S_t^{(i)}+D_t^{(i)}-S_{t-1}^{(i)}}{S_{t-1}^{(i)}}.
\end{equation}
We work with daily excess returns,
\begin{equation}
r_t^{(i),ex}=r_t^{(i)}-r_{f,t}^{(d)},
\end{equation}
where \(r_{f,t}^{(d)}\) is the daily risk-free rate. The forecasting task is to predict the next-day excess return \(r_{t+1}^{(i),ex}\) from a fixed historical window
\begin{equation}
\mathbf{r}_{t-C+1:t}^{(i),ex}=(r_{t-C+1}^{(i),ex},\ldots,r_t^{(i),ex}),
\end{equation}
where \(C\) is the input length. Our empirical design pools observations across assets to estimate a shared forecasting rule. A forecasting model is trained on pooled asset-level return histories and then applied cross-sectionally to all stocks in the investable universe. In the following, we suppress the asset index \(i\) when no confusion arises.

\subsection{Forecasting Interfaces}
The TSFMs introduced in \Cref{sec: ts_foundation_model} enter the financial task through their forecasting interface rather than through model-specific architectural details. For a given input window \(\mathbf{r}_{t-C+1:t}^{ex}\), a model can return either a predictive distribution or a point forecast. We use \(F_\theta\) to denote the point-forecast interface. It returns a point forecast obtained by model-specific aggregation of predictive samples or quantile outputs, or directly from the model when a point forecast is provided. The stochastic and deterministic formulations below describe the principal interfaces used in our experiments.

\textbf{Stochastic TSFMs.} Probabilistic models, such as Chronos-type TSFMs, generate samples from an estimated conditional distribution. Let \(G_\theta:\mathcal{Z}\times\mathbb{R}^{C}\rightarrow\mathbb{R}\) denote a conditional generator, where \(Z\in\mathcal{Z}\) is a random seed or latent noise variable. The model generates samples according to
\begin{equation}
G_\theta\!\left(Z,\mathbf{r}_{t-C+1:t}^{ex}\right)
\sim
\widehat{P}_\theta\!\left(\,\cdot\mid
\mathbf{r}_{t-C+1:t}^{ex}\right),
\end{equation}
where \(\widehat{P}_\theta\) is the estimated conditional distribution based on the supplied historical context. For models evaluated using sample-mean aggregation, the conditional mean forecast is estimated from \(N\) Monte Carlo samples by
\begin{equation}
\hat{r}_{t+1}^{ex}
=\frac{1}{N}\sum_{n=1}^{N}G_\theta\!\left(Z_n,\mathbf{r}_{t-C+1:t}^{ex}\right),
\qquad Z_n \overset{\mathrm{iid}}{\sim} P_Z.
\end{equation}
Here, $P_Z$ denotes the distribution of the latent random variable $Z$. The probability of a strictly positive excess return can be estimated as the fraction of sampled excess returns above zero.

\textbf{Deterministic TSFMs.} Point-forecast models, such as TimesFM-type TSFMs, directly output
\begin{equation}
\hat{r}_{t+1}^{ex}=f_\theta\!\left(\mathbf{r}_{t-C+1:t}^{ex}\right).
\end{equation}
The point forecast is used as the return prediction and as the ranking signal for portfolio construction. For directional evaluation, its sign provides the predicted direction of the next excess return.

\subsection{Time Series Foundation Models: Training and Inference}
We evaluate TSFMs under three regimes that differ in parameter initialization and the data used for adaptation. Let \(\mathcal{D}_{\mathrm{fin}}\) denote the financial return dataset used in this study, and let \(\mathcal{L}_{\theta}(X,Y)\) denote the model's native training loss, including any required target transformation, such as token cross-entropy for Chronos or mean squared error for point forecasts.

\subsubsection{Off-the-Shelf Pre-Trained Model (Zero-Shot Inference)} \label{TSFM_def_zero_shot}
Let \(\theta_{\mathrm{pre}}\) denote model parameters obtained from a large external pre-training corpus \(\mathcal{D}_{\mathrm{pre}}\), which may include heterogeneous real and synthetic time series. In the zero-shot regime, the model is applied directly to \(\mathcal{D}_{\mathrm{fin}}\) without any parameter updates:
\begin{equation}
\hat{r}_{t+1}^{ex}=F_{\theta_{\mathrm{pre}}}\!\left(\mathbf{r}_{t-C+1:t}^{ex}\right),
\end{equation}
where \(F_{\theta_{\mathrm{pre}}}\) returns the point forecast using the aggregation rule defined above. This setting tests whether generic temporal representations transfer to financial excess returns without domain-specific training. The ability to perform accurate zero-shot forecasting on unseen tasks is the principal premise underlying TSFMs, namely that large-scale pre-training yields transferable temporal representations that generalize across domains.

\subsubsection{Fine-Tuned Model} \label{TSFM_def_fine_tuned}
Fine-tuning starts from the pre-trained parameters \(\theta_{\mathrm{pre}}\) and updates them on the financial dataset:
\begin{equation}
\theta_{\mathrm{fin}}
=\arg\min_{\theta}
\mathbb{E}_{(X,Y)\sim\mathcal{D}_{\mathrm{fin}}}
\mathcal{L}_{\theta}(X,Y),
\end{equation}
where \(X=\mathbf{r}_{t-C+1:t}^{ex}\) and \(Y=r_{t+1}^{ex}\). In our experiments, all model parameters are trainable and the original loss function of each TSFM is retained. For all fine-tuning experiments, we use the hyperparameter settings recommended by the model developers without any modifications. This regime measures whether generic pre-training can provide a useful initialization for financial adaptation. This reflects the second central premise of TSFMs: even when zero-shot performance does not fully transfer to the downstream task, task-specific fine-tuning should improve performance and yield competitive results on downstream forecasting tasks.

\subsubsection{Models Pre-Trained from Scratch} \label{TSFM_def_pre_training}

The from-scratch regime uses the same financial training pairs but initializes the model randomly rather than from an external checkpoint:
\begin{equation}
\theta_{\mathrm{scratch}}
=\arg\min_{\theta}
\mathbb{E}_{(X,Y)\sim\mathcal{D}_{\mathrm{fin}}}
\mathcal{L}_{\theta}(X,Y),
\qquad \theta_0\sim\mathcal{P}_0.
\end{equation}
At a fixed architecture and training configuration, the defining distinction between fine-tuning and from-scratch pre-training is initialization: fine-tuning begins from \(\theta_{\mathrm{pre}}\), whereas from-scratch pre-training begins from \(\theta_0\). More importantly, training from scratch avoids reliance on public TSFM checkpoints whose pre-training corpora may overlap with our evaluation period. This regime is also consistent with the standard evaluation protocol in the asset pricing literature, in which models are estimated using only historical financial data available at each forecast origin and are subsequently evaluated on future out-of-sample observations, thereby avoiding potential look-ahead bias inherited from external pre-trained checkpoints.

\subsection{Evaluation Metrics}
We evaluate model performance along two dimensions: forecasting performance and portfolio performance. This distinction reflects that improvements in statistical fit need not translate into better investment decisions \citep{ait2001variable,brandt2009parametric}. A common portfolio-construction rule across models allows us to assess the economic usefulness of their cross-sectional ranking signals alongside their point-forecast accuracy.\footnote{Return predictability may reflect time-varying risk premia, and its implications for market efficiency depend on the maintained asset-pricing model \citep{fama1991efficient}. Our analysis evaluates the predictive and portfolio performance of alternative forecasting methods without distinguishing between risk compensation and mispricing as sources of predictable returns.} The forecasting performance metrics are described in \Cref{forcasting_per_metric_sec}, while the portfolio performance metrics are presented in \Cref{portfolio_per_metric_sec}.

\subsubsection{Forecasting Performance Metrics}
\label{forcasting_per_metric_sec}
Following \citet{gu2020empirical}, we use a zero-return benchmark to calculate the out-of-sample coefficient of determination. We compute this metric separately for each calendar year and report its equally weighted average across out-of-sample years. For each out-of-sample calendar year $y$, we compute
\begin{equation}
R^2_{\mathrm{OOS},y}
=
1-
\frac{
\sum_{(i,t)\in\mathcal{I}_y}
\left(r_{t+1}^{(i),ex}
-\hat{r}_{t+1}^{(i),ex}\right)^2
}{
\sum_{(i,t)\in\mathcal{I}_y}
\left(r_{t+1}^{(i),ex}\right)^2
},
\end{equation}
where $\mathcal{I}_y$ denotes the eligible stock--date observations assigned to out-of-sample evaluation year $y$, $r_{t+1}^{(i),ex}$ is the realized excess return, and $\hat{r}_{t+1}^{(i),ex}$ is the model forecast. Let $\mathcal{Y}_{\mathrm{OOS}}$ denote the set of out-of-sample evaluation years. We report their equally weighted average:
\begin{equation}
R^2_{\mathrm{OOS}}
=
\frac{1}{|\mathcal{Y}_{\mathrm{OOS}}|}
\sum_{y\in\mathcal{Y}_{\mathrm{OOS}}}
R^2_{\mathrm{OOS},y}.
\end{equation}
Reported values are expressed in percentage terms by multiplying $\RsqOOS$ by 100. A zero-return benchmark avoids estimating historical mean returns, which can be noisy for individual stocks.

For directional forecasting, an observation is classified as upward if its excess return is strictly positive and downward otherwise. The predicted class is determined by applying the same rule to the point forecast. We report overall accuracy, upward accuracy, downward accuracy, and macro-averaged $F_1$. Upward and downward accuracy denote class-specific recall: the proportion of actual observations in each class that are correctly classified. For each year $y$, the macro-$F_1$ score is the unweighted average of the two class-specific $F_1$ scores:
\begin{equation}
\mathrm{macro}\text{-}F_{1,y}
=
\frac{1}{2}
\sum_{c\in\{\mathrm{up},\mathrm{down}\}}
\frac{
2\,\mathrm{Precision}_{c,y}\,
\mathrm{Recall}_{c,y}
}{
\mathrm{Precision}_{c,y}
+\mathrm{Recall}_{c,y}
},
\end{equation}
where $\mathrm{Precision}_{c,y}$ is the fraction of predictions for class $c$ that are correct in year $y$, and $\mathrm{Recall}_{c,y}$ is the fraction of actual observations in class $c$ that are correctly identified. Precision and recall are each set to zero when their respective denominators are zero. A class-specific $F_1$ score is set to zero when the sum of its precision and recall is zero. All classification metrics are computed separately for each calendar year and then averaged with equal weight across the 23 out-of-sample years. Accuracy measures are reported in percentage terms, while macro-$F_1$ is reported on a scale from zero to one.

\subsubsection{Portfolio Performance Metrics} \label{portfolio_per_metric_sec}
We translate forecasts into a frictionless daily long--short portfolio sort designed to measure the economic strength of the models' cross-sectional ranking signals. At the end of day \(t\), stocks are ranked by their predicted next-day excess returns \(\hat{r}_{t+1}^{(i),ex}\). The investable universe is divided into deciles. Our baseline portfolio is a zero-cost, EW portfolio that is long the top decile and short the bottom decile, and is rebalanced daily. The long--short portfolio is scaled to unit gross exposure: one half of gross exposure is assigned to the long leg and one half to the short leg. Let \(R_{t+1}^{H}\) and \(R_{t+1}^{L}\) denote the standalone excess returns of the highest- and lowest-forecast deciles, respectively. The reported long--short return is therefore
\begin{equation}
r_{t+1}^{LS}=\frac{1}{2}R_{t+1}^{H}-\frac{1}{2}R_{t+1}^{L}.
\end{equation}
Accordingly, the spread portfolio rows reported use the same unit-gross-exposure convention and correspond to \(\frac{1}{2}(H-L)\).\footnote{To clarify the sign convention, $R^H_{t+1}$ and $R^L_{t+1}$ denote the standalone excess returns of the highest- and lowest-forecast decile portfolios, respectively, before applying portfolio-position signs. Hence, the long-leg contribution is $\frac{1}{2}R^H_{t+1}$, while the signed short-leg contribution is $-\frac{1}{2}R^L_{t+1}$. Thus, a negative return on the lowest-forecast decile generates a positive contribution from the short leg.} We also report a value-weighted (VW) version. For VW portfolios, stocks within each long or short decile are weighted by their market capitalization at the portfolio formation date, normalized so that weights sum to one within each leg before applying the one-half long and one-half short scaling. The VW long--short return is then one half of the VW return of the top decile minus one half of the VW return of the bottom decile.

The realized long--short return series $\{r_{t+1}^{LS}\}$ measures the economic value of the model's cross-sectional ranking signal. Unless reported by calendar year, portfolio metrics are computed from the full daily out-of-sample return series for 2001--2023. Annualized return, annualized standard deviation, and the Sharpe ratio are calculated as $252\mu_{LS}$, $\sqrt{252}\,s_{LS}$, and $\sqrt{252}\mu_{LS}/s_{LS}$, respectively, where $\mu_{LS}$ and $s_{LS}$ denote the sample mean and sample standard deviation of daily long--short returns. We also report average daily return in basis points, maximum drawdown, largest one-day loss, skewness, and excess kurtosis. Annualized return, annualized standard deviation, maximum drawdown, and largest one-day loss are reported in percentage terms.\footnote{All statistics are computed from unrounded values, so sums, differences, and ratios of reported entries may differ from reported results in the second decimal place.} For annual-performance figures, these metrics are computed separately within each calendar year.

% \notewg{To Eghbal: Please look at the definition here. The referee was confused with the signs of the long- short- and spread portfolio. Please check the caption of spread portfolio tables.}
\section{Data and Computational Resources}\label{sec:data}

\subsection{Data Sources and Training Universes} \label{data_source}
We construct a comprehensive panel of daily firm-level excess returns spanning 1990--2023 across 94 countries and regions.\footnote{We use daily rather than monthly returns because zero-shot evaluation, fine-tuning, and, in particular, from-scratch TSFM pre-training require substantially larger samples than are available at the monthly frequency. We also adopt a univariate setting, as many commonly used firm characteristics and predictive signals are not observed reliably at a daily frequency. A further implication is that the portfolio exercises involve daily rebalancing and therefore substantial turnover and trading costs. Accordingly, the frictionless portfolio results should be interpreted primarily as comparative measures of predictive and cross-sectional ranking performance across models, rather than as estimates of directly implementable investment performance.} Daily price and return information comes from two complementary sources. For the U.S., all equity data originate from the Center for Research in Security Prices (CRSP) database. For non-U.S. markets, we rely on Compustat global daily security files, which provide comparable firm identifiers, prices, and share counts. When a security is available in both CRSP and Compustat, the CRSP record is given priority because of its broader coverage of corporate actions, higher data quality, and inclusion of delisting returns.

We also use monthly firm-level characteristics from the JKP Global Factor Data of \citet{jensen2023there}, obtained through WRDS. We apply the dataset's filters for common stocks, main exchanges, primary listings, and one main observation per security and month. The characteristics are organized into thirteen conceptual clusters: investment, value, low risk, quality, seasonality, profit growth, leverage, profitability, momentum, debt issuance, accruals, short-term reversal, and size. The dataset contains up to 153 characteristics observed at the firm-month level across countries. We treat each firm--characteristic combination as a separate univariate time series and combine these monthly characteristic series with daily return series for joint pre-training. Following a common TSFM pre-training approach, we add JKP characteristics as separate univariate series, not as covariates for return forecasts.\footnote{The composition and sampling frequencies of the pre-training data vary across experiments, including daily series and mixtures of daily and monthly series. Regardless of the pre-training data configuration, the prediction target in all main forecasting and portfolio evaluations is the next-day excess return.} For each expanding-window training year, we retain only observations dated through the end of that year while preserving the chronological ordering and missing-value structure of each series. This yields three training universes based on observed financial data: U.S. excess returns, global excess returns, and global excess returns augmented with JKP characteristics.

As a fourth training universe, we augment global excess returns with synthetic time series. Following the synthetic data-generation framework used to pre-train Chronos \citep{ansari2024chronos}, we construct a synthetic panel matched in size and structure to the JKP augmentation panel for each expanding-window training year. Synthetic values are generated by randomly combining up to five Gaussian-process kernels drawn from periodic, linear, radial-basis-function, rational-quadratic, white-noise, and constant kernels and sampling one trajectory from the resulting Gaussian-process prior. Thus, the JKP- and synthetic-augmented samples are matched in size, structure, and missingness, while differing in the values and economic content of the augmentation series. These training universes allow us to assess the benefits of broader mixed-frequency financial information and, separately, the incremental value of synthetic augmentation commonly used in TSFM pre-training. \Cref{dataset_scale_summary} summarizes the dataset scale used throughout this study. Detailed annual and market-level coverage is reported in \Crefrange{Annual_breakdown_observations_1}{Annual_breakdown_securities_JPK_2} of the \hyperref[pp:internet-title]{Internet Appendix}.

% \notewg{To Eghbal: The referee asked about how JKP factors are used. This is how I understand, please verify this.}

\begin{table}
\centering
\begin{threeparttable}
\begin{adjustbox}{width=0.96\textwidth, center}
\scriptsize
\captionsetup{width=\linewidth}
\caption{Dataset Scale Summary}
\label{dataset_scale_summary}
\begin{minipage}{\linewidth}
\begin{tabular}{L{0.19\linewidth}L{0.19\linewidth}L{0.13\linewidth}ccL{0.29\linewidth}}
\toprule
\textbf{Dataset} & \textbf{Use} & \textbf{Period} & \textbf{Obs.} & \textbf{Sec.} & \textbf{Notes} \\
\midrule
U.S. excess returns & U.S.-only training universe & 1990--2022 & 131.22M & 53.65K & Daily excess returns; maximum expanding-window training sample in 2022. \\
Global excess returns & Scaled financial training universe & 1990--2022 & 444.70M & 194.62K & Daily excess returns pooled across markets. \\
JKP characteristics & Auxiliary pre-training series & 1990--2022 & 1,256.36M & 92.05K & Up to 153 monthly firm-level characteristics from \citet{jensen2023there}, treated as separate univariate series. \\
JKP-augmented total & Largest pre-training universe & 1990--2022 & 1,701.07M & 255.83K & Global excess returns combined with JKP characteristic series; securities are deduplicated across the two datasets. \\
U.S. OOS evaluation & Common test sample & 2001--2023 & 18.14M & 10.17K & CRSP ordinary common shares on NYSE, AMEX, and NASDAQ after the price and security filters. \\
\bottomrule
\end{tabular}
\vspace{0.1cm}
\begin{tablenotes}[para,flushleft]
\footnotesize
\textbf{Note:} Training counts refer to the largest expanding-window sample, covering 1990--2022. Observations are nonmissing scalar entries in the training panels, reported in millions. Security counts are distinct security identifiers with at least one nonmissing observation during this period, reported in thousands and deduplicated across datasets for the combined sample. Each security--characteristic combination contributes a separate univariate series. Cumulative data coverage, including observations preceding the training period, is shown in \Cref{Cumulative_obs_sec_us}, \Cref{Cumulative_obs_sec_global}, and \Cref{Cumulative_obs_sec_JPK_compare}. Country codes and data sizes in gigabytes are reported in \Cref{mapping_codes} and \Cref{data_size_mb_plot}, respectively, in the \hyperref[pp:internet-title]{Internet Appendix}.
\end{tablenotes}
\end{minipage}
\end{adjustbox}
\end{threeparttable}
\end{table}

% \Cref{dataset_scale_summary} summarizes the dataset scale used throughout the paper. Detailed annual and market-level coverage is reported in \Cref{Annual_breakdown_observations_1,Annual_breakdown_observations_2,Annual_breakdown_securities_1,Annual_breakdown_securities_2,Annual_breakdown_observations_JPK_1,Annual_breakdown_observations_JPK_2,Annual_breakdown_securities_JPK_1,Annual_breakdown_securities_JPK_2} of the Internet Appendix.

\subsection{Data Preprocessing}
Our empirical design draws on \citet{gu2020empirical}, particularly the zero-return forecasting benchmark and the cross-sectionally aggregated loss-differential approach to forecast comparison. The daily-data cleaning procedures, eligibility filters, return transformations, and missing-value treatment are described below. Before computing excess returns, the raw daily data are cleaned and harmonized across markets. We retain only listings on the main exchanges within each country. Observations belonging to the bottom five percent of the market capitalization distribution within each country-day are excluded to limit the influence of microcaps. When delisting information is available, the reported delisting return is incorporated on the final trading day. If a delisting is recorded but the associated return is missing, a delisting return of $-30\%$ is assigned as a uniform imputation assumption motivated by \citet{shumway1997delisting} to mitigate delisting bias. To maintain consistency between CRSP and Compustat records, daily returns from Compustat are winsorized using the CRSP return distribution as a benchmark: Compustat returns above the 99.9\textsuperscript{th} percentile or below the 0.1\textsuperscript{th} percentile of the corresponding CRSP distribution for the same day are capped at those values. Returns are measured in excess of the U.S. risk-free rate.\footnote{For an unlevered long position in a stock, the simple total return is bounded below by $-100\%$. With returns expressed in decimal form, $r_t^{ex}=r_t-r_{f,t}^{(d)}\geq -1-r_{f,t}^{(d)}$. Thus, excess returns can fall below $-100\%$ when the daily risk-free rate is positive.} Non-U.S. stock returns are expressed in U.S. dollars, incorporating exchange-rate movements, following \citet{jensen2023there}. For model estimation only, daily excess returns are capped at $-100\%$ and $+100\%$. This limits the disproportionate influence of extreme observations on model fitting, particularly under squared-error objectives, where a small number of unusually large returns can dominate the training loss. These wide bounds preserve variation within the interval while retaining more extreme observations at the respective boundaries rather than discarding them. The bounds are fixed across assets and dates. Out-of-sample realized returns used for forecast evaluation and portfolio performance calculations are not subject to this cap, ensuring that extreme outcomes retained after the other data-cleaning steps remain fully reflected in reported performance. After cleaning, all firm-level series are merged into a unified daily panel organized by calendar year,
\begin{equation}
\mathbf{r}_t^{ex} = \bigl(r_{t}^{(1),ex}, r_{t}^{(2),ex}, \ldots, r_{t}^{(M_t),ex}\bigr),
\end{equation}
\noindent with dates ordered in ascending sequence, where $M_t$ denotes the number of assets available at time $t$ and \(i \in \{1,\ldots,M_t\}\) indexes assets.

Missing excess returns are imputed within each country-day cross-section to preserve the institutional structure of national markets. For each trading date \(t\), if all firms' returns in a country are missing on that date, those values are left missing rather than imputed, as this may reflect a market-wide closure, holiday, or unavailable data. Otherwise, missing values within that country are replaced by the cross-sectional median of nonmissing observations from the same country on date \(t\). We use medians to reduce sensitivity to heavy-tailed or skewed return distributions. To avoid extrapolating returns beyond an asset's listing period, observations before the first valid trading date or after the last valid trading date of each asset are set to missing.
% \noteNi{can we give the citation to justify the preprocessing approach we chose?} Dates for which all asset excess returns are missing are removed from the panel.
\subsection{Training and Evaluation Protocol}\label{sec:evaluation_protocol}
We use an annual expanding-window split: the historical sample starts in 1990 and expands by one year at each re-estimation, while the following calendar year is held out for testing. Models indexed by $y=2000,\ldots,2022$ generate forecasts for year $y+1$, yielding 23 out-of-sample test years from 2001 to 2023. The same schedule applies to benchmark estimation, TSFM fine-tuning, and pre-training from scratch. Publicly available TSFMs evaluated in the zero-shot setting use a single fixed model throughout the test period.\footnote{Public pre-training corpora may overlap with the evaluation period; this potential source of look-ahead bias remains relevant to zero-shot and fine-tuned models, whereas our from-scratch models use only historical training data.} The input lengths of 5, 21, 252, and 512 trading days specify the past returns supplied to each forecast and are distinct from the expanding sample used for model estimation. For the U.S. portfolio analysis, we retain CRSP securities with share codes 11 and 12, an absolute price of at least \$5, and exchange codes 1--3 (NYSE, AMEX, and NASDAQ). This selection includes common shares of both U.S.- and foreign-incorporated corporations listed on these exchanges.

\subsection{Model Sizes and Computational Resources}
\begin{figure}
    \caption{Comparative Model Complexity}
    \label{Comparative_Model_Complexity}
    \centering
    \includegraphics[
        width=0.56\textwidth,
        clip,
        trim=0 0 0 8,
    ]{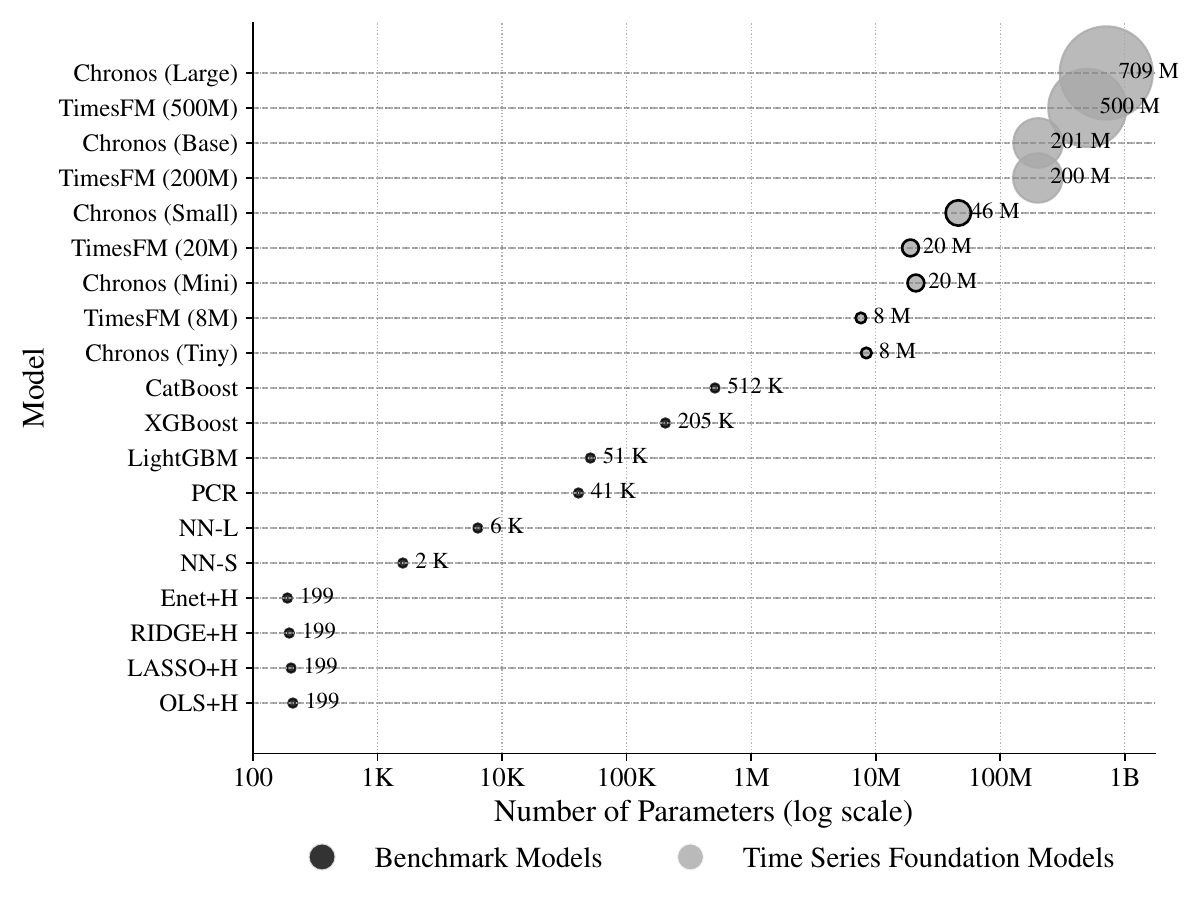}
    
    \begin{minipage}{0.8\textwidth}
        \scriptsize
        \textbf{Note:} Bubble area is proportional to parameter count. Black bubbles denote benchmark models; gray bubbles denote TSFMs. Black borders identify TSFM configurations used for from-scratch pre-training. `H' indicates Huber loss. For models whose parameter counts vary with the input window, the reported counts are averaged across the four input windows.
    \end{minipage}
\end{figure}

In this subsection, we summarize model scale and computational requirements. \Cref{Comparative_Model_Complexity} plots parameter counts on a log scale, contrasting benchmark models with Chronos (tiny, mini, small, base, and large)\footnote{Chronos variants contain approximately 8M, 20M, 46M, 200M, and 710M parameters for the tiny, mini, small, base, and large models, respectively.} and TimesFM (8M, 20M, 200M, and 500M). In the main zero-shot and fine-tuned experiments, we use Chronos (tiny, mini, small, base, and large) and TimesFM (200M and 500M). For from-scratch pre-training, computational constraints limit us to Chronos (tiny, mini, and small) and scaled-down TimesFM variants of approximately 8M and 20M parameters.\footnote{TimesFM (200M) refers to the model released under TimesFM 1, while TimesFM (500M) corresponds to TimesFM 2. The scaled-down TimesFM variants are derived from TimesFM 2 by proportionally reducing core Transformer hyperparameters while keeping the remaining settings consistent with the original specification. The restriction to smaller Chronos models and scaled-down TimesFM variants reflects the computational demands of annual from-scratch pre-training. Our expanding-window protocol requires fresh pre-training for every evaluation year and model configuration, substantially increasing the total resource requirements relative to a single training run. These smaller architectures make repeated pre-training feasible within the study's available computational resources.}

Inference, fine-tuning, and especially from-scratch pre-training are computationally intensive. All TSFM experiments were run on GPU servers, using approximately 50,000 GPU hours across pre-training, fine-tuning, and large-scale inference.\footnote{The economic importance of these resource requirements is underscored by \citet{bandi2026early}, who treat computational capacity as a scarce, capital-intensive input into AI production.} These computations include repeated annual pre-training across model sizes, training universes, and model configurations. Benchmark models rely primarily on CPU computation; because jobs were distributed across servers, exact usage was not recorded, but we estimate roughly 25,000 CPU hours for benchmark training and testing. The TSFM experiments used multi-node servers equipped with NVIDIA GH200 Grace Hopper GPUs, each integrating a Grace CPU with an H100 Tensor Core GPU and 96 GB of high-bandwidth memory.\footnote{A detailed account of the computational infrastructure and its underlying technical aspects is provided in \citet{mcintosh2024isambard}. Part of this research was also conducted using N8 Bede.} We initialize the pseudorandom number generators with the same fixed seed at the start of every benchmark training, TSFM fine-tuning, and from-scratch pre-training run.
\section{Numerical Results} \label{sec: numerical_results}
This section is organized around one question: can TSFMs become useful for financial forecasting once they are trained natively on financial data and allowed to scale with data size? We proceed in three steps. First, we establish a demanding empirical hurdle and summarize why off-the-shelf TSFMs do not transfer cleanly to daily excess return forecasting. Second, we show that finance-native pre-training substantially improves long-window portfolio performance relative to the principal off-the-shelf Chronos and TimesFM specifications, while improvements in forecasting accuracy depend on the metric and specification. Third, we examine whether these gains strengthen further under data scaling and whether they remain robust over time, under transaction costs, in seven major non-U.S. markets, and across alternative hyperparameter choices.

As a starting point, we evaluate a broad set of benchmark models to identify the most demanding baseline for TSFMs. Benchmark models include linear (OLS+H, LASSO+H, RIDGE+H, Enet+H, and PCR)\footnote{`H' denotes Huber-loss estimation. LASSO, Enet, and PCR denote least absolute shrinkage and selection operator, elastic net, and principal component regression, respectively. OLS+H denotes scale-adaptive Huber regression with $\ell_2$ regularization.}, ensemble (XGBoost, CatBoost, and LightGBM), and neural network (NN-S and NN-L)\footnote{NN-S refers to a single-hidden-layer neural network with 8 hidden units, while NN-L denotes a similar architecture comprising 32 hidden units. Further details are provided in \Cref{IA_Neural_Network} of the \hyperref[pp:internet-title]{Internet Appendix}.} models. These models span the main classes used in the financial forecasting literature, remain computationally feasible at our sample size, and include strong practical baselines, especially among tree-based ensembles.\footnote{A detailed description of the benchmark models is provided in \Cref{appendix_benchmark_models} of the \hyperref[pp:internet-title]{Internet Appendix}. Also, \Cref{tab:huber_hyperparams} of the \hyperref[pp:internet-title]{Internet Appendix} presents the hyperparameter settings for the benchmark models. The set of models is drawn from \citet{gu2020empirical} and \mbox{\citet{leippold2022MachineLearningChinese}}, subject to the additional requirement that they can be trained efficiently on the large-scale dataset employed in this study. For benchmark specifications with tuned hyperparameters, selection is performed in the initial estimation year (2000) by maximizing validation $R^2$, and the selected settings are held fixed in subsequent years. The validation sample is a chronologically held-out block within the historical estimation sample, preceding the corresponding out-of-sample test year; observations from the same date are assigned together to either training or validation. Specifications without hyperparameter tuning use fixed settings. This procedure ensures computational feasibility while enabling a fair comparison across different classes of models. The key hyperparameters that were optimized are reported under `Tuned', while those held constant throughout the analysis are reported under `Fixed'.} All experiments follow the annual expanding-window evaluation protocol in \Cref{sec:evaluation_protocol}.\footnote{Throughout this study, we use the terms `pre-training' and `fine-tuning' when referring to TSFMs, and the terms `training' or `estimation' when referring to other model classes.} All models are also evaluated with window sizes of 5, 21, 252, and 512 trading days to examine how the amount of historical information available to each model influences its performance. We vary the input window to test whether TSFMs exploit historical information differently from benchmark models. The window sizes of 5, 21, and 252 trading days correspond approximately to one week, one month, and one year of past excess returns, respectively. A window size of 512 trading days is additionally included, as it is a commonly used maximum input length in many TSFMs. Finally, our work focuses on univariate return forecasting, benchmarking all TSFMs in a univariate time series setting, although some recently proposed TSFMs are capable of handling multivariate data.

We present the empirical evidence in \Cref{primary_results}, \Cref{scaled_results}, \Cref{intenational_results}, and \Cref{hyperparameter_impact}. \Cref{primary_results} establishes the benchmark hurdle, documents the limited usefulness of off-the-shelf TSFMs, and then focuses on finance-native pre-training using U.S. data. \Cref{scaled_results} extends that pre-training analysis to larger and more heterogeneous datasets, and examines stability and transaction costs. \Cref{intenational_results} evaluates whether the same conclusions extend beyond the U.S. market. Finally, \Cref{hyperparameter_impact} examines sensitivity to model configuration.\footnote{All models pre-trained from scratch are available through our portal at \href{https://FinText.ai}{\texttt{FinText.ai}} and the Hugging Face repository at \href{https://huggingface.co/FinText}{\texttt{https://huggingface.co/FinText}}.}

\subsection{U.S. Evidence and Motivation for Finance-Native Pre-Training} \label{primary_results}
The U.S. evidence serves two purposes. First, it establishes CatBoost as a strong empirical hurdle. Second, it shows that zero-shot transfer and conventional fine-tuning do not consistently deliver performance competitive with the benchmark in this setting, which motivates the main focus of the section: finance-native pre-training from scratch on excess returns. In this subsection, benchmark estimation, TSFM fine-tuning, and from-scratch pre-training use U.S. data, and all models are evaluated on U.S. stocks. Zero-shot models retain their original external pre-training. Unless explicitly stated otherwise, TSFM hyperparameters follow the original model specifications. To facilitate comparison across training regimes, \Cref{FP_US_comparison} and \Cref{PP_US_comparison} report the selected representative specifications for forecasting and portfolio performance, respectively, together with CatBoost as the representative benchmark. The full results, including the complete set of models, additional performance metrics, and sample and portfolio breakdowns, are reported in \Cref{Appendix_Extended_Results}. These results are broadly consistent with the patterns documented here and do not alter the main conclusions.

\subsubsection{Benchmark and Empirical Hurdle} \label{benchmark_results}
The complete benchmark tables are reported in \Cref{FP_benchmark_full} and \Cref{PP_benchmark_full}. The forecasting table reports full-sample, large-cap, and small-cap results, where large-cap and small-cap stocks are defined by the top and bottom 25\% of market capitalization. The portfolio table reports the full-universe long--short portfolio and its signed long- and short-leg contributions; size-specific portfolio results are reported separately in \Cref{PP_US_cap_comparison}. Forecasting metrics are computed separately by calendar year using all stock-date observations in that year and then averaged over the out-of-sample period. Portfolio metrics are computed from daily decile portfolios sorted on predicted excess returns.

This benchmark step is important because TSFMs should not be compared only with generic time-series baselines. Daily equity return forecasting has a long empirical literature. We therefore use the benchmark exercise to identify the most demanding comparator before turning to TSFMs. The forecasting results in \Cref{FP_benchmark_full} establish a clear hurdle. Tree-based ensembles generally provide the strongest squared-error performance among the benchmarks, while the ranking depends on the metric and input window. CatBoost is a strong representative comparator, particularly for gross EW risk-adjusted portfolio performance at the longer windows. At the 512-day window, CatBoost attains an $\RsqOOS$ of $-0.03\%$, an overall directional accuracy of $51.16\%$, and a macro-$F_1$ score of $0.49$. XGBoost and LightGBM are close competitors: both deliver near-zero or mildly negative full-sample $\RsqOOS$ at the longer windows and directional accuracy slightly above 51\%, confirming that the benchmark strength is not unique to one tree-based implementation. Linear models are more mixed. OLS+H has relatively high directional accuracy, above 51.8\% across windows, but its $\RsqOOS$ is weaker than the ensemble models, while regularized linear models and PCR generally do not close the gap. Neural networks also produce mixed results: NN-S and NN-L sometimes improve with longer windows, but their $\RsqOOS$ remains substantially below the ensemble models. Small-cap stocks consistently exhibit stronger $\RsqOOS$ than large-cap stocks and generally exhibit higher overall directional accuracy. For CatBoost, averaging across all four window sizes in \Cref{FP_benchmark_full}, small-cap stocks have an average $\RsqOOS$ of about $0.57\%$ and average directional accuracy of about $52.29\%$, compared with about $-0.37\%$ and $50.80\%$ for large-cap stocks. A similar small-cap advantage appears for other benchmark classes, including linear models and neural networks. This pattern is consistent with the view that smaller stocks contain more exploitable return predictability, while large-cap returns are closer to the efficient-market benchmark.\footnote{We also report in \Cref{DM_test_Benchmark} the results of the modified Diebold--Mariano (DM) tests, following the approach of \citet{gu2020empirical}, which assess the statistical significance of out-of-sample forecasting performance differences among benchmark models. Given the number of configurations examined, individual pairwise significance results should also be interpreted in light of multiple-testing concerns \citep{harvey2016and}. Following \citet{gu2020empirical}, we compute, for each trading day, the cross-sectional average of the row model's squared forecast errors minus the column model's squared forecast errors, using observations common to both models. The modified DM statistic equals the time-series mean of this daily loss differential divided by its Newey--West standard error, computed with one daily lag. A positive and statistically significant DM statistic indicates that the model in the column outperforms the model in the corresponding row. Consistent with the benchmark performance results, XGBoost, CatBoost, and LightGBM exhibit significantly superior forecasting performance compared to traditional linear models such as OLS+H, LASSO+H, RIDGE+H, Enet+H, and PCR across most window sizes. Neural network models (NN-S and NN-L) show mixed results, occasionally outperforming linear models but generally lagging behind the ensemble models. Among the ensemble models, CatBoost and XGBoost most frequently deliver statistically significant improvements over competing models. Within-model DM comparisons do not show a uniform advantage for the longest input windows. For XGBoost and LightGBM, the 21-day specification significantly outperforms the 252- and 512-day specifications. For CatBoost, the differences among the 21-, 252-, and 512-day specifications are not statistically significant at the reported 5\% level.}

In terms of portfolio performance, \Cref{PP_benchmark_full} conveys the same message, although models other than CatBoost also generate economically meaningful signals. Linear models produce high annualized returns, with OLS+H around 45\% across windows, but their SRs are lower than those of the best ensemble models because their return volatility and drawdowns are larger. PCR is the weakest linear benchmark in economic terms. XGBoost and LightGBM remain very competitive: XGBoost reaches an annualized return of 47.42\% at the 252-day window, and LightGBM produces annualized returns above 45\% at the 21-, 252-, and 512-day windows. CatBoost delivers the highest benchmark SR at the 21-, 252-, and 512-day windows, with annualized returns above 46\% and SRs of 6.72, 6.72, and 6.39, respectively. At the 5-day window, XGBoost has a slightly higher SR than CatBoost, at 5.74 versus 5.70. Neural networks again trail the ensemble models in SRs, with generally lower annualized returns and less favorable drawdown profiles. The long leg consistently contributes more than the short leg, which matches the broader return-predictability evidence in \citet{gu2020empirical} and \citet{leippold2022MachineLearningChinese}.\footnote{\Cref{Spread_Portfolio_Benchmarks_table} reports increasing decile return profiles for the ensemble models, with particularly large long--short spreads at the longer input windows.} Taken together, these findings justify using CatBoost as the main hurdle for all subsequent TSFM comparisons. Accordingly, CatBoost is used as the representative benchmark model in \Cref{FP_US_comparison} and \Cref{PP_US_comparison}, reflecting its strong gross EW risk-adjusted portfolio performance, especially at the longer input windows.\footnote{In \Cref{PP_US_comparison}, we also report results for VW portfolios. Since VW portfolios generally exhibit weaker performance than their EW counterparts across the majority of our tests throughout the study, detailed VW portfolio results are omitted for brevity.}

\begin{table} % fixed ----------------------------
  \centering
  \begin{threeparttable}
    \begin{adjustbox}{max width=0.76\textwidth, center}
    \scriptsize
    \captionsetup{width=\linewidth}
    \caption{Representative U.S. Forecasting Comparison across Training Regimes}
    \label{FP_US_comparison}
    \begin{minipage}{\linewidth}
      \renewcommand{\arraystretch}{1.15}
      \setlength{\tabcolsep}{4pt}
      \begin{tabular}{l>{\raggedright\arraybackslash}p{0.42\linewidth}cccc}
        \toprule
        \textbf{Metric} & \textbf{Representative Model} & \textbf{5} & \textbf{21} & \textbf{252} & \textbf{512} \\
        \midrule
        \multirow{7}{*}{$\RsqOOS$}
          & Benchmark: CatBoost & -0.25 & -0.04 & -0.03 & -0.03 \\
          & Zero-shot: Chronos (Large) & -46.86 & -23.84 & -2.49 & -1.37\\
          & Zero-shot: TimesFM 2 (500M) & -27.96 & -11.81 & -3.86 & -2.80 \\
          & Fine-tuned: Chronos (Large) & -17.02 & -2.30 & -1.85 & -1.89 \\
          & Fine-tuned: TimesFM 2 (500M) & -529.52 & -562.05 & -175.47 & -146.79 \\
          & Finance-native: Chronos (Small) & -3.18 & -1.43 & -0.65 & -0.59 \\
          & Finance-native: TimesFM (20M) & -34.09 & -16.32 & -13.53 & -14.75 \\
        \midrule
        \multirow{7}{*}{Overall Acc.}
          & Benchmark: CatBoost & 50.94 & 51.12 & 51.08 & 51.16 \\
          & Zero-shot: Chronos (Large) & 48.70 & 49.13 & 50.49 & 51.01 \\
          & Zero-shot: TimesFM 2 (500M) & 48.34 & 48.70 & 49.59 & 49.82 \\
          & Fine-tuned: Chronos (Large) & 49.97 & 50.24 & 50.27 & 50.27 \\
          & Fine-tuned: TimesFM 2 (500M) & 48.93 & 49.47 & 50.00 & 49.97 \\
          & Finance-native: Chronos (Small) & 50.40 & 50.79 & 51.36 & 51.36 \\
          & Finance-native: TimesFM (20M) & 49.88 & 50.88 & 51.14 & 51.12 \\
        \bottomrule
      \end{tabular}
      \vspace{0.1cm}
      \begin{tablenotes}[para,flushleft]
        \footnotesize
        \textbf{Note:} This table reports full-sample forecasting results for representative models. The columns 5, 21, 252, and 512 denote the input window length in trading days. $\RsqOOS$ and `Overall Acc.' are reported in percentage terms. $\RsqOOS$ is averaged with equal weight across the 23 out-of-sample years. `Overall Acc.' is computed within each calendar year as the percentage of stock-date observations whose next-day excess return direction is correctly predicted, and is then averaged with equal weight across the 23 evaluation years. For brevity, the table reports only full-sample results; the complete benchmark, zero-shot, fine-tuned, and finance-native pre-trained results, including large-cap and small-cap breakdowns, are reported in \Cref{FP_benchmark_full}, \Cref{FP_TSFM_zero_shot_full}, \Cref{FP_TSFM_fine_tuned_full}, and \Cref{FP_TSFM_pre_trained_full}. Chronos (Large) is selected as the largest released public Chronos checkpoint and is retained for both the zero-shot and fine-tuned comparisons. This provides a fixed checkpoint size across these regimes; it does not imply that Chronos (Large) is the best-performing zero-shot Chronos size. Chronos (Mini) achieves higher zero-shot EW annualized returns and Sharpe ratios at the 252- and 512-day input windows.
      \end{tablenotes}
    \end{minipage}
    \end{adjustbox}
  \end{threeparttable}
\end{table}

\begin{table}[!htbp] % fixed ----------------------------
  \centering
  \begin{threeparttable}
    \begin{adjustbox}{max width=0.72\textwidth, center}
    \scriptsize
    \captionsetup{width=\linewidth}
    \caption{Representative U.S. Portfolio Comparison across Training Regimes}
    \label{PP_US_comparison}
    \begin{minipage}{\linewidth}
      \renewcommand{\arraystretch}{1.15}
      \setlength{\tabcolsep}{3pt}
      \begin{tabular}{clcccccccc}
        \toprule
        & & \multicolumn{4}{c}{\textbf{EW}} & \multicolumn{4}{c}{\textbf{VW}} \\
        \cmidrule(lr){3-6} \cmidrule(lr){7-10}
        \textbf{Metric} & \textbf{Representative Model} & \textbf{5} & \textbf{21} & \textbf{252} & \textbf{512} & \textbf{5} & \textbf{21} & \textbf{252} & \textbf{512} \\
        \midrule
        \multirow{7}{*}{AR}
          & Benchmark: CatBoost & 42.35 & 46.29 & 46.27 & 46.98 & 0.83 & 1.36 & 5.62 & 3.49 \\
          & Zero-shot: Chronos (Large) & -30.59 & -17.40 & 9.89 & 20.17 & -10.72 & -7.73 & 0.03 & 3.34 \\
          & Zero-shot: TimesFM 2 (500M) & -39.62 & -33.33 & -7.61 & -1.47 & -12.29 & -12.41 & -7.64 & -2.70 \\
          & Fine-tuned: Chronos (Large) & -2.11 & -0.69 & 0.23 & -0.94 & -1.18 & -2.17 & 0.64 & -1.15 \\
          & Fine-tuned: TimesFM 2 (500M) & -28.61 & -17.26 & -2.48 & -4.14 & -10.17 & -6.60 & 0.35 & -0.62 \\
          & Finance-native: Chronos (Small) & -0.32 & 13.99 & 33.61 & 36.84 & -3.02 & -0.44 & 8.69 & 11.05 \\
          & Finance-native: TimesFM (20M) & -18.10 & 10.48 & 30.24 & 30.12 & -9.75 & 0.31 & 6.34 & 6.01 \\
        \midrule
        \multirow{7}{*}{SD}
          & Benchmark: CatBoost & 7.43 & 6.89 & 6.88 & 7.35 & 10.63 & 10.36 & 10.18 & 10.82 \\
          & Zero-shot: Chronos (Large) & 7.66 & 6.90 & 7.62 & 6.91 & 10.32 & 9.46 & 11.27 & 11.51\\
          & Zero-shot: TimesFM 2 (500M) & 9.52 & 9.23 & 8.55 & 7.95 & 12.70 & 12.48 & 11.59 & 10.67 \\
          & Fine-tuned: Chronos (Large) & 2.62 & 2.76 & 3.10 & 2.97 & 5.59 & 6.38 & 6.53 & 6.27 \\
          & Fine-tuned: TimesFM 2 (500M) & 9.87 & 10.37 & 6.55 & 5.69 & 13.30 & 14.04 & 10.17 & 9.49 \\
          & Finance-native: Chronos (Small) & 4.83 & 5.47 & 6.87 & 6.79 & 7.31 & 7.58 & 10.76 & 10.64  \\
          & Finance-native: TimesFM (20M) & 9.95 & 8.50 & 8.71 & 8.34 & 13.37 & 11.87 & 12.17 & 11.75 \\
        \midrule
        \multirow{7}{*}{SR}
          & Benchmark: CatBoost & 5.70 & 6.72 & 6.72 & 6.39 & 0.08 & 0.13 & 0.55 & 0.32 \\
          & Zero-shot: Chronos (Large) & -3.99 & -2.52 & 1.30 & 2.92 & -1.04 & -0.82 & 0.00 & 0.29 \\
          & Zero-shot: TimesFM 2 (500M) & -4.16 & -3.61 & -0.89 & -0.18 & -0.97 & -0.99 & -0.66 & -0.25 \\
          & Fine-tuned: Chronos (Large) & -0.80 & -0.25 & 0.07 & -0.32 & -0.21 & -0.34 & 0.10 & -0.18 \\
          & Fine-tuned: TimesFM 2 (500M) & -2.90 & -1.67 & -0.38 & -0.73 & -0.77 & -0.47 & 0.03 & -0.07 \\
          & Finance-native: Chronos (Small) & -0.07 & 2.56 & 4.89 & 5.42 & -0.41 & -0.06 & 0.81 & 1.04 \\
          & Finance-native: TimesFM (20M) & -1.82 & 1.23 & 3.47 & 3.61 & -0.73 & 0.03 & 0.52 & 0.51 \\
        \bottomrule
      \end{tabular}
      \vspace{0.1cm}
      \begin{tablenotes}[para,flushleft]
        \footnotesize
        \textbf{Note:} This table reports long--short portfolio results for representative models. EW and VW denote equal-weighted and value-weighted portfolios, respectively. The columns 5, 21, 252, and 512 denote the input window length in trading days. AR, SD, and SR denote annualized return, annualized standard deviation, and Sharpe ratio, respectively. AR and SD are reported in percentage terms. For brevity, only long--short portfolio results are reported here; the complete benchmark, zero-shot, fine-tuned, and finance-native pre-trained portfolio results, including long- and short-leg contributions, are reported in \Cref{PP_benchmark_full}, \Cref{PP_TSFM_zero_shot_full}, \Cref{PP_TSFM_fine_tuned_full}, and \Cref{PP_TSFM_pre_trained_full}.
      \end{tablenotes}
    \end{minipage}
    \end{adjustbox}
  \end{threeparttable}
\end{table}

\subsubsection{Off-the-Shelf TSFMs: Zero-Shot Evidence} \label{zero_shot_results}
We report the zero-shot evidence mainly to establish an important result: generic off-the-shelf TSFMs do not transfer cleanly to daily excess return forecasting. This test is natural because zero-shot transfer is one of the central promises of TSFMs. If a model has learned general temporal representations from a broad pre-training corpus, it should be able to produce useful forecasts for a new time series without task-specific re-estimation. Daily excess returns provide a demanding test of that claim because the target is noisy, weakly predictable, heavy-tailed, and economically meaningful only if forecasts preserve cross-sectional ranking information. The primary comparison focuses on Chronos and TimesFM because they are widely used representative TSFMs and because their model families span two common design choices: discrete tokenization with probabilistic autoregressive decoding for Chronos and continuous patch-based forecasting for TimesFM. The \hyperref[pp:internet-title]{Internet Appendix} extends the zero-shot analysis to additional TSFMs, including Chronos-Bolt, TimesFM 2.5, Moirai, Kairos, MOMENT, Lag-Llama, TiRex, FlowState, TTM, Toto, and Sundial. This broader comparison matters because poor zero-shot performance for Chronos and TimesFM alone could reflect model-specific limitations rather than a more general transfer problem.\footnote{Point-forecast aggregation depends on the model's output interface. We average 20 predictive samples for Chronos and Sundial, and 100 for Lag-Llama and Toto. For sample-based Moirai specifications, we use the lower empirical median of 100 predictive draws, defined as the 50\textsuperscript{th} order statistic. For Kairos, we use the predicted 0.5 quantile. For Moirai 2, we take the median across its nine reported quantile values; this equals the predicted 0.5 quantile when the outputs are ordered but can differ when quantile crossing occurs. TTM supplies a direct deterministic point forecast. For the remaining models, we use the predictive median or direct point output, as applicable.}

The complete Chronos and TimesFM zero-shot results are reported in \Cref{FP_TSFM_zero_shot_full} and \Cref{PP_TSFM_zero_shot_full}. The zero-shot evidence is diagnostic rather than central. Across TSFMs, forecasting fit remains sharply below CatBoost, with the best Chronos checkpoint (Mini) reaching $-0.75\%$ $\RsqOOS$ at the 512-day window against CatBoost's $-0.03\%$. TimesFM performance depends strongly on the checkpoint. TimesFM 2 (500M) improves in $\RsqOOS$ as the input window lengthens, whereas TimesFM 1 (200M) exhibits increasingly extreme negative $\RsqOOS$ at longer windows. Directional accuracy is less uniformly poor than $\RsqOOS$, which is informative: TSFMs often fail to match the realized magnitude of daily returns, but some specifications still recover weak directional or ranking information. This distinction matters because portfolio performance depends more on discrimination across stocks than on point-forecast calibration. Portfolio results nevertheless show that off-the-shelf transfer is not sufficient. The best zero-shot economic specification in the representative table is Chronos (Large) at the 512-day window, with an annualized return of 20.17\% and an SR of 2.92. These values are positive, but they remain far below CatBoost's 46.98\% annualized return and 6.39 SR at the same window. TimesFM remains weaker, with its 512-day representative specification producing a negative long--short return.\footnote{\Cref{Cumulative_log_return_zero_shot_long_and_short} plots the cumulative log returns of zero-shot long--short portfolios. Chronos performance improves with longer input windows but remains below the benchmark, while both TimesFM specifications perform poorly across all window sizes. \Cref{Spread_Portfolio_zero_shot_table} also reports the full decile return profiles for the zero-shot TSFMs. Across all Chronos variants, the H--L spread is negative at the 5- and 21-day windows but turns positive at the 252- and 512-day windows, indicating that the inverted short-window rankings improve as the input window lengthens. The large negative short-window SRs indicate rankings that are informative but inverted, consistent with these models extrapolating recent returns where daily returns tend to reverse; we report signals as produced, since reversing them would require ex post knowledge of the sign. TimesFM 1 (200M) produces a positive unit-gross long--short return of 1.05\% at the 21-day window, but negative returns at the other three windows. TimesFM 2 (500M) produces negative long--short returns at all four windows.} The extended TSFM results in the \hyperref[pp:internet-title]{Internet Appendix} show substantial heterogeneity across architectures. At the 512-day window, zero-shot Toto attains a gross EW SR of 6.10, close to CatBoost's 6.39, despite an $\RsqOOS$ of $-110.78\%$. Thus, some off-the-shelf models produce competitive ranking signals even with poor point-forecast fit, but generic pre-training does not reliably deliver strong performance across models, windows, and evaluation metrics. Results for the additional TSFMs are reported in \Crefrange{FP_zero_shot_extended_1}{Spread_portfolio_zero_shot_extended_2} of the \hyperref[pp:internet-title]{Internet Appendix}. The selected representative zero-shot specifications, Chronos (Large) and TimesFM 2 (500M), are reported in \Cref{FP_US_comparison} and \Cref{PP_US_comparison} for the forecasting and portfolio performance comparisons, respectively.

\subsubsection{Off-the-Shelf TSFMs: Fine-Tuning Does Not Close the Gap} \label{fine_tuned_results}
Fine-tuning is included as a second diagnostic rather than as the center of this study. The question is whether adapting publicly released TSFMs to our financial dataset can close the performance gap left by zero-shot transfer. This is a natural next step because foundation-model workflows often treat fine-tuning as the default remedy when zero-shot performance is weak. We apply the same expanding-window discipline used for the benchmark models in \Cref{benchmark_results} so that the fine-tuned TSFMs are adapted only with data available before the relevant out-of-sample year. We also adopt the exact fine-tuning hyperparameter settings specified by the respective model developers in their studies. Full fine-tuning results for both Chronos and TimesFM are provided in \Cref{FP_TSFM_fine_tuned_full} and \Cref{PP_TSFM_fine_tuned_full}. The forecasting effects of fine-tuning are mixed across models and input windows. Fine-tuned Chronos (Large) records $\RsqOOS$ values of $-17.02\%$, $-2.30\%$, $-1.85\%$, and $-1.89\%$ at the 5-, 21-, 252-, and 512-day windows, respectively. These values improve on its zero-shot results at the first three windows but deteriorate at 512 days. Fine-tuned TimesFM 1 (200M) improves substantially on its extremely negative zero-shot $\RsqOOS$ values, whereas TimesFM 2 (500M) deteriorates across all four windows. All fine-tuned Chronos and TimesFM specifications retain negative full-sample annual-average $\RsqOOS$ values and remain below CatBoost on this metric. Overall directional accuracy also remains below CatBoost across all fine-tuned specifications. Thus, financial fine-tuning does not consistently improve forecasting performance or close the benchmark gap.

More importantly, the forecasting improvements observed for selected fine-tuned specifications do not translate into competitive gross portfolio performance. Across fine-tuned Chronos and TimesFM specifications, gross EW long--short SRs range from $-3.31$ to $0.07$, far below the benchmark. Relative to zero-shot inference, fine-tuning improves Chronos SRs at the 5- and 21-day windows but reduces them at the 252- and 512-day windows. It improves TimesFM 2 (500M) SRs at the 5-, 21-, and 252-day windows but reduces its SR at 512 days; all four SRs remain negative. Its effects on TimesFM 1 (200M) are mixed. The best representative fine-tuned configuration, Chronos (Large) at the 252-day window, has an annualized return of only 0.23\% and an SR of 0.07, compared with CatBoost's 46.27\% annualized return and 6.72 SR at the same window. This divergence illustrates the distinction between statistical prediction and portfolio choice emphasized by \citet{ait2001variable} and \citet{brandt2009parametric}. Improved squared-error performance need not improve the selection of stocks entering the extreme forecast deciles. Under the developer-recommended fine-tuning configurations evaluated here, adaptation improves point-forecast fit for selected specifications but does not close the portfolio-performance gap with the benchmark.\footnote{\Cref{Cumulative_log_return_fine_tuned_long_and_short} plots the cumulative log returns of fine-tuned long--short portfolios, showing that fine-tuned TSFMs consistently underperform the benchmark across all input windows. \Cref{Spread_Portfolio_fine_tuned_table} reports the spread portfolio performance of the fine-tuned TSFMs. A comparison with \Cref{Spread_Portfolio_zero_shot_table} shows that the effect of fine-tuning depends on the initial decile ordering. For short-window specifications with inverted zero-shot rankings, low-decile returns can fall and high-decile returns can rise, improving negative long--short spreads toward zero. At longer windows, fine-tuning can instead reduce positive spreads. Nevertheless, the fine-tuned portfolios remain substantially weaker than CatBoost.} The selected representative fine-tuned models are the same as those used under zero-shot inference, Chronos (Large) and TimesFM 2 (500M), and are reported in \Cref{FP_US_comparison} for forecasting performance and \Cref{PP_US_comparison} for portfolio performance.

The limited performance of the principal off-the-shelf Chronos and TimesFM specifications motivates the pre-training experiment below. Zero-shot inference asks generic TSFMs to transfer without adaptation; fine-tuning asks TSFMs to adapt after most representation learning has already occurred. Neither approach gives the model a financial return corpus as its native pre-training environment. \Cref{pre_trained_results} therefore asks whether TSFMs become more useful when their representations are learned from scratch on excess returns rather than inherited from broad, mostly non-financial time-series corpora. A finding that from-scratch financial pre-training outperforms adapted public checkpoints would run counter to the foundational premise of TSFMs, namely that broadly pre-trained temporal representations should transfer effectively across downstream domains.

\subsubsection{Finance-Native Pre-Training} \label{pre_trained_results}
Given the weak off-the-shelf evidence, the central question becomes whether TSFMs can recover performance once they are pre-trained directly on financial returns. We therefore construct Chronos and TimesFM models from scratch and pre-train them on the U.S. excess-return sample. The forecasting results show that finance-native pre-training helps these models recover predictive ability, particularly with longer input windows. Chronos delivers the clearest gains, with directional accuracy becoming competitive with the benchmark at the 252-day and 512-day windows, although $\RsqOOS$ remains negative across the pre-trained specifications. The portfolio results provide stronger evidence of improvement: pre-training produces more useful cross-sectional ranking signals, yielding higher long--short returns and SRs than the representative off-the-shelf Chronos and TimesFM specifications at longer windows. Among the representative finance-native specifications, Chronos (Small) outperforms TimesFM (20M) in gross EW long--short annualized returns and SRs across all four input windows, although both remain below CatBoost on these metrics. This ranking does not extend to all portfolio constructions: finance-native Chronos (Small) exceeds CatBoost in VW annualized returns and SRs at the 252- and 512-day windows. The detailed evidence in \Cref{complete_pre_trained_tsfm_results} therefore shows that finance-native pre-training improves selected forecasting metrics and long-window portfolio performance, with the most pronounced gains in the economic usefulness of the forecasts.

More specifically, \Cref{FP_TSFM_pre_trained_full} reports directional accuracy of 51.36\% for Chronos (Small) at both the 252-day and 512-day windows, compared with CatBoost's corresponding values of 51.08\% and 51.16\%. At the 512-day window, however, its $\RsqOOS$ is $-0.59\%$, compared with $-0.03\%$ for CatBoost, illustrating the distinction between directional prediction and point-forecast fit.\footnote{The modified Diebold--Mariano tests in \Cref{DM_test_Pre_Trained} indicate that CatBoost retains its advantage in point-forecast accuracy.} In portfolio terms, \Cref{PP_TSFM_pre_trained_full} reports an annualized return of 36.84\% and an SR of 5.42 for Chronos (Small) at the 512-day window, compared with 30.12\% and 3.61 for TimesFM (20M). Within the Chronos family, the small specification outperforms the tiny and mini variants in portfolio performance at both the 252-day and 512-day windows.\footnote{The decile returns in \Cref{Spread_Portfolio_pre_trained_table} show clearer separation between stocks with high and low predicted returns at longer input windows.} Based on their gross EW long--short portfolio performance at the longer windows, we select Chronos (Small) and TimesFM (20M) as representative finance-native pre-trained specifications from their respective model families, with their forecasting performance reported in \Cref{FP_US_comparison} and their portfolio performance reported in \Cref{PP_US_comparison}.

% \Cref{Cumulative_log_return_pre_trained_long_and_short} displays the cumulative performance of the long and short portfolio legs separately.

\begin{figure}
    \caption{Cumulative Log Returns of Finance-Native Pre-Trained TSFMs: Long--Short Portfolios}
    \label{Cumulative_log_return_pre_trained_long_short}
    \centering
    \includegraphics[
        width=0.71\textwidth,
        clip,
        trim=0 35 0 28, % left bottom right top
    ]{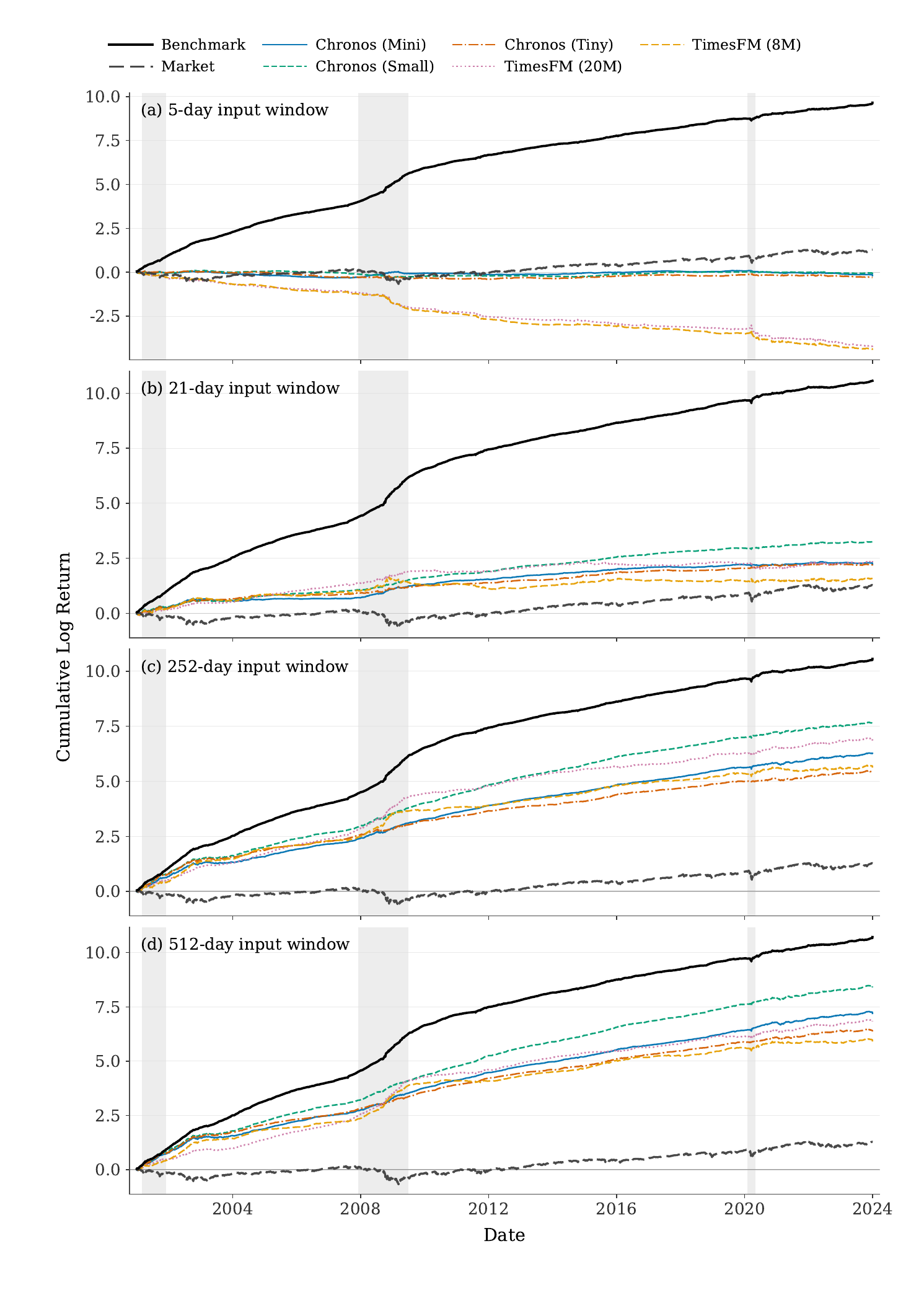} % Adjust the width and trims
    \begin{minipage}{0.8\textwidth}
        \footnotesize
        \textbf{Note:} This figure displays the cumulative log returns of long--short portfolios constructed from various forecasting models with input windows of 5 (a), 21 (b), 252 (c), and 512 (d) trading days. The benchmark model is CatBoost, the representative benchmark. The time series foundation models (TSFMs) include Chronos (tiny, mini, and small) and TimesFM (with 8 million and 20 million parameters). Forecasts are generated using the finance-native pre-trained models. Each subplot corresponds to a specific input window length, as indicated by the text labels in the upper-left corners. The benchmark model (CatBoost) is highlighted in black with bold lines. The dashed black line represents the cumulative log return of the market (S\&P 500). Shaded areas indicate U.S. recession periods, as defined by the National Bureau of Economic Research (NBER). The model-based long--short portfolios use equal weighting within each leg and are scaled to unit gross exposure. The S\&P 500 is shown separately as a market reference.
    \end{minipage}
\end{figure}

\Cref{FP_US_comparison} compares forecasting performance across training regimes. Finance-native pre-training improves directional accuracy relative to the representative zero-shot and fine-tuned models, particularly at longer windows. At the 512-day window, finance-native Chronos (Small) achieves directional accuracy of 51.36\%, compared with 51.01\% for zero-shot Chronos (Large), 50.27\% for fine-tuned Chronos (Large), and 51.16\% for CatBoost. Its $\RsqOOS$ of $-0.59\%$ improves on both zero-shot Chronos (Large), at $-1.37\%$, and fine-tuned Chronos (Large), at $-1.89\%$, but remains below CatBoost at $-0.03\%$. For TimesFM, pre-training likewise improves directional accuracy but yields a lower $\RsqOOS$ than zero-shot transfer across all four windows. Thus, finance-native pre-training improves directional accuracy for both representative model families, while its effect on squared-error performance differs by family: finance-native Chronos (Small) outperforms both representative off-the-shelf Chronos specifications across all four windows, whereas finance-native TimesFM (20M) remains below its zero-shot counterpart on $\RsqOOS$. The portfolio evidence in \Cref{PP_US_comparison} shows clearer gains at the 252-day and 512-day windows. At the 512-day window, finance-native Chronos (Small) achieves an EW annualized return of 36.84\% and an SR of 5.42, compared with 20.17\% and 2.92 for zero-shot Chronos and $-0.94\%$ and $-0.32$ for fine-tuned Chronos. These results substantially narrow the gap to CatBoost's 46.98\% and 6.39. Finance-native TimesFM (20M) also improves on its off-the-shelf counterparts, reaching an annualized return of 30.12\% and an SR of 3.61, although Chronos remains stronger among the finance-native specifications. For VW portfolios, finance-native Chronos achieves 11.05\% and 1.04, exceeding CatBoost's 3.49\% and 0.32 at the same window, while absolute performance remains below that of EW portfolios. \Cref{Cumulative_log_return_pre_trained_long_short} presents the same portfolio evidence in terms of cumulative log returns, reinforcing the stronger TSFM performance at longer input windows and the benchmark's remaining advantage. Overall, smaller finance-native models deliver stronger long-window portfolio performance than their much larger off-the-shelf Chronos and TimesFM counterparts, suggesting that the pre-training domain matters alongside model size. The economic gains are consistent with improved ranking ability despite the remaining point-forecast fit gap.

The pervasiveness of return predictability across firm sizes is an important consideration in evaluating financial forecasting models \citep{fama2008dissecting,10.1093/rfs/hhy131}. We therefore examine large-cap and small-cap stocks separately in \Cref{FP_US_cap_comparison} and compare their portfolio performance in \Cref{PP_US_cap_comparison}. Together with the equal- and value-weighted results in \Cref{PP_US_comparison}, these comparisons assess whether model performance extends beyond smaller stocks. For the benchmark, CatBoost, the small-cap advantage appears in both squared-error performance and directional accuracy across all four input windows. At the 512-day window, its $\RsqOOS$ is $0.59\%$ for small-cap stocks, compared with $-0.28\%$ for large-cap stocks, while directional accuracy is $52.39\%$ versus $50.95\%$. Among TSFMs, however, this advantage is less consistent. Both representative zero-shot models perform worse on small-cap stocks on both forecasting metrics at the 5- and 21-day windows. Fine-tuned TimesFM 2 (500M) also has more negative small-cap than large-cap $\RsqOOS$ at the 5-, 252-, and 512-day windows, but the ordering reverses at 21 days. Fine-tuned Chronos (Large) records negative $\RsqOOS$ in both size groups, with more negative values for small-cap stocks across all four windows. At the 512-day window, its $\RsqOOS$ is $-2.40\%$ for small-cap stocks and $-1.41\%$ for large-cap stocks. Its directional accuracy remains below CatBoost in both size groups across all windows. At the 512-day window, zero-shot Chronos (Large) achieves small-cap directional accuracy of $52.46\%$, marginally exceeding CatBoost's $52.39\%$ despite retaining negative $\RsqOOS$. Thus, the stronger small-cap predictability evident for the benchmark does not automatically translate into better TSFM performance across both forecasting metrics.

\begin{table}[!htbp]
  \centering
  \begin{threeparttable}
    \begin{adjustbox}{max width=0.80\textwidth, center}
    \scriptsize
    \captionsetup{width=\linewidth}
    \caption{Representative U.S. Forecasting Comparison by Market Capitalization}
    \label{FP_US_cap_comparison}
    \begin{minipage}{\linewidth}
      \renewcommand{\arraystretch}{1.15}
      \setlength{\tabcolsep}{3pt}
      \begin{tabular}{llcccccccc}
        \toprule
        & & \multicolumn{4}{c}{\textbf{Large-Cap}} & \multicolumn{4}{c}{\textbf{Small-Cap}} \\
        \cmidrule(lr){3-6} \cmidrule(lr){7-10}
        \textbf{Metric} & \textbf{Representative Model} & \textbf{5} & \textbf{21} & \textbf{252} & \textbf{512} & \textbf{5} & \textbf{21} & \textbf{252} & \textbf{512} \\
        \midrule
        \multirow{7}{*}{$\RsqOOS$}
          & Benchmark: CatBoost & -0.50 & -0.38 & -0.32 & -0.28 & 0.39 & 0.66 & 0.62 & 0.59 \\
          & Zero-shot: Chronos (Large) & -35.50 & -12.58 & -1.69 & -1.03 & -60.59 & -35.61 & -3.30 & -1.70 \\
          & Zero-shot: TimesFM 2 (500M) & -27.27 & -10.90 & -3.78 & -2.90 & -29.93 & -13.18 & -3.96 & -2.64 \\
          & Fine-tuned: Chronos (Large) & -16.42 & -2.01 & -1.35 & -1.41 & -18.34 & -2.74 & -2.31 & -2.40 \\
          & Fine-tuned: TimesFM 2 (500M) & -520.53 & -561.35 & -141.03 & -124.07 & -560.53 & -554.68 & -176.16 & -145.57 \\
          & Finance-native: Chronos (Small) & -3.17 & -1.40 & -0.92 & -0.84 & -3.20 & -1.33 & -0.17 & -0.15 \\
          & Finance-native: TimesFM (20M) & -32.89 & -16.98 & -15.65 & -17.60 & -37.39 & -15.35 & -11.86 & -12.61 \\
        \midrule
        \multirow{7}{*}{Overall Acc.}
          & Benchmark: CatBoost & 50.70 & 50.74 & 50.80 & 50.95 & 51.96 & 52.47 & 52.33 & 52.39 \\
          & Zero-shot: Chronos (Large) & 49.58 & 49.91 & 50.39 & 50.61 & 46.94 & 47.85 & 51.35 & 52.46 \\
          & Zero-shot: TimesFM 2 (500M) & 49.25 & 49.44 & 49.69 & 50.01 & 46.35 & 47.19 & 49.72 & 49.94 \\
          & Fine-tuned: Chronos (Large) & 49.98 & 49.85 & 49.83 & 49.79 & 50.02 & 50.86 & 50.98 & 50.98 \\
          & Fine-tuned: TimesFM 2 (500M) & 49.41 & 50.00 & 50.38 & 50.29 & 47.82 & 48.59 & 49.55 & 49.53 \\
          & Finance-native: Chronos (Small) & 49.58 & 50.08 & 50.86 & 50.84 & 51.65 & 52.27 & 53.12 & 53.20 \\
          & Finance-native: TimesFM (20M) & 49.84 & 50.66 & 50.95 & 50.98 & 49.77 & 51.53 & 51.82 & 51.82 \\
        \bottomrule
      \end{tabular}
      \vspace{0.1cm}
      \begin{tablenotes}[para,flushleft]
        \footnotesize
        \textbf{Note:} This table reports forecasting results separately for large-cap and small-cap stocks for representative models from each training regime. The columns 5, 21, 252, and 512 denote the input window length in trading days. $\RsqOOS$ and `Overall Acc.' are reported in percentage terms. `Overall Acc.' denotes overall directional accuracy. Large-cap and small-cap denote the top and bottom 25\% of stocks by market capitalization, respectively. The complete results, including the full-sample values, are reported in \Cref{FP_benchmark_full}, \Cref{FP_TSFM_zero_shot_full}, \Cref{FP_TSFM_fine_tuned_full}, and \Cref{FP_TSFM_pre_trained_full}.
      \end{tablenotes}
    \end{minipage}
    \end{adjustbox}
  \end{threeparttable}

   \vspace{1.20cm} 

  \begin{threeparttable}
    \begin{adjustbox}{max width=0.69\textwidth, center}
    \scriptsize
    \captionsetup{width=\linewidth}
    \caption{Representative U.S. Portfolio Comparison by Market Capitalization}
    \label{PP_US_cap_comparison}
    \begin{minipage}{\linewidth}
      \renewcommand{\arraystretch}{1.15}
      \setlength{\tabcolsep}{3pt}
      \begin{tabular}{llcccccccc}
        \toprule
        & & \multicolumn{4}{c}{\textbf{Large-Cap}}
            & \multicolumn{4}{c}{\textbf{Small-Cap}} \\
        \cmidrule(lr){3-6} \cmidrule(lr){7-10}
        \textbf{Metric} & \textbf{Representative Model}
          & \textbf{5} & \textbf{21} & \textbf{252} & \textbf{512}
          & \textbf{5} & \textbf{21} & \textbf{252} & \textbf{512} \\
        \midrule
        \multirow{7}{*}{AR}
          & Benchmark: CatBoost & -1.23 & -1.18 & 2.31 & 1.41 & 119.06 & 126.97 & 125.39 & 126.33 \\
          & Zero-shot: Chronos (Large) & -10.73 & -9.24 & -1.61 & -0.85 & -72.05 & -37.83 & 33.29 & 63.23 \\
          & Zero-shot: TimesFM 2 (500M) & -11.20 & -11.69 & -8.51 & -3.41 & -93.75 & -72.63 & -1.82 & 10.57 \\
          & Fine-tuned: Chronos (Large) & -0.28 & -1.24 & 0.41 & 0.36 & -5.97 & -0.93 & 2.66 & 1.26 \\
          & Fine-tuned: TimesFM 2 (500M) & -8.09 & -4.56 & 0.41 & -1.10 & -64.53 & -37.66 & -5.95 & -9.07 \\
          & Finance-native: Chronos (Small) & -1.56 & -0.80 & 4.99 & 7.36 & 5.99 & 45.34 & 91.18 & 94.60 \\
          & Finance-native: TimesFM (20M) & -7.87 & 0.46 & 7.36 & 5.55 & -36.82 & 29.74 & 75.39 & 78.23 \\
        \midrule
        \multirow{7}{*}{SD}
          & Benchmark: CatBoost & 8.42 & 7.97 & 7.94 & 8.57 & 9.90 & 9.87 & 9.47 & 9.95 \\
          & Zero-shot: Chronos (Large) & 8.38 & 8.04 & 9.22 & 9.11 & 9.25 & 8.77 & 8.77 & 7.95 \\
          & Zero-shot: TimesFM 2 (500M) & 10.85 & 10.60 & 9.69 & 8.78 & 11.06 & 10.63 & 9.37 & 9.00 \\
          & Fine-tuned: Chronos (Large) & 3.53 & 4.12 & 4.00 & 3.70 & 6.14 & 5.87 & 5.82 & 5.82 \\
          & Fine-tuned: TimesFM 2 (500M) & 11.37 & 11.78 & 8.72 & 7.90 & 11.01 & 10.89 & 7.78 & 7.35 \\
          & Finance-native: Chronos (Small) & 5.78 & 6.28 & 8.90 & 8.93 & 7.14 & 7.34 & 9.17 & 8.61 \\
          & Finance-native: TimesFM (20M) & 11.56 & 10.03 & 10.23 & 9.80 & 10.70 & 9.60 & 10.07 & 10.01 \\
        \midrule
        \multirow{7}{*}{SR}
          & Benchmark: CatBoost & -0.15 & -0.15 & 0.29 & 0.16 & 12.03 & 12.86 & 13.24 & 12.69 \\
          & Zero-shot: Chronos (Large) & -1.28 & -1.15 & -0.17 & -0.09 & -7.79 & -4.32 & 3.80 & 7.95 \\
          & Zero-shot: TimesFM 2 (500M) & -1.03 & -1.10 & -0.88 & -0.39 & -8.47 & -6.83 & -0.19 & 1.17 \\
          & Fine-tuned: Chronos (Large) & -0.08 & -0.30 & 0.10 & 0.10 & -0.97 & -0.16 & 0.46 & 0.22 \\
          & Fine-tuned: TimesFM 2 (500M) & -0.71 & -0.39 & 0.05 & -0.14 & -5.86 & -3.46 & -0.77 & -1.23 \\
          & Finance-native: Chronos (Small) & -0.27 & -0.13 & 0.56 & 0.82 & 0.84 & 6.18 & 9.94 & 10.99 \\
          & Finance-native: TimesFM (20M) & -0.68 & 0.05 & 0.72 & 0.57 & -3.44 & 3.10 & 7.49 & 7.82 \\
        \bottomrule
      \end{tabular}
      \vspace{0.1cm}
      \begin{tablenotes}[para,flushleft]
        \footnotesize
        \textbf{Note:} This table reports gross EW long--short portfolio results split by market capitalization for representative models from each training regime. The columns 5, 21, 252, and 512 denote the input window length in trading days. AR, SD, and SR denote annualized return, annualized standard deviation, and Sharpe ratio, respectively. AR and SD are reported in percentage terms. Large-cap and small-cap denote the top and bottom 25\% of stocks by market capitalization, respectively. Within each capitalization group, stocks are sorted daily into forecast deciles to construct the EW long--short portfolio. Group-days with fewer than ten eligible stocks are excluded. Full-universe portfolio results, without the capitalization split, are reported in \Cref{PP_benchmark_full}, \Cref{PP_TSFM_zero_shot_full}, \Cref{PP_TSFM_fine_tuned_full}, and \Cref{PP_TSFM_pre_trained_full}.
      \end{tablenotes}
    \end{minipage}
    \end{adjustbox}
  \end{threeparttable}
\end{table}

Against this benchmark, finance-native pre-training produces its clearest directional gains among smaller stocks. At the 512-day window, finance-native Chronos (Small) exceeds CatBoost's small-cap directional accuracy by 0.81 percentage points, but remains 0.11 percentage points below CatBoost for large-cap stocks. Its small-cap--large-cap accuracy gap is consequently wider than CatBoost's, at 2.36 versus 1.44 percentage points. This concentration is also evident at the 252-day window, where Chronos's advantage over CatBoost is 0.79 percentage points for small-cap stocks but only 0.06 percentage points for large-cap stocks. At both longer windows, its small-cap $\RsqOOS$ is also closer to zero than its large-cap value, although CatBoost retains better squared-error performance in both groups. Finance-native TimesFM (20M) likewise exhibits higher small-cap than large-cap directional accuracy at longer windows, but remains below CatBoost in the small-cap segment while slightly exceeding it among large-cap stocks. The size breakdown therefore shows that the directional benefits of finance-native pre-training vary across both market segments and model families, with Chronos's advantage over the benchmark concentrated primarily in small-cap stocks.

The portfolio evidence in \Cref{PP_US_cap_comparison} also helps interpret the high gross SRs reported in this study. At the 512-day window, CatBoost achieves an annualized return of $126.33\%$ and an SR of 12.69 for small-cap stocks, compared with $1.41\%$ and 0.16 for large-cap stocks. Finance-native Chronos (Small) exhibits a similar concentration, with corresponding returns of $94.60\%$ and $7.36\%$ and SRs of 10.99 and 0.82. These high small-cap SRs combine large gross mean returns with annualized portfolio standard deviations of $9.95\%$ for CatBoost and $8.61\%$ for Chronos. Equal weighting also gives smaller firms greater influence than value weighting, making performance in this segment particularly relevant to the aggregate EW results.\footnote{\citet{jiang2023re} report an annualized gross SR of 7.15 for a weekly equal-weighted strategy, compared with 1.49 under value weighting and 1.02 when the equal-weighted strategy is restricted to the largest 500 stocks. This attenuation is consistent with the importance of smaller stocks to our gross performance. \citet{nagel2012evaporating} reports an unconditional annualized SR of 8.44 for a daily transaction-price reversal strategy and 9.58 after hedging conditional market exposure, with the fitted conditional annualized SR peaking at approximately 15 during the financial crisis. Rebalancing frequency matters economically: daily trading can capture short-lived predictive gains that less frequent rebalancing misses, while the trading required to capture those gains can substantially erode net performance \citep{barber2001can}. \citet{avramov2006liquidity} further show that the largest short-run reversal profits occur in high-turnover, illiquid stocks but remain below estimated transaction costs. Although differences in signals, stock universes, holding periods, and conditional versus unconditional measurement limit direct numerical comparisons, this evidence supports our size-related interpretation: the highest gross SRs are concentrated in smaller stocks, where strong measured return predictability can coexist with unprofitable implementation.} Nevertheless, the model ranking differs by segment at the 252- and 512-day windows: CatBoost remains stronger for small-cap portfolios, whereas finance-native Chronos delivers higher large-cap returns and SRs. At the 512-day window, Chronos's stronger large-cap portfolio performance occurs despite its slightly lower directional accuracy, illustrating that directional accuracy alone does not determine portfolio performance. The large-cap portfolio ranking at both longer windows is consistent with Chronos's stronger VW performance in \Cref{PP_US_comparison}.

Translating these gross returns into implementable investment performance is particularly challenging for small-cap portfolios, where trading costs and price impact per dollar traded tend to be higher. Our baseline evaluation omits these costs, which can substantially erode the gains from daily rebalancing. \citet{novy2016taxonomy} document that transaction costs reduce anomaly profitability, that relatively few high-turnover strategies retain significant net returns even after cost mitigation, and that price impact constrains strategy capacity. These findings are particularly relevant to our daily rebalanced portfolios, which repeatedly incur execution costs and may become more costly to implement as investment size increases. Short positions introduce additional constraints: \citet{d2002market} documents borrowing fees, limited share availability, and recall risk, with borrowing difficulties disproportionately affecting small, illiquid stocks. The size breakdown is consistent with these implementation concerns, although it does not identify the mechanisms underlying the high gross SRs. Given these limitations, the portfolio exercises in this study should be interpreted primarily as an economic horse race among forecasting models, assessing the usefulness of their cross-sectional rankings under a common frictionless framework. The reported gross SRs provide comparative evidence on model performance rather than establish implementable trading strategies; achievable investment performance depends on turnover, execution costs, borrowing conditions, and portfolio capacity.

\begin{figure}
    \caption{Training Time by Model}
    \label{training_time_by_model_plot}
    \centering
    \includegraphics[
        width=0.65\textwidth,
        clip,
        trim=0 4 0 16, % left bottom right top
    ]{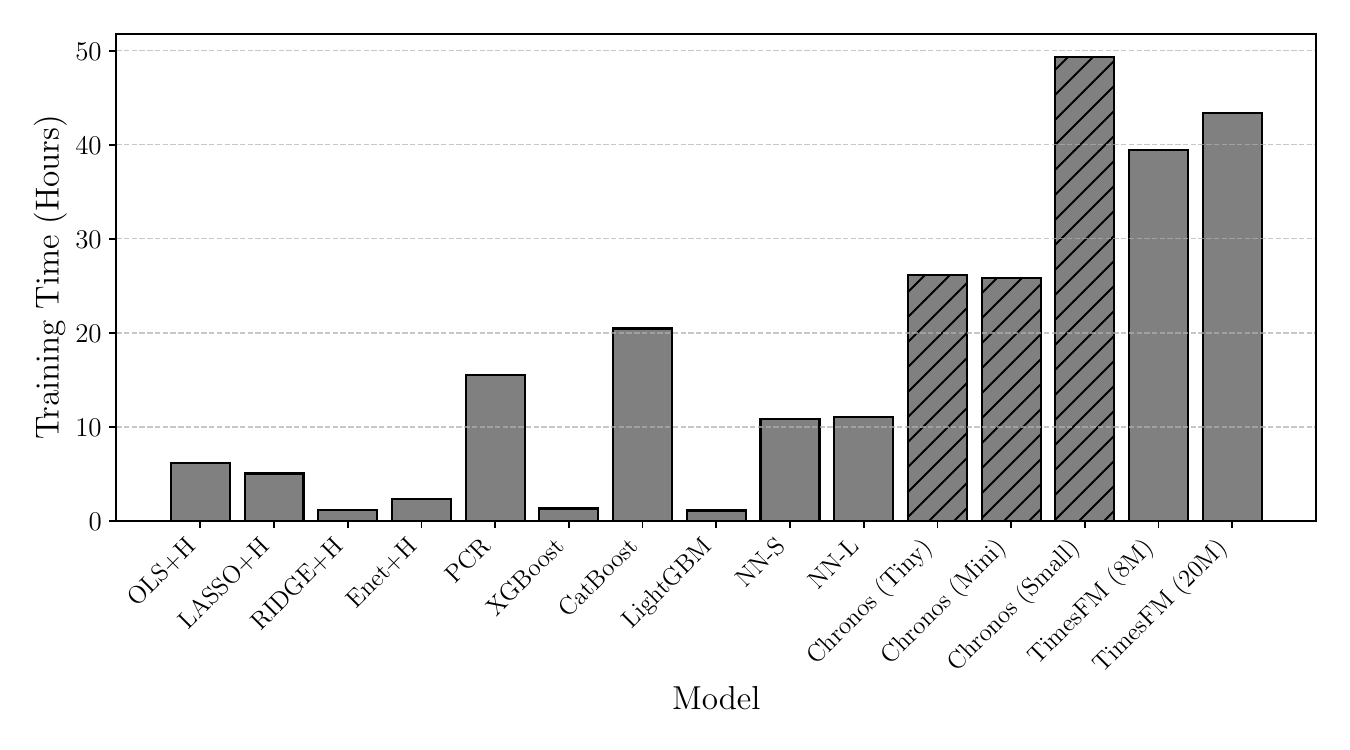} % Adjust the width and trims
    \begin{minipage}{0.8\textwidth}
        \footnotesize
        \textbf{Note:} This figure reports the total training time (in hours) required for each forecasting model. Benchmark models (OLS+H, LASSO+H, RIDGE+H, Enet+H, PCR, XGBoost, CatBoost, LightGBM, NN-S, and NN-L) are compared with time series foundation models (TSFMs, including Chronos and TimesFM, indicated by hatched bars). `H' indicates that the model is estimated using the Huber loss. Training times reflect the implementations and hardware used in this study: benchmarks were trained on CPUs and TSFMs on GPUs. They are practical runtime comparisons rather than hardware-standardized measures of computational efficiency. Reported values correspond to the final model trained for the year 2022, which uses the maximum available U.S. excess return data. All values are approximate, and small differences in training times should not be interpreted as meaningful.
    \end{minipage}
\end{figure}

These forecasting and portfolio gains also involve a substantial computational trade-off. \Cref{training_time_by_model_plot} reports conventional benchmark training times of up to about 20 hours, compared with approximately 25--50 hours for finance-native Chronos models and about 40--45 hours for TimesFM variants.\footnote{These figures reflect the study's implementation, including CPU training for conventional benchmarks and GPU training for TSFMs, and should consequently be interpreted as practical resource requirements rather than hardware-standardized comparisons.} This limitation is particularly important when moving to pre-training from scratch, because the model must learn its financial return representations directly rather than reuse those embedded in a public checkpoint. The computational burden extends beyond a single training run: repeated estimation, hyperparameter selection, and comparisons across model configurations multiply the resource requirements. It can therefore constrain experimentation and make frequent retraining less practical, especially as the training dataset or model size increases.

\subsection{Scaling Finance-Native Pre-Training} \label{scaled_results} 
The original TSFMs were pre-trained on substantially larger and more diverse corpora than our initial finance-native sample. The next question is whether the gains from finance-native pre-training continue to strengthen as the data become larger and more heterogeneous, helping assess whether limited training scale contributes to the remaining performance gap with the benchmarks. We therefore treat data scaling as an extension of the pre-training results in \Cref{primary_results}. The comparison begins with benchmark models, where global pooling generally weakens gross portfolio performance despite improvements in some forecasting metrics, and then turns to pre-trained TSFMs, where scaling improves selected specifications.

We construct three scaled datasets to answer distinct questions, with their endpoint sizes summarized in \Cref{dataset_scale_summary}. First, we expand the training data from U.S. excess returns to global excess returns, increasing the number of observations from approximately 131.22 million to 444.70 million. Second, we augment the global data with monthly JKP characteristic series, increasing the TSFM pre-training corpus to approximately 1.70 billion scalar observations for the 2022 models. This extension adds lower-frequency financial information, motivated by the use of multiple sampling frequencies in the original TSFM pre-training corpora. Third, we replace the JKP characteristics with synthetic time series of identical dimensionality and panel structure to assess whether comparable gains can arise without the additional financial content. Since TSFMs incorporate synthetic data in their original pre-training, this comparison tests whether synthetic augmentation also benefits financial return forecasting. It further helps assess whether any gains from JKP augmentation depend on its financial content and relevance to return prediction, or can also arise from synthetic data matched in size and panel structure. For each variation and year, we pre-train the TSFMs from scratch again.\footnote{For the augmented specifications, $\mathcal{D}_{\mathrm{fin}}$ in the pre-training formulation in \Cref{TSFM_def_pre_training} includes a mixture of global excess-return series and monthly JKP characteristic series or synthetic series, as described in \Cref{data_source}. At inference, the forecasting task remains as defined in \Cref{problem_steup}: each model receives a historical window of excess returns and predicts the next-day excess return.}

% \Cref{benchmark_results_scaled} provides the benchmark comparison, while \Cref{TSFM_results_scaled} contains the main scaling evidence for TSFMs. \Cref{performance_over_time} examines whether the resulting gains are stable over time, and \Cref{transaction_cost} evaluates their robustness to transaction costs.

%\subsubsection{Data Scaling} \label{benchmark_results_scaled}
Before turning to scaled TSFMs, we briefly document how benchmark models respond to the same data expansion. This comparison matters because it shows that simply adding more heterogeneous data is not enough on its own; the benefit depends on the model class. We therefore re-estimate all benchmark models on global excess returns and compare them directly with their U.S.-only counterparts. \Cref{FP_scaling_representative} and \Cref{PP_scaling_representative} summarize the representative changes, while the complete benchmark scaling comparisons are reported in \Cref{impact_benchmark_scale_forecasting} and \Cref{impact_benchmark_scale_portfolio}. The reported values are differences relative to the corresponding U.S.-only specification, with the columns referring to the 5-, 21-, 252-, and 512-day input windows. Thus, a positive value indicates an increase in the relevant performance metric, whereas a negative value indicates a decrease. For the representative benchmark, CatBoost, global scaling weakens performance across all four windows. The forecasting changes are negative across all four windows: $\RsqOOS$ falls by 0.16, 0.14, 0.05, and 0.07 percentage points, while overall accuracy falls by 1.19, 1.14, 0.78, and 1.10 percentage points. The economic comparison points in the same direction. Annualized returns fall by 19.77, 17.06, 9.91, and 17.36 percentage points, and SRs fall by 2.14, 2.10, 0.84, and 1.72.\footnote{It should be noted that all benchmark models, model complexities, and hyperparameter search environments are kept consistent with those used in \Cref{benchmark_results}, ensuring a fair and controlled comparison between the results reported here and those based on U.S.-only data. Although it is possible to further tune each model specifically for the larger and more heterogeneous global dataset, such optimization lies beyond the scope of this study. The objective is to isolate the effect of data scaling itself, rather than to maximize the performance of individual models through additional calibration. This design choice allows for a direct assessment of how increasing the data scale while holding model configurations constant affects both forecasting and portfolio performance, reflecting the performance of each model in an out-of-the-box setting with minimal modification.}

%\subsubsection{Data Scaling in Pre-Trained TSFMs} \label{TSFM_results_scaled}
We now turn to the TSFM scaling result. The question is whether pre-trained TSFMs benefit from more finance-native data even when benchmark models do not benefit uniformly. The comparison covers four pre-training regimes: U.S.-only, global, JKP-augmented, and synthetic-augmented. The representative benchmark remains CatBoost trained on U.S. data, as we have shown that global scaling weakens, rather than strengthens, the benchmark. The representative forecasting and portfolio changes are summarized in \Cref{FP_scaling_representative} and \Cref{PP_scaling_representative}, while the complete TSFM scaling comparisons are reported in \Cref{impact_TSFM_scale_forecasting} and \Cref{impact_TSFM_scale_portfolio}.\footnote{Following \citet{ansari2024chronos}, synthetic series are drawn from a zero-mean Gaussian process prior on a fixed uniform grid. For each series a small set of base kernels is sampled with replacement from a bank that includes periodic, linear, radial basis, rational quadratic, white noise, and constant components. These kernels are composed by repeatedly applying random sums or products to form a covariance function, from which a single trajectory is sampled. Although synthetic series could alternatively be generated using finance-specific procedures calibrated to reproduce stylized facts of asset returns, we adopt the generic Gaussian-process construction used in the TSFM literature to remain consistent with the pre-training framework motivating this exercise. Accordingly, the synthetic augmentation is not designed to reproduce financial-market structure directly. Instead, it allows us to assess whether increasing pre-training scale and heterogeneity through generic synthetic time series improves subsequent financial return forecasting.} Moving from U.S.-only to global pre-training does not help every TSFM equally, but it already departs from the benchmark pattern. In \Cref{FP_scaling_representative}, Chronos (Small) gives up modest $\RsqOOS$ at the longer windows but improves overall accuracy by 0.15 percentage points at both the 252- and 512-day windows. In \Cref{PP_scaling_representative}, the same global expansion increases Chronos (Small)'s annualized return by 4.67 and 4.13 percentage points and its SR by 0.70 and 0.62 at the 252- and 512-day windows, respectively, while annualized volatility remains largely unchanged. TimesFM is less stable under global-only pre-training, but becomes much stronger once the global sample is augmented. Relative to U.S.-only pre-training, JKP-augmented TimesFM (20M) gains 11.95 and 13.47 percentage points in $\RsqOOS$ at the 252- and 512-day windows, raises annualized returns by 16.68 and 19.85 percentage points, and increases SRs by 3.19 and 4.00 as annualized volatility declines. At the 512-day window, JKP-augmented TimesFM (20M) reaches an SR of 7.62, above CatBoost's 6.39. Chronos also shows economic gains at the longer windows, but TimesFM (20M) exhibits the largest long-window SR gains under JKP augmentation.\footnote{We also examine how scaling the fine-tuning data affects the performance of TSFMs. \Cref{impact_TSFM_scale_fine_tuned_forecasting} and \Cref{impact_TSFM_scale_fine_tuned_portfolio} of the \hyperref[pp:internet-title]{Internet Appendix} illustrate the effects of expanding the fine-tuning dataset from U.S.-only to global and subsequently to JKP-augmented configurations. The effects of expanding the fine-tuning sample depend on the model, input window, and metric. TimesFM 2 (500M) shows large improvements in $\RsqOOS$, whereas TimesFM 1 (200M) has mixed changes. Several Chronos specifications also show substantial improvements in squared-error fit under global fine-tuning, while JKP augmentation can reverse these gains. Directional-accuracy changes are generally small. Portfolio SR changes are mixed, and most configurations remain negative and well below CatBoost. Larger fine-tuning datasets therefore do not consistently close the economic-performance gap documented in \Cref{fine_tuned_results}.}

A similar pattern emerges under synthetic augmentation, which delivers results broadly comparable to JKP augmentation and, in some cases, stronger performance. In \Cref{FP_scaling_representative}, synthetic-augmented global pre-training gives TimesFM (20M) $\RsqOOS$ gains of 11.68 and 13.18 percentage points relative to U.S.-only pre-training at the 252- and 512-day windows, while overall accuracy improves by 0.25 and 0.34 percentage points, compared with $-0.02$ and 0.04 under JKP augmentation. In \Cref{PP_scaling_representative}, it raises the TimesFM (20M) annualized return by 18.16 and 20.71 percentage points and its SR by 3.14 and 3.46 at those windows, while giving Chronos (Small) larger SR improvements than JKP augmentation. That pattern suggests that, conditional on a finance-native core of excess-return data, the gains are not driven solely by the financial content of the added JKP characteristics; instead, TSFMs also appear to benefit from greater scale and heterogeneity in the augmented pre-training sample. In that sense, the scaling result points to the joint importance of domain-relevant core data and pre-training scale: a finance-native excess-return base, combined with larger and more heterogeneous augmentation, allows the stronger TSFMs, especially TimesFM, to translate model capacity into better economic performance.\footnote{\Cref{DM_test_TSFMs_Scaling} reports the modified DM test results. The table compares pre-trained TSFMs under two augmentation regimes: JKP-augmented and synthetic-augmented (models labeled `Syn.'). The relative performance of synthetic and JKP augmentation depends on model size and input window. For example, at matched 252- and 512-day windows, the DM statistics favor synthetic augmentation for TimesFM (8M), at 7.87 and 7.83, but favor JKP augmentation for TimesFM (20M), at $-2.22$ and $-2.69$; all four differences are significant at the reported 5\% level. Thus, the tests do not establish general superiority of synthetic augmentation. CatBoost has significantly lower daily cross-sectional squared error than all augmented TSFM specifications in the reported comparisons.} These model-dependent gains complement \citet{kelly2024virtue} by highlighting the role of training-data scale and composition alongside model complexity. While \citet{hellum2023global} find that global models outperform local models in characteristic-based return prediction, our return-history results provide mixed support for international pooling, with gains depending on the model and input window.

\begin{table}[!htbp]
  \centering
  \begin{threeparttable}
    \begin{adjustbox}{max width=0.63\textwidth, center}
    \scriptsize
    \captionsetup{width=\linewidth}
    \caption{Representative Forecasting Impact of Data Scaling}
    \label{FP_scaling_representative}
    \begin{minipage}{\linewidth}
      \renewcommand{\arraystretch}{1.15}
      \setlength{\tabcolsep}{4pt}
      \begin{tabular}{llcccc}
        \toprule
        \textbf{Metric} & \makebox[9em][l]{\textbf{Model}}\textbf{Data-scale comparison} & \textbf{5} & \textbf{21} & \textbf{252} & \textbf{512} \\
        \midrule
        \multirow{7}{*}{$\RsqOOS$}
          & \makebox[9em][l]{Benchmark}Global vs.\ U.S. & -0.16 & -0.14 & -0.05 & -0.07 \\
          & \makebox[9em][l]{Chronos (Small)}Global vs.\ U.S. & 0.44 & -0.11 & -0.07 & -0.14 \\
          & \makebox[9em][l]{}JKP+Global vs.\ U.S. & -6.86 & -5.15 & -0.31 & -0.32 \\
          & \makebox[9em][l]{}Synthetic+Global vs.\ U.S. & -31.93 & -4.23 & -0.14 & -0.10 \\
          & \makebox[9em][l]{TimesFM (20M)}Global vs.\ U.S. & -24.61 & -48.32 & -26.45 & -30.54 \\
          & \makebox[9em][l]{}JKP+Global vs.\ U.S. & 6.50 & 5.51 & 11.95 & 13.47 \\
          & \makebox[9em][l]{}Synthetic+Global vs.\ U.S. & 8.51 & 8.45 & 11.68 & 13.18 \\
        \midrule
        \multirow{7}{*}{Overall Acc.}
          & \makebox[9em][l]{Benchmark}Global vs.\ U.S. & -1.19 & -1.14 & -0.78 & -1.10 \\
          & \makebox[9em][l]{Chronos (Small)}Global vs.\ U.S. & -0.48 & 0.02 & 0.15 & 0.15 \\
          & \makebox[9em][l]{}JKP+Global vs.\ U.S. & -1.17 & -0.69 & 0.28 & 0.34 \\
          & \makebox[9em][l]{}Synthetic+Global vs.\ U.S. & -0.80 & -0.50 & 0.24 & 0.38 \\
          & \makebox[9em][l]{TimesFM (20M)}Global vs.\ U.S. & -0.26 & -0.52 & -0.41 & -0.36 \\
          & \makebox[9em][l]{}JKP+Global vs.\ U.S. & -0.54 & -0.15 & -0.02 & 0.04 \\
          & \makebox[9em][l]{}Synthetic+Global vs.\ U.S. & -0.44 & -0.46 & 0.25 & 0.34 \\
        \bottomrule
      \end{tabular}
      \vspace{0.1cm}
      \begin{tablenotes}[flushleft]
        \footnotesize
        \item \textbf{Note:} This table reports representative changes in forecasting performance from the complete scaling comparison. Changes in $\RsqOOS$ and overall accuracy are reported in percentage points. Each entry equals the metric under the stated training-data regime minus the corresponding U.S.-only baseline for the same model and input window, computed from unrounded values. The benchmark is CatBoost. The training-data regimes are: Global (expanding from U.S. to global markets), JKP+Global (Global data augmented with JKP characteristics), and Synthetic+Global (Global data augmented with synthetic series). Positive differences indicate improved performance, whereas negative differences indicate deterioration. Values are reported across input window lengths of 5, 21, 252, and 512 trading days. The full visual comparison is reported in \Cref{appendix:scaling_figures}.
      \end{tablenotes}
    \end{minipage}
    \end{adjustbox}
  \end{threeparttable}

  \vspace{1.20cm}

  \begin{threeparttable}
    \begin{adjustbox}{max width=0.63\textwidth, center}
    \scriptsize
    \captionsetup{width=\linewidth}
    \caption{Representative Portfolio Impact of Data Scaling}
    \label{PP_scaling_representative}
    \begin{minipage}{\linewidth}
      \renewcommand{\arraystretch}{1.15}
      \setlength{\tabcolsep}{4pt}
      \begin{tabular}{llcccc}
        \toprule
        \textbf{Metric} & \makebox[9em][l]{\textbf{Model}}\textbf{Data-scale comparison} & \textbf{5} & \textbf{21} & \textbf{252} & \textbf{512} \\
        \midrule
        \multirow{7}{*}{AR}
          & \makebox[9em][l]{Benchmark}Global vs.\ U.S. & -19.77 & -17.06 & -9.91 & -17.36 \\
          & \makebox[9em][l]{Chronos (Small)}Global vs.\ U.S. & -2.73 & 4.37 & 4.67 & 4.13 \\
          & \makebox[9em][l]{}JKP+Global vs.\ U.S. & -21.80 & -15.27 & 4.21 & 4.45 \\
          & \makebox[9em][l]{}Synthetic+Global vs.\ U.S. & -9.35 & -12.29 & 5.52 & 4.88 \\
          & \makebox[9em][l]{TimesFM (20M)}Global vs.\ U.S. & -2.23 & -8.47 & -5.24 & -4.83 \\
          & \makebox[9em][l]{}JKP+Global vs.\ U.S. & -4.61 & 2.36 & 16.68 & 19.85 \\
          & \makebox[9em][l]{}Synthetic+Global vs.\ U.S. & -5.35 & -5.83 & 18.16 & 20.71 \\
        \midrule
        \multirow{7}{*}{SD}
          & \makebox[9em][l]{Benchmark}Global vs.\ U.S. & -1.09 & -0.56 & -0.70 & -1.01 \\
          & \makebox[9em][l]{Chronos (Small)}Global vs.\ U.S. & 0.04 & -0.76 & -0.03 & -0.01 \\
          & \makebox[9em][l]{}JKP+Global vs.\ U.S. & 1.92 & 0.98 & 0.03 & -0.05 \\
          & \makebox[9em][l]{}Synthetic+Global vs.\ U.S. & 1.36 & -0.47 & -0.55 & -0.58 \\
          & \makebox[9em][l]{TimesFM (20M)}Global vs.\ U.S. & -0.64 & 0.37 & -1.09 & -0.67 \\
          & \makebox[9em][l]{}JKP+Global vs.\ U.S. & -0.33 & -0.07 & -1.66 & -1.78 \\
          & \makebox[9em][l]{}Synthetic+Global vs.\ U.S. & -0.24 & 0.36 & -1.39 & -1.15 \\
        \midrule
        \multirow{7}{*}{SR}
          & \makebox[9em][l]{Benchmark}Global vs.\ U.S. & -2.14 & -2.10 & -0.84 & -1.72 \\
          & \makebox[9em][l]{Chronos (Small)}Global vs.\ U.S. & -0.56 & 1.34 & 0.70 & 0.62 \\
          & \makebox[9em][l]{}JKP+Global vs.\ U.S. & -3.21 & -2.75 & 0.59 & 0.70 \\
          & \makebox[9em][l]{}Synthetic+Global vs.\ U.S. & -1.50 & -2.22 & 1.30 & 1.29 \\
          & \makebox[9em][l]{TimesFM (20M)}Global vs.\ U.S. & -0.36 & -1.01 & -0.19 & -0.31 \\
          & \makebox[9em][l]{}JKP+Global vs.\ U.S. & -0.54 & 0.29 & 3.19 & 4.00 \\
          & \makebox[9em][l]{}Synthetic+Global vs.\ U.S. & -0.60 & -0.71 & 3.14 & 3.46 \\
        \bottomrule
      \end{tabular}
      \vspace{0.1cm}
      \begin{tablenotes}[flushleft]
        \footnotesize
        \item \textbf{Note:} This table reports representative changes in long--short portfolio performance from the complete scaling comparison. Each entry equals the metric under the stated training-data regime minus the corresponding U.S.-only baseline for the same model and input window, computed from unrounded values. The benchmark is CatBoost. The training-data regimes are: Global (expanding from U.S. to global markets), JKP+Global (Global data augmented with JKP characteristics), and Synthetic+Global (Global data augmented with synthetic series). Annualized return (AR) and annualized standard deviation (SD) changes are in percentage points. Positive differences in AR and SR indicate improvement, whereas negative differences indicate deterioration. For SD, negative differences indicate lower volatility, whereas positive differences indicate higher volatility. Stocks are equally weighted within each decile, and the unit-gross long--short return is $\frac{1}{2}(H-L)$, where $H$ and $L$ denote the highest- and lowest-forecast decile returns, respectively. The full visual comparison is reported in \Cref{appendix:scaling_figures}.
      \end{tablenotes}
    \end{minipage}
    \end{adjustbox}
  \end{threeparttable}
\end{table}
\vspace*{\fill}

\begin{figure} %% Fixed -----------------------------------------------
    \caption{Yearly SRs of Long--Short Portfolios}
    \label{yearly_sharpe_long_short}
    \centering
    \includegraphics[
        width=0.75\textwidth,
        clip,
        trim=0 0 0 5, % left bottom right top
    ]{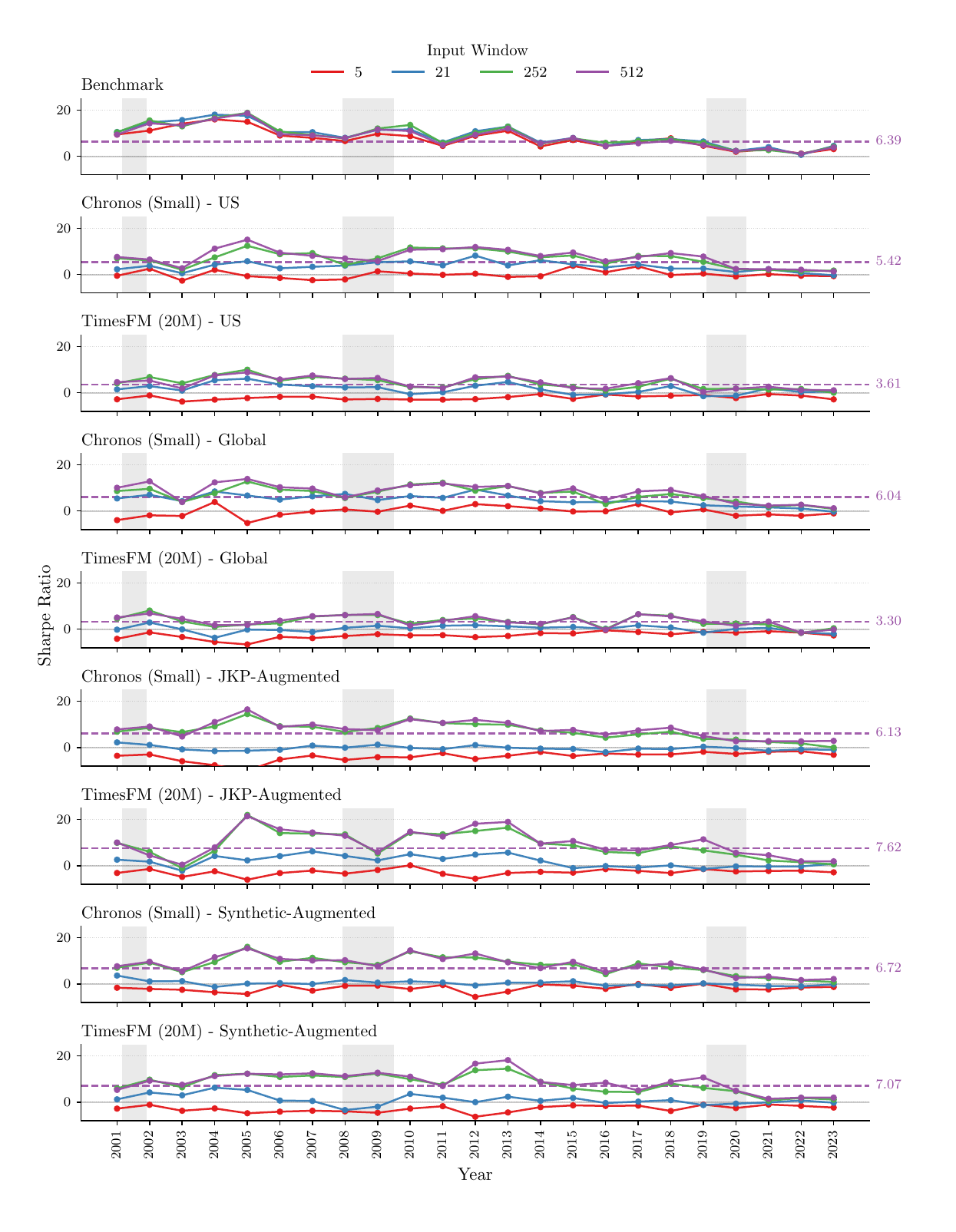} % Adjust the width and trims
    
    \begin{minipage}{0.8\textwidth}
        \footnotesize
        \textbf{Note:} The figure reports annual SRs from 2001 to 2023 for long--short EW portfolios. Results are presented by model across multiple window sizes (5, 21, 252, and 512 trading days). The set of models includes a benchmark, the CatBoost model, that is trained using U.S. excess returns, as well as the largest configurations used in our from-scratch experiments: Chronos and TimesFM (with parameter sizes of small and 20M, respectively), both trained from scratch. We also report results for TSFMs pre-trained on different datasets: U.S.-only excess returns, global excess returns, global excess returns augmented with JKP characteristics (as defined in \citet{jensen2023there}, denoted JKP-Augmented), and global excess returns augmented with synthetic series generated following \citet{ansari2024chronos} and matched in size and panel structure to the JKP augmentation sample (Synthetic-Augmented). The dashed horizontal line in each panel denotes the full-sample SR of the corresponding 512-day specification, with its value reported on the right-hand side of the panel. Shaded regions correspond to U.S. recession periods as identified by the National Bureau of Economic Research (NBER).
    \end{minipage}
\end{figure}

Examining performance over time is important because full-sample portfolio statistics can conceal substantial variation and do not establish whether strong aggregate results extend across many years or are concentrated in a few favorable episodes. To assess this distinction, \Cref{yearly_sharpe_long_short} reports annual SRs of EW long--short portfolios over the out-of-sample period for CatBoost and representative finance-native Chronos and TimesFM models across alternative input windows and pre-training datasets. The figure shows substantial time variation in risk-adjusted performance across all model classes, together with a general weakening toward the later part of the out-of-sample period. The relative ranking nevertheless remains informative: CatBoost is generally stronger at shorter windows, whereas the augmented TSFMs perform more favorably at the 252-day and 512-day windows. CatBoost’s annual SR profiles are closely aligned across all four input windows, while the TSFMs exhibit a pronounced separation between their short- and long-window specifications. In particular, the shortest-window TSFM specifications frequently generate negative annual SRs, and augmentation does not systematically eliminate weak short-window performance. By contrast, the two longer-window profiles generally move closely together, indicating that extending the input history beyond 252 days does not improve performance uniformly in every year. The response to scaling also differs across model families. For Chronos, global pre-training and augmentation largely preserve the temporal pattern already evident under U.S.-only pre-training, while strengthening long-window performance overall. TimesFM exhibits a more pronounced dependence on augmentation: global expansion alone does not deliver a comparable improvement, whereas both JKP and synthetic augmentation substantially strengthen its long-window profiles. This distinction is consistent with the preceding evidence that additional scale and heterogeneity particularly benefit TimesFM.

The yearly profiles also show that the high full-sample SRs of the augmented TSFMs are accompanied by several periods of exceptionally strong performance, particularly during the earlier and middle parts of the sample, rather than uniformly high SRs in every year. At the same time, the gains are not confined to a single isolated year, with positive long-window SRs observed across a broader set of years. Both augmentation regimes exhibit this pattern, consistent with the preceding evidence that the benefits are not driven solely by the financial content of the JKP characteristics. Their relative performance nevertheless varies over time, with neither augmentation method uniformly dominating the other across years. The shaded recession periods also do not coincide with a uniform collapse in performance: many long-window specifications remain positive during both the early-2000s recession and the global financial crisis. Nevertheless, the broad weakening near the end of the sample affects both the benchmark and the TSFMs, with augmented long-window SRs generally falling below their full-sample reference lines and their advantage over shorter windows narrowing. Overall, the temporal evidence reinforces the benefits of finance-native pre-training and data augmentation at longer windows, while showing that these gains are economically meaningful but distinctly time-varying. The broad late-sample deterioration across both CatBoost and the TSFMs further suggests that model choice affects the level of risk-adjusted performance but does not eliminate the decline shared across the evaluated specifications.\footnote{We also report the annual SR results for all models evaluated in \Cref{benchmark_results}, trained on U.S. excess returns. The corresponding outcomes are presented in \Cref{yearly_sharpe_long_short_benchmarks}. The findings indicate that all evaluated models exhibit a broadly similar pattern of performance deterioration toward the end of the sample, although some deviations arise in specific years. The generally superior performance of ensemble models also remains evident, with the performance gap narrowing toward the end of the sample. Moreover, the general upward or downward movements are broadly consistent across models, and the annual profiles are closely aligned across input windows for most benchmarks. While model choice affects the magnitude of the resulting SRs, it does not eliminate the broad late-sample deterioration.} This deterioration, observed across TSFMs and a broad range of conventional model families, echoes the decline in return predictability documented by \citet{mclean2016does} and the real-time weakening of return-based anomalies reported by \citet{marrow2024real}. Whereas these studies examine the
evolution of individual predictors and anomalies, our evidence documents late-sample weakening across forecasting architectures and training regimes.

\begin{table}[tbp]\centering
\begin{threeparttable}
\begin{adjustbox}{max width=0.86\textwidth, center}
\scriptsize
\captionsetup{width=\linewidth}
\caption{Long--Short Portfolio SRs under Position-Based Cost Scenarios}
\label{sharpe_ratio_with_transaction_costs}
\begin{minipage}{\linewidth}
\renewcommand{\arraystretch}{1.2}%
\setlength{\tabcolsep}{3pt}
\begin{tabular}{l|*{16}{c}}
\toprule
\multirow{2}{*}{\textbf{Model}}
& \multicolumn{4}{c}{0 bps}
& \multicolumn{4}{c}{20 bps}
& \multicolumn{4}{c}{40 bps}
& \multicolumn{4}{c}{Mixed} \\
\cmidrule(lr){2-5}
\cmidrule(lr){6-9}
\cmidrule(lr){10-13}
\cmidrule(lr){14-17}
& 5 & 21 & 252 & 512
& 5 & 21 & 252 & 512
& 5 & 21 & 252 & 512
& 5 & 21 & 252 & 512 \\
\midrule

\multicolumn{17}{l}{\textbf{U.S.}} \\
Benchmark
& 5.70 & 6.72 & 6.72 & 6.39
& -1.08 & -0.60 & -0.60 & -0.47
& -7.86 & -7.91 & -7.92 & -7.32
& -0.41 & 0.09 & 0.09 & 0.16 \\
Chronos (Small)
& -0.07 & 2.56 & 4.89 & 5.42
& -10.51 & -6.65 & -2.44 & -2.00
& -20.95 & -15.86 & -9.78 & -9.42
& -9.19 & -5.64 & -1.70 & -1.34 \\
TimesFM (20M)
& -1.82 & 1.23 & 3.47 & 3.61
& -6.88 & -4.70 & -2.32 & -2.43
& -11.95 & -10.63 & -8.10 & -8.48
& -6.30 & -4.02 & -1.68 & -1.76 \\
\midrule

\multicolumn{17}{l}{\textbf{Global}} \\
Chronos (Small)
& -0.63 & 3.89 & 5.59 & 6.04
& -10.99 & -6.80 & -1.77 & -1.39
& -21.35 & -17.49 & -9.14 & -8.83
& -9.58 & -5.61 & -1.01 & -0.69 \\
TimesFM (20M)
& -2.18 & 0.23 & 3.28 & 3.30
& -7.60 & -5.46 & -3.34 & -3.28
& -13.01 & -11.15 & -9.95 & -9.85
& -6.97 & -4.79 & -2.58 & -2.53 \\
\midrule

\multicolumn{17}{l}{\textbf{JKP-Augmented}} \\
Chronos (Small)
& -3.28 & -0.20 & 5.48 & 6.13
& -10.74 & -8.01 & -1.82 & -1.35
& -18.21 & -15.82 & -9.12 & -8.83
& -9.79 & -7.06 & -1.02 & -0.61 \\
TimesFM (20M)
& -2.36 & 1.52 & 6.66 & 7.62
& -7.60 & -4.46 & -0.49 & -0.07
& -12.83 & -10.44 & -7.65 & -7.75
& -7.01 & -3.82 & 0.14 & 0.60 \\
\midrule

\multicolumn{17}{l}{\textbf{Synthetic-Augmented}} \\
Chronos (Small)
& -1.56 & 0.34 & 6.19 & 6.72
& -9.71 & -9.74 & -1.78 & -1.40
& -17.86 & -19.82 & -9.75 & -9.51
& -8.62 & -8.28 & -0.95 & -0.58 \\
TimesFM (20M)
& -2.41 & 0.52 & 6.62 & 7.07
& -7.60 & -5.16 & -0.27 & 0.06
& -12.79 & -10.85 & -7.17 & -6.95
& -7.02 & -4.54 & 0.37 & 0.71 \\
\bottomrule
\end{tabular}

\vspace{0.1cm}
\begin{tablenotes}[para,flushleft]\footnotesize
\textbf{Note:} This table reports annualized SRs of EW long--short portfolios for representative models at input windows of 5, 21, 252, and 512 trading days. Performance is reported after hypothetical daily position-based deductions of 0, 20, or 40 bps, and after size-dependent deductions using cost inputs from \citet{frazzini2012trading}; the mixed costs correspond roughly to 21.27 bps for small-cap stocks and 11.21 bps for large-cap stocks. The benchmark is CatBoost, reported here only under U.S.-only training. The TSFMs are Chronos (Small) and TimesFM (20M), pre-trained from scratch on U.S., global, JKP-augmented global, or synthetic-augmented global data. All portfolios are evaluated on U.S. stocks. Costs are charged daily on the absolute weight of every selected position, including retained positions, with total gross exposure normalized to one. The fixed-cost scenarios therefore deduct 20 or 40 bps from the portfolio return each day. These are position-based cost scenarios, not turnover-based estimates.
\end{tablenotes}
\end{minipage}
\end{adjustbox}
\end{threeparttable}
\end{table}

% \subsubsection{Pre-Trained TSFMs under Transaction Costs} \label{transaction_cost}
As noted earlier, transaction costs may substantially reduce the economic value of these strategies because they trade daily across a broad cross-section and are therefore exposed to potentially substantial trading costs. \Cref{sharpe_ratio_with_transaction_costs} reports SRs for the benchmark and representative finance-native pre-trained TSFMs before and after hypothetical daily position-based cost deductions across the 5-, 21-, 252-, and 512-day input windows, considering fixed-cost and mixed-cost specifications.\footnote{The fixed-cost specifications apply a daily charge of 20 or 40 bps to the absolute portfolio weight of every selected position, including positions retained from the previous day. Charges are aggregated across both legs under the unit-gross-exposure convention, giving a total portfolio-level deduction of 20 or 40 bps per day, respectively. The mixed-cost specification applies the same convention using size-dependent charges. These are position-based cost scenarios rather than turnover-based estimates. The zero-cost specification provides the frictionless benchmark. The mixed-cost specification uses the market-impact cost estimates of \citet{frazzini2012trading}. For transaction-cost assignment, stocks are classified as large-cap or small-cap relative to the daily median market capitalization of NYSE-listed stocks, with costs of 11.21 bps and 21.27 bps applied to large-cap and small-cap positions, respectively. This NYSE-based cutoff is specific to the transaction-cost exercise and differs from the size-group definitions used in
\Cref{FP_US_cap_comparison} and \Cref{PP_US_cap_comparison}.} These hypothetical cost deductions substantially reduce SRs, although some specifications retain positive cost-adjusted performance. Among the representative models in \Cref{sharpe_ratio_with_transaction_costs}, CatBoost achieves the highest net SRs at the 5- and 21-day windows under all three nonzero-cost specifications. At the 252- and 512-day windows, synthetic-augmented TimesFM (20M) achieves the highest net SRs under all three nonzero-cost specifications. Its SRs are $-0.27$ and $0.06$ under 20 bps, $-7.17$ and $-6.95$ under 40 bps, and $0.37$ and $0.71$ under mixed costs, respectively. JKP-augmented TimesFM (20M) also retains positive SRs under mixed costs at these longer windows, at $0.14$ and $0.60$. CatBoost records positive mixed-cost SRs of $0.09$, $0.09$, and $0.16$ at the 21-, 252-, and 512-day windows, respectively. All representative specifications have negative SRs under 40 bps. Thus, the relative advantage of synthetic-augmented TimesFM (20M) at longer windows persists under these hypothetical position-based deductions, but positive cost-adjusted performance remains limited to selected specifications and assumptions. This result complements evidence that economic restrictions attenuate machine-learning profitability \citep{avramov2023machine} and underscores the distinction between gross predictive performance and implementable investment opportunities \citep{jensen2026machine}.\footnote{\Cref{sharpe_ratio_with_transaction_costs_benchmarks} presents the corresponding results for the benchmark models. All benchmark specifications have negative net SRs under the 20 bps and 40 bps scenarios. Under mixed costs, U.S.-trained OLS+H has positive reported SRs at 21 and 252 days, while XGBoost and CatBoost have positive reported SRs at 21, 252, and 512 days. LightGBM has positive reported SRs at 252 and 512 days, with its 21-day value rounded to 0.00. OLS+H has a 5-day value rounded to $-0.00$ and a 512-day value of $-0.01$, so positive net performance cannot be claimed at all four windows. Among U.S.-trained models, OLS+H has the highest net SR at every window under 40 bps. Under both 20 bps and mixed costs, OLS+H ranks highest at 5 days and XGBoost ranks highest at the remaining windows. All globally trained benchmark specifications have negative net SRs under the three nonzero-cost scenarios. Broader training universes therefore do not systematically improve net risk-adjusted performance in this exercise.} These results apply to the daily portfolio rule evaluated here. Allowing trading intensity to depend on signal persistence and transaction costs \citep{garleanu2013dynamic} may alter net performance, but is beyond the scope of our comparison.

% \Cref{international_results_benchmarks} presents the international results for the benchmark models, and \Cref{international_results_tsfm} presents the corresponding results for the TSFMs.

\subsection{International Evidence} \label{intenational_results}
All preceding out-of-sample results focus on the U.S. market. We extend the analysis to seven major markets: Hong Kong (HKG), Taiwan (TWN), South Korea (KOR), Germany (DEU), the United Kingdom (GBR), India (IND), and Australia (AUS), selected based on market size, data reliability, and the availability of short-selling mechanisms, subject to market-specific and time-varying restrictions.\footnote{The reported long--short results
represent hypothetical portfolio performance and do not establish implementability under historical short-selling restrictions.} The international out-of-sample evaluation spans 2001--2023, matching the U.S. evaluation period. The benchmark models are trained on global data and evaluated separately in each market. We report TSFM results under global, JKP-augmented, and synthetic-augmented pre-training.

\begin{landscape}
\thispagestyle{landscape}
\begin{table}[p]\centering
\begin{threeparttable}
\begin{adjustbox}{width=0.90\linewidth, center}
\scriptsize
\captionsetup{width=\linewidth}
\caption{Representative International Results across Input Windows}
\label{international_representative_comparison}
\begin{minipage}{\linewidth}
\renewcommand{\arraystretch}{1.2}%
\setlength{\tabcolsep}{3pt}
\begin{tabular}{l|*{28}{c}}
\toprule
\multirow{2}{*}{\textbf{Market}}
& \multicolumn{4}{c}{\textbf{Benchmark}}
& \multicolumn{4}{c}{\makecell{\textbf{Chronos (Small)}\\\textbf{Global}}}
& \multicolumn{4}{c}{\makecell{\textbf{Chronos (Small)}\\\textbf{JKP}}}
& \multicolumn{4}{c}{\makecell{\textbf{Chronos (Small)}\\\textbf{Synthetic}}}
& \multicolumn{4}{c}{\makecell{\textbf{TimesFM (20M)}\\\textbf{Global}}}
& \multicolumn{4}{c}{\makecell{\textbf{TimesFM (20M)}\\\textbf{JKP}}}
& \multicolumn{4}{c}{\makecell{\textbf{TimesFM (20M)}\\\textbf{Synthetic}}} \\
\cmidrule(lr){2-5}
\cmidrule(lr){6-9}
\cmidrule(lr){10-13}
\cmidrule(lr){14-17}
\cmidrule(lr){18-21}
\cmidrule(lr){22-25}
\cmidrule(lr){26-29}
& 5 & 21 & 252 & 512
& 5 & 21 & 252 & 512
& 5 & 21 & 252 & 512
& 5 & 21 & 252 & 512
& 5 & 21 & 252 & 512
& 5 & 21 & 252 & 512
& 5 & 21 & 252 & 512 \\
\midrule

\multicolumn{29}{l}{\textbf{Panel A: }\boldmath$\RsqOOS$} \\
HKG
& 0.65 & 1.07 & 1.11 & 1.08
& -1.11 & -0.43 & -0.48 & -0.88
& -5.59 & -2.16 & -0.96 & -1.26
& -23.89 & -3.48 & -0.45 & -0.46
& -43.24 & -51.53 & -32.18 & -34.42
& -19.95 & -6.61 & -0.06 & 0.13
& -18.09 & -4.92 & -0.06 & -0.05 \\
TWN
& -0.35 & 0.15 & 0.27 & 0.16
& -1.65 & -1.60 & -4.67 & -4.74
& -5.64 & -3.40 & -4.38 & -4.63
& -23.57 & -5.80 & -2.61 & -2.78
& -42.39 & -50.72 & -34.00 & -38.71
& -20.53 & -10.50 & -1.83 & -1.54
& -19.64 & -7.74 & -2.45 & -2.31 \\
KOR
& -0.70 & -0.60 & -0.30 & -0.24
& -1.40 & -1.39 & -1.67 & -2.30
& -4.69 & -2.84 & -1.83 & -2.62
& -26.36 & -6.08 & -0.83 & -1.07
& -40.20 & -46.77 & -26.77 & -37.65
& -20.73 & -11.87 & -1.58 & -2.22
& -20.12 & -9.25 & -3.09 & -3.39 \\
DEU
& 4.77 & 6.06 & 6.56 & 6.30
& -0.73 & 1.82 & 3.17 & 2.91
& -5.44 & -0.95 & 3.90 & 3.53
& -23.48 & -1.69 & 3.01 & 2.56
& -46.81 & -49.51 & -18.79 & -17.75
& -22.07 & -1.75 & 7.29 & 7.69
& -20.39 & -0.80 & 7.20 & 7.49 \\
GBR
& -0.23 & 0.25 & 0.40 & 0.28
& -1.14 & -0.60 & -0.59 & -0.77
& -6.20 & -3.85 & -0.69 & -0.74
& -29.62 & -3.30 & -0.58 & -0.69
& -43.15 & -49.79 & -25.59 & -30.12
& -21.04 & -8.68 & -1.09 & -0.97
& -19.69 & -6.33 & -1.53 & -1.54 \\
IND
& 3.11 & 4.23 & 4.71 & 4.23
& 0.50 & 3.66 & 4.55 & 4.41
& -3.74 & -0.56 & 4.01 & 3.60
& -19.95 & -3.74 & 3.71 & 3.07
& -41.00 & -48.93 & -29.95 & -27.73
& -18.36 & -5.61 & 4.29 & 4.65
& -16.77 & -4.07 & 3.23 & 3.52 \\
AUS
& 1.98 & 2.47 & 2.66 & 2.54
& -0.97 & 0.32 & 0.75 & 0.72
& -5.05 & -1.45 & 0.79 & 0.95
& -27.64 & -1.69 & 0.45 & 0.80
& -46.18 & -52.62 & -26.34 & -28.02
& -21.57 & -5.12 & 2.16 & 2.49
& -20.10 & -3.64 & 1.93 & 2.07 \\
\midrule

\multicolumn{29}{l}{\textbf{Panel B: SR}} \\
HKG
& 1.75 & 1.86 & 1.93 & 1.92
& 0.22 & 0.81 & 1.41 & 1.37
& -0.13 & 0.27 & 1.09 & 1.16
& -0.42 & 0.26 & 1.12 & 1.33
& -0.78 & 0.12 & 0.80 & 0.82
& -0.96 & 0.21 & 1.42 & 1.58
& -1.09 & 0.27 & 1.28 & 1.72 \\
TWN
& 0.42 & 1.50 & 1.59 & 1.15
& 0.42 & 0.63 & 1.29 & 1.30
& 0.53 & 1.14 & 1.77 & 1.88
& 0.82 & 0.82 & 1.29 & 1.37
& 0.36 & 0.07 & -0.42 & -0.58
& 0.17 & 0.07 & 1.13 & 1.16
& 0.38 & 0.58 & 0.85 & 0.96 \\
KOR
& 0.19 & 0.30 & 0.28 & 0.29
& -0.07 & 0.21 & 0.64 & 0.75
& -0.24 & 0.22 & 0.92 & 0.32
& -0.44 & -0.22 & 0.65 & 0.53
& -0.08 & -0.11 & 0.35 & 0.21
& -0.16 & -0.08 & 0.69 & 0.66
& -0.14 & -0.22 & 0.23 & 0.32 \\
DEU
& 5.59 & 6.08 & 5.92 & 5.90
& 0.29 & 2.19 & 3.42 & 3.40
& -2.19 & 0.58 & 3.20 & 3.27
& -1.81 & 0.70 & 3.10 & 3.20
& -2.70 & 0.69 & 3.11 & 3.48
& -2.63 & 2.35 & 5.56 & 5.77
& -3.48 & 1.45 & 5.65 & 5.89 \\
GBR
& 1.27 & 1.94 & 2.35 & 1.88
& -0.18 & 0.81 & 1.47 & 1.77
& -0.96 & -0.04 & 1.41 & 1.55
& -0.79 & 0.30 & 1.46 & 1.88
& -0.85 & -0.19 & 0.89 & 0.86
& -0.98 & 0.18 & 1.89 & 1.81
& -0.74 & 0.23 & 1.73 & 1.61 \\
IND
& 5.49 & 6.44 & 6.74 & 6.16
& 3.08 & 5.60 & 6.71 & 6.40
& 1.74 & 3.47 & 6.11 & 5.77
& 1.63 & 3.18 & 5.53 & 5.34
& 1.00 & 1.44 & 1.12 & 1.58
& 0.72 & 3.14 & 6.31 & 6.39
& 0.81 & 2.79 & 5.24 & 4.87 \\
AUS
& 5.89 & 6.16 & 6.02 & 6.16
& 0.17 & 2.26 & 2.65 & 2.99
& -2.32 & 0.54 & 2.59 & 2.77
& -2.41 & 0.45 & 2.35 & 2.55
& -3.84 & -0.12 & 2.49 & 2.78
& -4.75 & 1.29 & 5.39 & 5.84
& -4.56 & 0.33 & 5.31 & 5.57 \\
\bottomrule
\end{tabular}

\vspace{0.1cm}
\begin{tablenotes}[para,flushleft]\footnotesize
\textbf{Note:} This table reports market-level representative results at input windows of 5, 21, 252, and 512 trading days. Panel A reports $\RsqOOS$ in percentage terms, and Panel B reports annualized SRs of EW long--short portfolios. The benchmark is CatBoost trained on global data. Global, JKP, and Synthetic indicate global, JKP-augmented global, and synthetic-augmented global TSFM pre-training datasets, respectively.
\end{tablenotes}
\end{minipage}
\end{adjustbox}
\end{threeparttable}
\end{table}
\end{landscape}

Before turning to the international TSFM results, we summarize the benchmark evidence in \Cref{international_benchmark_forecasting_table} and \Cref{international_benchmark_portfolio_table}, which report the complete forecasting and portfolio results. Consistent with the U.S. results, tree-based ensembles generally deliver the strongest performance, while linear models, PCR, and neural networks are less competitive in most markets and windows. The main exceptions occur at the 5-day window: in Germany and Australia, OLS+H and RIDGE+H attain higher SRs than all three ensembles; in South Korea, NN-L does so (0.36 versus at most 0.19); and in Hong Kong, NN-S attains a higher $\RsqOOS$ (0.73\%) than all three ensembles. CatBoost provides a strong common comparator and serves as the representative benchmark in \Cref{international_representative_comparison}, which reports results across the 5-, 21-, 252-, and 512-day input windows. CatBoost delivers the clearest forecasting gains in Germany, India, and Australia, with $\RsqOOS$ values ranging from 4.77\% to 6.56\%, 3.11\% to 4.71\%, and 1.98\% to 2.66\%, respectively, and corresponding SRs ranging from 5.59 to 6.08, 5.49 to 6.74, and 5.89 to 6.16. Hong Kong shows consistently positive forecasting and portfolio performance, with $\RsqOOS$ values of 0.65\%--1.11\% and SRs of 1.75--1.93. The U.K.\ shows more modest forecasting performance, with $\RsqOOS$ ranging from $-0.23\%$ to 0.40\%, although SRs remain positive at 1.27--2.35. Taiwan remains only weakly predictable for CatBoost, and South Korea is its most challenging market, with negative $\RsqOOS$ across all windows and SRs of only 0.19--0.30. These patterns establish the market-level hurdle for the TSFM comparison below.

We next examine international TSFM performance using \Cref{international_representative_comparison}, which compares Chronos (Small) and TimesFM (20M) across all four input windows under global, JKP-augmented, and synthetic-augmented pre-training. \Cref{international_tsfm_forecasting_table} reports forecasting results for all TSFM specifications, while \Cref{international_tsfm_portfolio_table} reports their corresponding portfolio SRs. Performance is generally strongest at longer windows, with augmentation particularly benefiting TimesFM (20M). Under global-only pre-training, Chronos (Small) is competitive in India and achieves positive long-window $\RsqOOS$ in Germany and Australia, whereas TimesFM (20M) records strongly negative $\RsqOOS$ across all markets and windows. Augmentation substantially improves TimesFM's forecasting and portfolio performance, especially in Germany, India, and Australia. At the 512-day window, JKP-augmented TimesFM (20M) achieves $\RsqOOS$ values of 7.69\% in Germany and 4.65\% in India, exceeding CatBoost's 6.30\% and 4.23\%, respectively. Its portfolio SR is close to CatBoost's in Germany and exceeds it in India, while both forecasting and portfolio performance remain close to the benchmark in Australia. Synthetic augmentation delivers broadly similar improvements, although its performance is weaker in India. Augmentation does not uniformly benefit Chronos, whose global-only specification remains strongest in India. Gains are also uneven elsewhere: forecasting fit generally trails CatBoost, but some TSFM specifications achieve higher SRs in Taiwan and South Korea despite negative $\RsqOOS$, reinforcing the distinction between point-forecast accuracy and economically useful rankings. Short-window performance remains considerably weaker, particularly at the 5-day window. Overall, the international evidence supports the U.S. finding that augmented finance-native pre-training can make TSFMs competitive with strong benchmarks at longer windows, while showing that these gains depend on the market, model, and evaluation metric.\footnote{\Cref{dis_evidence_appdedix} complements these aggregate results with a distributional illustration and discusses the limitations of generic TSFM transfer to financial forecasting. \Cref{chronos_excess_return_30_days_forecast} compares predictive intervals at horizons of 1--30 trading days for selected U.S. stocks from zero-shot, fine-tuned, and globally pre-trained finance-native Chronos (Small) models using a 512-day historical context. The examples illustrate that generic pre-training does not ensure alignment with financial return dynamics and that fine-tuning does not necessarily improve this alignment. Finance-native global pre-training appears to bring forecasts closer to the realized return level in these selected examples, although substantial uncertainty remains.}

%%%%% Fixed ---------------------------------------------
\begin{table}[tbp]\centering
\begin{threeparttable}
\begin{adjustbox}{max width=0.80\textwidth, center}
\scriptsize
\captionsetup{width=\linewidth}
\caption{Representative Forecasting and Portfolio Impact of Model Configuration}
\label{model_configuration_representative}
\begin{minipage}{\linewidth}
\renewcommand{\arraystretch}{1.15}
\setlength{\tabcolsep}{4pt}
\begin{tabular}{llcccc}
\toprule
\textbf{Metric}
& \makebox[9em][l]{\textbf{Model}}\textbf{Configuration}
& \textbf{5} & \textbf{21} & \textbf{252} & \textbf{512} \\
\midrule

\multirow{4}{*}{$\RsqOOS$}
& \makebox[9em][l]{Chronos (Small)}Restricted quantization
& -11.26 & -8.32 & -5.91 & -5.63 \\
& \makebox[9em][l]{}Dynamic quantization
& -1.38 & -1.52 & -1.69 & -1.80 \\
& \makebox[9em][l]{TimesFM (20M)}Patch length 8
& -300.02 & -277.18 & -\num{600745.71} & -\num{1589013.38} \\
& \makebox[9em][l]{}Patch length 128
& 36.00 & 58.76 & 39.27 & 44.63 \\
\midrule

\multirow{4}{*}{Overall Acc.}
& \makebox[9em][l]{Chronos (Small)}Restricted quantization
& -0.32 & -0.68 & -0.99 & -0.91 \\
& \makebox[9em][l]{}Dynamic quantization
& -0.17 & -0.55 & -0.71 & -0.70 \\
& \makebox[9em][l]{TimesFM (20M)}Patch length 8
& -0.41 & -0.07 & -0.65 & -0.66 \\
& \makebox[9em][l]{}Patch length 128
& -0.67 & -0.75 & -0.70 & -0.83 \\
\midrule

\multirow{4}{*}{SR}
& \makebox[9em][l]{Chronos (Small)}Restricted quantization
& -1.21 & -1.63 & 1.67 & 1.29 \\
& \makebox[9em][l]{}Dynamic quantization
& -0.93 & -1.38 & 2.05 & 0.92 \\
& \makebox[9em][l]{TimesFM (20M)}Patch length 8
& -0.26 & -0.31 & -3.52 & -3.56 \\
& \makebox[9em][l]{}Patch length 128
& -0.84 & -1.99 & -3.34 & -3.66 \\
\bottomrule
\end{tabular}

\vspace{0.1cm}
\begin{tablenotes}[para,flushleft]\footnotesize
\textbf{Note:} This table reports representative changes in forecasting and portfolio performance relative to each model's default configuration under global pre-training. Columns denote input window lengths in trading days. For Chronos, restricted quantization replaces the default range of $[-15,15]$ with $[-2,2]$, while dynamic quantization recalibrates the bounds annually to the 1\textsuperscript{st} and 99\textsuperscript{th} percentiles of pooled pre-training excess returns normalized by their pooled mean absolute value. For TimesFM, the input patch length changes from the default of 32 to 8 or 128. Changes in $\RsqOOS$ and overall directional accuracy are reported in percentage points; SR changes are in Sharpe units and refer to gross EW long--short portfolios evaluated on U.S. stocks. The complete forecasting and portfolio comparisons are reported in \Cref{impact_TSFM_HP_forecasting} and \Cref{impact_TSFM_HP_performance}, respectively. Values with five or more integer digits are shown in scientific notation.
\end{tablenotes}
\end{minipage}
\end{adjustbox}
\end{threeparttable}
\end{table}

\subsection{The Role of Model Configuration} \label{hyperparameter_impact}
So far, TSFM pre-training has used the default training hyperparameters for the model configurations considered. In contrast, as discussed in \Cref{sec: numerical_results}, benchmark specifications with tunable hyperparameters underwent selection during the initial estimation year (2000), with the selected configurations reused in subsequent years; specifications without hyperparameter tuning, including the neural networks, used fixed settings. Although hyperparameter tuning is standard practice in financial ML studies, our baseline TSFM experiments retain the developers' default training settings, potentially limiting their performance in this financial forecasting task. This raises the question of how sensitive TSFMs are to settings originally developed for general-purpose forecasting on large, heterogeneous datasets. We therefore conduct a controlled experiment using global pre-training data, varying selected hyperparameters while holding the remaining settings fixed. For Chronos, described in \Cref{subsec: chronos}, we compare the default tokenizer quantization range of $[-15,15]$ with a restricted range of $[-2,2]$ and a dynamic strategy whose bounds are recalibrated annually to the 1\textsuperscript{st} and 99\textsuperscript{th} percentiles of pooled pre-training excess returns normalized by their pooled mean absolute value. For TimesFM, described in \Cref{subsec: TimesFM}, we change the default input patch length from 32 to either 8 or 128. \Cref{model_configuration_representative} summarizes changes in forecasting and portfolio performance for representative models, while \Cref{impact_TSFM_HP_forecasting} and \Cref{impact_TSFM_HP_performance} report the complete comparisons. Because all TSFM configurations in this experiment are pre-trained on global data, the benchmark shown in these two figures is CatBoost trained on global excess returns, rather than the U.S.-trained CatBoost used as the representative benchmark in the other U.S. comparisons. \Cref{model_configuration_representative} reports changes relative to each model's default global pre-training configuration across the four input windows. Positive values indicate improvements and negative values indicate deterioration; changes in $\RsqOOS$ and directional accuracy are in percentage points, while SR changes are in Sharpe units.\footnote{We focus on the quantization range for Chronos and the input patch length for TimesFM because these hyperparameters govern key aspects of the models' representations. The exercise is illustrative rather than exhaustive: tuning other hyperparameters or jointly optimizing model configurations may yield further gains, but such an analysis lies beyond the scope of this study.}

The results in \Cref{model_configuration_representative} highlight substantial sensitivity to model configuration and a distinction between forecasting accuracy and portfolio performance. For Chronos (Small), restricted and dynamic quantization worsen $\RsqOOS$ and overall directional accuracy across all four input windows. Nevertheless, both adjustments improve SRs at the 252- and 512-day windows: restricted quantization increases SRs by 1.67 and 1.29, respectively, while dynamic quantization produces improvements of 2.05 and 0.92. Both adjustments reduce SRs at shorter windows. For TimesFM (20M), reducing the input patch length from 32 to 8 worsens $\RsqOOS$ across all windows, with particularly severe deterioration at the 252- and 512-day windows. Increasing the patch length to 128 improves $\RsqOOS$ across all windows. However, both adjustments reduce directional accuracy and portfolio SRs. At the 512-day window, extending the patch length improves $\RsqOOS$ by 44.63 percentage points but reduces the SR by 3.66. Thus, configuration changes can improve point-forecast fit without improving portfolio performance, or improve risk-adjusted portfolio performance despite weaker forecasting metrics. The benefits of adjustment therefore depend on the model, input window, and evaluation metric.
\section{Conclusion}\label{sec:conclusion}

This study offers a large-scale empirical assessment of TSFMs for excess return forecasting across U.S. and global markets. Off-the-shelf, generic pre-trained TSFMs generally underperform strong benchmarks, especially tree-based ensembles, although selected models produce competitive gross portfolio signals despite poor point-forecast fit. Fine-tuning yields mixed results relative to zero-shot settings and does not close the portfolio performance gap with benchmark models. Finance-native pre-training substantially improves gross long--short portfolio performance relative to the principal off-the-shelf Chronos and TimesFM specifications, especially at longer input windows. Improvements in forecasting performance depend on the metric and model: directional accuracy can improve even when $\RsqOOS$ remains below the benchmark. Conventional models retain a substantial gross portfolio advantage at shorter windows. Improvements in goodness-of-fit vary across models, training regimes, and markets, and do not consistently translate into stronger directional accuracy or portfolio performance.

%\notewg{WG: @Eghbal, when you say point estimate here, what is the other option? I think we are everywhere using the mean prediction as the next step, and never investigate the distrbution of prediction? Hence, when you say weaker, which metric are you looking at?}

Selected smaller TSFMs pre-trained from scratch on financial data outperform the principal off-the-shelf Chronos and TimesFM specifications on selected forecasting metrics and long-window gross portfolio performance. While global pooling weakens the gross portfolio performance of tree-based benchmarks in our U.S. evaluation, the benefits of broader and more diverse pre-training sets for TSFMs depend on the model and input window. Augmentation with monthly characteristics or synthetic time series improves selected long-window specifications. Across international markets, tree-based ensembles remain the most reliable baseline; however, scaled, finance-specialized TSFMs become increasingly competitive. Their performance, though, is sensitive to pre-training composition, window length, and hyperparameter choices. Performance varies substantially over time, with broad late-sample deterioration affecting both the benchmarks and the TSFMs. Augmentation improves some long-window specifications but does not eliminate this deterioration. The gains from finance-native pre-training also come with substantial computational costs, particularly for from-scratch pre-training, underscoring the need for more efficient, finance-oriented architectures. Overall, TSFMs represent a promising paradigm for financial forecasting when domain alignment and computational feasibility are jointly managed. Some off-the-shelf models also produce comparatively strong gross ranking signals: for example, Toto attains an SR of 6.10 at the 512-day window despite strongly negative $\RsqOOS$. Such cases reinforce the need to distinguish point-forecast accuracy from cross-sectional ranking performance. Taken together, the findings support evaluating finance-native pre-training alongside strong conventional benchmarks, rather than assuming that either generic transfer or domain-specific pre-training guarantees reliable forecasts or implementable net returns.

The broad transfer capabilities demonstrated by LLMs have motivated the expectation that large-scale pre-training can yield similarly generalizable representations for time series. Our findings suggest that this expectation requires qualification in financial forecasting: generalization across heterogeneous domains does not automatically imply effective transfer to financial applications. The usefulness of the TSFMs examined here depends on domain alignment, while establishing their value requires rigorous empirical evaluation rather than reliance on model scale, leaderboard rankings, or performance on general forecasting benchmarks. This distinction matters for both industry adoption and the growing academic literature applying AI to finance. Treating model sophistication or success in other domains as evidence of financial usefulness risks overstating predictive ability and investment value. Without chronological evaluation, strong benchmark comparisons, and economic assessment, apparent gains may support misleading research conclusions and unreliable investment decisions. Rigorous financial validation is therefore essential before claims of AI generalization are translated into practical use, where unwarranted confidence can result in substantial financial losses.

Building on these findings, several avenues for future research emerge. Probabilistic TSFMs generate predictive distributions, enabling analysis of calibration, tail risk, and higher-order features. Extending the analysis to longer forecasting horizons would further illuminate how predictive signals evolve and decay over time. Furthermore, a natural progression is to test multivariate extensions of TSFMs that capture cross-asset dependencies, market factors, and macro-financial linkages. Beyond returns, applying TSFMs to other forms of financial data such as volatility, order flow, or limit order book information could enhance their applicability to risk management and market microstructure analysis. Future research should conduct a more comprehensive analysis of how hyperparameter configurations and architectural design choices influence model performance, with the objective of developing finance-oriented TSFMs specifically optimized for distinct predictive and decision-making applications. Finally, while this study focuses on comparing forecasting methods and their cross-sectional ranking signals, future research could optimize portfolio construction and rebalancing to assess whether predictive gains can be retained while reducing turnover and
implementation costs.

\clearpage

\clearpage
\clearpage
\putbib[bib_list]

\clearpage
\singlespacing

% Main-paper appendix
\appendix
\doublespacing

% ------------------------------------------------
% Appendix A: no subsections
% ------------------------------------------------
\counterwithin{figure}{section}
\counterwithin{table}{section}
\counterwithin{equation}{section}

\section{Data Overview}

\vspace*{\fill}
\begin{figure}[H]
    \centering
    \caption{Cumulative Number of Observations and Unique Securities (U.S.)}
     \label{Cumulative_obs_sec_us}
    % First figure
    \includegraphics[
        width=0.6\textwidth,
        trim=0 0 0 10, % left bottom right top
        clip
    ]{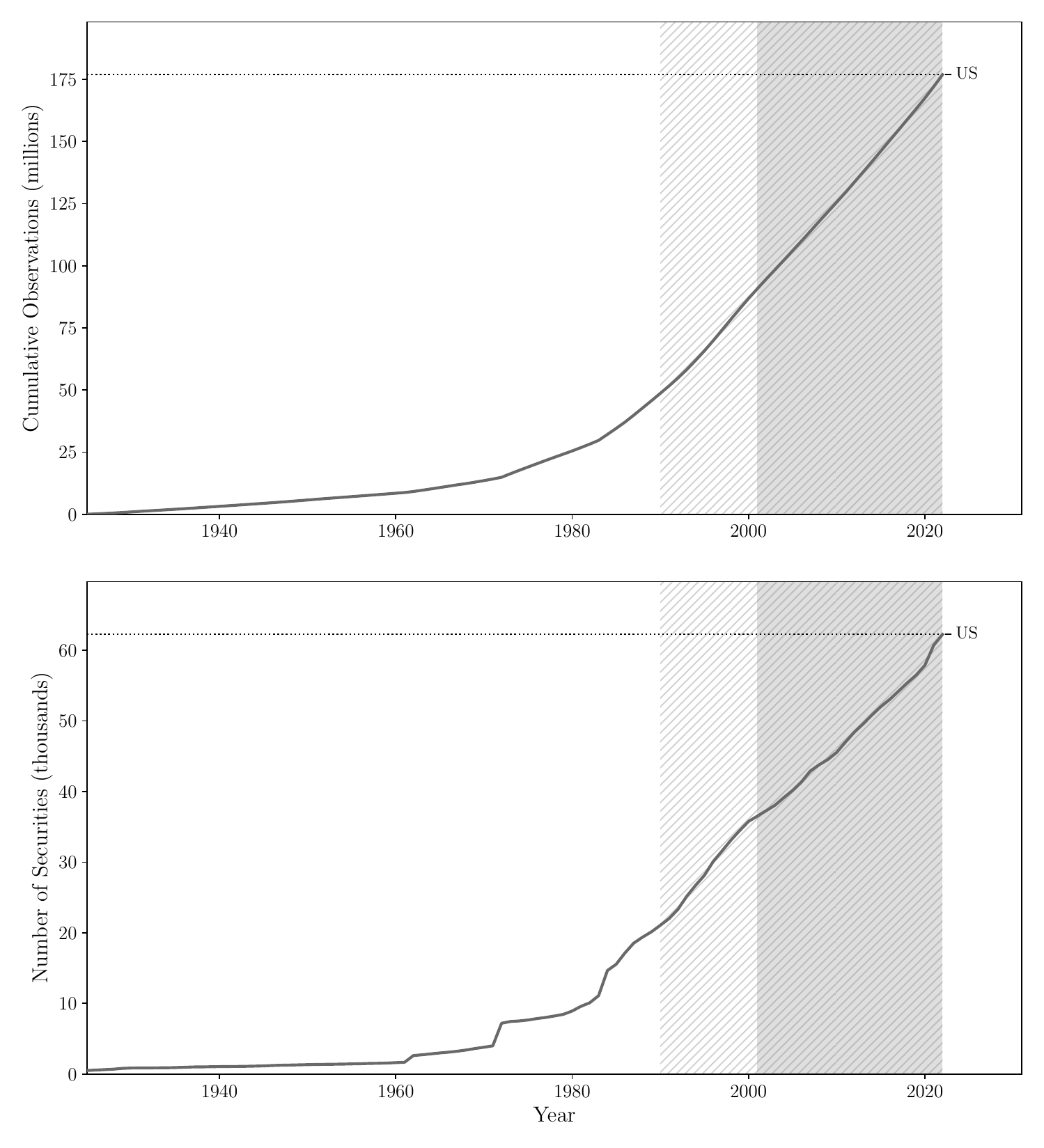}
    % Note under both figures
    \begin{minipage}{0.80\textwidth}
        \footnotesize
        \textbf{Note:} The top panel displays the cumulative number of U.S. excess return observations (in millions) from 1925 to 2022. The bottom panel displays the cumulative number of unique U.S. securities (in thousands) over the same period. In each panel, the horizontal dotted line marks the end-of-2022 value. The light-shaded region begins in 1990, the start of the expanding-window training sample, while the darker region covers 2001--2022, excluding the initial training endpoint of 2000. Each annual training sample is used for the following year's evaluation. The panels show historical data availability from 1925 onward; only observations from 1990 onward enter the study.
    \end{minipage}
\end{figure}
\vspace*{\fill}

\clearpage

\vspace*{\fill}
\begin{figure}[H]
    \centering
    \caption{Cumulative Number of Observations and Unique Securities by Market}
    \label{Cumulative_obs_sec_global}
    % First figure
    \includegraphics[
        width=0.6\textwidth,
        trim=0 0 0 10, % left bottom right top
        clip
    ]{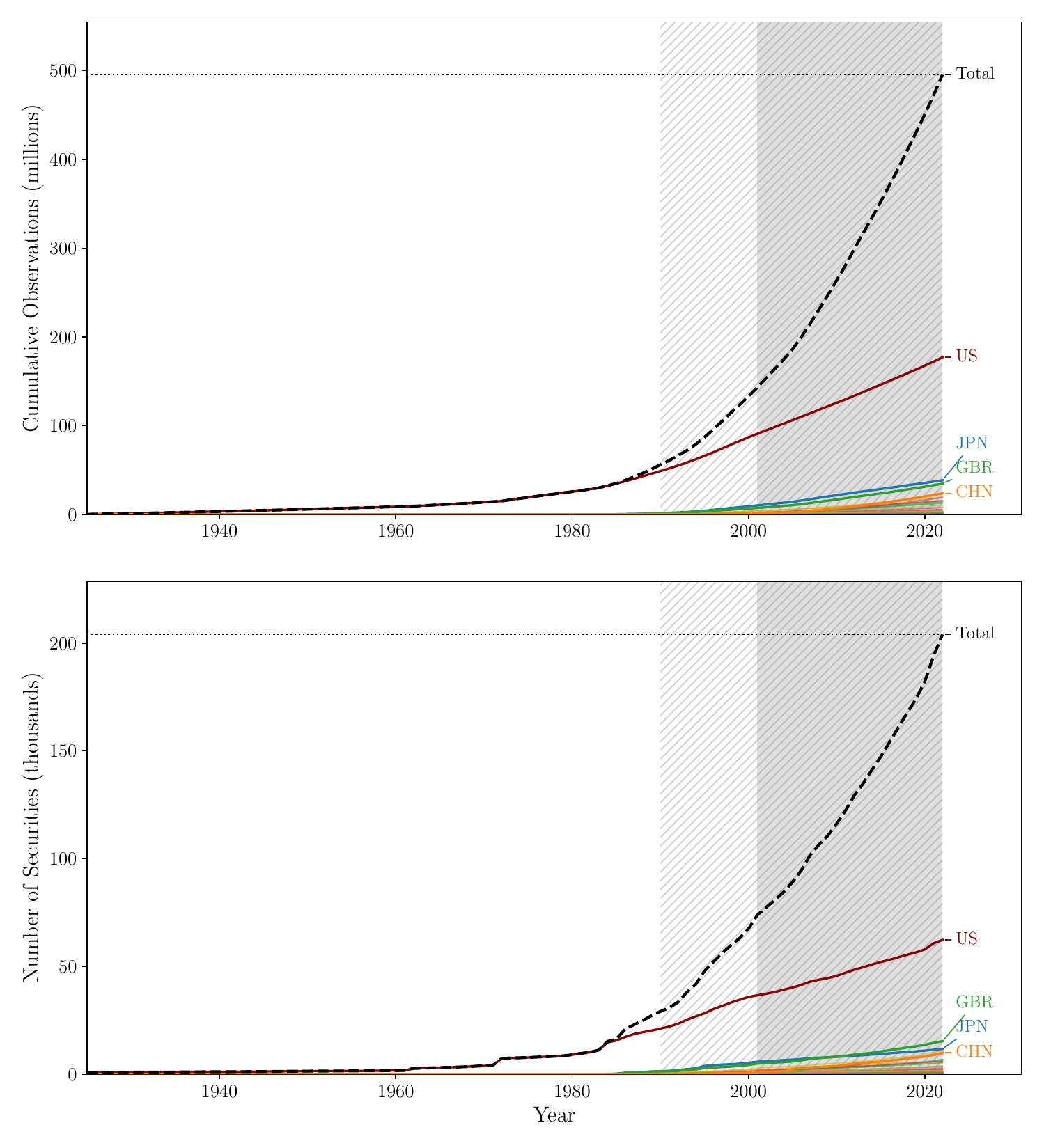}

    % Note under both figures
    \begin{minipage}{0.80\textwidth}
        \footnotesize
        \textbf{Note:} The top panel displays the cumulative number of excess return observations (in millions) across markets from 1925 to 2022. The bottom panel displays the cumulative number of unique securities (in thousands) over the same period. The United States, Japan, the United Kingdom, and China are highlighted in both panels. In each panel, the dashed curve represents the aggregate total across markets, and the horizontal dotted line marks its end-of-2022 value. The light-shaded region begins in 1990, the start of the expanding-window training sample, while the darker region covers 2001--2022, excluding the initial training endpoint of 2000. Each annual training sample is used for the following year's evaluation. The panels show historical data availability from 1925 onward; only observations from 1990 onward enter the study. Uruguay is excluded, while multiple-market listings classified as MUL are included.
    \end{minipage}
\end{figure}
\vspace*{\fill}

\clearpage

\vspace*{\fill}
\begin{figure}[H]
    \centering
    \caption{Cumulative Number of Observations and Unique Securities: Excess Returns vs. JKP}
    \label{Cumulative_obs_sec_JPK_compare}

    \includegraphics[
        width=0.6\textwidth,
        trim=0 0 0 10,
        clip
    ]{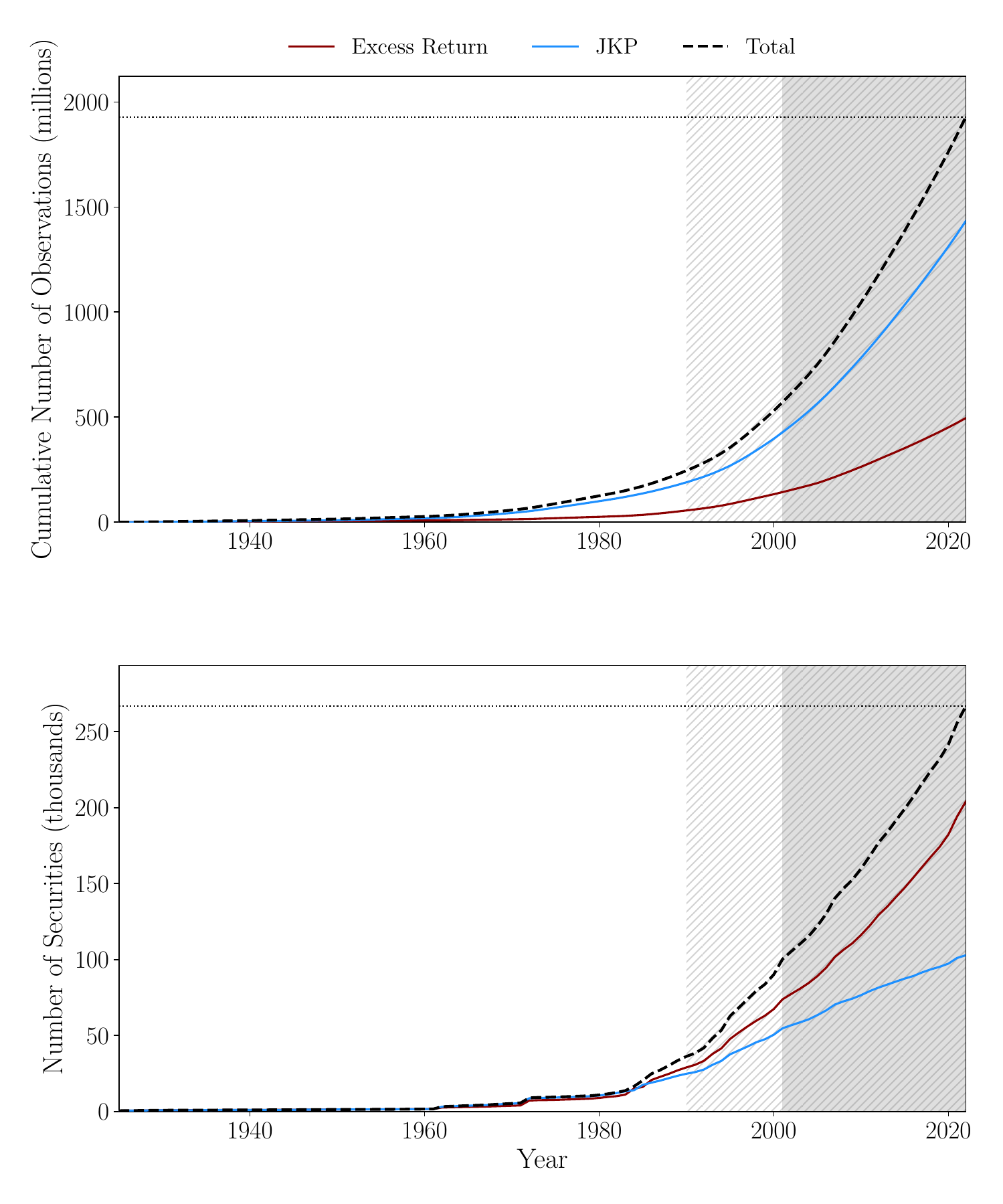}

    \begin{minipage}{0.80\textwidth}
        \footnotesize
        \textbf{Note:} The top panel displays the cumulative number of excess return observations, JKP characteristic observations from \citet{jensen2023there}, and their combined total (in millions) across markets from 1925 to 2022. The bottom panel displays the cumulative number of unique securities (in thousands), with the combined count deduplicated across the return and JKP datasets. In each panel, the dashed curve represents the combined total, while the horizontal dotted line marks its end-of-2022 value. The light-shaded region begins in 1990, the start of the expanding-window training sample, while the darker region covers 2001--2022, excluding the initial training endpoint of 2000. Each annual training sample is used for the following year's evaluation. The panels show historical data availability from 1925 onward; only observations from 1990 onward enter the study. Uruguay is excluded, while multiple-market listings classified as MUL are included.
    \end{minipage}
\end{figure}
\vspace*{\fill}

\clearpage

\newcolumntype{G}{c c c c}

\section{Extended Results} \label{Appendix_Extended_Results}
This appendix reports the detailed empirical results that support the condensed evidence in \Cref{sec: numerical_results}. The material follows the order of the main text. We first provide the complete benchmark, zero-shot TSFM, fine-tuned TSFM, and finance-native pre-trained TSFM results. We then report detailed scaling figures, modified Diebold--Mariano tests, temporal stability, and position-based cost comparisons. Finally, we report the complete international benchmark and TSFM results, a distributional illustration, and detailed model-configuration figures. Expanded neural-network benchmarks and fine-tuning-scale comparisons are reported in \Cref{ia:larger_neural_networks} and \Cref{ia:scaling_fine_tuning} of the Internet Appendix. Benchmark descriptions and the broader TSFM model summary are also provided in the \hyperref[pp:internet-title]{Internet Appendix}.

\subsection{Complete Benchmark Results} \label{complete_benchmark_results}
This subsection presents complete benchmark results for both forecasting performance and portfolio performance across all models and window sizes.

\vspace*{\fill}
\begin{table}[h]
  \centering
  \begin{threeparttable}
    \begin{adjustbox}{width=1.05\textwidth, center}
    \scriptsize
    \captionsetup{width=\linewidth}
    \caption{Benchmark Models - Forecasting Performance (Complete)}
    \label{FP_benchmark_full}
    \begin{minipage}{\linewidth}
        \renewcommand{\arraystretch}{1.3}
      \begin{tabular}{c|cccc|cccc|cccc|cccc|cccc}
        \toprule
        \textbf{Model} & \multicolumn{4}{c|}{\textbf{OLS+H}} & \multicolumn{4}{c|}{\textbf{LASSO+H}} & \multicolumn{4}{c|}{\textbf{RIDGE+H}} & \multicolumn{4}{c|}{\textbf{Enet+H}} & \multicolumn{4}{c}{\textbf{PCR}} \\
        \textbf{Window Size} & 5 & 21 & 252 & 512 & 5 & 21 & 252 & 512 & 5 & 21 & 252 & 512 & 5 & 21 & 252 & 512 & 5 & 21 & 252 & 512 \\
        \midrule
        $\RsqOOS$ & \makecell{-0.41 \\ -0.88 \\ 0.26} & \makecell{-0.42 \\ -0.88 \\ 0.26} & \makecell{-0.48 \\ -0.96 \\ 0.22} & \makecell{-0.58 \\ -1.07 \\ 0.15} &
        \makecell{-2.84 \\ -3.75 \\ -1.50} & \makecell{-1.40 \\ -2.12 \\ -0.38} & \makecell{-1.63 \\ -2.41 \\ -0.54} & \makecell{-1.79 \\ -2.59 \\ -0.68} &
        \makecell{-1.30 \\ -1.97 \\ -0.29} & \makecell{-1.34 \\ -2.01 \\ -0.33} & \makecell{-1.63 \\ -2.38 \\ -0.55} & \makecell{-1.77 \\ -2.51 \\ -0.68} &
        \makecell{-1.35 \\ -2.04 \\ -0.32} & \makecell{-1.35 \\ -2.04 \\ -0.31} & \makecell{-1.30 \\ -2.01 \\ -0.29} & \makecell{-1.08 \\ -1.75 \\ -0.15} &
        \makecell{-1.31 \\ -1.85 \\ -0.61} & \makecell{-1.22 \\ -1.73 \\ -0.57} & \makecell{-1.47 \\ -2.02 \\ -0.77} & \makecell{-2.66 \\ -3.56 \\ -1.50} \\
        \addlinespace[3pt]
        Overall Acc. & \makecell{51.92 \\ 50.59 \\ 54.66} & \makecell{51.93 \\ 50.62 \\ 54.64} & \makecell{51.86 \\ 50.67 \\ 54.40} & \makecell{51.82 \\ 50.69 \\ 54.27} &
        \makecell{51.34 \\ 51.09 \\ 52.38} & \makecell{51.24 \\ 51.18 \\ 51.89} & \makecell{51.15 \\ 51.15 \\ 51.71} & \makecell{51.18 \\ 51.20 \\ 51.75} &
        \makecell{51.24 \\ 51.20 \\ 51.93} & \makecell{51.27 \\ 51.14 \\ 52.11} & \makecell{51.27 \\ 51.28 \\ 51.97} & \makecell{51.32 \\ 51.21 \\ 52.16} &
        \makecell{51.22 \\ 51.18 \\ 51.94} & \makecell{51.28 \\ 51.21 \\ 52.00} & \makecell{51.15 \\ 51.20 \\ 51.65} & \makecell{51.12 \\ 51.18 \\ 51.57} &
        \makecell{50.99 \\ 51.00 \\ 51.37} & \makecell{50.92 \\ 50.96 \\ 51.31} & \makecell{50.90 \\ 50.97 \\ 51.29} & \makecell{51.04 \\ 50.77 \\ 52.00} \\

        \addlinespace[3pt]
        Up Acc. & \makecell{45.56 \\ 42.63 \\ 49.23} & \makecell{45.92 \\ 43.01 \\ 49.68} & \makecell{47.50 \\ 44.79 \\ 51.22} & \makecell{47.77 \\ 45.21 \\ 51.38} &
        \makecell{63.70 \\ 63.90 \\ 66.90} & \makecell{67.15 \\ 68.11 \\ 69.53} & \makecell{68.03 \\ 69.20 \\ 70.32} & \makecell{67.94 \\ 69.09 \\ 70.26} &
        \makecell{67.34 \\ 68.29 \\ 70.04} & \makecell{66.17 \\ 66.89 \\ 68.96} & \makecell{67.26 \\ 68.42 \\ 69.36} & \makecell{65.96 \\ 66.79 \\ 68.19} &
        \makecell{66.87 \\ 67.80 \\ 69.47} & \makecell{67.31 \\ 68.26 \\ 69.88} & \makecell{69.12 \\ 70.46 \\ 71.37} & \makecell{69.33 \\ 70.63 \\ 71.43} &
        \makecell{66.11 \\ 68.00 \\ 66.83} & \makecell{65.95 \\ 68.13 \\ 66.34} & \makecell{66.00 \\ 68.32 \\ 66.21} & \makecell{62.33 \\ 63.48 \\ 63.29} \\

        \addlinespace[3pt]
        Down Acc. & \makecell{58.19 \\ 59.01 \\ 59.48} & \makecell{57.87 \\ 58.70 \\ 59.05} & \makecell{56.19 \\ 56.92 \\ 57.23} & \makecell{55.81 \\ 56.49 \\ 56.81} &
        \makecell{39.38 \\ 37.80 \\ 39.63} & \makecell{35.61 \\ 33.40 \\ 36.16} & \makecell{34.74 \\ 32.41 \\ 35.26} & \makecell{34.82 \\ 32.49 \\ 35.36} &
        \makecell{35.56 \\ 33.36 \\ 35.94} & \makecell{36.78 \\ 34.77 \\ 37.20} & \makecell{35.66 \\ 33.29 \\ 36.60} & \makecell{37.04 \\ 34.92 \\ 37.97} &
        \makecell{36.02 \\ 33.86 \\ 36.50} & \makecell{35.53 \\ 33.31 \\ 36.04} & \makecell{33.68 \\ 31.19 \\ 34.22} & \makecell{33.32 \\ 30.85 \\ 33.96} &
        \makecell{36.22 \\ 33.19 \\ 37.67} & \makecell{36.20 \\ 32.96 \\ 37.95} & \makecell{36.15 \\ 32.82 \\ 38.06} & \makecell{39.99 \\ 37.45 \\ 41.97} \\

        \addlinespace[3pt]
        F1 & \makecell{0.52 \\ 0.50 \\ 0.54} & \makecell{0.52 \\ 0.50 \\ 0.54} & \makecell{0.52 \\ 0.51 \\ 0.54} & \makecell{0.52 \\ 0.51 \\ 0.54} &
        \makecell{0.51 \\ 0.50 \\ 0.52} & \makecell{0.50 \\ 0.49 \\ 0.51} & \makecell{0.50 \\ 0.49 \\ 0.51} & \makecell{0.50 \\ 0.49 \\ 0.51} &
        \makecell{0.50 \\ 0.49 \\ 0.51} & \makecell{0.50 \\ 0.50 \\ 0.51} & \makecell{0.50 \\ 0.49 \\ 0.51} & \makecell{0.50 \\ 0.50 \\ 0.51} &
        \makecell{0.50 \\ 0.49 \\ 0.51} & \makecell{0.50 \\ 0.49 \\ 0.51} & \makecell{0.50 \\ 0.49 \\ 0.50} & \makecell{0.49 \\ 0.49 \\ 0.50} &
        \makecell{0.50 \\ 0.49 \\ 0.51} & \makecell{0.50 \\ 0.49 \\ 0.51} & \makecell{0.50 \\ 0.49 \\ 0.51} & \makecell{0.50 \\ 0.50 \\ 0.52} \\
        \addlinespace[3pt]

        \midrule \midrule
        \textbf{Model} & \multicolumn{4}{c|}{\textbf{XGBoost}} & \multicolumn{4}{c|}{\textbf{CatBoost}} & \multicolumn{4}{c|}{\textbf{LightGBM}} & \multicolumn{4}{c|}{\textbf{NN-S}} & \multicolumn{4}{c}{\textbf{NN-L}} \\
        \textbf{Window Size} & 5 & 21 & 252 & 512 & 5 & 21 & 252 & 512 & 5 & 21 & 252 & 512 & 5 & 21 & 252 & 512 & 5 & 21 & 252 & 512 \\
        \midrule
        $\RsqOOS$ & \makecell{-0.22 \\ -0.45 \\ 0.38} & \makecell{0.01 \\ -0.28 \\ 0.62} & \makecell{-0.09 \\ -0.34 \\ 0.50} & \makecell{-0.12 \\ -0.32 \\ 0.46} &
        \makecell{-0.25 \\ -0.50 \\ 0.39} & \makecell{-0.04 \\ -0.38 \\ 0.66} & \makecell{-0.03 \\ -0.32 \\ 0.62} & \makecell{-0.03 \\ -0.28 \\ 0.59} &
        \makecell{-0.22 \\ -0.43 \\ 0.37} & \makecell{-0.01 \\ -0.27 \\ 0.59} & \makecell{-0.08 \\ -0.31 \\ 0.50} & \makecell{-0.13 \\ -0.34 \\ 0.46} &
        \makecell{-3.04 \\ -4.48 \\ -1.86} & \makecell{-3.27 \\ -5.12 \\ -1.96} & \makecell{-1.87 \\ -2.54 \\ -0.90} & \makecell{-2.31 \\ -2.79 \\ -1.32} &
        \makecell{-3.28 \\ -5.13 \\ -1.95} & \makecell{-2.26 \\ -3.05 \\ -1.31} & \makecell{-2.24 \\ -2.82 \\ -1.28} & \makecell{-2.54 \\ -3.39 \\ -1.53} \\
        \addlinespace[3pt]

        Overall Acc. & \makecell{51.23 \\ 50.90 \\ 52.52} & \makecell{51.23 \\ 51.01 \\ 52.44} & \makecell{51.27 \\ 51.15 \\ 52.30} & \makecell{51.34 \\ 51.22 \\ 52.46} &
        \makecell{50.94 \\ 50.70 \\ 51.96} & \makecell{51.12 \\ 50.74 \\ 52.47} & \makecell{51.08 \\ 50.80 \\ 52.33} & \makecell{51.16 \\ 50.95 \\ 52.39} &
        \makecell{51.17 \\ 50.95 \\ 52.34} & \makecell{51.18 \\ 51.01 \\ 52.29} & \makecell{51.24 \\ 51.13 \\ 52.34} & \makecell{51.38 \\ 51.20 \\ 52.58} &
        \makecell{50.94 \\ 50.52 \\ 51.94} & \makecell{50.97 \\ 50.54 \\ 52.06} & \makecell{50.96 \\ 50.81 \\ 51.74} & \makecell{51.32 \\ 50.98 \\ 52.33} &
        \makecell{51.25 \\ 50.48 \\ 52.79} & \makecell{51.28 \\ 51.04 \\ 52.24} & \makecell{51.04 \\ 50.35 \\ 52.34} & \makecell{50.85 \\ 50.30 \\ 52.04} \\
        \addlinespace[3pt]

        Up Acc. & \makecell{68.63 \\ 71.26 \\ 67.39} & \makecell{69.44 \\ 73.96 \\ 66.84} & \makecell{70.53 \\ 76.27 \\ 67.79} & \makecell{69.42 \\ 74.88 \\ 66.86} &
        \makecell{68.96 \\ 72.05 \\ 68.00} & \makecell{67.82 \\ 72.97 \\ 65.11} & \makecell{66.87 \\ 71.06 \\ 64.65} & \makecell{69.14 \\ 74.39 \\ 66.07} &
        \makecell{69.61 \\ 72.42 \\ 68.75} & \makecell{70.05 \\ 74.80 \\ 67.93} & \makecell{70.05 \\ 75.86 \\ 67.49} & \makecell{68.14 \\ 73.46 \\ 66.00} &
        \makecell{48.28 \\ 48.02 \\ 49.12} & \makecell{46.43 \\ 45.96 \\ 47.59} & \makecell{62.42 \\ 62.93 \\ 63.69} & \makecell{56.89 \\ 56.89 \\ 58.32} &
        \makecell{46.58 \\ 46.08 \\ 47.64} & \makecell{59.45 \\ 59.19 \\ 60.17} & \makecell{53.18 \\ 51.98 \\ 55.50} & \makecell{49.67 \\ 48.62 \\ 51.40} \\
        \addlinespace[3pt]

        Down Acc. & \makecell{33.97 \\ 29.26 \\ 38.91} & \makecell{33.15 \\ 26.64 \\ 39.27} & \makecell{32.25 \\ 24.64 \\ 38.27} & \makecell{33.36 \\ 26.09 \\ 39.21} &
        \makecell{33.14 \\ 28.11 \\ 37.42} & \makecell{34.60 \\ 27.25 \\ 41.01} & \makecell{35.59 \\ 29.59 \\ 41.24} & \makecell{33.37 \\ 26.24 \\ 39.86} &
        \makecell{32.95 \\ 28.21 \\ 37.40} & \makecell{32.53 \\ 25.86 \\ 38.10} & \makecell{32.73 \\ 25.10 \\ 38.66} & \makecell{34.73 \\ 27.60 \\ 40.27} &
        \makecell{53.19 \\ 52.60 \\ 54.21} & \makecell{55.28 \\ 54.98 \\ 55.94} & \makecell{39.64 \\ 37.84 \\ 41.04} & \makecell{45.35 \\ 44.22 \\ 46.46} &
        \makecell{55.32 \\ 54.81 \\ 56.49} & \makecell{42.51 \\ 41.45 \\ 44.37} & \makecell{49.16 \\ 49.15 \\ 49.53} & \makecell{52.30 \\ 52.32 \\ 52.93} \\
        \addlinespace[3pt]

        F1 & \makecell{0.49 \\ 0.47 \\ 0.51} & \makecell{0.49 \\ 0.46 \\ 0.51} & \makecell{0.49 \\ 0.46 \\ 0.51} & \makecell{0.49 \\ 0.46 \\ 0.51} &
        \makecell{0.49 \\ 0.47 \\ 0.51} & \makecell{0.49 \\ 0.47 \\ 0.52} & \makecell{0.50 \\ 0.48 \\ 0.52} & \makecell{0.49 \\ 0.47 \\ 0.51} &
        \makecell{0.49 \\ 0.47 \\ 0.51} & \makecell{0.49 \\ 0.47 \\ 0.51} & \makecell{0.49 \\ 0.47 \\ 0.51} & \makecell{0.49 \\ 0.47 \\ 0.51} &
        \makecell{0.43 \\ 0.41 \\ 0.44} & \makecell{0.43 \\ 0.41 \\ 0.45} & \makecell{0.47 \\ 0.46 \\ 0.49} & \makecell{0.48 \\ 0.46 \\ 0.50} &
        \makecell{0.44 \\ 0.42 \\ 0.46} & \makecell{0.45 \\ 0.44 \\ 0.47} & \makecell{0.48 \\ 0.46 \\ 0.50} & \makecell{0.47 \\ 0.45 \\ 0.49} \\
        \addlinespace[3pt]

        \bottomrule
      \end{tabular}
      \vspace{0.1cm}
      \begin{tablenotes}[para,flushleft]
        \footnotesize
        \textbf{Note:} This table presents each metric as a set of three values, ordered from top to bottom: full sample, top 25\% of firms by market capitalization (large-cap), and bottom 25\% (small-cap) for various predictive models across different window sizes (5, 21, 252, and 512 trading days). Metrics are first computed separately for each calendar year using all stock-date observations within that year. The reported values represent the average of these yearly statistics. Metrics include out-of-sample $R^2$ ($\RsqOOS$), overall directional accuracy, upward and downward classification accuracy, and macro-averaged F1 score. Benchmark models include linear (OLS+H, LASSO+H, RIDGE+H, Enet+H, and PCR), ensemble (XGBoost, CatBoost, and LightGBM), and neural network (NN-S and NN-L) models. `Overall Acc.' denotes overall directional accuracy, `Up Acc.' and `Down Acc.' represent the model's accuracy in predicting upward and downward excess returns respectively, and `F1' refers to the macro-averaged F1 score. Accuracy measures and $\RsqOOS$ are reported in percentage terms; macro-$F_1$ is reported on a zero-to-one scale. `H' indicates that the model is estimated using the Huber loss.
      \end{tablenotes}
    \end{minipage}
    \end{adjustbox}
  \end{threeparttable}
\end{table}
\vspace*{\fill}

\vspace*{\fill}
\begin{table}
  \centering
  \begin{threeparttable}
    \begin{adjustbox}{width=1.05\textwidth, center}
    \scriptsize
    \captionsetup{width=\linewidth}
    \caption{Benchmark Models - Portfolio Performance (Complete)}
    \label{PP_benchmark_full}
    \begin{minipage}{\linewidth}
      \renewcommand{\arraystretch}{1.2}
      \begin{tabular}{c|cccc|cccc|cccc|cccc|cccc}
        \toprule
        \textbf{Model} & \multicolumn{4}{c|}{\textbf{OLS+H}} & \multicolumn{4}{c|}{\textbf{LASSO+H}} & \multicolumn{4}{c|}{\textbf{RIDGE+H}} & \multicolumn{4}{c|}{\textbf{Enet+H}} & \multicolumn{4}{c}{\textbf{PCR}} \\
        \textbf{Window Size} & 5 & 21 & 252 & 512 & 5 & 21 & 252 & 512 & 5 & 21 & 252 & 512 & 5 & 21 & 252 & 512 & 5 & 21 & 252 & 512 \\
        \midrule
        AR & \makecell{45.03 \\ 29.28 \\ 15.75} & \makecell{45.47 \\ 29.59 \\ 15.89} & \makecell{45.63 \\ 29.51 \\ 16.13} & \makecell{45.05 \\ 29.01 \\ 16.04} &
        \makecell{44.54 \\ 28.96 \\ 15.58} & \makecell{43.44 \\ 28.16 \\ 15.28} & \makecell{43.61 \\ 28.29 \\ 15.31} & \makecell{43.55 \\ 28.25 \\ 15.30} &
        \makecell{44.66 \\ 29.01 \\ 15.64} & \makecell{44.22 \\ 28.97 \\ 15.25} & \makecell{44.51 \\ 28.64 \\ 15.87} & \makecell{43.99 \\ 28.40 \\ 15.59} &
        \makecell{44.04 \\ 28.71 \\ 15.34} & \makecell{44.06 \\ 28.66 \\ 15.40} & \makecell{44.06 \\ 28.67 \\ 15.39} & \makecell{43.60 \\ 28.30 \\ 15.30} &
        \makecell{29.96 \\ 19.52 \\ 10.44} & \makecell{29.28 \\ 19.42 \\ 9.85} & \makecell{28.37 \\ 18.49 \\ 9.88} & \makecell{33.18 \\ 21.82 \\ 11.36} \\ \addlinespace[3pt]

        SD & \makecell{9.35 \\ 13.90 \\ 11.86} & \makecell{9.37 \\ 13.92 \\ 11.85} & \makecell{9.37 \\ 13.91 \\ 11.86} & \makecell{9.35 \\ 13.89 \\ 11.87} &
        \makecell{9.29 \\ 13.84 \\ 11.89} & \makecell{9.03 \\ 13.66 \\ 11.90} & \makecell{9.04 \\ 13.67 \\ 11.90} & \makecell{9.10 \\ 13.68 \\ 11.91} &
        \makecell{9.25 \\ 13.84 \\ 11.87} & \makecell{9.27 \\ 13.80 \\ 11.92} & \makecell{9.39 \\ 13.78 \\ 12.02} & \makecell{9.38 \\ 13.80 \\ 12.00} &
        \makecell{9.21 \\ 13.79 \\ 11.90} & \makecell{9.20 \\ 13.79 \\ 11.90} & \makecell{9.17 \\ 13.76 \\ 11.90} & \makecell{9.05 \\ 13.67 \\ 11.90} &
        \makecell{8.40 \\ 13.11 \\ 12.15} & \makecell{8.27 \\ 13.04 \\ 12.09} & \makecell{8.16 \\ 12.99 \\ 12.01} & \makecell{8.51 \\ 13.17 \\ 12.06} \\ \addlinespace[3pt]

        SR & \makecell{4.82 \\ 2.11 \\ 1.33} & \makecell{4.85 \\ 2.12 \\ 1.34} & \makecell{4.87 \\ 2.12 \\ 1.36} & \makecell{4.82 \\ 2.09 \\ 1.35} &
        \makecell{4.80 \\ 2.09 \\ 1.31} & \makecell{4.81 \\ 2.06 \\ 1.28} & \makecell{4.82 \\ 2.07 \\ 1.29} & \makecell{4.79 \\ 2.06 \\ 1.28} &
        \makecell{4.83 \\ 2.10 \\ 1.32} & \makecell{4.77 \\ 2.10 \\ 1.28} & \makecell{4.74 \\ 2.08 \\ 1.32} & \makecell{4.69 \\ 2.06 \\ 1.30} &
        \makecell{4.78 \\ 2.08 \\ 1.29} & \makecell{4.79 \\ 2.08 \\ 1.29} & \makecell{4.80 \\ 2.08 \\ 1.29} & \makecell{4.82 \\ 2.07 \\ 1.29} &
        \makecell{3.57 \\ 1.49 \\ 0.86} & \makecell{3.54 \\ 1.49 \\ 0.82} & \makecell{3.48 \\ 1.42 \\ 0.82} & \makecell{3.90 \\ 1.66 \\ 0.94} \\ \addlinespace[3pt]

        Daily Return & \makecell{17.87 \\ 11.62 \\ 6.25} & \makecell{18.04 \\ 11.74 \\ 6.30} & \makecell{18.11 \\ 11.71 \\ 6.40} & \makecell{17.88 \\ 11.51 \\ 6.37} &
        \makecell{17.68 \\ 11.49 \\ 6.18} & \makecell{17.24 \\ 11.18 \\ 6.06} & \makecell{17.30 \\ 11.23 \\ 6.08} & \makecell{17.28 \\ 11.21 \\ 6.07} &
        \makecell{17.72 \\ 11.51 \\ 6.21} & \makecell{17.55 \\ 11.49 \\ 6.05} & \makecell{17.66 \\ 11.36 \\ 6.30} & \makecell{17.46 \\ 11.27 \\ 6.19} &
        \makecell{17.48 \\ 11.39 \\ 6.09} & \makecell{17.48 \\ 11.37 \\ 6.11} & \makecell{17.48 \\ 11.38 \\ 6.11} & \makecell{17.30 \\ 11.23 \\ 6.07} &
        \makecell{11.89 \\ 7.75 \\ 4.14} & \makecell{11.62 \\ 7.71 \\ 3.91} & \makecell{11.26 \\ 7.34 \\ 3.92} & \makecell{13.17 \\ 8.66 \\ 4.51} \\ \addlinespace[3pt]

        Max DD & \makecell{17.92 \\ 32.58 \\ 30.89} & \makecell{17.69 \\ 32.96 \\ 30.32} & \makecell{17.68 \\ 32.86 \\ 29.46} & \makecell{19.40 \\ 33.12 \\ 29.06} &
        \makecell{17.36 \\ 32.34 \\ 31.34} & \makecell{14.89 \\ 31.39 \\ 30.82} & \makecell{14.89 \\ 31.39 \\ 30.82} & \makecell{15.33 \\ 31.55 \\ 32.45} &
        \makecell{17.51 \\ 32.42 \\ 31.52} & \makecell{17.31 \\ 32.30 \\ 31.27} & \makecell{17.76 \\ 32.13 \\ 27.95} & \makecell{18.18 \\ 32.14 \\ 28.23} &
        \makecell{17.35 \\ 32.20 \\ 32.40} & \makecell{17.20 \\ 32.15 \\ 32.32} & \makecell{16.33 \\ 32.00 \\ 32.57} & \makecell{15.02 \\ 31.41 \\ 30.82} &
        \makecell{15.86 \\ 27.11 \\ 33.25} & \makecell{16.64 \\ 27.59 \\ 33.44} & \makecell{16.27 \\ 27.46 \\ 34.81} & \makecell{15.07 \\ 28.61 \\ 36.78} \\ \addlinespace[3pt]

        Max 1-day Loss & \makecell{5.60 \\ 8.74 \\ 5.11} & \makecell{5.66 \\ 8.79 \\ 5.08} & \makecell{5.68 \\ 8.77 \\ 4.94} & \makecell{5.68 \\ 8.76 \\ 4.95} &
        \makecell{5.59 \\ 8.75 \\ 5.25} & \makecell{5.06 \\ 8.40 \\ 5.48} & \makecell{5.06 \\ 8.40 \\ 5.48} & \makecell{5.14 \\ 8.44 \\ 5.47} &
        \makecell{5.59 \\ 8.77 \\ 5.23} & \makecell{5.57 \\ 8.75 \\ 5.31} & \makecell{5.50 \\ 8.64 \\ 5.19} & \makecell{5.39 \\ 8.50 \\ 5.30} &
        \makecell{5.53 \\ 8.68 \\ 5.33} & \makecell{5.49 \\ 8.69 \\ 5.29} & \makecell{5.42 \\ 8.62 \\ 5.33} & \makecell{5.06 \\ 8.40 \\ 5.48} &
        \makecell{3.24 \\ 7.45 \\ 4.64} & \makecell{2.94 \\ 7.16 \\ 4.88} & \makecell{3.65 \\ 7.59 \\ 4.92} & \makecell{4.21 \\ 7.85 \\ 6.13} \\ \addlinespace[3pt]

        Skew & \makecell{1.26 \\ -0.09 \\ 0.53} & \makecell{1.27 \\ -0.08 \\ 0.54} & \makecell{1.22 \\ -0.10 \\ 0.54} & \makecell{1.25 \\ -0.09 \\ 0.55} &
        \makecell{1.22 \\ -0.11 \\ 0.50} & \makecell{1.08 \\ -0.08 \\ 0.44} & \makecell{1.09 \\ -0.07 \\ 0.45} & \makecell{1.06 \\ -0.08 \\ 0.44} &
        \makecell{1.18 \\ -0.11 \\ 0.49} & \makecell{1.18 \\ -0.08 \\ 0.48} & \makecell{1.07 \\ -0.10 \\ 0.44} & \makecell{1.08 \\ -0.09 \\ 0.43} &
        \makecell{1.16 \\ -0.10 \\ 0.47} & \makecell{1.16 \\ -0.10 \\ 0.47} & \makecell{1.13 \\ -0.10 \\ 0.46} & \makecell{1.09 \\ -0.07 \\ 0.45} &
        \makecell{0.85 \\ -0.01 \\ 0.42} & \makecell{0.84 \\ -0.08 \\ 0.38} & \makecell{0.76 \\ -0.10 \\ 0.40} & \makecell{0.78 \\ -0.18 \\ 0.31} \\ \addlinespace[3pt]

        Kurt & \makecell{17.24 \\ 12.02 \\ 6.36} & \makecell{17.62 \\ 12.22 \\ 6.33} & \makecell{17.51 \\ 12.13 \\ 6.35} & \makecell{17.57 \\ 12.12 \\ 6.29} &
        \makecell{16.65 \\ 11.85 \\ 6.15} & \makecell{15.05 \\ 11.07 \\ 5.96} & \makecell{15.06 \\ 11.06 \\ 6.01} & \makecell{15.05 \\ 11.16 \\ 6.01} &
        \makecell{16.59 \\ 11.86 \\ 6.11} & \makecell{16.56 \\ 11.82 \\ 6.22} & \makecell{16.20 \\ 11.61 \\ 6.16} & \makecell{16.14 \\ 11.56 \\ 6.14} &
        \makecell{16.43 \\ 11.70 \\ 6.09} & \makecell{16.31 \\ 11.71 \\ 6.07} & \makecell{16.12 \\ 11.64 \\ 6.02} & \makecell{14.99 \\ 11.06 \\ 6.00} &
        \makecell{12.35 \\ 10.10 \\ 6.09} & \makecell{12.31 \\ 9.79 \\ 6.06} & \makecell{11.06 \\ 10.02 \\ 5.78} & \makecell{12.38 \\ 10.49 \\ 6.10} \\ \addlinespace[3pt]

        \midrule \midrule

        \textbf{Model} & \multicolumn{4}{c|}{\textbf{XGBoost}} & \multicolumn{4}{c|}{\textbf{CatBoost}} & \multicolumn{4}{c|}{\textbf{LightGBM}} & \multicolumn{4}{c|}{\textbf{NN-S}} & \multicolumn{4}{c}{\textbf{NN-L}} \\
        \textbf{Window Size} & 5 & 21 & 252 & 512 & 5 & 21 & 252 & 512 & 5 & 21 & 252 & 512 & 5 & 21 & 252 & 512 & 5 & 21 & 252 & 512 \\
        \midrule
        AR & \makecell{44.27 \\ 27.40 \\ 16.87} & \makecell{46.66 \\ 29.04 \\ 17.62} & \makecell{47.42 \\ 30.34 \\ 17.08} & \makecell{47.17 \\ 30.28 \\ 16.89} &
        \makecell{42.35 \\ 26.80 \\ 15.54} & \makecell{46.29 \\ 28.47 \\ 17.82} & \makecell{46.27 \\ 29.18 \\ 17.09} & \makecell{46.98 \\ 29.43 \\ 17.55} &
        \makecell{43.43 \\ 26.98 \\ 16.44} & \makecell{45.70 \\ 28.75 \\ 16.95} & \makecell{46.25 \\ 29.75 \\ 16.50} & \makecell{46.30 \\ 29.92 \\ 16.38} &
        \makecell{40.30 \\ 27.48 \\ 12.82} & \makecell{42.36 \\ 27.63 \\ 14.74} & \makecell{39.91 \\ 25.63 \\ 14.28} & \makecell{39.87 \\ 25.72 \\ 14.15} &
        \makecell{43.77 \\ 28.54 \\ 15.23} & \makecell{42.73 \\ 27.83 \\ 14.89} & \makecell{40.55 \\ 26.28 \\ 14.27} & \makecell{38.51 \\ 25.01 \\ 13.50} \\ \addlinespace[3pt]

        SD & \makecell{7.71 \\ 14.15 \\ 12.07} & \makecell{7.20 \\ 13.91 \\ 12.43} & \makecell{8.18 \\ 13.95 \\ 12.55} & \makecell{8.33 \\ 13.99 \\ 12.55} &
        \makecell{7.43 \\ 14.15 \\ 12.11} & \makecell{6.89 \\ 13.89 \\ 12.38} & \makecell{6.88 \\ 13.72 \\ 12.32} & \makecell{7.35 \\ 13.82 \\ 12.32} &
        \makecell{7.76 \\ 14.24 \\ 12.08} & \makecell{7.54 \\ 14.00 \\ 12.43} & \makecell{8.26 \\ 14.05 \\ 12.58} & \makecell{8.43 \\ 14.11 \\ 12.53} &
        \makecell{9.32 \\ 14.00 \\ 11.78} & \makecell{9.22 \\ 13.85 \\ 11.97} & \makecell{9.01 \\ 13.62 \\ 11.93} & \makecell{8.96 \\ 13.71 \\ 11.86} &
        \makecell{9.25 \\ 13.85 \\ 11.93} & \makecell{9.32 \\ 13.83 \\ 11.90} & \makecell{8.87 \\ 13.70 \\ 11.86} & \makecell{8.91 \\ 13.57 \\ 11.92} \\ \addlinespace[3pt]

        SR & \makecell{5.74 \\ 1.94 \\ 1.40} & \makecell{6.48 \\ 2.09 \\ 1.42} & \makecell{5.80 \\ 2.18 \\ 1.36} & \makecell{5.66 \\ 2.16 \\ 1.35} &
        \makecell{5.70 \\ 1.89 \\ 1.28} & \makecell{6.72 \\ 2.05 \\ 1.44} & \makecell{6.72 \\ 2.13 \\ 1.39} & \makecell{6.39 \\ 2.13 \\ 1.42} &
        \makecell{5.60 \\ 1.90 \\ 1.36} & \makecell{6.06 \\ 2.05 \\ 1.36} & \makecell{5.60 \\ 2.12 \\ 1.31} & \makecell{5.49 \\ 2.12 \\ 1.31} &
        \makecell{4.32 \\ 1.96 \\ 1.09} & \makecell{4.59 \\ 1.99 \\ 1.23} & \makecell{4.43 \\ 1.88 \\ 1.20} & \makecell{4.45 \\ 1.88 \\ 1.19} &
        \makecell{4.73 \\ 2.06 \\ 1.28} & \makecell{4.58 \\ 2.01 \\ 1.25} & \makecell{4.57 \\ 1.92 \\ 1.20} & \makecell{4.32 \\ 1.84 \\ 1.13} \\ \addlinespace[3pt]

        Daily Return & \makecell{17.57 \\ 10.87 \\ 6.69} & \makecell{18.52 \\ 11.52 \\ 6.99} & \makecell{18.82 \\ 12.04 \\ 6.78} & \makecell{18.72 \\ 12.02 \\ 6.70} &
        \makecell{16.81 \\ 10.64 \\ 6.17} & \makecell{18.37 \\ 11.30 \\ 7.07} & \makecell{18.36 \\ 11.58 \\ 6.78} & \makecell{18.64 \\ 11.68 \\ 6.97} &
        \makecell{17.23 \\ 10.71 \\ 6.53} & \makecell{18.14 \\ 11.41 \\ 6.73} & \makecell{18.35 \\ 11.81 \\ 6.55} & \makecell{18.37 \\ 11.87 \\ 6.50} &
        \makecell{15.99 \\ 10.91 \\ 5.09} & \makecell{16.81 \\ 10.96 \\ 5.85} & \makecell{15.84 \\ 10.17 \\ 5.67} & \makecell{15.82 \\ 10.21 \\ 5.62} &
        \makecell{17.37 \\ 11.33 \\ 6.04} & \makecell{16.95 \\ 11.04 \\ 5.91} & \makecell{16.09 \\ 10.43 \\ 5.66} & \makecell{15.28 \\ 9.92 \\ 5.36} \\ \addlinespace[3pt]

        Max DD & \makecell{12.19 \\ 32.06 \\ 31.51} & \makecell{12.60 \\ 32.07 \\ 31.01} & \makecell{15.56 \\ 33.44 \\ 32.74} & \makecell{16.02 \\ 33.63 \\ 31.84} &
        \makecell{12.03 \\ 32.49 \\ 31.98} & \makecell{13.14 \\ 32.85 \\ 30.00} & \makecell{13.67 \\ 31.85 \\ 31.75} & \makecell{14.46 \\ 32.62 \\ 29.58} &
        \makecell{14.00 \\ 33.30 \\ 32.29} & \makecell{16.16 \\ 33.86 \\ 34.18} & \makecell{16.85 \\ 33.94 \\ 32.78} & \makecell{17.17 \\ 33.70 \\ 31.31} &
        \makecell{18.99 \\ 33.34 \\ 32.34} & \makecell{18.70 \\ 33.36 \\ 32.58} & \makecell{16.78 \\ 32.25 \\ 33.58} & \makecell{17.26 \\ 32.68 \\ 33.09} &
        \makecell{17.08 \\ 32.37 \\ 32.35} & \makecell{18.60 \\ 33.21 \\ 31.50} & \makecell{14.56 \\ 31.97 \\ 32.68} & \makecell{17.63 \\ 31.64 \\ 31.11} \\ \addlinespace[3pt]

        Max 1-day Loss & \makecell{4.97 \\ 8.56 \\ 5.51} & \makecell{3.41 \\ 8.06 \\ 4.74} & \makecell{4.36 \\ 8.23 \\ 5.63} & \makecell{4.56 \\ 8.19 \\ 5.12} &
        \makecell{4.86 \\ 8.28 \\ 5.56} & \makecell{3.75 \\ 8.21 \\ 4.49} & \makecell{4.15 \\ 8.19 \\ 5.03} & \makecell{4.93 \\ 8.19 \\ 5.27} &
        \makecell{5.10 \\ 8.59 \\ 5.66} & \makecell{4.26 \\ 8.33 \\ 5.36} & \makecell{4.96 \\ 8.28 \\ 5.56} & \makecell{5.17 \\ 8.32 \\ 5.52} &
        \makecell{5.62 \\ 8.81 \\ 6.31} & \makecell{5.47 \\ 8.54 \\ 5.53} & \makecell{4.80 \\ 8.21 \\ 4.98} & \makecell{5.15 \\ 8.37 \\ 5.28} &
        \makecell{5.51 \\ 8.67 \\ 5.58} & \makecell{5.61 \\ 8.80 \\ 5.63} & \makecell{5.49 \\ 8.67 \\ 5.18} & \makecell{5.29 \\ 8.32 \\ 5.19} \\ \addlinespace[3pt]

        Skew & \makecell{1.77 \\ -0.05 \\ 0.35} & \makecell{2.07 \\ 0.01 \\ 0.36} & \makecell{1.53 \\ -0.01 \\ 0.32} & \makecell{1.45 \\ -0.03 \\ 0.36} &
        \makecell{2.20 \\ -0.01 \\ 0.35} & \makecell{2.15 \\ 0.00 \\ 0.38} & \makecell{2.21 \\ 0.02 \\ 0.37} & \makecell{1.67 \\ -0.05 \\ 0.35} &
        \makecell{1.75 \\ -0.08 \\ 0.33} & \makecell{1.69 \\ -0.06 \\ 0.35} & \makecell{1.53 \\ -0.04 \\ 0.31} & \makecell{1.52 \\ -0.04 \\ 0.32} &
        \makecell{1.27 \\ -0.12 \\ 0.48} & \makecell{1.27 \\ -0.15 \\ 0.50} & \makecell{1.16 \\ -0.09 \\ 0.50} & \makecell{1.62 \\ -0.01 \\ 0.51} &
        \makecell{1.20 \\ -0.11 \\ 0.51} & \makecell{1.33 \\ -0.11 \\ 0.53} & \makecell{1.23 \\ -0.14 \\ 0.44} & \makecell{1.31 \\ -0.13 \\ 0.48} \\ \addlinespace[3pt]

        Kurt & \makecell{26.41 \\ 11.42 \\ 6.12} & \makecell{27.96 \\ 11.49 \\ 5.20} & \makecell{20.76 \\ 12.11 \\ 5.67} & \makecell{19.69 \\ 11.96 \\ 5.62} &
        \makecell{32.84 \\ 11.60 \\ 6.15} & \makecell{31.26 \\ 12.29 \\ 5.15} & \makecell{32.14 \\ 12.51 \\ 5.39} & \makecell{26.37 \\ 12.53 \\ 5.49} &
        \makecell{26.97 \\ 11.35 \\ 6.32} & \makecell{25.50 \\ 12.15 \\ 5.78} & \makecell{22.71 \\ 12.23 \\ 5.91} & \makecell{22.20 \\ 12.17 \\ 5.81} &
        \makecell{17.24 \\ 11.95 \\ 7.02} & \makecell{16.59 \\ 11.60 \\ 6.58} & \makecell{15.34 \\ 11.68 \\ 6.12} & \makecell{20.55 \\ 12.27 \\ 6.18} &
        \makecell{16.43 \\ 11.44 \\ 6.71} & \makecell{17.24 \\ 11.66 \\ 6.64} & \makecell{17.19 \\ 11.96 \\ 5.99} & \makecell{17.73 \\ 11.76 \\ 6.46} \\ \addlinespace[3pt]

        \bottomrule
      \end{tabular}
      \vspace{0.1cm}
      \begin{tablenotes}[para,flushleft]
        \footnotesize
        \textbf{Note:} This table reports portfolio performance metrics computed from the full daily out-of-sample excess return series across different input window lengths (5, 21, 252, and 512 trading days) for each model. Each cell displays three values from top to bottom: the long--short portfolio, the long-leg contribution, and the short-leg contribution. The long leg equals one half of the top-decile return, while the short leg is the signed contribution from shorting the bottom decile and equals minus one half of the bottom-decile return. Metrics include annualized return (AR), standard deviation (SD), Sharpe ratio (SR), daily return (in basis points), maximum drawdown (Max DD), maximum one-day loss (Max 1-day Loss), skewness, and excess kurtosis of portfolio returns. AR, SD, maximum drawdown, and maximum one-day loss are reported in percentage terms. Benchmark models include linear (OLS+H, LASSO+H, RIDGE+H, Enet+H, and PCR), ensemble (XGBoost, CatBoost, and LightGBM), and neural network (NN-S and NN-L) models. Portfolios are formed using decile sorting based on model forecasts, with equal weighting across stocks. `H' indicates that the model is estimated using the Huber loss.
      \end{tablenotes}
    \end{minipage}
    \end{adjustbox}
  \end{threeparttable}
\end{table}
\vspace*{\fill}

\begin{landscape} %% Fixed -----------------------------------------------
\thispagestyle{landscape}
\begin{table}[p]
\vspace*{\fill} % push it down\
\centering
\caption{Comparison of Daily Out-of-Sample Forecasting Performance of Benchmark Models Using Diebold--Mariano Test}
\label{DM_test_Benchmark}
\begin{adjustbox}{max width=1.46\textwidth}
\begin{threeparttable}
\renewcommand{\arraystretch}{1.2}
\setlength{\tabcolsep}{3pt}

 \\
\bottomrule
\end{tabular}
\begin{tablenotes}[para,flushleft]
\large
\textbf{Note:} This table reports Diebold--Mariano test statistics based on the daily cross-sectional mean of squared-error differences (row minus column), following the forecast-comparison approach of \citet{gu2020empirical}. Positive statistics favor the column specification; negative statistics favor the row specification. All comparisons use the same stock-date observations. Heteroskedasticity- and autocorrelation-consistent (HAC) standard errors use one lag. An asterisk denotes significance at the 5\% level using a two-sided asymptotic standard normal test, without adjustment for multiple comparisons. Input windows are measured in trading days. `H' denotes estimation using Huber loss.
\end{tablenotes}
\end{threeparttable}
\end{adjustbox}
\end{table}
\end{landscape}

\clearpage

\vspace*{\fill}
\begin{table}[h]
  \centering
  \begin{threeparttable}
    \begin{adjustbox}{width=1.05\textwidth, center}
    \scriptsize
    \captionsetup{width=\linewidth}
    \caption{Benchmark Models - Spread Portfolio Performance}
    \label{Spread_Portfolio_Benchmarks_table}
    \begin{minipage}{\linewidth}
      \renewcommand{\arraystretch}{1.2}
      \begin{tabular}{c|cccc|cccc|cccc|cccc|cccc}
        \toprule
        \textbf{Model} & \multicolumn{4}{c|}{\textbf{OLS+H}} & \multicolumn{4}{c|}{\textbf{LASSO+H}} & \multicolumn{4}{c|}{\textbf{RIDGE+H}} & \multicolumn{4}{c|}{\textbf{Enet+H}} & \multicolumn{4}{c}{\textbf{PCR}} \\
        \textbf{Window Size} & 5 & 21 & 252 & 512 & 5 & 21 & 252 & 512 & 5 & 21 & 252 & 512 & 5 & 21 & 252 & 512 & 5 & 21 & 252 & 512 \\
        \midrule
        Low & -31.50 & -31.77 & -32.25 & -32.09 & -31.17 & -30.56 & -30.63 & -30.59 & -31.29 & -30.50 & -31.75 & -31.18 & -30.68 & -30.80 & -30.78 & -30.60 & -20.88 & -19.71 & -19.77 & -22.72 \\
        \addlinespace[3pt]
        2 & -8.88 & -8.73 & -8.72 & -8.51 & -8.02 & -4.57 & -4.84 & -4.86 & -7.73 & -8.39 & -7.81 & -7.76 & -7.13 & -6.80 & -6.28 & -4.97 & 1.40 & 1.00 & 1.66 & -0.94 \\
        \addlinespace[3pt]
        3 & -0.54 & -0.80 & -0.93 & -1.16 & 0.09 & 2.35 & 2.27 & 2.24 & 0.37 & 1.19 & 0.57 & 0.80 & 0.95 & 1.06 & 1.58 & 2.18 & 6.89 & 6.70 & 6.13 & 5.17 \\
        \addlinespace[3pt]
        4 & 4.50 & 4.03 & 4.57 & 4.87 & 4.44 & 6.87 & 6.49 & 6.38 & 4.60 & 4.21 & 4.87 & 4.69 & 5.20 & 5.18 & 5.88 & 6.60 & 7.79 & 8.82 & 9.09 & 7.96 \\
        \addlinespace[3pt]
        5 & 7.90 & 8.05 & 8.14 & 8.09 & 8.51 & 9.11 & 9.04 & 8.99 & 8.09 & 8.52 & 8.55 & 9.19 & 8.81 & 8.66 & 8.54 & 8.99 & 10.87 & 10.60 & 11.21 & 10.02 \\
        \addlinespace[3pt]
        6 & 12.62 & 12.11 & 12.04 & 12.27 & 12.22 & 11.85 & 12.09 & 11.87 & 12.58 & 12.15 & 12.13 & 11.82 & 12.52 & 12.42 & 12.14 & 12.00 & 12.86 & 12.34 & 12.60 & 13.14 \\
        \addlinespace[3pt]
        7 & 16.47 & 16.98 & 16.63 & 16.81 & 16.43 & 14.81 & 14.73 & 15.04 & 16.16 & 15.88 & 16.40 & 16.45 & 15.26 & 15.28 & 15.05 & 14.88 & 14.38 & 14.36 & 15.44 & 15.51 \\
        \addlinespace[3pt]
        8 & 22.38 & 22.47 & 22.45 & 22.85 & 21.58 & 19.27 & 19.51 & 19.47 & 21.74 & 21.75 & 22.01 & 22.89 & 20.75 & 20.73 & 20.54 & 19.42 & 18.03 & 17.84 & 17.83 & 17.85 \\
        \addlinespace[3pt]
        9 & 31.56 & 31.59 & 32.18 & 31.94 & 31.10 & 27.64 & 27.84 & 28.06 & 30.57 & 30.36 & 30.87 & 29.41 & 30.00 & 30.04 & 29.10 & 28.00 & 22.71 & 22.29 & 21.93 & 23.46 \\
        \addlinespace[3pt]
        High & 58.57 & 59.17 & 59.01 & 58.02 & 57.92 & 56.33 & 56.59 & 56.50 & 58.03 & 57.93 & 57.27 & 56.80 & 57.41 & 57.32 & 57.34 & 56.61 & 39.04 & 38.85 & 36.97 & 43.65 \\
        \addlinespace[3pt]
        $\tfrac{1}{2}(H-L)$ & 45.03 & 45.47 & 45.63 & 45.05 & 44.54 & 43.44 & 43.61 & 43.55 & 44.66 & 44.22 & 44.51 & 43.99 & 44.04 & 44.06 & 44.06 & 43.60 & 29.96 & 29.28 & 28.37 & 33.18 \\   

        \midrule \midrule

        \textbf{Model} & \multicolumn{4}{c|}{\textbf{XGBoost}} & \multicolumn{4}{c|}{\textbf{CatBoost}} & \multicolumn{4}{c|}{\textbf{LightGBM}} & \multicolumn{4}{c|}{\textbf{NN-S}} & \multicolumn{4}{c}{\textbf{NN-L}} \\
        \textbf{Window Size} & 5 & 21 & 252 & 512 & 5 & 21 & 252 & 512 & 5 & 21 & 252 & 512 & 5 & 21 & 252 & 512 & 5 & 21 & 252 & 512 \\
        \midrule
        Low & -33.74 & -35.24 & -34.17 & -33.78 & -31.09 & -35.64 & -34.19 & -35.11 & -32.89 & -33.91 & -33.00 & -32.77 & -25.64 & -29.47 & -28.56 & -28.31 & -30.45 & -29.79 & -28.53 & -27.00 \\
        \addlinespace[3pt]
        2 & -3.45 & -4.41 & -6.08 & -5.27 & -0.80 & -2.89 & -3.83 & -3.98 & -3.09 & -4.67 & -6.37 & -6.98 & -6.98 & -7.43 & -5.24 & -5.11 & -8.50 & -8.27 & -5.82 & -4.83 \\
        \addlinespace[3pt]
        3 & 4.48 & 4.23 & 2.15 & 2.12 & 4.33 & 4.56 & 2.80 & 3.33 & 3.12 & 1.95 & 1.91 & 1.17 & 0.50 & 0.63 & 1.46 & 1.94 & 0.01 & -0.27 & 1.51 & 2.72 \\
        \addlinespace[3pt]
        4 & 6.22 & 8.35 & 6.82 & 6.43 & 7.51 & 8.36 & 6.40 & 6.87 & 6.03 & 6.49 & 6.00 & 5.91 & 4.19 & 5.17 & 6.09 & 6.42 & 4.31 & 4.10 & 5.26 & 6.75 \\
        \addlinespace[3pt]
        5 & 10.43 & 9.76 & 9.83 & 8.78 & 9.40 & 10.27 & 9.88 & 9.40 & 10.02 & 9.81 & 9.30 & 9.41 & 7.99 & 9.21 & 10.44 & 9.97 & 8.65 & 8.60 & 8.82 & 8.98 \\
        \addlinespace[3pt]
        6 & 13.00 & 12.02 & 11.61 & 11.34 & 11.57 & 11.44 & 12.00 & 12.43 & 12.37 & 12.35 & 11.35 & 12.05 & 11.48 & 12.56 & 12.70 & 12.52 & 12.27 & 12.69 & 13.14 & 13.01 \\
        \addlinespace[3pt]
        7 & 15.10 & 14.89 & 14.55 & 14.68 & 15.27 & 15.26 & 15.66 & 14.08 & 15.53 & 14.87 & 15.28 & 15.11 & 15.31 & 15.85 & 16.17 & 15.22 & 16.43 & 16.70 & 16.53 & 15.71 \\
        \addlinespace[3pt]
        8 & 19.34 & 18.54 & 20.01 & 19.70 & 18.24 & 18.67 & 18.31 & 19.43 & 20.76 & 20.37 & 20.80 & 20.31 & 21.05 & 21.58 & 20.51 & 20.05 & 22.56 & 21.85 & 20.92 & 20.15 \\
        \addlinespace[3pt]
        9 & 26.93 & 26.90 & 27.69 & 28.54 & 25.06 & 26.11 & 27.71 & 27.79 & 27.29 & 28.35 & 28.33 & 29.05 & 30.22 & 29.75 & 28.27 & 28.96 & 30.74 & 31.82 & 28.71 & 27.59 \\
        \addlinespace[3pt]
        High & 54.80 & 58.08 & 60.68 & 60.56 & 53.61 & 56.94 & 58.36 & 58.85 & 53.97 & 57.49 & 59.50 & 59.84 & 54.97 & 55.25 & 51.26 & 51.43 & 57.09 & 55.66 & 52.56 & 50.02 \\
        \addlinespace[3pt]
        $\tfrac{1}{2}(H-L)$ & 44.27 & 46.66 & 47.42 & 47.17 & 42.35 & 46.29 & 46.27 & 46.98 & 43.43 & 45.70 & 46.25 & 46.30 & 40.30 & 42.36 & 39.91 & 39.87 & 43.77 & 42.73 & 40.55 & 38.51 \\  
        \bottomrule
      \end{tabular}
      \vspace{0.1cm}
      \begin{tablenotes}[para,flushleft]
        \footnotesize
        \textbf{Note:} This table presents the annualized average returns of decile spread portfolios formed using forecasts from various predictive models across different window sizes (5, 21, 252, and 512 trading days). For each model and window size, we report standalone decile returns from Low to High. Here, $H$ and $L$ denote the standalone returns of the highest- and lowest-forecast deciles. The final row reports the unit-gross long--short return, $\frac{1}{2}(H-L)$, with long- and short-leg contributions of $\frac{1}{2}H$ and $-\frac{1}{2}L$, respectively. Models evaluated include linear (OLS+H, LASSO+H, RIDGE+H, Enet+H, and PCR), ensemble (XGBoost, CatBoost, and LightGBM), and neural network (NN-S and NN-L) models. `H' indicates that the model is estimated using the Huber loss. Portfolios are constructed using equal-weighted (EW) decile sorting based on predicted returns.
      \end{tablenotes}
    \end{minipage}
    \end{adjustbox}
  \end{threeparttable}
\end{table}
\vspace*{\fill}

\clearpage
\subsection{Complete Zero-Shot TSFM Results} \label{complete_zero_shot_tsfm_results}
This subsection presents complete zero-shot TSFM results for both forecasting performance and portfolio performance across all models and window sizes.
\vspace*{\fill}
\begin{table}[h] %% Fixed -------------------------------------------------------------------------
  \centering
  \begin{threeparttable}
    \begin{adjustbox}{width=0.96\textwidth, center}
    \scriptsize
    \captionsetup{width=\linewidth}
    \caption{Zero-Shot TSFMs - Forecasting Performance (Complete)}
    \label{FP_TSFM_zero_shot_full}
    \begin{minipage}{\linewidth}
        \renewcommand{\arraystretch}{1.3}
      \begin{tabular}{c|cccc|cccc|cccc|cccc}
        \toprule
        \textbf{Model} & \multicolumn{4}{c|}{\textbf{Benchmark}} & \multicolumn{4}{c|}{\textbf{Chronos (Tiny)}} & \multicolumn{4}{c|}{\textbf{Chronos (Mini)}} & \multicolumn{4}{c}{\textbf{Chronos (Small)}}  \\
        \textbf{Window Size} & 5 & 21 & 252 & 512 & 5 & 21 & 252 & 512 & 5 & 21 & 252 & 512 & 5 & 21 & 252 & 512   \\

        \midrule
        $\RsqOOS$ & \makecell{-0.25 \\ -0.50 \\ 0.39} & \makecell{-0.04 \\ -0.38 \\ 0.66} & \makecell{-0.03 \\ -0.32 \\ 0.62} & \makecell{-0.03 \\ -0.28 \\ 0.59} &
        \makecell{-75.87 \\ -68.38 \\ -85.21} & \makecell{-8.34 \\ -8.27 \\ -8.87} & \makecell{-1.96 \\ -2.66 \\ -1.37} & \makecell{-1.77 \\ -2.50 \\ -1.15} &
        \makecell{-43.14 \\ -40.16 \\ -47.34} & \makecell{-11.60 \\ -9.68 \\ -13.96} & \makecell{-1.04 \\ -1.32 \\ -0.83} & \makecell{-0.75 \\ -1.13 \\ -0.40} &
        \makecell{-77.07 \\ -52.28 \\ -104.53} & \makecell{-13.04 \\ -8.01 \\ -18.67} & \makecell{-1.30 \\ -1.42 \\ -1.21} & \makecell{-1.27 \\ -1.34 \\ -1.16} \\

        \addlinespace[3pt]

        Overall Acc. & \makecell{50.94 \\ 50.70 \\ 51.96} & \makecell{51.12 \\ 50.74 \\ 52.47} & \makecell{51.08 \\ 50.80 \\ 52.33} & \makecell{51.16 \\ 50.95 \\ 52.39} &
        \makecell{48.39 \\ 49.41 \\ 46.26} & \makecell{49.31 \\ 49.88 \\ 48.35} & \makecell{50.30 \\ 50.73 \\ 50.18} & \makecell{50.46 \\ 50.86 \\ 50.60} &
        \makecell{48.60 \\ 49.36 \\ 46.95} & \makecell{49.32 \\ 49.68 \\ 48.75} & \makecell{50.54 \\ 50.60 \\ 51.13} & \makecell{50.54 \\ 50.68 \\ 51.06} &
        \makecell{48.54 \\ 49.22 \\ 46.98} & \makecell{49.50 \\ 49.46 \\ 49.53} & \makecell{50.87 \\ 50.39 \\ 52.25} & \makecell{50.99 \\ 50.25 \\ 52.62} \\

        \addlinespace[3pt]

        Up Acc. & \makecell{68.96 \\ 72.05 \\ 68.00} & \makecell{67.82 \\ 72.97 \\ 65.11} & \makecell{66.87 \\ 71.06 \\ 64.65} & \makecell{69.14 \\ 74.39 \\ 66.07} &
        \makecell{54.53 \\ 55.72 \\ 52.81} & \makecell{59.51 \\ 62.19 \\ 56.75} & \makecell{76.02 \\ 79.80 \\ 71.08} & \makecell{78.02 \\ 82.42 \\ 70.99} &
        \makecell{51.67 \\ 53.54 \\ 48.62} & \makecell{54.39 \\ 58.34 \\ 48.76} & \makecell{66.93 \\ 73.00 \\ 59.45} & \makecell{68.43 \\ 74.41 \\ 60.84} &
        \makecell{47.69 \\ 49.71 \\ 44.33} & \makecell{44.87 \\ 49.24 \\ 38.35} & \makecell{39.93 \\ 52.61 \\ 24.45} & \makecell{28.51 \\ 41.28 \\ 15.37} \\

        \addlinespace[3pt]

        Down Acc. & \makecell{33.14 \\ 28.11 \\ 37.42} & \makecell{34.60 \\ 27.25 \\ 41.01} & \makecell{35.59 \\ 29.59 \\ 41.24} & \makecell{33.37 \\ 26.24 \\ 39.86} &
        \makecell{42.32 \\ 42.71 \\ 40.38} & \makecell{39.23 \\ 36.80 \\ 40.84} & \makecell{24.91 \\ 20.04 \\ 31.37} & \makecell{23.25 \\ 17.55 \\ 32.21} &
        \makecell{45.50 \\ 44.86 \\ 45.39} & \makecell{44.21 \\ 40.39 \\ 48.64} & \makecell{34.27 \\ 26.82 \\ 43.59} & \makecell{32.87 \\ 25.56 \\ 42.22} &
        \makecell{49.28 \\ 48.56 \\ 49.27} & \makecell{53.82 \\ 49.35 \\ 59.38} & \makecell{61.13 \\ 47.41 \\ 76.83} & \makecell{72.60 \\ 58.95 \\ 85.76} \\

        \addlinespace[3pt]

        F1 & \makecell{0.49 \\ 0.47 \\ 0.51} & \makecell{0.49 \\ 0.47 \\ 0.52} & \makecell{0.50 \\ 0.48 \\ 0.52} & \makecell{0.49 \\ 0.47 \\ 0.51} &
        \makecell{0.48 \\ 0.49 \\ 0.46} & \makecell{0.49 \\ 0.49 \\ 0.48} & \makecell{0.47 \\ 0.45 \\ 0.48} & \makecell{0.46 \\ 0.44 \\ 0.49} &
        \makecell{0.49 \\ 0.49 \\ 0.47} & \makecell{0.49 \\ 0.49 \\ 0.49} & \makecell{0.49 \\ 0.47 \\ 0.51} & \makecell{0.49 \\ 0.47 \\ 0.51} &
        \makecell{0.48 \\ 0.49 \\ 0.47} & \makecell{0.49 \\ 0.49 \\ 0.49} & \makecell{0.50 \\ 0.49 \\ 0.48} & \makecell{0.48 \\ 0.49 \\ 0.44} \\

        \addlinespace[3pt]

        \midrule \midrule
        \textbf{Model} & \multicolumn{4}{c|}{\textbf{Chronos (Base)}} & \multicolumn{4}{c|}{\textbf{Chronos (Large)}} & \multicolumn{4}{c|}{\textbf{TimesFM 1 (200M)}} & \multicolumn{4}{c}{\textbf{TimesFM 2 (500M)}}  \\
        \textbf{Window Size} & 5 & 21 & 252 & 512 & 5 & 21 & 252 & 512 & 5 & 21 & 252 & 512 & 5 & 21 & 252 & 512  \\
        \midrule
        $\RsqOOS$ &  \makecell{-57.59 \\ -38.80 \\ -78.51} & \makecell{-21.47 \\ -13.21 \\ -30.26} & \makecell{-1.97 \\ -1.79 \\ -2.29} & \makecell{-1.35 \\ -1.60 \\ -1.22} &
        \makecell{-46.86 \\ -35.50 \\ -60.59} & \makecell{-23.84 \\ -12.58 \\ -35.61} & \makecell{-2.49 \\ -1.69 \\ -3.30} & \makecell{-1.37 \\ -1.03 \\ -1.70} &
        \makecell{-1469.95 \\ -610.91 \\ -3497.22} & \makecell{-5004.66 \\ -517.06 \\ -9803.15} & \makecell{-\num{23311.78} \\ -\num{20779.55} \\ -\num{30920.77}} & \makecell{-\num{62603.87} \\ -\num{26993.51} \\ -\num{120644.10}} &
        \makecell{-27.96 \\ -27.27 \\ -29.93} & \makecell{-11.81 \\ -10.90 \\ -13.18} & \makecell{-3.86 \\ -3.78 \\ -3.96} & \makecell{-2.80 \\ -2.90 \\ -2.64} \\

        \addlinespace[3pt]

        Overall Acc. & \makecell{48.59 \\ 49.42 \\ 46.84} & \makecell{49.19 \\ 49.61 \\ 48.45} & \makecell{50.57 \\ 49.97 \\ 51.85} & \makecell{50.80 \\ 49.86 \\ 52.61} &
        \makecell{48.70 \\ 49.58 \\ 46.94} & \makecell{49.13 \\ 49.91 \\ 47.85} & \makecell{50.49 \\ 50.39 \\ 51.35} & \makecell{51.01 \\ 50.61 \\ 52.46} &
        \makecell{49.55 \\ 50.24 \\ 48.53} & \makecell{50.00 \\ 49.86 \\ 50.15} & \makecell{49.87 \\ 49.62 \\ 50.24} & \makecell{49.60 \\ 49.74 \\ 49.36} &
        \makecell{48.34 \\ 49.25 \\ 46.35} & \makecell{48.70 \\ 49.44 \\ 47.19} & \makecell{49.59 \\ 49.69 \\ 49.72} & \makecell{49.82 \\ 50.01 \\ 49.94} \\

        \addlinespace[3pt]

        Up Acc. &  \makecell{53.94 \\ 55.68 \\ 51.04} & \makecell{52.98 \\ 56.87 \\ 47.25} & \makecell{31.53 \\ 39.94 \\ 23.36} & \makecell{24.55 \\ 30.55 \\ 21.48} &
        \makecell{59.70 \\ 61.30 \\ 57.24} & \makecell{64.13 \\ 66.96 \\ 60.71} & \makecell{51.52 \\ 60.25 \\ 42.21} & \makecell{49.60 \\ 59.93 \\ 37.30} &
        \makecell{73.64 \\ 73.80 \\ 73.55} & \makecell{49.94 \\ 49.14 \\ 51.28} & \makecell{45.46 \\ 44.42 \\ 47.35} & \makecell{54.59 \\ 53.29 \\ 56.82} &
        \makecell{50.14 \\ 52.65 \\ 46.22} & \makecell{52.74 \\ 56.01 \\ 48.61} & \makecell{52.16 \\ 56.29 \\ 47.40} & \makecell{56.11 \\ 61.13 \\ 50.53} \\

        \addlinespace[3pt]

        Down Acc. &  \makecell{43.27 \\ 42.75 \\ 43.02} & \makecell{45.31 \\ 41.76 \\ 49.33} & \makecell{68.83 \\ 59.80 \\ 77.18} & \makecell{76.18 \\ 69.42 \\ 80.34} &
        \makecell{37.85 \\ 37.20 \\ 37.70} & \makecell{34.35 \\ 31.87 \\ 36.35} & \makecell{49.16 \\ 39.47 \\ 59.58} & \makecell{52.04 \\ 40.18 \\ 66.00} &
        \makecell{26.02 \\ 25.68 \\ 26.26} & \makecell{50.01 \\ 50.51 \\ 49.15} & \makecell{54.08 \\ 54.86 \\ 52.79} & \makecell{44.72 \\ 46.00 \\ 42.76} &
        \makecell{46.46 \\ 45.49 \\ 46.38} & \makecell{44.56 \\ 42.28 \\ 45.81} & \makecell{46.81 \\ 42.34 \\ 51.66} & \makecell{43.44 \\ 38.02 \\ 49.30} \\

        \addlinespace[3pt]

        F1 & \makecell{0.48 \\ 0.49 \\ 0.47} & \makecell{0.49 \\ 0.49 \\ 0.48} & \makecell{0.48 \\ 0.49 \\ 0.47} & \makecell{0.47 \\ 0.47 \\ 0.47} &
        \makecell{0.48 \\ 0.49 \\ 0.47} & \makecell{0.48 \\ 0.48 \\ 0.47} & \makecell{0.50 \\ 0.49 \\ 0.50} & \makecell{0.50 \\ 0.49 \\ 0.51} &
        \makecell{0.47 \\ 0.47 \\ 0.46} & \makecell{0.50 \\ 0.50 \\ 0.50} & \makecell{0.50 \\ 0.49 \\ 0.50} & \makecell{0.49 \\ 0.49 \\ 0.49} &
        \makecell{0.48 \\ 0.49 \\ 0.46} & \makecell{0.48 \\ 0.49 \\ 0.47} & \makecell{0.49 \\ 0.49 \\ 0.49} & \makecell{0.49 \\ 0.49 \\ 0.50} \\

        \addlinespace[3pt]
        \bottomrule
      \end{tabular}
      \vspace{0.1cm}
      \begin{tablenotes}[para,flushleft]
        \footnotesize
        \textbf{Note:} This table presents each metric as a set of three values, ordered from top to bottom: full sample, top 25\% of firms by market capitalization (large-cap), and bottom 25\% (small-cap) for various predictive models across different window sizes (5, 21, 252, and 512 trading days). The benchmark model is CatBoost, the representative benchmark. The time series foundation models (TSFMs) include Chronos (tiny, mini, small, base, and large) and TimesFM (version 1 with 200 million and version 2 with 500 million parameters). Zero-shot inference is performed using the pre-trained models released by the respective authors. Metrics are first computed separately for each calendar year using all stock-date observations within that year. The reported values represent the average of these yearly statistics. Metrics include out-of-sample $R^2$ ($\RsqOOS$), overall directional accuracy, upward and downward classification accuracy, and macro-averaged F1 score. `Overall Acc.' denotes overall directional accuracy, `Up Acc.' and `Down Acc.' represent the model's accuracy in predicting upward and downward excess returns respectively, and `F1' refers to the macro-averaged F1 score. Accuracy measures and $\RsqOOS$ are reported in percentage terms; macro-$F_1$ is reported on a zero-to-one scale. Values with five or more integer digits are shown in scientific notation.
      \end{tablenotes}
    \end{minipage}
    \end{adjustbox}
  \end{threeparttable}
\end{table}
\vspace*{\fill}

\begin{table} %% Fixed -------------------------------------------------------------------------
  \centering
  \begin{threeparttable}
    \begin{adjustbox}{width=0.96\textwidth, center}
    \scriptsize
    \captionsetup{width=\linewidth}
    \caption{Zero-Shot TSFMs - Portfolio Performance (Complete)}
    \label{PP_TSFM_zero_shot_full}
    \begin{minipage}{\linewidth}
      \renewcommand{\arraystretch}{1.2}
      \begin{tabular}{c|cccc|cccc|cccc|cccc}
        \toprule
        \textbf{Model} & \multicolumn{4}{c|}{\textbf{Benchmark}} & \multicolumn{4}{c|}{\textbf{Chronos (Tiny)}} & \multicolumn{4}{c|}{\textbf{Chronos (Mini)}} & \multicolumn{4}{c}{\textbf{Chronos (Small)}} \\
        \textbf{Window Size} & 5 & 21 & 252 & 512 & 5 & 21 & 252 & 512 & 5 & 21 & 252 & 512 & 5 & 21 & 252 & 512 \\
        \midrule
    AR & \makecell{42.35 \\ 26.80 \\ 15.54} & \makecell{46.29 \\ 28.47 \\ 17.82} & \makecell{46.27 \\ 29.18 \\ 17.09} & \makecell{46.98 \\ 29.43 \\ 17.55} &
    \makecell{-37.07 \\ -12.51 \\ -24.56} & \makecell{-17.31 \\ -3.68 \\ -13.63} & \makecell{9.15 \\ 9.42 \\ -0.27} & \makecell{14.25 \\ 11.20 \\ 3.05} &
    \makecell{-32.85 \\ -9.23 \\ -23.62} & \makecell{-15.50 \\ -1.13 \\ -14.37} & \makecell{20.17 \\ 14.44 \\ 5.73} & \makecell{21.09 \\ 15.40 \\ 5.69} &
    \makecell{-36.73 \\ -12.20 \\ -24.53} & \makecell{-14.36 \\ -1.27 \\ -13.10} & \makecell{6.53 \\ 7.63 \\ -1.09} & \makecell{7.64 \\ 8.98 \\ -1.34} \\ \addlinespace[3pt]

    SD & \makecell{7.43 \\ 14.15 \\ 12.11} & \makecell{6.89 \\ 13.89 \\ 12.38} & \makecell{6.88 \\ 13.72 \\ 12.32} & \makecell{7.35 \\ 13.82 \\ 12.32} &
    \makecell{8.11 \\ 11.79 \\ 13.34} & \makecell{6.07 \\ 12.00 \\ 13.52} & \makecell{5.42 \\ 12.69 \\ 13.58} & \makecell{5.28 \\ 12.86 \\ 13.45} &
    \makecell{8.00 \\ 11.91 \\ 13.73} & \makecell{6.69 \\ 12.05 \\ 13.75} & \makecell{5.44 \\ 12.50 \\ 13.74} & \makecell{5.57 \\ 12.51 \\ 13.52} &
    \makecell{8.13 \\ 11.75 \\ 13.70} & \makecell{6.79 \\ 11.62 \\ 14.46} & \makecell{7.56 \\ 11.48 \\ 15.53} & \makecell{8.14 \\ 10.63 \\ 15.62} \\ \addlinespace[3pt]

    SR & \makecell{5.70 \\ 1.89 \\ 1.28} & \makecell{6.72 \\ 2.05 \\ 1.44} & \makecell{6.72 \\ 2.13 \\ 1.39} & \makecell{6.39 \\ 2.13 \\ 1.42} &
    \makecell{-4.57 \\ -1.06 \\ -1.84} & \makecell{-2.85 \\ -0.31 \\ -1.01} & \makecell{1.69 \\ 0.74 \\ -0.02} & \makecell{2.70 \\ 0.87 \\ 0.23} &
    \makecell{-4.10 \\ -0.77 \\ -1.72} & \makecell{-2.32 \\ -0.09 \\ -1.05} & \makecell{3.71 \\ 1.16 \\ 0.42} & \makecell{3.79 \\ 1.23 \\ 0.42} &
    \makecell{-4.52 \\ -1.04 \\ -1.79} & \makecell{-2.12 \\ -0.11 \\ -0.91} & \makecell{0.86 \\ 0.66 \\ -0.07} & \makecell{0.94 \\ 0.84 \\ -0.09} \\ \addlinespace[3pt]

    Daily Return & \makecell{16.81 \\ 10.64 \\ 6.17} & \makecell{18.37 \\ 11.30 \\ 7.07} & \makecell{18.36 \\ 11.58 \\ 6.78} & \makecell{18.64 \\ 11.68 \\ 6.97} &
    \makecell{-14.71 \\ -4.97 \\ -9.74} & \makecell{-6.87 \\ -1.46 \\ -5.41} & \makecell{3.63 \\ 3.74 \\ -0.11} & \makecell{5.66 \\ 4.45 \\ 1.21} &
    \makecell{-13.03 \\ -3.66 \\ -9.37} & \makecell{-6.15 \\ -0.45 \\ -5.70} & \makecell{8.00 \\ 5.73 \\ 2.27} & \makecell{8.37 \\ 6.11 \\ 2.26} &
    \makecell{-14.57 \\ -4.84 \\ -9.73} & \makecell{-5.70 \\ -0.50 \\ -5.20} & \makecell{2.59 \\ 3.03 \\ -0.43} & \makecell{3.03 \\ 3.56 \\ -0.53} \\ \addlinespace[3pt]

    Max DD & \makecell{12.03 \\ 32.49 \\ 31.98} & \makecell{13.14 \\ 32.85 \\ 30.00} & \makecell{13.67 \\ 31.85 \\ 31.75} & \makecell{14.46 \\ 32.62 \\ 29.58} &
    \makecell{99.98 \\ 95.91 \\ 99.70} & \makecell{98.16 \\ 69.10 \\ 96.45} & \makecell{19.64 \\ 35.16 \\ 51.28} & \makecell{12.82 \\ 34.74 \\ 46.49} &
    \makecell{99.95 \\ 91.15 \\ 99.64} & \makecell{97.25 \\ 55.02 \\ 97.06} & \makecell{10.57 \\ 27.87 \\ 46.91} & \makecell{12.40 \\ 29.08 \\ 46.08} &
    \makecell{99.98 \\ 95.53 \\ 99.71} & \makecell{96.42 \\ 54.45 \\ 96.12} & \makecell{23.19 \\ 35.18 \\ 64.16} & \makecell{24.63 \\ 29.40 \\ 62.50} \\ \addlinespace[3pt]

    Max 1-day Loss & \makecell{4.86 \\ 8.28 \\ 5.56} & \makecell{3.75 \\ 8.21 \\ 4.49} & \makecell{4.15 \\ 8.19 \\ 5.03} & \makecell{4.93 \\ 8.19 \\ 5.27} &
    \makecell{5.34 \\ 6.31 \\ 7.84} & \makecell{4.91 \\ 6.85 \\ 7.68} & \makecell{5.00 \\ 6.96 \\ 7.60} & \makecell{5.45 \\ 6.98 \\ 7.40} &
    \makecell{12.18 \\ 7.23 \\ 11.65} & \makecell{4.67 \\ 6.55 \\ 7.99} & \makecell{4.89 \\ 6.85 \\ 7.87} & \makecell{12.24 \\ 6.64 \\ 11.65} &
    \makecell{12.11 \\ 6.78 \\ 11.65} & \makecell{4.98 \\ 6.82 \\ 7.81} & \makecell{4.29 \\ 7.06 \\ 7.25} & \makecell{4.31 \\ 6.64 \\ 6.64} \\ \addlinespace[3pt]

    Skew & \makecell{2.20 \\ -0.01 \\ 0.35} & \makecell{2.15 \\ 0.00 \\ 0.38} & \makecell{2.21 \\ 0.02 \\ 0.37} & \makecell{1.67 \\ -0.05 \\ 0.35} &
    \makecell{-0.77 \\ -0.41 \\ 0.15} & \makecell{-1.11 \\ -0.42 \\ 0.12} & \makecell{-1.49 \\ -0.46 \\ 0.06} & \makecell{-1.36 \\ -0.44 \\ 0.12} &
    \makecell{-2.97 \\ -0.44 \\ -0.29} & \makecell{-0.81 \\ -0.41 \\ 0.17} & \makecell{-1.05 \\ -0.43 \\ 0.12} & \makecell{-7.78 \\ -0.31 \\ -0.27} &
    \makecell{-2.77 \\ -0.47 \\ -0.26} & \makecell{-1.12 \\ -0.52 \\ 0.03} & \makecell{-0.66 \\ -0.53 \\ 0.04} & \makecell{-0.50 \\ -0.45 \\ 0.10} \\ \addlinespace[3pt]

    Kurt & \makecell{32.84 \\ 11.60 \\ 6.15} & \makecell{31.26 \\ 12.29 \\ 5.15} & \makecell{32.14 \\ 12.51 \\ 5.39} & \makecell{26.37 \\ 12.53 \\ 5.49} &
    \makecell{14.18 \\ 5.66 \\ 11.65} & \makecell{20.52 \\ 6.45 \\ 9.70} & \makecell{19.51 \\ 5.47 \\ 9.22} & \makecell{20.58 \\ 5.74 \\ 9.81} &
    \makecell{67.69 \\ 6.34 \\ 15.73} & \makecell{18.09 \\ 5.08 \\ 10.44} & \makecell{17.60 \\ 5.24 \\ 8.75} & \makecell{266.68 \\ 5.64 \\ 13.62} &
    \makecell{61.70 \\ 5.99 \\ 15.40} & \makecell{15.74 \\ 6.18 \\ 8.06} & \makecell{7.27 \\ 6.17 \\ 5.93} & \makecell{7.09 \\ 7.73 \\ 5.32} \\ \addlinespace[3pt]

        \midrule \midrule

        \textbf{Model} & \multicolumn{4}{c|}{\textbf{Chronos (Base)}} & \multicolumn{4}{c|}{\textbf{Chronos (Large)}} & \multicolumn{4}{c|}{\textbf{TimesFM 1 (200M)}} & \multicolumn{4}{c}{\textbf{TimesFM 2 (500M)}}  \\
        \textbf{Window Size} & 5 & 21 & 252 & 512 & 5 & 21 & 252 & 512 & 5 & 21 & 252 & 512 & 5 & 21 & 252 & 512 \\
        \midrule

    AR & \makecell{-31.94 \\ -10.04 \\ -21.90} & \makecell{-16.69 \\ -2.99 \\ -13.70} & \makecell{7.37 \\ 8.21 \\ -0.84} & \makecell{14.35 \\ 13.50 \\ 0.85} &
    \makecell{-30.59 \\ -8.95 \\ -21.64} & \makecell{-17.40 \\ -3.05 \\ -14.35} & \makecell{9.89 \\ 8.82 \\ 1.07} & \makecell{20.17 \\ 13.13 \\ 7.04} &
    \makecell{-6.83 \\ -1.42 \\ -5.41} & \makecell{1.05 \\ 6.66 \\ -5.61} & \makecell{-1.40 \\ 5.00 \\ -6.41} & \makecell{-4.85 \\ 3.06 \\ -7.91} &
    \makecell{-39.62 \\ -13.04 \\ -26.58} & \makecell{-33.33 \\ -11.11 \\ -22.22} & \makecell{-7.61 \\ 1.02 \\ -8.64} & \makecell{-1.47 \\ 4.73 \\ -6.19} \\ \addlinespace[3pt]

    SD & \makecell{7.65 \\ 12.07 \\ 13.61} & \makecell{6.74 \\ 11.94 \\ 14.00} & \makecell{8.31 \\ 10.58 \\ 15.63} & \makecell{8.52 \\ 10.10 \\ 15.63} &
    \makecell{7.66 \\ 12.20 \\ 13.61} & \makecell{6.90 \\ 11.90 \\ 13.71} & \makecell{7.62 \\ 10.96 \\ 14.97} & \makecell{6.91 \\ 10.72 \\ 14.27} &
    \makecell{6.40 \\ 12.98 \\ 11.37} & \makecell{6.49 \\ 11.85 \\ 13.27} & \makecell{5.06 \\ 11.06 \\ 12.88} & \makecell{5.33 \\ 11.26 \\ 12.63} &
    \makecell{9.52 \\ 11.79 \\ 14.32} & \makecell{9.23 \\ 11.57 \\ 14.53} & \makecell{8.55 \\ 11.36 \\ 14.89} & \makecell{7.95 \\ 11.57 \\ 14.86} \\ \addlinespace[3pt]

    SR & \makecell{-4.18 \\ -0.83 \\ -1.61} & \makecell{-2.48 \\ -0.25 \\ -0.98} & \makecell{0.89 \\ 0.78 \\ -0.05} & \makecell{1.68 \\ 1.34 \\ 0.05} &
    \makecell{-3.99 \\ -0.73 \\ -1.59} & \makecell{-2.52 \\ -0.26 \\ -1.05} & \makecell{1.30 \\ 0.80 \\ 0.07} & \makecell{2.92 \\ 1.22 \\ 0.49} &
    \makecell{-1.07 \\ -0.11 \\ -0.48} & \makecell{0.16 \\ 0.56 \\ -0.42} & \makecell{-0.28 \\ 0.45 \\ -0.50} & \makecell{-0.91 \\ 0.27 \\ -0.63} &
    \makecell{-4.16 \\ -1.11 \\ -1.86} & \makecell{-3.61 \\ -0.96 \\ -1.53} & \makecell{-0.89 \\ 0.09 \\ -0.58} & \makecell{-0.18 \\ 0.41 \\ -0.42} \\ \addlinespace[3pt]

    Daily Return & \makecell{-12.68 \\ -3.98 \\ -8.69} & \makecell{-6.62 \\ -1.19 \\ -5.44} & \makecell{2.92 \\ 3.26 \\ -0.33} & \makecell{5.69 \\ 5.36 \\ 0.34} &
    \makecell{-12.14 \\ -3.55 \\ -8.59} & \makecell{-6.91 \\ -1.21 \\ -5.69} & \makecell{3.92 \\ 3.50 \\ 0.43} & \makecell{8.00 \\ 5.21 \\ 2.79} &
    \makecell{-2.71 \\ -0.56 \\ -2.15} & \makecell{0.42 \\ 2.64 \\ -2.23} & \makecell{-0.56 \\ 1.99 \\ -2.54} & \makecell{-1.93 \\ 1.22 \\ -3.14} &
    \makecell{-15.72 \\ -5.18 \\ -10.55} & \makecell{-13.23 \\ -4.41 \\ -8.82} & \makecell{-3.02 \\ 0.41 \\ -3.43} & \makecell{-0.58 \\ 1.88 \\ -2.46} \\ \addlinespace[3pt]

    Max DD & \makecell{99.94 \\ 92.83 \\ 99.46} & \makecell{97.90 \\ 64.37 \\ 96.52} & \makecell{23.95 \\ 31.04 \\ 65.64} & \makecell{19.72 \\ 27.80 \\ 53.41} &
    \makecell{99.92 \\ 90.86 \\ 99.42} & \makecell{98.22 \\ 64.45 \\ 96.99} & \makecell{19.87 \\ 38.92 \\ 53.10} & \makecell{15.80 \\ 28.55 \\ 45.85} &
    \makecell{81.06 \\ 55.51 \\ 76.29} & \makecell{31.36 \\ 32.75 \\ 81.59} & \makecell{32.50 \\ 35.81 \\ 81.81} & \makecell{70.13 \\ 41.90 \\ 87.24} &
    \makecell{99.99 \\ 96.33 \\ 99.82} & \makecell{99.96 \\ 94.25 \\ 99.50} & \makecell{84.86 \\ 54.11 \\ 90.14} & \makecell{46.40 \\ 45.57 \\ 84.32} \\ \addlinespace[3pt]

    Max 1-day Loss & \makecell{4.99 \\ 6.99 \\ 8.34} & \makecell{4.98 \\ 6.44 \\ 8.17} & \makecell{3.93 \\ 6.18 \\ 7.42} & \makecell{5.97 \\ 6.01 \\ 6.25} &
    \makecell{12.08 \\ 6.97 \\ 11.65} & \makecell{4.60 \\ 6.39 \\ 7.72} & \makecell{5.05 \\ 5.63 \\ 7.53} & \makecell{4.31 \\ 5.42 \\ 6.21} &
    \makecell{3.62 \\ 6.92 \\ 6.02} & \makecell{3.97 \\ 7.07 \\ 6.99} & \makecell{3.84 \\ 5.59 \\ 7.15} & \makecell{4.83 \\ 5.66 \\ 7.86} &
    \makecell{7.12 \\ 7.85 \\ 9.23} & \makecell{7.88 \\ 6.71 \\ 9.22} & \makecell{11.93 \\ 6.29 \\ 11.65} & \makecell{12.19 \\ 6.28 \\ 11.65} \\ \addlinespace[3pt]

    Skew & \makecell{-0.79 \\ -0.43 \\ 0.12} & \makecell{-0.97 \\ -0.48 \\ 0.08} & \makecell{-0.57 \\ -0.52 \\ 0.02} & \makecell{-0.54 \\ -0.48 \\ 0.08} &
    \makecell{-3.18 \\ -0.41 \\ -0.30} & \makecell{-1.03 \\ -0.50 \\ 0.12} & \makecell{-0.84 \\ -0.55 \\ 0.08} & \makecell{-0.72 \\ -0.43 \\ 0.15} &
    \makecell{-0.08 \\ -0.30 \\ 0.37} & \makecell{-0.45 \\ -0.21 \\ 0.12} & \makecell{-0.34 \\ -0.43 \\ 0.21} & \makecell{-0.14 \\ -0.37 \\ 0.14} &
    \makecell{-1.51 \\ -0.59 \\ 0.02} & \makecell{-1.68 \\ -0.65 \\ 0.01} & \makecell{-3.27 \\ -0.66 \\ -0.30} & \makecell{-3.91 \\ -0.56 \\ -0.29} \\ \addlinespace[3pt]

    Kurt & \makecell{13.79 \\ 6.12 \\ 10.77} & \makecell{15.52 \\ 5.80 \\ 9.53} & \makecell{6.44 \\ 7.22 \\ 5.53} & \makecell{6.74 \\ 9.55 \\ 4.58} &
    \makecell{74.25 \\ 6.04 \\ 15.77} & \makecell{16.83 \\ 5.61 \\ 10.66} & \makecell{11.78 \\ 5.34 \\ 7.17} & \makecell{9.52 \\ 5.51 \\ 6.42} &
    \makecell{8.62 \\ 6.50 \\ 8.65} & \makecell{11.93 \\ 8.34 \\ 7.15} & \makecell{17.94 \\ 5.81 \\ 8.65} & \makecell{35.83 \\ 6.21 \\ 9.81} &
    \makecell{19.07 \\ 6.41 \\ 12.69} & \makecell{21.97 \\ 5.95 \\ 12.29} & \makecell{59.69 \\ 5.33 \\ 14.20} & \makecell{81.67 \\ 5.03 \\ 14.08} \\ \addlinespace[3pt]
        \bottomrule
      \end{tabular}
      \vspace{0.1cm}
      \begin{tablenotes}[para,flushleft]
        \footnotesize
        \textbf{Note:} This table reports portfolio performance metrics computed from the full daily out-of-sample excess return series across different input window lengths (5, 21, 252, and 512 trading days) for each model. The benchmark model is CatBoost, the representative benchmark. The time series foundation models (TSFMs) include Chronos (tiny, mini, small, base, and large) and TimesFM (version 1 with 200 million and version 2 with 500 million parameters). Zero-shot inference is performed using the pre-trained models released by the respective authors. Each cell displays three values from top to bottom: the long--short portfolio, the long-leg contribution, and the short-leg contribution. The long leg equals one half of the top-decile return, while the short leg is the signed contribution from shorting the bottom decile and equals minus one half of the bottom-decile return. Metrics include annualized return (AR), standard deviation (SD), Sharpe ratio (SR), daily return (in basis points), maximum drawdown (Max DD), maximum one-day loss (Max 1-day Loss), skewness, and excess kurtosis of portfolio returns. AR, SD, maximum drawdown, and maximum one-day loss are reported in percentage terms. Portfolios are formed using decile sorting based on model forecasts, with equal weighting across stocks.
      \end{tablenotes}
    \end{minipage}
    \end{adjustbox}
  \end{threeparttable}
\end{table}

\clearpage
\vspace*{\fill} % Fixed ---------------------------------------
\begin{figure}[ht!]
    \caption{Cumulative Log Returns of Zero-Shot TSFMs: Long--Short Portfolios}
    \label{Cumulative_log_return_zero_shot_long_and_short}
    \centering
    \includegraphics[
        width=0.71\textwidth,
        clip,
        trim=0 35 0 28, % left bottom right top
    ]{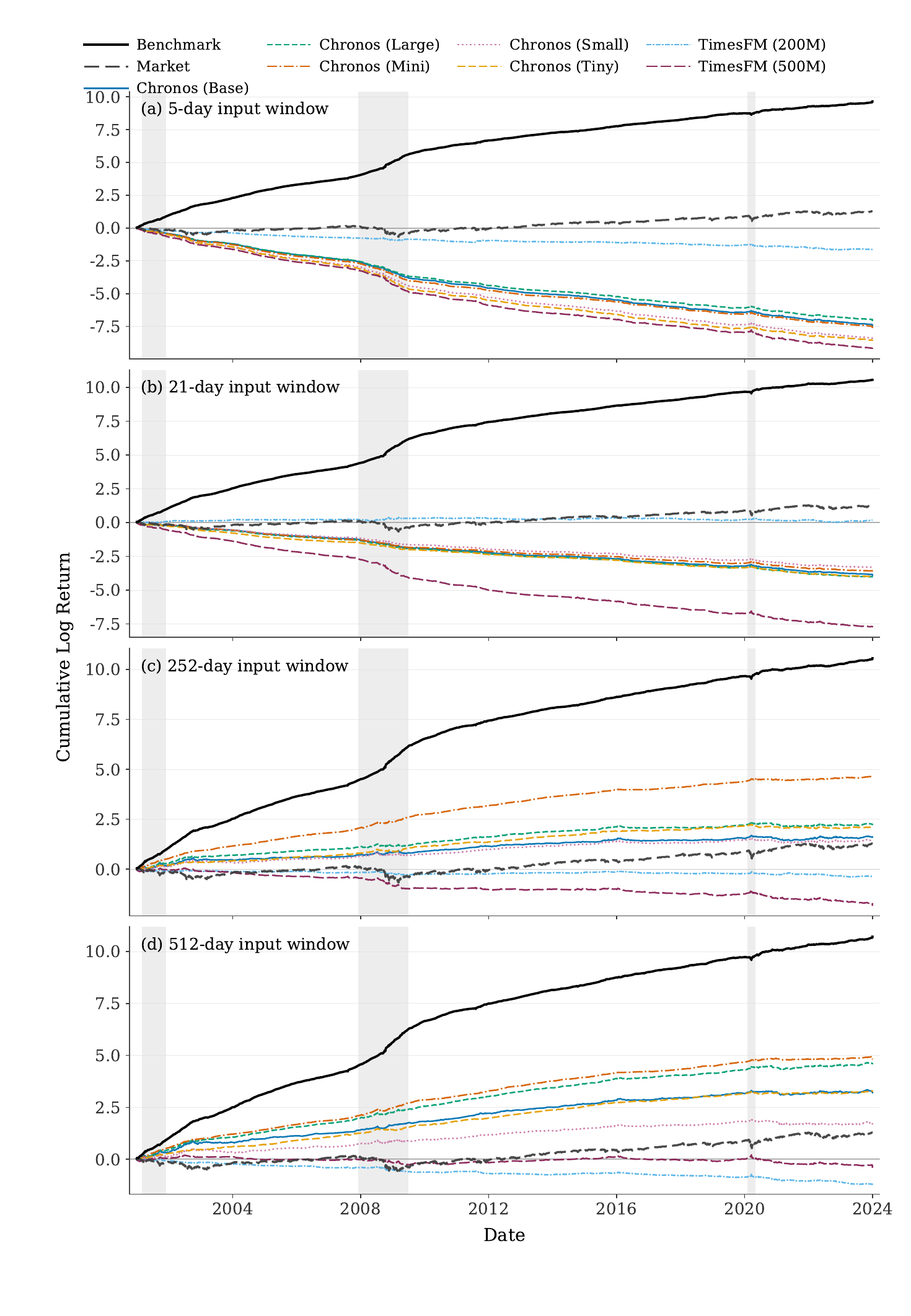} % Adjust the width and trims
    \begin{minipage}{0.8\textwidth}
        \footnotesize
        \textbf{Note:} This figure displays the cumulative log returns of long--short portfolios constructed from various forecasting models with input windows of 5 (a), 21 (b), 252 (c), and 512 (d) trading days. The benchmark model is CatBoost, the representative benchmark. The time series foundation models (TSFMs) include Chronos (tiny, mini, small, base, and large) and TimesFM (version 1 with 200 million and version 2 with 500 million parameters). Zero-shot inference is performed using the pre-trained models released by the respective authors. Each subplot corresponds to a specific input window length, as indicated by the text labels in the upper-left corners. The benchmark model (CatBoost) is highlighted in black with bold lines. The dashed black line represents the cumulative log return of the market (S\&P 500). Shaded areas indicate U.S. recession periods, as defined by the National Bureau of Economic Research (NBER). The model-based long--short portfolios use equal weighting within each leg and are scaled to unit gross exposure. The S\&P 500 is shown separately as a market reference.

    \end{minipage}
\end{figure}
\vspace*{\fill}
\clearpage

\vspace*{\fill}
\begin{table}[htbp]
  \centering
  \begin{threeparttable}
    \begin{adjustbox}{width=0.96\textwidth, center}
    \scriptsize
    \captionsetup{width=\linewidth}
    \caption{Zero-Shot TSFMs - Spread Portfolio Performance}
    \label{Spread_Portfolio_zero_shot_table}
    \begin{minipage}{\linewidth}
      \renewcommand{\arraystretch}{1.2}
      \begin{tabular}{c|cccc|cccc|cccc|cccc}
        \toprule
        \textbf{Model} & \multicolumn{4}{c|}{\textbf{Benchmark}} & \multicolumn{4}{c|}{\textbf{Chronos (Tiny)}} & \multicolumn{4}{c|}{\textbf{Chronos (Mini)}} & \multicolumn{4}{c}{\textbf{Chronos (Small)}}\\
        \textbf{Window Size} & 5 & 21 & 252 & 512 & 5 & 21 & 252 & 512 & 5 & 21 & 252 & 512 & 5 & 21 & 252 & 512 \\
        \midrule
        Low & -31.09 & -35.64 & -34.19 & -35.11 & 49.11 & 27.27 & 0.54 & -6.10 & 47.24 & 28.75 & -11.45 & -11.38 & 49.05 & 26.20 & 2.19 & 2.69  \\
        \addlinespace[3pt]
        2 &  -0.80 & -2.89 & -3.83 & -3.98 & 25.05 & 17.80 & 5.40 & 4.73 & 24.98 & 19.14 & 4.34 & 3.80 & 25.48 & 19.25 & 8.55 & 7.73  \\
        \addlinespace[3pt]
        3 & 4.33 & 4.56 & 2.80 & 3.33 & 18.50 & 14.73 & 8.36 & 9.44 & 17.84 & 15.43 & 8.18 & 8.65 & 19.28 & 15.83 & 10.24 & 8.67 \\
        \addlinespace[3pt]
        4 & 7.51 & 8.36 & 6.40 & 6.87 & 14.27 & 13.41 & 11.19 & 10.17 & 13.14 & 13.44 & 10.22 & 9.47 & 14.75 & 14.38 & 11.22 & 10.71 \\
        \addlinespace[3pt]
        5 & 9.40 & 10.27 & 9.88 & 9.40 & 13.13 & 13.08 & 12.20 & 12.07 & 11.26 & 11.95 & 11.25 & 10.44 & 12.34 & 12.16 & 11.70 & 10.76 \\
        \addlinespace[3pt]
        6 & 11.57 & 11.44 & 12.00 & 12.43 & 9.56 & 12.18 & 11.84 & 13.62 & 9.13 & 8.98 & 13.17 & 11.94 & 9.03 & 10.01 & 13.46 & 12.42 \\
        \addlinespace[3pt]
        7 & 15.27 & 15.26 & 15.66 & 14.08 & 7.78 & 9.34 & 14.21 & 13.78 & 6.37 & 7.39 & 14.17 & 13.97 & 7.32 & 8.06 & 12.76 & 12.91 \\
        \addlinespace[3pt]
        8 & 18.24 & 18.67 & 18.31 & 19.43 & 4.42 & 8.14 & 14.36 & 15.56 & 3.54 & 6.16 & 15.74 & 15.69 & 3.29 & 5.25 & 13.96 & 14.63\\
        \addlinespace[3pt]
        9 & 25.06 & 26.11 & 27.71 & 27.79 & -3.73 & 4.49 & 16.13 & 17.41 & -2.00 & 4.09 & 18.60 & 19.73 & -3.08 & 4.46 & 13.74 & 14.60  \\
        \addlinespace[3pt]
        High & 53.61 & 56.94 & 58.36 & 58.85 & -25.03 & -7.36 & 18.84 & 22.41 & -18.45 & -2.26 & 28.89 & 30.80 & -24.40 & -2.53 & 15.26 & 17.97 \\
        \addlinespace[3pt]
        $\tfrac{1}{2}(H-L)$ & 42.35 & 46.29 & 46.27 & 46.98 & -37.07 & -17.31 & 9.15 & 14.25 & -32.85 & -15.50 & 20.17 & 21.09 & -36.73 & -14.36 & 6.53 & 7.64 \\

        \midrule \midrule

        \textbf{Model} & \multicolumn{4}{c|}{\textbf{Chronos (Base)}} & \multicolumn{4}{c|}{\textbf{Chronos (Large)}} & \multicolumn{4}{c|}{\textbf{TimesFM 1 (200M)}} & \multicolumn{4}{c}{\textbf{TimesFM 2 (500M)}} \\
        \textbf{Window Size} & 5 & 21 & 252 & 512 & 5 & 21 & 252 & 512 & 5 & 21 & 252 & 512 & 5 & 21 & 252 & 512  \\
        \midrule
        Low & 43.81 & 27.40 & 1.68 & -1.70 & 43.28 & 28.69 & -2.15 & -14.07 & 10.83 & 11.23 & 12.81 & 15.83 & 53.16 & 44.44 & 17.27 & 12.38 \\
        \addlinespace[3pt]
        2 & 23.50 & 21.61 & 10.90 & 7.02 & 22.98 & 19.83 & 7.65 & 4.86 & 15.88 & 10.68 & 12.12 & 14.54 & 31.51 & 29.18 & 19.41 & 15.22  \\
        \addlinespace[3pt]
        3 & 18.66 & 16.62 & 10.84 & 8.70 & 17.16 & 16.90 & 10.79 & 9.43 & 15.39 & 10.96 & 11.90 & 14.32 & 22.96 & 22.12 & 16.66 & 15.10   \\
        \addlinespace[3pt]
        4 & 15.58 & 14.36 & 11.88 & 10.14 & 14.23 & 14.21 & 11.85 & 12.10 & 14.94 & 9.92 & 11.45 & 12.71 & 16.82 & 17.38 & 15.04 & 14.50  \\
        \addlinespace[3pt]
        5 & 12.65 & 11.92 & 11.86 & 10.38 & 12.13 & 12.11 & 12.38 & 12.42 & 14.42 & 9.95 & 11.41 & 12.19 & 11.94 & 13.17 & 13.36 & 11.67 \\
        \addlinespace[3pt]
        6 & 9.54 & 11.16 & 12.06 & 10.28 & 9.55 & 10.63 & 13.08 & 13.68 & 14.05 & 11.36 & 11.19 & 10.11 & 7.91 & 9.33 & 10.49 & 10.99  \\
        \addlinespace[3pt]
        7 & 7.90 & 8.13 & 11.66 & 11.62 & 8.00 & 8.66 & 13.28 & 14.13 & 12.43 & 10.77 & 11.05 & 10.40 & 2.98 & 5.03 & 9.02 & 9.37  \\
        \addlinespace[3pt]
        8 & 3.66 & 5.39 & 12.37 & 13.26 & 4.88 & 5.99 & 13.17 & 15.68 & 11.10 & 11.94 & 11.00 & 9.37 & -0.71 & 0.77 & 6.10 & 7.70 \\
        \addlinespace[3pt]
        9 & -2.14 & 2.46 & 13.43 & 16.39 & -1.24 & 2.15 & 15.42 & 18.63 & 6.91 & 12.95 & 10.14 & 7.47 & -7.43 & -6.40 & 3.68 & 6.70  \\
        \addlinespace[3pt]
        High & -20.08 & -5.98 & 16.42 & 27.00 & -17.91 & -6.11 & 17.63 & 26.26 & -2.83 & 13.32 & 10.01 & 6.12 & -26.09 & -22.22 & 2.05 & 9.45  \\
        \addlinespace[3pt]
        $\tfrac{1}{2}(H-L)$ & -31.94 & -16.69 & 7.37 & 14.35 & -30.59 & -17.40 & 9.89 & 20.17 & -6.83 & 1.05 & -1.40 & -4.85 & -39.62 & -33.33 & -7.61 & -1.47  \\
        \bottomrule
      \end{tabular}
      \vspace{0.1cm}
      \begin{tablenotes}[para,flushleft]
        \footnotesize
        \textbf{Note:} This table presents the annualized average returns of decile spread portfolios formed using forecasts from various predictive models across different window sizes (5, 21, 252, and 512 trading days). The benchmark model is CatBoost, the representative benchmark. The time series foundation models (TSFMs) include Chronos (tiny, mini, small, base, and large) and TimesFM (version 1 with 200 million and version 2 with 500 million parameters). Zero-shot inference is performed using the pre-trained models released by the respective authors. For each model and window size, we report standalone decile returns from Low to High. Here, $H$ and $L$ denote the standalone returns of the highest- and lowest-forecast deciles. The final row reports the unit-gross long--short return, $\frac{1}{2}(H-L)$, with long- and short-leg contributions of $\frac{1}{2}H$ and $-\frac{1}{2}L$, respectively. Portfolios are constructed using equal-weighted (EW) decile sorting based on predicted returns.
      \end{tablenotes}
    \end{minipage}
    \end{adjustbox}
  \end{threeparttable}
\end{table}
\vspace*{\fill}

\clearpage
\subsection{Complete Fine-Tuned TSFM Results} \label{complete_fine_tuned_tsfm_results}
This subsection presents complete fine-tuned TSFM results for both forecasting performance and portfolio performance across all models and window sizes.
\vspace*{\fill}
\begin{table}[h] %% Fixed -------------------------------------------------------------------------
  \centering
  \begin{threeparttable}
    \begin{adjustbox}{width=0.96\textwidth, center}
    \scriptsize
    \captionsetup{width=\linewidth}
    \caption{Fine-Tuned TSFMs - Forecasting Performance (Complete)}
    \label{FP_TSFM_fine_tuned_full}
    \begin{minipage}{\linewidth}
        \renewcommand{\arraystretch}{1.3}
      \begin{tabular}{c|cccc|cccc|cccc|cccc}
        \toprule
        \textbf{Model} & \multicolumn{4}{c|}{\textbf{Benchmark}} & \multicolumn{4}{c|}{\textbf{Chronos (Tiny)}} & \multicolumn{4}{c|}{\textbf{Chronos (Mini)}} & \multicolumn{4}{c}{\textbf{Chronos (Small)}} \\
        \textbf{Window Size} & 5 & 21 & 252 & 512 & 5 & 21 & 252 & 512 & 5 & 21 & 252 & 512 & 5 & 21 & 252 & 512   \\

        \midrule
        $\RsqOOS$ & \makecell{-0.25 \\ -0.50 \\ 0.39} & \makecell{-0.04 \\ -0.38 \\ 0.66} & \makecell{-0.03 \\ -0.32 \\ 0.62} & \makecell{-0.03 \\ -0.28 \\ 0.59} &
        \makecell{-284.11 \\ -286.64 \\ -290.52} & \makecell{-260.19 \\ -293.73 \\ -233.98} & \makecell{-204.18 \\ -282.37 \\ -140.10} & \makecell{-206.36 \\ -300.83 \\ -134.05} &
        \makecell{-562.12 \\ -565.53 \\ -578.12} & \makecell{-127.03 \\ -149.73 \\ -108.14} & \makecell{-75.19 \\ -115.26 \\ -47.11} & \makecell{-71.96 \\ -117.61 \\ -43.00} &
        \makecell{-288.93 \\ -268.76 \\ -322.02} & \makecell{-86.10 \\ -97.87 \\ -75.73} & \makecell{-58.67 \\ -78.13 \\ -42.60} & \makecell{-51.89 \\ -70.01 \\ -38.35} \\

        \addlinespace[3pt]

        Overall Acc. & \makecell{50.94 \\ 50.70 \\ 51.96} & \makecell{51.12 \\ 50.74 \\ 52.47} & \makecell{51.08 \\ 50.80 \\ 52.33} & \makecell{51.16 \\ 50.95 \\ 52.39} &
        \makecell{49.69 \\ 50.53 \\ 48.60} & \makecell{49.71 \\ 50.83 \\ 48.30} & \makecell{49.77 \\ 50.92 \\ 48.45} & \makecell{49.82 \\ 50.95 \\ 48.54} &
        \makecell{49.86 \\ 50.55 \\ 49.05} & \makecell{49.86 \\ 50.81 \\ 48.81} & \makecell{50.02 \\ 50.81 \\ 49.32} & \makecell{50.04 \\ 50.80 \\ 49.38} &
        \makecell{49.92 \\ 50.40 \\ 49.46} & \makecell{49.79 \\ 50.78 \\ 48.65} & \makecell{49.86 \\ 50.82 \\ 48.79} & \makecell{49.84 \\ 50.76 \\ 48.79} \\

        \addlinespace[3pt]

        Up Acc. & \makecell{68.96 \\ 72.05 \\ 68.00} & \makecell{67.82 \\ 72.97 \\ 65.11} & \makecell{66.87 \\ 71.06 \\ 64.65} & \makecell{69.14 \\ 74.39 \\ 66.07} &
        \makecell{76.67 \\ 77.47 \\ 75.25} & \makecell{87.92 \\ 89.46 \\ 84.75} & \makecell{90.35 \\ 94.07 \\ 83.13} & \makecell{90.30 \\ 94.42 \\ 82.36} &
        \makecell{72.87 \\ 74.20 \\ 70.44} & \makecell{84.18 \\ 87.64 \\ 77.64} & \makecell{79.70 \\ 87.99 \\ 66.93} & \makecell{78.87 \\ 87.85 \\ 65.46} &
        \makecell{67.87 \\ 70.14 \\ 63.73} & \makecell{85.31 \\ 88.00 \\ 80.14} & \makecell{84.45 \\ 88.76 \\ 77.32} & \makecell{82.83 \\ 86.77 \\ 76.24} \\

        \addlinespace[3pt]

        Down Acc. & \makecell{33.14 \\ 28.11 \\ 37.42} & \makecell{34.60 \\ 27.25 \\ 41.01} & \makecell{35.59 \\ 29.59 \\ 41.24} & \makecell{33.37 \\ 26.24 \\ 39.86} &
        \makecell{23.35 \\ 22.48 \\ 24.80} & \makecell{12.34 \\ 10.52 \\ 15.68} & \makecell{10.09 \\ 5.93 \\ 17.36} & \makecell{10.20 \\ 5.60 \\ 18.18} &
        \makecell{27.17 \\ 25.68 \\ 29.63} & \makecell{16.25 \\ 12.35 \\ 22.88} & \makecell{21.01 \\ 12.10 \\ 33.54} & \makecell{21.90 \\ 12.25 \\ 35.02} &
        \makecell{32.25 \\ 29.72 \\ 36.44} & \makecell{15.03 \\ 11.97 \\ 20.31} & \makecell{16.02 \\ 11.30 \\ 23.10} & \makecell{17.57 \\ 13.27 \\ 24.10} \\

        \addlinespace[3pt]

        F1 & \makecell{0.49 \\ 0.47 \\ 0.51} & \makecell{0.49 \\ 0.47 \\ 0.52} & \makecell{0.50 \\ 0.48 \\ 0.52} & \makecell{0.49 \\ 0.47 \\ 0.51} &
        \makecell{0.46 \\ 0.46 \\ 0.46} & \makecell{0.42 \\ 0.41 \\ 0.42} & \makecell{0.40 \\ 0.38 \\ 0.43} & \makecell{0.40 \\ 0.38 \\ 0.43} &
        \makecell{0.47 \\ 0.47 \\ 0.47} & \makecell{0.43 \\ 0.42 \\ 0.45} & \makecell{0.45 \\ 0.42 \\ 0.48} & \makecell{0.46 \\ 0.42 \\ 0.48} &
        \makecell{0.48 \\ 0.48 \\ 0.49} & \makecell{0.43 \\ 0.42 \\ 0.44} & \makecell{0.43 \\ 0.42 \\ 0.45} & \makecell{0.44 \\ 0.42 \\ 0.46} \\

        \addlinespace[3pt]

        \midrule \midrule
        \textbf{Model} & \multicolumn{4}{c|}{\textbf{Chronos (Base)}} & \multicolumn{4}{c|}{\textbf{Chronos (Large)}} & \multicolumn{4}{c|}{\textbf{TimesFM 1 (200M)}} & \multicolumn{4}{c}{\textbf{TimesFM 2 (500M)}} \\
        \textbf{Window Size} & 5 & 21 & 252 & 512 & 5 & 21 & 252 & 512 & 5 & 21 & 252 & 512 & 5 & 21 & 252 & 512  \\
        \midrule
        $\RsqOOS$ & \makecell{-1507.22 \\ -1496.04 \\ -1548.20} & \makecell{-11.16 \\ -11.72 \\ -10.78} & \makecell{-4.81 \\ -5.02 \\ -4.67} & \makecell{-5.42 \\ -5.74 \\ -5.22} &
        \makecell{-17.02 \\ -16.42 \\ -18.34} & \makecell{-2.30 \\ -2.01 \\ -2.74} & \makecell{-1.85 \\ -1.35 \\ -2.31} & \makecell{-1.89 \\ -1.41 \\ -2.40} &
        \makecell{-29.46 \\ -29.60 \\ -30.56} & \makecell{-15.74 \\ -14.54 \\ -17.09} & \makecell{-5.19 \\ -5.53 \\ -4.99} & \makecell{-5.02 \\ -5.63 \\ -4.59} &
        \makecell{-529.52 \\ -520.53 \\ -560.53} & \makecell{-562.05 \\ -561.35 \\ -554.68} & \makecell{-175.47 \\ -141.03 \\ -176.16} & \makecell{-146.79 \\ -124.07 \\ -145.57} \\

        \addlinespace[3pt]

        Overall Acc. & \makecell{50.03 \\ 50.00 \\ 50.14} & \makecell{49.95 \\ 50.19 \\ 49.64} & \makecell{49.99 \\ 49.84 \\ 50.10} & \makecell{49.99 \\ 49.94 \\ 49.95} &
        \makecell{49.97 \\ 49.98 \\ 50.02} & \makecell{50.24 \\ 49.85 \\ 50.86} & \makecell{50.27 \\ 49.83 \\ 50.98} & \makecell{50.27 \\ 49.79 \\ 50.98} &
        \makecell{48.72 \\ 49.28 \\ 47.34} & \makecell{49.21 \\ 49.78 \\ 48.19} & \makecell{49.82 \\ 50.23 \\ 49.38} & \makecell{49.88 \\ 50.19 \\ 49.44} &
        \makecell{48.93 \\ 49.41 \\ 47.82} & \makecell{49.47 \\ 50.00 \\ 48.59} & \makecell{50.00 \\ 50.38 \\ 49.55} & \makecell{49.97 \\ 50.29 \\ 49.53} \\

        \addlinespace[3pt]

        Up Acc. & \makecell{49.62 \\ 50.66 \\ 47.75} & \makecell{58.73 \\ 59.56 \\ 57.33} & \makecell{43.83 \\ 41.55 \\ 47.21} & \makecell{48.48 \\ 46.29 \\ 51.17} &
        \makecell{49.72 \\ 50.37 \\ 48.71} & \makecell{36.13 \\ 36.72 \\ 35.53} & \makecell{35.33 \\ 36.89 \\ 33.22} & \makecell{35.02 \\ 36.41 \\ 33.17} &
        \makecell{50.95 \\ 53.11 \\ 47.67} & \makecell{56.60 \\ 59.11 \\ 54.09} & \makecell{59.55 \\ 61.65 \\ 57.84} & \makecell{59.89 \\ 62.52 \\ 57.49} &
        \makecell{47.30 \\ 49.37 \\ 44.31} & \makecell{55.10 \\ 57.45 \\ 52.64} & \makecell{57.48 \\ 59.77 \\ 55.43} & \makecell{57.58 \\ 60.01 \\ 55.66} \\

        \addlinespace[3pt]

        Down Acc. & \makecell{50.22 \\ 49.12 \\ 51.94} & \makecell{41.31 \\ 40.38 \\ 42.67} & \makecell{55.96 \\ 58.39 \\ 52.62} & \makecell{51.35 \\ 53.61 \\ 48.73} &
        \makecell{50.01 \\ 49.30 \\ 50.85} & \makecell{63.85 \\ 63.26 \\ 64.38} & \makecell{64.72 \\ 63.06 \\ 66.71} & \makecell{65.02 \\ 63.53 \\ 66.75} &
        \makecell{46.44 \\ 45.10 \\ 46.97} & \makecell{41.99 \\ 39.93 \\ 43.07} & \makecell{40.56 \\ 38.55 \\ 42.11} & \makecell{40.34 \\ 37.56 \\ 42.49} &
        \makecell{50.32 \\ 49.10 \\ 50.83} & \makecell{43.67 \\ 41.74 \\ 44.84} & \makecell{42.62 \\ 40.41 \\ 44.37} & \makecell{42.46 \\ 39.97 \\ 44.09} \\

        \addlinespace[3pt]

        F1 & \makecell{0.49 \\ 0.49 \\ 0.49} & \makecell{0.50 \\ 0.50 \\ 0.49} & \makecell{0.50 \\ 0.49 \\ 0.50} & \makecell{0.50 \\ 0.50 \\ 0.50} &
        \makecell{0.49 \\ 0.49 \\ 0.49} & \makecell{0.49 \\ 0.49 \\ 0.49} & \makecell{0.49 \\ 0.48 \\ 0.49} & \makecell{0.49 \\ 0.48 \\ 0.49} &
        \makecell{0.49 \\ 0.49 \\ 0.47} & \makecell{0.49 \\ 0.49 \\ 0.48} & \makecell{0.49 \\ 0.49 \\ 0.49} & \makecell{0.49 \\ 0.49 \\ 0.49} &
        \makecell{0.48 \\ 0.48 \\ 0.46} & \makecell{0.47 \\ 0.48 \\ 0.47} & \makecell{0.48 \\ 0.48 \\ 0.48} & \makecell{0.48 \\ 0.48 \\ 0.47} \\

        \addlinespace[3pt]
        \bottomrule
      \end{tabular}
      \vspace{0.1cm}
      \begin{tablenotes}[para,flushleft]
        \footnotesize
        \textbf{Note:} This table presents each metric as a set of three values, ordered from top to bottom: full sample, top 25\% of firms by market capitalization (large-cap), and bottom 25\% (small-cap) for various predictive models across different window sizes (5, 21, 252, and 512 trading days). The benchmark model is CatBoost, the representative benchmark. The time series foundation models (TSFMs) include Chronos (tiny, mini, small, base, and large) and TimesFM (version 1 with 200 million and version 2 with 500 million parameters). The models released by the respective authors are fine-tuned on an annual basis. Metrics are first computed separately for each calendar year using all stock-date observations within that year. The reported values represent the average of these yearly statistics. Metrics include out-of-sample $R^2$ ($\RsqOOS$), overall directional accuracy, upward and downward classification accuracy, and macro-averaged F1 score. `Overall Acc.' denotes overall directional accuracy, `Up Acc.' and `Down Acc.' represent the model's accuracy in predicting upward and downward excess returns respectively, and `F1' refers to the macro-averaged F1 score. Accuracy measures and $\RsqOOS$ are reported in percentage terms; macro-$F_1$ is reported on a zero-to-one scale.
      \end{tablenotes}
    \end{minipage}
    \end{adjustbox}
  \end{threeparttable}
\end{table}
\vspace*{\fill}

\begin{table} %% Fixed -------------------------------------------------------------------------
  \centering
  \begin{threeparttable}
    \begin{adjustbox}{width=0.96\textwidth, center}
    \scriptsize
    \captionsetup{width=\linewidth}
    \caption{Fine-Tuned TSFMs - Portfolio Performance (Complete)}
    \label{PP_TSFM_fine_tuned_full}
    \begin{minipage}{\linewidth}
      \renewcommand{\arraystretch}{1.2}
      \begin{tabular}{c|cccc|cccc|cccc|cccc}
        \toprule
        \textbf{Model} & \multicolumn{4}{c|}{\textbf{Benchmark}} & \multicolumn{4}{c|}{\textbf{Chronos (Tiny)}} & \multicolumn{4}{c|}{\textbf{Chronos (Mini)}} & \multicolumn{4}{c}{\textbf{Chronos (Small)}} \\
        \textbf{Window Size} & 5 & 21 & 252 & 512 & 5 & 21 & 252 & 512 & 5 & 21 & 252 & 512 & 5 & 21 & 252 & 512 \\
        \midrule
    AR & \makecell{42.35 \\ 26.80 \\ 15.54} & \makecell{46.29 \\ 28.47 \\ 17.82} & \makecell{46.27 \\ 29.18 \\ 17.09} & \makecell{46.98 \\ 29.43 \\ 17.55} &
    \makecell{-3.43 \\ 3.59 \\ -7.01} & \makecell{-2.59 \\ 3.42 \\ -6.01} & \makecell{-2.46 \\ 3.69 \\ -6.15} & \makecell{-1.01 \\ 4.57 \\ -5.58} &
    \makecell{-3.34 \\ 3.73 \\ -7.07} & \makecell{-1.84 \\ 3.94 \\ -5.78} & \makecell{-0.32 \\ 4.39 \\ -4.71} & \makecell{-0.86 \\ 4.52 \\ -5.38} &
    \makecell{-2.87 \\ 3.13 \\ -6.00} & \makecell{-2.72 \\ 3.94 \\ -6.66} & \makecell{-1.36 \\ 4.61 \\ -5.97} & \makecell{-1.67 \\ 4.67 \\ -6.34} \\ \addlinespace[3pt]

    SD & \makecell{7.43 \\ 14.15 \\ 12.11} & \makecell{6.89 \\ 13.89 \\ 12.38} & \makecell{6.88 \\ 13.72 \\ 12.32} & \makecell{7.35 \\ 13.82 \\ 12.32} &
    \makecell{3.71 \\ 13.74 \\ 11.77} & \makecell{5.84 \\ 14.73 \\ 10.12} & \makecell{7.17 \\ 14.79 \\ 8.79} & \makecell{6.82 \\ 14.41 \\ 8.80} &
    \makecell{3.58 \\ 13.64 \\ 11.55} & \makecell{5.77 \\ 14.65 \\ 10.09} & \makecell{6.21 \\ 14.48 \\ 9.56} & \makecell{5.63 \\ 13.99 \\ 9.64} &
    \makecell{3.51 \\ 13.47 \\ 11.37} & \makecell{5.71 \\ 14.42 \\ 10.06} & \makecell{5.84 \\ 14.12 \\ 9.67} & \makecell{5.07 \\ 13.63 \\ 9.79} \\ \addlinespace[3pt]

    SR & \makecell{5.70 \\ 1.89 \\ 1.28} & \makecell{6.72 \\ 2.05 \\ 1.44} & \makecell{6.72 \\ 2.13 \\ 1.39} & \makecell{6.39 \\ 2.13 \\ 1.42} &
    \makecell{-0.92 \\ 0.26 \\ -0.60} & \makecell{-0.44 \\ 0.23 \\ -0.59} & \makecell{-0.34 \\ 0.25 \\ -0.70} & \makecell{-0.15 \\ 0.32 \\ -0.63} &
    \makecell{-0.93 \\ 0.27 \\ -0.61} & \makecell{-0.32 \\ 0.27 \\ -0.57} & \makecell{-0.05 \\ 0.30 \\ -0.49} & \makecell{-0.15 \\ 0.32 \\ -0.56} &
    \makecell{-0.82 \\ 0.23 \\ -0.53} & \makecell{-0.48 \\ 0.27 \\ -0.66} & \makecell{-0.23 \\ 0.33 \\ -0.62} & \makecell{-0.33 \\ 0.34 \\ -0.65} \\ \addlinespace[3pt]

    Daily Return & \makecell{16.81 \\ 10.64 \\ 6.17} & \makecell{18.37 \\ 11.30 \\ 7.07} & \makecell{18.36 \\ 11.58 \\ 6.78} & \makecell{18.64 \\ 11.68 \\ 6.97} &
    \makecell{-1.36 \\ 1.42 \\ -2.78} & \makecell{-1.03 \\ 1.36 \\ -2.38} & \makecell{-0.97 \\ 1.46 \\ -2.44} & \makecell{-0.40 \\ 1.81 \\ -2.21} &
    \makecell{-1.32 \\ 1.48 \\ -2.81} & \makecell{-0.73 \\ 1.56 \\ -2.29} & \makecell{-0.13 \\ 1.74 \\ -1.87} & \makecell{-0.34 \\ 1.79 \\ -2.13} &
    \makecell{-1.14 \\ 1.24 \\ -2.38} & \makecell{-1.08 \\ 1.56 \\ -2.64} & \makecell{-0.54 \\ 1.83 \\ -2.37} & \makecell{-0.66 \\ 1.85 \\ -2.51} \\ \addlinespace[3pt]

    Max DD & \makecell{12.03 \\ 32.49 \\ 31.98} & \makecell{13.14 \\ 32.85 \\ 30.00} & \makecell{13.67 \\ 31.85 \\ 31.75} & \makecell{14.46 \\ 32.62 \\ 29.58} &
    \makecell{56.25 \\ 43.19 \\ 83.96} & \makecell{52.31 \\ 41.73 \\ 79.55} & \makecell{52.83 \\ 43.97 \\ 78.45} & \makecell{35.19 \\ 42.05 \\ 76.23} &
    \makecell{55.21 \\ 42.14 \\ 84.08} & \makecell{44.49 \\ 41.51 \\ 78.12} & \makecell{28.12 \\ 40.52 \\ 72.25} & \makecell{27.81 \\ 41.40 \\ 75.02} &
    \makecell{50.94 \\ 44.79 \\ 79.97} & \makecell{50.14 \\ 42.37 \\ 81.08} & \makecell{31.62 \\ 42.41 \\ 77.73} & \makecell{38.69 \\ 37.93 \\ 79.76} \\ \addlinespace[3pt]

    Max 1-day Loss & \makecell{4.86 \\ 8.28 \\ 5.56} & \makecell{3.75 \\ 8.21 \\ 4.49} & \makecell{4.15 \\ 8.19 \\ 5.03} & \makecell{4.93 \\ 8.19 \\ 5.27} &
    \makecell{5.06 \\ 7.12 \\ 5.80} & \makecell{2.07 \\ 7.33 \\ 5.07} & \makecell{3.20 \\ 6.87 \\ 4.49} & \makecell{2.32 \\ 7.13 \\ 4.08} &
    \makecell{2.14 \\ 7.34 \\ 5.34} & \makecell{2.37 \\ 7.25 \\ 5.00} & \makecell{3.04 \\ 6.73 \\ 3.87} & \makecell{3.50 \\ 6.99 \\ 6.20} &
    \makecell{1.45 \\ 6.90 \\ 5.66} & \makecell{6.79 \\ 7.41 \\ 6.44} & \makecell{6.95 \\ 6.82 \\ 6.52} & \makecell{2.60 \\ 6.93 \\ 6.20} \\ \addlinespace[3pt]

    Skew & \makecell{2.20 \\ -0.01 \\ 0.35} & \makecell{2.15 \\ 0.00 \\ 0.38} & \makecell{2.21 \\ 0.02 \\ 0.37} & \makecell{1.67 \\ -0.05 \\ 0.35} &
    \makecell{-1.63 \\ -0.21 \\ 0.24} & \makecell{0.20 \\ -0.13 \\ 0.45} & \makecell{0.19 \\ -0.18 \\ 0.63} & \makecell{0.21 \\ -0.18 \\ 0.68} &
    \makecell{-0.05 \\ -0.26 \\ 0.30} & \makecell{1.09 \\ -0.15 \\ 0.44} & \makecell{1.13 \\ -0.12 \\ 0.62} & \makecell{0.20 \\ -0.20 \\ 0.36} &
    \makecell{0.13 \\ -0.22 \\ 0.36} & \makecell{-0.96 \\ -0.18 \\ 0.19} & \makecell{-0.94 \\ -0.18 \\ 0.26} & \makecell{0.08 \\ -0.21 \\ 0.28} \\ \addlinespace[3pt]

    Kurt & \makecell{32.84 \\ 11.60 \\ 6.15} & \makecell{31.26 \\ 12.29 \\ 5.15} & \makecell{32.14 \\ 12.51 \\ 5.39} & \makecell{26.37 \\ 12.53 \\ 5.49} &
    \makecell{40.29 \\ 6.45 \\ 7.94} & \makecell{3.72 \\ 5.97 \\ 8.93} & \makecell{3.25 \\ 4.70 \\ 10.64} & \makecell{2.95 \\ 4.86 \\ 10.44} &
    \makecell{4.62 \\ 6.88 \\ 7.83} & \makecell{18.07 \\ 6.00 \\ 8.18} & \makecell{17.92 \\ 5.35 \\ 7.81} & \makecell{7.02 \\ 5.42 \\ 8.66} &
    \makecell{2.98 \\ 6.51 \\ 8.31} & \makecell{24.64 \\ 6.03 \\ 10.20} & \makecell{24.78 \\ 5.19 \\ 10.17} & \makecell{4.55 \\ 5.68 \\ 9.56} \\ \addlinespace[3pt]

        \midrule \midrule

        \textbf{Model} & \multicolumn{4}{c|}{\textbf{Chronos (Base)}} & \multicolumn{4}{c|}{\textbf{Chronos (Large)}} & \multicolumn{4}{c|}{\textbf{TimesFM 1 (200M)}} & \multicolumn{4}{c}{\textbf{TimesFM 2 (500M)}} \\
        \textbf{Window Size} & 5 & 21 & 252 & 512 & 5 & 21 & 252 & 512 & 5 & 21 & 252 & 512 & 5 & 21 & 252 & 512 \\
        \midrule

    AR & \makecell{-2.00 \\ 3.79 \\ -5.79} & \makecell{-0.61 \\ 4.58 \\ -5.20} & \makecell{-1.44 \\ 4.45 \\ -5.89} & \makecell{-1.03 \\ 4.77 \\ -5.80} &
    \makecell{-2.11 \\ 4.47 \\ -6.57} & \makecell{-0.69 \\ 4.78 \\ -5.47} & \makecell{0.23 \\ 5.10 \\ -4.87} & \makecell{-0.94 \\ 4.69 \\ -5.63} &
    \makecell{-31.91 \\ -9.56 \\ -22.35} & \makecell{-20.27 \\ -4.67 \\ -15.59} & \makecell{-1.83 \\ 3.77 \\ -5.61} & \makecell{-2.64 \\ 3.72 \\ -6.36} &
    \makecell{-28.61 \\ -7.59 \\ -21.02} & \makecell{-17.26 \\ -3.64 \\ -13.62} & \makecell{-2.48 \\ 3.78 \\ -6.25} & \makecell{-4.14 \\ 3.29 \\ -7.42} \\ \addlinespace[3pt]

    SD & \makecell{2.80 \\ 12.59 \\ 12.87} & \makecell{2.50 \\ 12.99 \\ 12.17} & \makecell{2.98 \\ 11.34 \\ 12.22} & \makecell{2.98 \\ 11.15 \\ 11.93} &
    \makecell{2.62 \\ 12.39 \\ 12.53} & \makecell{2.76 \\ 11.99 \\ 12.60} & \makecell{3.10 \\ 11.85 \\ 11.62} & \makecell{2.97 \\ 11.74 \\ 11.66} &
    \makecell{9.64 \\ 11.80 \\ 14.24} & \makecell{9.84 \\ 11.66 \\ 14.51} & \makecell{6.05 \\ 12.55 \\ 12.02} & \makecell{5.56 \\ 12.45 \\ 11.70} &
    \makecell{9.87 \\ 10.83 \\ 14.52} & \makecell{10.37 \\ 10.99 \\ 14.60} & \makecell{6.55 \\ 12.12 \\ 11.99} & \makecell{5.69 \\ 12.04 \\ 11.66}  \\ \addlinespace[3pt]

    SR & \makecell{-0.72 \\ 0.30 \\ -0.45} & \makecell{-0.25 \\ 0.35 \\ -0.43} & \makecell{-0.48 \\ 0.39 \\ -0.48} & \makecell{-0.35 \\ 0.43 \\ -0.49} &
    \makecell{-0.80 \\ 0.36 \\ -0.52} & \makecell{-0.25 \\ 0.40 \\ -0.43} & \makecell{0.07 \\ 0.43 \\ -0.42} & \makecell{-0.32 \\ 0.40 \\ -0.48} &
    \makecell{-3.31 \\ -0.81 \\ -1.57} & \makecell{-2.06 \\ -0.40 \\ -1.07} & \makecell{-0.30 \\ 0.30 \\ -0.47} & \makecell{-0.47 \\ 0.30 \\ -0.54} &
    \makecell{-2.90 \\ -0.70 \\ -1.45} & \makecell{-1.67 \\ -0.33 \\ -0.93} & \makecell{-0.38 \\ 0.31 \\ -0.52} & \makecell{-0.73 \\ 0.27 \\ -0.64} \\ \addlinespace[3pt]

    Daily Return & \makecell{-0.79 \\ 1.50 \\ -2.30} & \makecell{-0.24 \\ 1.82 \\ -2.06} & \makecell{-0.57 \\ 1.77 \\ -2.34} & \makecell{-0.41 \\ 1.89 \\ -2.30} &
    \makecell{-0.84 \\ 1.77 \\ -2.61} & \makecell{-0.28 \\ 1.90 \\ -2.17} & \makecell{0.09 \\ 2.02 \\ -1.93} & \makecell{-0.37 \\ 1.86 \\ -2.23} &
    \makecell{-12.66 \\ -3.80 \\ -8.87} & \makecell{-8.04 \\ -1.85 \\ -6.19} & \makecell{-0.73 \\ 1.50 \\ -2.22} & \makecell{-1.05 \\ 1.48 \\ -2.52} &
    \makecell{-11.35 \\ -3.01 \\ -8.34} & \makecell{-6.85 \\ -1.45 \\ -5.40} & \makecell{-0.98 \\ 1.50 \\ -2.48} & \makecell{-1.64 \\ 1.30 \\ -2.95} \\ \addlinespace[3pt]

    Max DD & \makecell{42.03 \\ 44.33 \\ 81.53} & \makecell{16.90 \\ 41.18 \\ 78.51} & \makecell{28.74 \\ 37.91 \\ 79.68} & \makecell{21.45 \\ 41.15 \\ 79.34} &
    \makecell{42.76 \\ 42.81 \\ 83.02} & \makecell{20.85 \\ 39.57 \\ 80.12} & \makecell{12.22 \\ 37.87 \\ 77.33} & \makecell{21.40 \\ 40.26 \\ 79.02} &
    \makecell{99.94 \\ 91.88 \\ 99.52} & \makecell{99.12 \\ 75.11 \\ 97.79} & \makecell{50.51 \\ 55.78 \\ 82.26} & \makecell{49.89 \\ 54.14 \\ 81.15} &
    \makecell{99.87 \\ 86.93 \\ 99.35} & \makecell{98.27 \\ 67.48 \\ 96.58} & \makecell{56.77 \\ 41.25 \\ 84.33} & \makecell{64.01 \\ 40.59 \\ 84.51} \\ \addlinespace[3pt]

    Max 1-day Loss & \makecell{1.12 \\ 7.22 \\ 6.07} & \makecell{1.09 \\ 6.80 \\ 5.55} & \makecell{5.12 \\ 6.87 \\ 4.94} & \makecell{5.59 \\ 6.30 \\ 5.21} &
    \makecell{1.13 \\ 6.74 \\ 5.96} & \makecell{1.54 \\ 6.97 \\ 5.78} & \makecell{1.00 \\ 6.72 \\ 5.05} & \makecell{5.97 \\ 6.81 \\ 5.37} &
    \makecell{7.96 \\ 5.94 \\ 8.13} & \makecell{12.01 \\ 5.53 \\ 11.65} & \makecell{3.02 \\ 7.05 \\ 5.87} & \makecell{6.53 \\ 6.88 \\ 6.22} &
    \makecell{8.00 \\ 5.28 \\ 8.15} & \makecell{12.01 \\ 5.50 \\ 11.65} & \makecell{3.73 \\ 6.99 \\ 5.43} & \makecell{6.53 \\ 6.78 \\ 6.22} \\ \addlinespace[3pt]

    Skew & \makecell{1.96 \\ -0.31 \\ 0.20} & \makecell{-0.01 \\ -0.23 \\ 0.29} & \makecell{-3.53 \\ -0.37 \\ 0.27} & \makecell{-4.81 \\ -0.39 \\ 0.25} &
    \makecell{-0.07 \\ -0.28 \\ 0.24} & \makecell{-0.52 \\ -0.39 \\ 0.23} & \makecell{9.30 \\ -0.25 \\ 0.34} & \makecell{-5.54 \\ -0.36 \\ 0.27} &
    \makecell{-1.57 \\ -0.47 \\ -0.07} & \makecell{-2.81 \\ -0.54 \\ -0.45} & \makecell{-0.35 \\ -0.48 \\ 0.16} & \makecell{-1.73 \\ -0.45 \\ 0.15} &
    \makecell{-1.27 \\ -0.41 \\ 0.04} & \makecell{-1.99 \\ -0.38 \\ -0.17} & \makecell{-0.26 \\ -0.33 \\ 0.35} & \makecell{-1.28 \\ -0.33 \\ 0.30}  \\ \addlinespace[3pt]

    Kurt & \makecell{51.11 \\ 7.04 \\ 7.15} & \makecell{2.63 \\ 6.57 \\ 7.07} & \makecell{96.45 \\ 6.97 \\ 6.21} & \makecell{138.03 \\ 6.45 \\ 6.61} &
    \makecell{2.87 \\ 7.17 \\ 6.90} & \makecell{6.48 \\ 7.97 \\ 6.88} & \makecell{351.97 \\ 7.93 \\ 6.05} & \makecell{179.60 \\ 7.76 \\ 6.24} &
    \makecell{18.46 \\ 5.52 \\ 12.47} & \makecell{43.80 \\ 5.25 \\ 16.49} & \makecell{5.71 \\ 6.79 \\ 5.19} & \makecell{28.34 \\ 6.18 \\ 7.03} &
    \makecell{17.07 \\ 4.29 \\ 11.98} & \makecell{37.97 \\ 4.95 \\ 17.08} & \makecell{7.59 \\ 6.45 \\ 5.99} & \makecell{25.51 \\ 6.11 \\ 7.41} \\ \addlinespace[3pt]

        \bottomrule
      \end{tabular}
      \vspace{0.1cm}
      \begin{tablenotes}[para,flushleft]
        \footnotesize
        \textbf{Note:} This table reports portfolio performance metrics computed from the full daily out-of-sample excess return series across different input window lengths (5, 21, 252, and 512 trading days) for each model. The benchmark model is CatBoost, the representative benchmark. The time series foundation models (TSFMs) include Chronos (tiny, mini, small, base, and large) and TimesFM (version 1 with 200 million and version 2 with 500 million parameters). The models released by the respective authors are fine-tuned on an annual basis. Each cell displays three values from top to bottom: the long--short portfolio, the long-leg contribution, and the short-leg contribution. The long leg equals one half of the top-decile return, while the short leg is the signed contribution from shorting the bottom decile and equals minus one half of the bottom-decile return. Metrics include annualized return (AR), standard deviation (SD), Sharpe ratio (SR), daily return (in basis points), maximum drawdown (Max DD), maximum one-day loss (Max 1-day Loss), skewness, and excess kurtosis of portfolio returns. AR, SD, maximum drawdown, and maximum one-day loss are reported in percentage terms. Portfolios are formed using decile sorting based on model forecasts, with equal weighting across stocks.
      \end{tablenotes}
    \end{minipage}
    \end{adjustbox}
  \end{threeparttable}
\end{table}

\vspace*{\fill}
\begin{figure}[h] %%%%%%% Fixed ------------------------------------------------------------
    \caption{Cumulative Log Returns of Fine-Tuned TSFMs: Long--Short Portfolios}
    \label{Cumulative_log_return_fine_tuned_long_and_short}
    \centering
    \includegraphics[
        width=0.67\textwidth,
        clip,
        trim=0 35 0 28, % left bottom right top
    ]{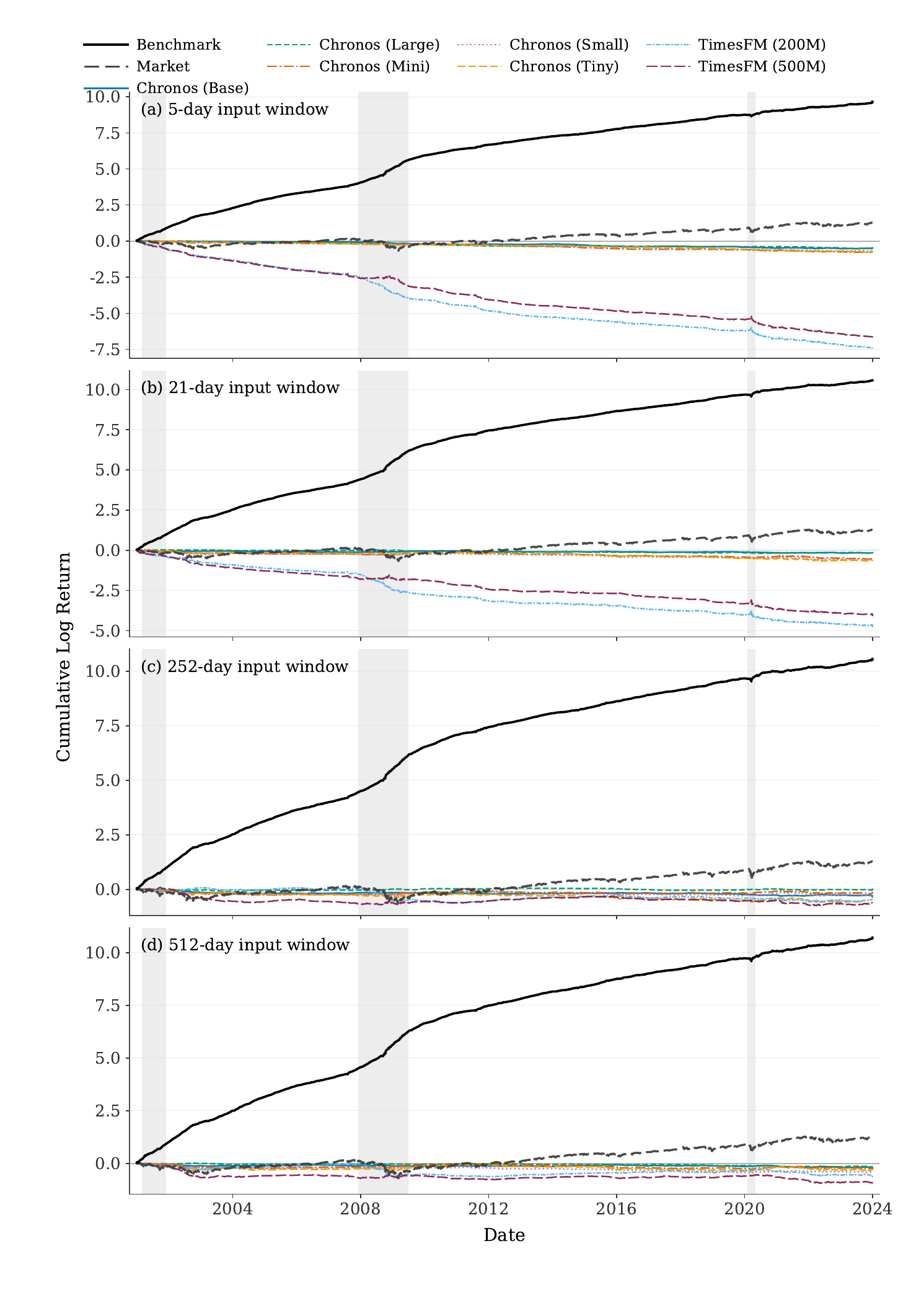} % Adjust the width and trims
    \begin{minipage}{0.8\textwidth}
        \footnotesize
        \textbf{Note:} This figure displays the cumulative log returns of long--short portfolios constructed from various forecasting models with input windows of 5 (a), 21 (b), 252 (c), and 512 (d) trading days. The benchmark model is CatBoost, the representative benchmark. The time series foundation models (TSFMs) include Chronos (tiny, mini, small, base, and large) and TimesFM (version 1 with 200 million and version 2 with 500 million parameters). The models released by the respective authors are fine-tuned on an annual basis. Each subplot corresponds to a specific input window length, as indicated by the text labels in the upper-left corners. The benchmark model (CatBoost) is highlighted in black with bold lines. The dashed black line represents the cumulative log return of the market (S\&P 500). Shaded areas indicate U.S. recession periods, as defined by the National Bureau of Economic Research (NBER). The model-based long--short portfolios use equal weighting within each leg and are scaled to unit gross exposure. The S\&P 500 is shown separately as a market reference.

    \end{minipage}
\end{figure}
\vspace*{\fill}
\clearpage

\vspace*{\fill}
\begin{table}[h]
  \centering
  \begin{threeparttable}
    \begin{adjustbox}{width=0.96\textwidth, center}
    \scriptsize
    \captionsetup{width=\linewidth}
    \caption{Fine-Tuned TSFMs - Spread Portfolio Performance}
    \label{Spread_Portfolio_fine_tuned_table}
    \begin{minipage}{\linewidth}
      \renewcommand{\arraystretch}{1.2}
      \begin{tabular}{c|cccc|cccc|cccc|cccc}
        \toprule
        \textbf{Model} & \multicolumn{4}{c|}{\textbf{Benchmark}} & \multicolumn{4}{c|}{\textbf{Chronos (Tiny)}} & \multicolumn{4}{c|}{\textbf{Chronos (Mini)}} & \multicolumn{4}{c}{\textbf{Chronos (Small)}} \\
        \textbf{Window Size} & 5 & 21 & 252 & 512 & 5 & 21 & 252 & 512 & 5 & 21 & 252 & 512 & 5 & 21 & 252 & 512 \\
        \midrule
        Low & -31.09 & -35.64 & -34.19 & -35.11 & 14.03 & 12.01 & 12.29 & 11.16 & 14.14 & 11.55 & 9.43 & 10.76 & 12.00 & 13.31 & 11.94 & 12.67 \\
        \addlinespace[3pt]
        2 &  -0.80 & -2.89 & -3.83 & -3.98 & 12.67 & 12.36 & 11.74 & 12.20 & 14.32 & 11.58 & 11.83 & 11.58 & 14.99 & 12.14 & 10.64 & 11.34 \\ 
        \addlinespace[3pt]
        3 & 4.33 & 4.56 & 2.80 & 3.33 & 11.27 & 11.63 & 10.77 & 11.26 & 11.99 & 12.31 & 12.30 & 11.79 & 13.39 & 11.33 & 12.79 & 11.69 \\ 
        \addlinespace[3pt]
        4 & 7.51 & 8.36 & 6.40 & 6.87 & 12.31 & 11.89 & 12.13 & 12.24 & 11.45 & 12.22 & 12.20 & 12.39 & 11.71 & 11.46 & 12.34 & 11.70 \\ 
        \addlinespace[3pt]
        5 & 9.40 & 10.27 & 9.88 & 9.40 & 11.01 & 12.25 & 12.45 & 11.44 & 10.36 & 11.61 & 11.80 & 12.14 & 11.14 & 11.44 & 11.93 & 10.98 \\
        \addlinespace[3pt]
        6 & 11.57 & 11.44 & 12.00 & 12.43 & 10.90 & 11.54 & 12.19 & 11.61 & 11.35 & 11.97 & 12.25 & 11.86 & 11.07 & 11.95 & 11.56 & 12.65 \\ 
        \addlinespace[3pt]
        7 & 15.27 & 15.26 & 15.66 & 14.08 & 10.88 & 12.65 & 11.87 & 11.98 & 11.12 & 11.33 & 11.22 & 10.88 & 11.18 & 11.45 & 10.58 & 11.08 \\ 
        \addlinespace[3pt]
        8 & 18.24 & 18.67 & 18.31 & 19.43 & 11.21 & 11.83 & 11.79 & 11.22 & 11.17 & 11.15 & 11.60 & 11.48 & 10.47 & 11.58 & 11.94 & 10.86 \\ 
        \addlinespace[3pt]
        9 & 25.06 & 26.11 & 27.71 & 27.79 & 11.63 & 10.09 & 10.48 & 10.83 & 9.72 & 11.47 & 11.68 & 11.18 & 10.87 & 10.54 & 10.16 & 10.78 \\
        \addlinespace[3pt]
        High & 53.61 & 56.94 & 58.36 & 58.85 & 7.17 & 6.84 & 7.38 & 9.14 & 7.47 & 7.88 & 8.78 & 9.04 & 6.26 & 7.88 & 9.22 & 9.34 \\ 
        \addlinespace[3pt]
        $\tfrac{1}{2}(H-L)$ & 42.35 & 46.29 & 46.27 & 46.98 & -3.43 & -2.59 & -2.46 & -1.01 & -3.34 & -1.84 & -0.32 & -0.86 & -2.87 & -2.72 & -1.36 & -1.67 \\ 
        \midrule \midrule

        \textbf{Model} & \multicolumn{4}{c|}{\textbf{Chronos (Base)}} & \multicolumn{4}{c|}{\textbf{Chronos (Large)}} & \multicolumn{4}{c|}{\textbf{TimesFM 1 (200M)}} & \multicolumn{4}{c}{\textbf{TimesFM 2 (500M)}} \\
        \textbf{Window Size} & 5 & 21 & 252 & 512 & 5 & 21 & 252 & 512 & 5 & 21 & 252 & 512 & 5 & 21 & 252 & 512 \\
        \midrule
        Low & 11.58 & 10.39 & 11.78 & 11.60 & 13.14 & 10.94 & 9.74 & 11.26 & 44.70 & 31.19 & 11.21 & 12.73 & 42.04 & 27.24 & 12.50 & 14.85  \\
        \addlinespace[3pt]
        2 & 12.17 & 12.26 & 10.90 & 10.56 & 14.58 & 11.36 & 11.79 & 11.92 & 29.17 & 22.43 & 11.26 & 11.41 & 27.92 & 21.54 & 12.17 & 12.96 \\
        \addlinespace[3pt]
        3 & 14.38 & 11.87 & 11.19 & 11.73 & 11.71 & 10.92 & 11.18 & 12.00 & 21.37 & 18.83 & 11.68 & 12.15 & 19.74 & 17.49 & 12.67 & 13.49 \\
        \addlinespace[3pt]
        4 & 13.35 & 11.72 & 12.01 & 11.93 & 12.03 & 11.30 & 11.88 & 11.03 & 16.18 & 16.22 & 13.62 & 12.46 & 15.81 & 15.25 & 13.25 & 12.68 \\
        \addlinespace[3pt]
        5 & 11.29 & 10.58 & 12.23 & 12.63 & 11.09 & 11.60 & 11.57 & 11.80 & 12.26 & 12.53 & 12.33 & 12.30 & 12.30 & 12.69 & 11.88 & 11.46 \\
        \addlinespace[3pt]
        6 & 12.01 & 12.37 & 12.78 & 11.07 & 10.66 & 11.72 & 11.74 & 11.56 & 8.40 & 10.44 & 11.33 & 11.17 & 8.36 & 9.73 & 11.34 & 10.56 \\
        \addlinespace[3pt]
        7 & 10.54 & 12.04 & 11.67 & 12.38 & 9.92 & 12.76 & 11.35 & 11.92 & 4.41 & 6.85 & 11.11 & 10.73 & 4.33 & 8.27 & 10.66 & 10.95 \\
        \addlinespace[3pt]
        8 & 10.66 & 11.12 & 10.84 & 10.97 & 10.70 & 10.75 & 12.45 & 11.49 & 0.55 & 4.10 & 11.98 & 11.89 & 1.72 & 5.94 & 11.27 & 10.19 \\
        \addlinespace[3pt]
        9 & 9.52 & 11.56 & 10.78 & 10.65 & 10.31 & 12.18 & 11.19 & 10.73 & -4.85 & -0.17 & 11.03 & 10.81 & -3.97 & 2.22 & 9.78 & 9.36 \\
        \addlinespace[3pt]
        High & 7.58 & 9.17 & 8.90 & 9.55 & 8.93 & 9.55 & 10.20 & 9.38 & -19.13 & -9.35 & 7.54 & 7.44 & -15.18 & -7.28 & 7.55 & 6.57 \\
        \addlinespace[3pt]
        $\tfrac{1}{2}(H-L)$ & -2.00 & -0.61 & -1.44 & -1.03 & -2.11 & -0.69 & 0.23 & -0.94 & -31.91 & -20.27 & -1.83 & -2.64 & -28.61 & -17.26 & -2.48 & -4.14 \\
        \bottomrule
      \end{tabular}
      \vspace{0.1cm}
      \begin{tablenotes}[para,flushleft]
        \footnotesize
        \textbf{Note:} This table presents the annualized average returns of decile spread portfolios formed using forecasts from various predictive models across different window sizes (5, 21, 252, and 512 trading days). The benchmark model is CatBoost, the representative benchmark. The time series foundation models (TSFMs) include Chronos (tiny, mini, small, base, and large) and TimesFM (version 1 with 200 million and version 2 with 500 million parameters). The models released by the respective authors are fine-tuned on an annual basis. For each model and window size, we report standalone decile returns from Low to High. Here, $H$ and $L$ denote the standalone returns of the highest- and lowest-forecast deciles. The final row reports the unit-gross long--short return, $\frac{1}{2}(H-L)$, with long- and short-leg contributions of $\frac{1}{2}H$ and $-\frac{1}{2}L$, respectively. Portfolios are constructed using equal-weighted (EW) decile sorting based on predicted returns.
      \end{tablenotes}
    \end{minipage}
    \end{adjustbox}
  \end{threeparttable}
\end{table}
\vspace*{\fill}

\clearpage
\subsection{Complete Finance-Native Pre-Trained TSFM Results} \label{complete_pre_trained_tsfm_results}
This subsection presents complete finance-native pre-trained TSFM results for both forecasting performance and portfolio performance across all models and window sizes.
\vspace*{\fill}
\begin{table}[h] %% Fixed -------------------------------------------------------------------------
  \centering
  \begin{threeparttable}
    \begin{adjustbox}{width=0.82\textwidth, center}
    \scriptsize
    \captionsetup{width=\linewidth}
    \caption{Finance-Native Pre-Trained TSFMs - Forecasting Performance (Complete)}
    \label{FP_TSFM_pre_trained_full}
    \begin{minipage}{\linewidth}
        \renewcommand{\arraystretch}{1.3}
      \begin{tabular}{c|cccc|cccc|cccc}
        \toprule
        \textbf{Model} & \multicolumn{4}{c|}{\textbf{Benchmark}} & \multicolumn{4}{c|}{\textbf{Chronos (Tiny)}} & \multicolumn{4}{c}{\textbf{Chronos (Mini)}} \\
        \textbf{Window Size} & 5 & 21 & 252 & 512 & 5 & 21 & 252 & 512 & 5 & 21 & 252 & 512 \\

        \midrule
        $\RsqOOS$ & \makecell{-0.25 \\ -0.50 \\ 0.39} & \makecell{-0.04 \\ -0.38 \\ 0.66} & \makecell{-0.03 \\ -0.32 \\ 0.62} & \makecell{-0.03 \\ -0.28 \\ 0.59} &
        \makecell{-3.77 \\ -3.85 \\ -3.74} & \makecell{-1.09 \\ -1.24 \\ -0.83} & \makecell{-0.65 \\ -0.87 \\ -0.30} & \makecell{-0.61 \\ -0.89 \\ -0.24} &
        \makecell{-4.50 \\ -4.43 \\ -4.61} & \makecell{-1.26 \\ -1.53 \\ -0.87} & \makecell{-0.60 \\ -0.78 \\ -0.26} & \makecell{-0.61 \\ -0.83 \\ -0.23} \\

        \addlinespace[3pt]

        Overall Acc. & \makecell{50.94 \\ 50.70 \\ 51.96} & \makecell{51.12 \\ 50.74 \\ 52.47} & \makecell{51.08 \\ 50.80 \\ 52.33} & \makecell{51.16 \\ 50.95 \\ 52.39} &
        \makecell{50.32 \\ 49.29 \\ 51.77} & \makecell{50.65 \\ 49.85 \\ 52.11} & \makecell{51.27 \\ 50.81 \\ 52.80} & \makecell{51.32 \\ 50.87 \\ 52.91} &
        \makecell{50.24 \\ 49.31 \\ 51.53} & \makecell{50.57 \\ 49.86 \\ 51.96} & \makecell{51.31 \\ 50.90 \\ 52.88} & \makecell{51.34 \\ 50.87 \\ 53.01} \\

        \addlinespace[3pt]

        Up Acc. & \makecell{68.96 \\ 72.05 \\ 68.00} & \makecell{67.82 \\ 72.97 \\ 65.11} & \makecell{66.87 \\ 71.06 \\ 64.65} & \makecell{69.14 \\ 74.39 \\ 66.07} &
        \makecell{18.50 \\ 17.97 \\ 19.80} & \makecell{31.78 \\ 32.02 \\ 33.38} & \makecell{46.82 \\ 55.26 \\ 41.48} & \makecell{47.66 \\ 55.40 \\ 42.68} &
        \makecell{22.99 \\ 22.31 \\ 24.63} & \makecell{33.22 \\ 33.20 \\ 35.01} & \makecell{47.43 \\ 55.85 \\ 42.63} & \makecell{46.58 \\ 53.22 \\ 43.56} \\

        \addlinespace[3pt]

        Down Acc. & \makecell{33.14 \\ 28.11 \\ 37.42} & \makecell{34.60 \\ 27.25 \\ 41.01} & \makecell{35.59 \\ 29.59 \\ 41.24} & \makecell{33.37 \\ 26.24 \\ 39.86} &
        \makecell{81.36 \\ 81.84 \\ 80.29} & \makecell{68.79 \\ 68.00 \\ 68.62} & \makecell{55.20 \\ 45.48 \\ 62.76} & \makecell{54.54 \\ 45.51 \\ 61.90} &
        \makecell{76.82 \\ 77.43 \\ 75.47} & \makecell{67.25 \\ 66.80 \\ 66.88} & \makecell{54.67 \\ 45.02 \\ 61.85} & \makecell{55.61 \\ 47.72 \\ 61.26} \\

        \addlinespace[3pt]

        F1 & \makecell{0.49 \\ 0.47 \\ 0.51} & \makecell{0.49 \\ 0.47 \\ 0.52} & \makecell{0.50 \\ 0.48 \\ 0.52} & \makecell{0.49 \\ 0.47 \\ 0.51} &
        \makecell{0.44 \\ 0.43 \\ 0.45} & \makecell{0.48 \\ 0.47 \\ 0.50} & \makecell{0.50 \\ 0.48 \\ 0.52} & \makecell{0.50 \\ 0.48 \\ 0.52} &
        \makecell{0.46 \\ 0.45 \\ 0.47} & \makecell{0.48 \\ 0.47 \\ 0.50} & \makecell{0.50 \\ 0.49 \\ 0.52} & \makecell{0.50 \\ 0.49 \\ 0.52} \\

        \addlinespace[3pt]

        \midrule \midrule
        \textbf{Model} & \multicolumn{4}{c|}{\textbf{Chronos (Small)}} & \multicolumn{4}{c|}{\textbf{TimesFM (8M)}} & \multicolumn{4}{c}{\textbf{TimesFM (20M)}} \\
        \textbf{Window Size} & 5 & 21 & 252 & 512 & 5 & 21 & 252 & 512 & 5 & 21 & 252 & 512  \\
        \midrule
        $\RsqOOS$ & \makecell{-3.18 \\ -3.17 \\ -3.20} & \makecell{-1.43 \\ -1.40 \\ -1.33} & \makecell{-0.65 \\ -0.92 \\ -0.17} & \makecell{-0.59 \\ -0.84 \\ -0.15} &
        \makecell{-32.17 \\ -32.35 \\ -33.93} & \makecell{-16.29 \\ -17.83 \\ -14.55} & \makecell{-22.06 \\ -25.74 \\ -19.90} & \makecell{-27.58 \\ -32.83 \\ -24.91} &
        \makecell{-34.09 \\ -32.89 \\ -37.39} & \makecell{-16.32 \\ -16.98 \\ -15.35} & \makecell{-13.53 \\ -15.65 \\ -11.86} & \makecell{-14.75 \\ -17.60 \\ -12.61} \\
        \addlinespace[3pt]

        Overall Acc. & \makecell{50.40 \\ 49.58 \\ 51.65} & \makecell{50.79 \\ 50.08 \\ 52.27} & \makecell{51.36 \\ 50.86 \\ 53.12} & \makecell{51.36 \\ 50.84 \\ 53.20} &
        \makecell{49.73 \\ 49.20 \\ 50.19} & \makecell{50.80 \\ 49.94 \\ 52.28} & \makecell{51.14 \\ 50.20 \\ 52.82} & \makecell{51.15 \\ 50.26 \\ 52.77} &
        \makecell{49.88 \\ 49.84 \\ 49.77} & \makecell{50.88 \\ 50.66 \\ 51.53} & \makecell{51.14 \\ 50.95 \\ 51.82} & \makecell{51.12 \\ 50.98 \\ 51.82} \\

        \addlinespace[3pt]

        Up Acc. & \makecell{26.10 \\ 25.60 \\ 27.57} & \makecell{36.59 \\ 37.04 \\ 37.75} & \makecell{45.39 \\ 52.00 \\ 42.71} & \makecell{45.96 \\ 51.52 \\ 43.93} &
        \makecell{39.75 \\ 41.50 \\ 37.11} & \makecell{34.48 \\ 36.12 \\ 32.77} & \makecell{34.98 \\ 36.26 \\ 34.21} & \makecell{35.52 \\ 36.80 \\ 34.48} &
        \makecell{49.57 \\ 51.06 \\ 47.46} & \makecell{54.82 \\ 55.89 \\ 54.05} & \makecell{57.46 \\ 57.98 \\ 57.70} & \makecell{57.60 \\ 58.45 \\ 57.38} \\

        \addlinespace[3pt]

        Down Acc. & \makecell{73.85 \\ 74.16 \\ 72.89} & \makecell{64.36 \\ 63.16 \\ 65.01} & \makecell{56.85 \\ 48.97 \\ 62.33} & \makecell{56.35 \\ 49.52 \\ 61.38} &
        \makecell{59.15 \\ 56.93 \\ 61.47} & \makecell{66.32 \\ 63.98 \\ 69.11} & \makecell{66.39 \\ 64.48 \\ 68.59} & \makecell{65.84 \\ 64.00 \\ 68.22} &
        \makecell{49.34 \\ 47.46 \\ 51.17} & \makecell{46.20 \\ 44.04 \\ 48.71} & \makecell{44.34 \\ 42.74 \\ 46.17} & \makecell{44.14 \\ 42.26 \\ 46.43} \\

        \addlinespace[3pt]

        F1 & \makecell{0.47 \\ 0.46 \\ 0.48} & \makecell{0.49 \\ 0.48 \\ 0.51} & \makecell{0.50 \\ 0.49 \\ 0.52} & \makecell{0.50 \\ 0.49 \\ 0.52} &
        \makecell{0.47 \\ 0.47 \\ 0.47} & \makecell{0.45 \\ 0.45 \\ 0.46} & \makecell{0.43 \\ 0.42 \\ 0.44} & \makecell{0.43 \\ 0.43 \\ 0.44} &
        \makecell{0.47 \\ 0.47 \\ 0.47} & \makecell{0.47 \\ 0.46 \\ 0.47} & \makecell{0.45 \\ 0.45 \\ 0.46} & \makecell{0.45 \\ 0.45 \\ 0.46} \\

        \addlinespace[3pt]

        \bottomrule
      \end{tabular}
      \vspace{0.1cm}
      \begin{tablenotes}[para,flushleft]
        \footnotesize
        \textbf{Note:} This table presents each metric as a set of three values, ordered from top to bottom: full sample, top 25\% of firms by market capitalization (large-cap), and bottom 25\% (small-cap) for various predictive models across different window sizes (5, 21, 252, and 512 trading days). The benchmark model is CatBoost, the representative benchmark. The time series foundation models (TSFMs) include Chronos (tiny, mini, and small) and TimesFM (with 8 million and 20 million parameters). The models are pre-trained from scratch using U.S. excess return data on an annual basis. Metrics are first computed separately for each calendar year using all stock-date observations within that year. The reported values represent the average of these yearly statistics. Metrics include out-of-sample $R^2$ ($\RsqOOS$), overall directional accuracy, upward and downward classification accuracy, and macro-averaged F1 score. `Overall Acc.' denotes overall directional accuracy, `Up Acc.' and `Down Acc.' represent the model's accuracy in predicting upward and downward excess returns respectively, and `F1' refers to the macro-averaged F1 score. Accuracy measures and $\RsqOOS$ are reported in percentage terms; macro-$F_1$ is reported on a zero-to-one scale.
      \end{tablenotes}
    \end{minipage}
    \end{adjustbox}
  \end{threeparttable}
\end{table}
\vspace*{\fill}

\begin{table} %% Fixed -------------------------------------------------------------------------
  \centering
  \begin{threeparttable}
    \begin{adjustbox}{width=0.82\textwidth, center}
    \scriptsize
    \captionsetup{width=\linewidth}
    \caption{Finance-Native Pre-Trained TSFMs - Portfolio Performance (Complete)}
    \label{PP_TSFM_pre_trained_full}
    \begin{minipage}{\linewidth}
      \renewcommand{\arraystretch}{1.2}
      \begin{tabular}{c|cccc|cccc|cccc}
        \toprule
        \textbf{Model} & \multicolumn{4}{c|}{\textbf{Benchmark}} & \multicolumn{4}{c|}{\textbf{Chronos (Tiny)}} & \multicolumn{4}{c}{\textbf{Chronos (Mini)}} \\
        \textbf{Window Size} & 5 & 21 & 252 & 512 & 5 & 21 & 252 & 512 & 5 & 21 & 252 & 512   \\
        \midrule
    AR & \makecell{42.35 \\ 26.80 \\ 15.54} & \makecell{46.29 \\ 28.47 \\ 17.82} & \makecell{46.27 \\ 29.18 \\ 17.09} & \makecell{46.98 \\ 29.43 \\ 17.55} &
    \makecell{-1.14 \\ 6.33 \\ -7.47} & \makecell{9.47 \\ 11.63 \\ -2.17} & \makecell{23.69 \\ 19.41 \\ 4.27} & \makecell{28.00 \\ 21.82 \\ 6.18} &
    \makecell{-0.70 \\ 7.31 \\ -8.01} & \makecell{9.84 \\ 11.76 \\ -1.92} & \makecell{27.55 \\ 21.01 \\ 6.54} & \makecell{31.63 \\ 23.80 \\ 7.83} \\ \addlinespace[3pt]

    SD & \makecell{7.43 \\ 14.15 \\ 12.11} & \makecell{6.89 \\ 13.89 \\ 12.38} & \makecell{6.88 \\ 13.72 \\ 12.32} & \makecell{7.35 \\ 13.82 \\ 12.32} &
    \makecell{5.56 \\ 10.42 \\ 14.02} & \makecell{6.36 \\ 11.07 \\ 14.40} & \makecell{7.12 \\ 11.77 \\ 14.18} & \makecell{6.83 \\ 11.96 \\ 13.90} &
    \makecell{4.89 \\ 11.13 \\ 13.67} & \makecell{5.89 \\ 11.40 \\ 13.95} & \makecell{6.80 \\ 12.20 \\ 13.64} & \makecell{6.96 \\ 11.98 \\ 13.67} \\ \addlinespace[3pt]

    SR & \makecell{5.70 \\ 1.89 \\ 1.28} & \makecell{6.72 \\ 2.05 \\ 1.44} & \makecell{6.72 \\ 2.13 \\ 1.39} & \makecell{6.39 \\ 2.13 \\ 1.42} &
    \makecell{-0.21 \\ 0.61 \\ -0.53} & \makecell{1.49 \\ 1.05 \\ -0.15} & \makecell{3.33 \\ 1.65 \\ 0.30} & \makecell{4.10 \\ 1.82 \\ 0.44} &
    \makecell{-0.14 \\ 0.66 \\ -0.59} & \makecell{1.67 \\ 1.03 \\ -0.14} & \makecell{4.05 \\ 1.72 \\ 0.48} & \makecell{4.54 \\ 1.99 \\ 0.57}  \\ \addlinespace[3pt]

    Daily Return & \makecell{16.81 \\ 10.64 \\ 6.17} & \makecell{18.37 \\ 11.30 \\ 7.07} & \makecell{18.36 \\ 11.58 \\ 6.78} & \makecell{18.64 \\ 11.68 \\ 6.97} &
    \makecell{-0.45 \\ 2.51 \\ -2.97} & \makecell{3.76 \\ 4.62 \\ -0.86} & \makecell{9.40 \\ 7.70 \\ 1.70} & \makecell{11.11 \\ 8.66 \\ 2.45} &
    \makecell{-0.28 \\ 2.90 \\ -3.18} & \makecell{3.90 \\ 4.67 \\ -0.76} & \makecell{10.93 \\ 8.34 \\ 2.60} & \makecell{12.55 \\ 9.44 \\ 3.11} \\ \addlinespace[3pt]

    Max DD & \makecell{12.03 \\ 32.49 \\ 31.98} & \makecell{13.14 \\ 32.85 \\ 30.00} & \makecell{13.67 \\ 31.85 \\ 31.75} & \makecell{14.46 \\ 32.62 \\ 29.58} &
    \makecell{38.04 \\ 34.73 \\ 86.71} & \makecell{12.58 \\ 27.08 \\ 63.71} & \makecell{10.35 \\ 30.70 \\ 46.14} & \makecell{10.68 \\ 28.47 \\ 42.83} &
    \makecell{31.70 \\ 25.49 \\ 87.79} & \makecell{12.77 \\ 28.61 \\ 60.81} & \makecell{10.07 \\ 30.38 \\ 42.73} & \makecell{11.73 \\ 29.23 \\ 40.11} \\ \addlinespace[3pt]

    Max 1-day Loss & \makecell{4.86 \\ 8.28 \\ 5.56} & \makecell{3.75 \\ 8.21 \\ 4.49} & \makecell{4.15 \\ 8.19 \\ 5.03} & \makecell{4.93 \\ 8.19 \\ 5.27} &
    \makecell{3.68 \\ 6.02 \\ 7.44} & \makecell{7.04 \\ 6.47 \\ 6.59} & \makecell{4.36 \\ 7.63 \\ 6.15} & \makecell{4.31 \\ 7.31 \\ 5.80} &
    \makecell{4.98 \\ 7.13 \\ 6.35} & \makecell{6.13 \\ 7.05 \\ 5.75} & \makecell{3.70 \\ 8.13 \\ 4.93} & \makecell{5.42 \\ 8.14 \\ 5.80} \\ \addlinespace[3pt]

    Skew & \makecell{2.20 \\ -0.01 \\ 0.35} & \makecell{2.15 \\ 0.00 \\ 0.38} & \makecell{2.21 \\ 0.02 \\ 0.37} & \makecell{1.67 \\ -0.05 \\ 0.35} &
    \makecell{-0.73 \\ -0.47 \\ 0.07} & \makecell{-1.36 \\ -0.32 \\ 0.01} & \makecell{-0.08 \\ -0.37 \\ 0.18} & \makecell{0.19 \\ -0.19 \\ 0.16} &
    \makecell{-1.16 \\ -0.41 \\ 0.08} & \makecell{-1.02 \\ -0.29 \\ 0.03} & \makecell{0.17 \\ -0.18 \\ 0.23} & \makecell{0.03 \\ -0.16 \\ 0.20} \\ \addlinespace[3pt]

    Kurt & \makecell{32.84 \\ 11.60 \\ 6.15} & \makecell{31.26 \\ 12.29 \\ 5.15} & \makecell{32.14 \\ 12.51 \\ 5.39} & \makecell{26.37 \\ 12.53 \\ 5.49} &
    \makecell{10.22 \\ 8.05 \\ 7.47} & \makecell{24.30 \\ 10.49 \\ 6.07} & \makecell{6.63 \\ 14.61 \\ 5.46} & \makecell{7.72 \\ 14.09 \\ 5.55} &
    \makecell{18.03 \\ 11.33 \\ 6.53} & \makecell{23.67 \\ 13.20 \\ 5.01} & \makecell{10.41 \\ 17.99 \\ 4.39} & \makecell{12.04 \\ 18.09 \\ 5.22} \\ \addlinespace[3pt]

        \midrule \midrule

        \textbf{Model} & \multicolumn{4}{c|}{\textbf{Chronos (Small)}} & \multicolumn{4}{c|}{\textbf{TimesFM (8M)}} & \multicolumn{4}{c}{\textbf{TimesFM (20M)}}  \\
        \textbf{Window Size} & 5 & 21 & 252 & 512 & 5 & 21 & 252 & 512 & 5 & 21 & 252 & 512 \\
        \midrule

    AR & \makecell{-0.32 \\ 6.72 \\ -7.04} & \makecell{13.99 \\ 14.08 \\ -0.09} & \makecell{33.61 \\ 25.27 \\ 8.34} & \makecell{36.84 \\ 26.76 \\ 10.08} &
    \makecell{-18.77 \\ -3.42 \\ -15.35} & \makecell{7.33 \\ 9.55 \\ -2.21} & \makecell{24.88 \\ 19.17 \\ 5.71} & \makecell{26.07 \\ 19.95 \\ 6.12} &
    \makecell{-18.10 \\ -3.17 \\ -14.93} & \makecell{10.48 \\ 11.37 \\ -0.89} & \makecell{30.24 \\ 21.44 \\ 8.80} & \makecell{30.12 \\ 21.57 \\ 8.55} \\ \addlinespace[3pt]

    SD & \makecell{4.83 \\ 11.13 \\ 13.74} & \makecell{5.47 \\ 11.60 \\ 13.76} & \makecell{6.87 \\ 11.72 \\ 13.83} & \makecell{6.79 \\ 11.77 \\ 13.55} &
    \makecell{9.89 \\ 11.15 \\ 14.36} & \makecell{9.74 \\ 10.85 \\ 14.33} & \makecell{8.81 \\ 12.34 \\ 12.69} & \makecell{8.18 \\ 12.60 \\ 12.20} &
    \makecell{9.95 \\ 11.26 \\ 14.33} & \makecell{8.50 \\ 11.83 \\ 13.54} & \makecell{8.71 \\ 13.24 \\ 12.25} & \makecell{8.34 \\ 13.33 \\ 11.82} \\ \addlinespace[3pt]

    SR & \makecell{-0.07 \\ 0.60 \\ -0.51} & \makecell{2.56 \\ 1.21 \\ -0.01} & \makecell{4.89 \\ 2.16 \\ 0.60} & \makecell{5.42 \\ 2.27 \\ 0.74} &
    \makecell{-1.90 \\ -0.31 \\ -1.07} & \makecell{0.75 \\ 0.88 \\ -0.15} & \makecell{2.83 \\ 1.55 \\ 0.45} & \makecell{3.19 \\ 1.58 \\ 0.50} &
    \makecell{-1.82 \\ -0.28 \\ -1.04} & \makecell{1.23 \\ 0.96 \\ -0.07} & \makecell{3.47 \\ 1.62 \\ 0.72} & \makecell{3.61 \\ 1.62 \\ 0.72} \\ \addlinespace[3pt]

    Daily Return & \makecell{-0.13 \\ 2.67 \\ -2.79} & \makecell{5.55 \\ 5.59 \\ -0.03} & \makecell{13.34 \\ 10.03 \\ 3.31} & \makecell{14.62 \\ 10.62 \\ 4.00} &
    \makecell{-7.45 \\ -1.36 \\ -6.09} & \makecell{2.91 \\ 3.79 \\ -0.88} & \makecell{9.87 \\ 7.61 \\ 2.27} & \makecell{10.35 \\ 7.92 \\ 2.43} &
    \makecell{-7.18 \\ -1.26 \\ -5.93} & \makecell{4.16 \\ 4.51 \\ -0.36} & \makecell{12.00 \\ 8.51 \\ 3.49} & \makecell{11.95 \\ 8.56 \\ 3.39} \\ \addlinespace[3pt]

    Max DD & \makecell{34.48 \\ 42.31 \\ 84.76} & \makecell{9.82 \\ 28.55 \\ 47.64} & \makecell{8.17 \\ 29.16 \\ 44.34} & \makecell{9.91 \\ 29.28 \\ 41.58} &
    \makecell{98.79 \\ 65.75 \\ 97.72} & \makecell{41.74 \\ 23.62 \\ 75.34} & \makecell{17.17 \\ 31.61 \\ 41.79} & \makecell{14.41 \\ 32.20 \\ 36.67} &
    \makecell{98.59 \\ 64.13 \\ 97.47} & \makecell{28.85 \\ 28.38 \\ 68.23} & \makecell{10.96 \\ 30.98 \\ 35.47} & \makecell{11.14 \\ 30.52 \\ 36.48} \\ \addlinespace[3pt]

    Max 1-day Loss & \makecell{4.91 \\ 6.74 \\ 7.26} & \makecell{6.95 \\ 7.18 \\ 6.52} & \makecell{3.48 \\ 8.14 \\ 5.73} & \makecell{6.59 \\ 8.50 \\ 6.36} &
    \makecell{7.19 \\ 7.49 \\ 8.56} & \makecell{5.34 \\ 5.72 \\ 8.64} & \makecell{7.39 \\ 8.37 \\ 6.05} & \makecell{7.22 \\ 8.53 \\ 6.05} &
    \makecell{8.22 \\ 7.37 \\ 8.90} & \makecell{5.52 \\ 5.73 \\ 8.16} & \makecell{7.60 \\ 8.06 \\ 6.44} & \makecell{7.40 \\ 7.98 \\ 6.44} \\ \addlinespace[3pt]

    Skew & \makecell{-1.67 \\ -0.52 \\ 0.01} & \makecell{-1.45 \\ -0.29 \\ 0.04} & \makecell{0.22 \\ -0.25 \\ 0.19} & \makecell{-0.29 \\ -0.25 \\ 0.11} &
    \makecell{-1.42 \\ -0.60 \\ -0.16} & \makecell{-0.46 \\ -0.33 \\ 0.06} & \makecell{0.11 \\ -0.34 \\ 0.21} & \makecell{0.62 \\ -0.25 \\ 0.32} &
    \makecell{-1.69 \\ -0.54 \\ -0.23} & \makecell{-0.72 \\ -0.42 \\ -0.01} & \makecell{0.14 \\ -0.22 \\ 0.17} & \makecell{0.60 \\ -0.14 \\ 0.29} \\ \addlinespace[3pt]

    Kurt & \makecell{22.02 \\ 9.58 \\ 6.96} & \makecell{41.52 \\ 12.62 \\ 5.97} & \makecell{8.91 \\ 17.58 \\ 5.73} & \makecell{20.43 \\ 18.00 \\ 5.93} &
    \makecell{16.68 \\ 7.43 \\ 11.93} & \makecell{8.85 \\ 8.25 \\ 8.54} & \makecell{19.29 \\ 16.21 \\ 5.85} & \makecell{25.54 \\ 16.34 \\ 6.30} &
    \makecell{19.62 \\ 6.41 \\ 12.57} & \makecell{9.34 \\ 6.71 \\ 9.14} & \makecell{20.64 \\ 12.30 \\ 6.60} & \makecell{23.65 \\ 11.95 \\ 7.14} \\ \addlinespace[3pt]

        \bottomrule
      \end{tabular}
      \vspace{0.1cm}
      \begin{tablenotes}[para,flushleft]
        \footnotesize
        \textbf{Note:} This table reports portfolio performance metrics computed from the full daily out-of-sample excess return series across different input window lengths (5, 21, 252, and 512 trading days) for each model. The benchmark model is CatBoost, the representative benchmark. The time series foundation models (TSFMs) include Chronos (tiny, mini, and small) and TimesFM (with 8 million and 20 million parameters). Inference is performed using the finance-native pre-trained models. Each cell displays three values from top to bottom: the long--short portfolio, the long-leg contribution, and the short-leg contribution. The long leg equals one half of the top-decile return, while the short leg is the signed contribution from shorting the bottom decile and equals minus one half of the bottom-decile return. Metrics include annualized return (AR), standard deviation (SD), Sharpe ratio (SR), daily return (in basis points), maximum drawdown (Max DD), maximum one-day loss (Max 1-day Loss), skewness, and excess kurtosis of portfolio returns. AR, SD, maximum drawdown, and maximum one-day loss are reported in percentage terms. Portfolios are formed using decile sorting based on model forecasts, with equal weighting across stocks.
      \end{tablenotes}
    \end{minipage}
    \end{adjustbox}
  \end{threeparttable}
\end{table}

\begin{landscape}
\vspace*{\fill}
\thispagestyle{landscape}
\begin{table}[h]
\vspace*{\fill} % push it down\
\centering
\caption{Comparison of Daily Out-of-Sample Forecasting Performance of Finance-Native Pre-Trained TSFMs Using Diebold--Mariano Test}
\label{DM_test_Pre_Trained}
\begin{adjustbox}{max width=1.3\textwidth}
\begin{threeparttable}
\renewcommand{\arraystretch}{1.2}
\setlength{\tabcolsep}{3pt}

\begin{tabular}{c c| *{5}{G}}
\toprule
 & & \multicolumn{4}{c}{Chronos (Tiny)} & \multicolumn{4}{c}{Chronos (Mini)} & \multicolumn{4}{c}{Chronos (Small)} & \multicolumn{4}{c}{TimesFM (8M)} & \multicolumn{4}{c}{TimesFM (20M)} \\
 & & 5 & 21 & 252 & 512 & 5 & 21 & 252 & 512 & 5 & 21 & 252 & 512 & 5 & 21 & 252 & 512 & 5 & 21 & 252 & 512 \\
\midrule
\addlinespace[0.5em]
\rotatebox[origin=c]{90}{Benchmark} & \begin{tabular}{r}5 \\ 21 \\ 252 \\ 512\end{tabular} & \begin{tabular}{c}-8.97* \\ -9.85* \\ -9.87* \\ -9.95*\end{tabular} & \begin{tabular}{c}-4.47* \\ -6.42* \\ -6.57* \\ -6.60*\end{tabular} & \begin{tabular}{c}-2.89* \\ -5.02* \\ -5.44* \\ -5.20*\end{tabular} & \begin{tabular}{c}-3.26* \\ -5.40* \\ -5.93* \\ -5.68*\end{tabular} & \begin{tabular}{c}-12.61* \\ -13.67* \\ -13.76* \\ -13.81*\end{tabular} & \begin{tabular}{c}-5.95* \\ -8.41* \\ -8.74* \\ -8.65*\end{tabular} & \begin{tabular}{c}-2.11* \\ -3.42* \\ -3.57* \\ -3.46*\end{tabular} & \begin{tabular}{c}-2.44* \\ -3.71* \\ -3.89* \\ -3.78*\end{tabular} & \begin{tabular}{c}-10.80* \\ -12.46* \\ -12.66* \\ -12.65*\end{tabular} & \begin{tabular}{c}-5.36* \\ -6.84* \\ -6.99* \\ -6.98*\end{tabular} & \begin{tabular}{c}-2.55* \\ -3.98* \\ -4.12* \\ -4.03*\end{tabular} & \begin{tabular}{c}-2.52* \\ -3.88* \\ -4.04* \\ -3.94*\end{tabular} & \begin{tabular}{c}-28.46* \\ -28.87* \\ -28.84* \\ -28.89*\end{tabular} & \begin{tabular}{c}-20.27* \\ -20.61* \\ -20.43* \\ -20.61*\end{tabular} & \begin{tabular}{c}-32.34* \\ -32.67* \\ -32.33* \\ -32.60*\end{tabular} & \begin{tabular}{c}-27.55* \\ -27.75* \\ -27.64* \\ -27.76*\end{tabular} & \begin{tabular}{c}-29.14* \\ -29.58* \\ -29.58* \\ -29.61*\end{tabular} & \begin{tabular}{c}-22.92* \\ -23.24* \\ -23.09* \\ -23.25*\end{tabular} & \begin{tabular}{c}-26.52* \\ -26.72* \\ -26.42* \\ -26.81*\end{tabular} & \begin{tabular}{c}-25.42* \\ -25.57* \\ -25.37* \\ -25.65*\end{tabular} \\
\noalign{\vskip 0.6em}
\rotatebox[origin=c]{90}{Chronos (Tiny)} & \begin{tabular}{r}5 \\ 21 \\ 252 \\ 512\end{tabular} & \begin{tabular}{c}{} \\ {} \\ {} \\ {}\end{tabular} & \begin{tabular}{c}{} \\ {} \\ {} \\ {}\end{tabular} & \begin{tabular}{c}{} \\ {} \\ {} \\ {}\end{tabular} & \begin{tabular}{c}{} \\ {} \\ {} \\ {}\end{tabular} & \begin{tabular}{c}-1.61 \\ -11.94* \\ -12.23* \\ -11.72*\end{tabular} & \begin{tabular}{c}8.00* \\ -2.36* \\ -5.54* \\ -3.93*\end{tabular} & \begin{tabular}{c}7.22* \\ 1.89 \\ 0.22 \\ 0.42\end{tabular} & \begin{tabular}{c}6.77* \\ 1.40 \\ -0.46 \\ -0.26\end{tabular} & \begin{tabular}{c}2.49* \\ -16.05* \\ -11.88* \\ -10.08*\end{tabular} & \begin{tabular}{c}6.82* \\ -2.15* \\ -4.02* \\ -3.43*\end{tabular} & \begin{tabular}{c}7.80* \\ 1.95 \\ -0.40 \\ -0.11\end{tabular} & \begin{tabular}{c}7.35* \\ 1.68 \\ -0.49 \\ -0.19\end{tabular} & \begin{tabular}{c}-31.28* \\ -30.73* \\ -29.13* \\ -28.56*\end{tabular} & \begin{tabular}{c}-19.23* \\ -21.68* \\ -19.61* \\ -19.03*\end{tabular} & \begin{tabular}{c}-23.52* \\ -30.99* \\ -30.92* \\ -30.92*\end{tabular} & \begin{tabular}{c}-21.36* \\ -26.10* \\ -26.48* \\ -26.65*\end{tabular} & \begin{tabular}{c}-32.23* \\ -31.64* \\ -30.13* \\ -29.38*\end{tabular} & \begin{tabular}{c}-19.63* \\ -23.73* \\ -22.41* \\ -21.60*\end{tabular} & \begin{tabular}{c}-12.87* \\ -22.23* \\ -23.03* \\ -22.87*\end{tabular} & \begin{tabular}{c}-13.30* \\ -21.38* \\ -22.31* \\ -22.40*\end{tabular} \\
\noalign{\vskip 0.6em}
\rotatebox[origin=c]{90}{Chronos (Mini)} & \begin{tabular}{r}5 \\ 21 \\ 252 \\ 512\end{tabular} & \begin{tabular}{c}{} \\ {} \\ {} \\ {}\end{tabular} & \begin{tabular}{c}{} \\ {} \\ {} \\ {}\end{tabular} & \begin{tabular}{c}{} \\ {} \\ {} \\ {}\end{tabular} & \begin{tabular}{c}{} \\ {} \\ {} \\ {}\end{tabular} & \begin{tabular}{c}{} \\ {} \\ {} \\ {}\end{tabular} & \begin{tabular}{c}{} \\ {} \\ {} \\ {}\end{tabular} & \begin{tabular}{c}{} \\ {} \\ {} \\ {}\end{tabular} & \begin{tabular}{c}{} \\ {} \\ {} \\ {}\end{tabular} & \begin{tabular}{c}4.40* \\ -12.17* \\ -9.18* \\ -8.20*\end{tabular} & \begin{tabular}{c}9.52* \\ -0.94 \\ -3.25* \\ -2.71*\end{tabular} & \begin{tabular}{c}11.15* \\ 3.51* \\ -0.80 \\ 0.11\end{tabular} & \begin{tabular}{c}10.83* \\ 3.07* \\ -1.42 \\ 0.05\end{tabular} & \begin{tabular}{c}-27.09* \\ -29.66* \\ -28.05* \\ -27.55*\end{tabular} & \begin{tabular}{c}-15.40* \\ -20.54* \\ -18.44* \\ -17.94*\end{tabular} & \begin{tabular}{c}-23.01* \\ -31.00* \\ -30.23* \\ -30.01*\end{tabular} & \begin{tabular}{c}-21.09* \\ -26.06* \\ -26.22* \\ -26.22*\end{tabular} & \begin{tabular}{c}-27.86* \\ -30.50* \\ -29.17* \\ -28.58*\end{tabular} & \begin{tabular}{c}-16.65* \\ -22.71* \\ -21.20* \\ -20.61*\end{tabular} & \begin{tabular}{c}-12.33* \\ -22.00* \\ -21.66* \\ -21.24*\end{tabular} & \begin{tabular}{c}-12.92* \\ -21.23* \\ -21.25* \\ -21.03*\end{tabular} \\
\noalign{\vskip 0.6em}
\rotatebox[origin=c]{90}{Chronos (Small)} & \begin{tabular}{r}5 \\ 21 \\ 252 \\ 512\end{tabular} & \begin{tabular}{c}{} \\ {} \\ {} \\ {}\end{tabular} & \begin{tabular}{c}{} \\ {} \\ {} \\ {}\end{tabular} & \begin{tabular}{c}{} \\ {} \\ {} \\ {}\end{tabular} & \begin{tabular}{c}{} \\ {} \\ {} \\ {}\end{tabular} & \begin{tabular}{c}{} \\ {} \\ {} \\ {}\end{tabular} & \begin{tabular}{c}{} \\ {} \\ {} \\ {}\end{tabular} & \begin{tabular}{c}{} \\ {} \\ {} \\ {}\end{tabular} & \begin{tabular}{c}{} \\ {} \\ {} \\ {}\end{tabular} & \begin{tabular}{c}{} \\ {} \\ {} \\ {}\end{tabular} & \begin{tabular}{c}{} \\ {} \\ {} \\ {}\end{tabular} & \begin{tabular}{c}{} \\ {} \\ {} \\ {}\end{tabular} & \begin{tabular}{c}{} \\ {} \\ {} \\ {}\end{tabular} & \begin{tabular}{c}-30.40* \\ -28.99* \\ -28.83* \\ -28.35*\end{tabular} & \begin{tabular}{c}-18.92* \\ -19.59* \\ -19.49* \\ -18.79*\end{tabular} & \begin{tabular}{c}-26.05* \\ -29.45* \\ -30.77* \\ -30.45*\end{tabular} & \begin{tabular}{c}-22.86* \\ -25.48* \\ -26.26* \\ -26.19*\end{tabular} & \begin{tabular}{c}-31.60* \\ -29.91* \\ -30.10* \\ -29.49*\end{tabular} & \begin{tabular}{c}-20.85* \\ -21.81* \\ -22.40* \\ -21.60*\end{tabular} & \begin{tabular}{c}-15.82* \\ -20.55* \\ -22.09* \\ -21.57*\end{tabular} & \begin{tabular}{c}-15.83* \\ -20.17* \\ -21.39* \\ -21.10*\end{tabular} \\
\noalign{\vskip 0.6em}
\rotatebox[origin=c]{90}{TimesFM (8M)} & \begin{tabular}{r}5 \\ 21 \\ 252 \\ 512\end{tabular} & \begin{tabular}{c}{} \\ {} \\ {} \\ {}\end{tabular} & \begin{tabular}{c}{} \\ {} \\ {} \\ {}\end{tabular} & \begin{tabular}{c}{} \\ {} \\ {} \\ {}\end{tabular} & \begin{tabular}{c}{} \\ {} \\ {} \\ {}\end{tabular} & \begin{tabular}{c}{} \\ {} \\ {} \\ {}\end{tabular} & \begin{tabular}{c}{} \\ {} \\ {} \\ {}\end{tabular} & \begin{tabular}{c}{} \\ {} \\ {} \\ {}\end{tabular} & \begin{tabular}{c}{} \\ {} \\ {} \\ {}\end{tabular} & \begin{tabular}{c}{} \\ {} \\ {} \\ {}\end{tabular} & \begin{tabular}{c}{} \\ {} \\ {} \\ {}\end{tabular} & \begin{tabular}{c}{} \\ {} \\ {} \\ {}\end{tabular} & \begin{tabular}{c}{} \\ {} \\ {} \\ {}\end{tabular} & \begin{tabular}{c}{} \\ {} \\ {} \\ {}\end{tabular} & \begin{tabular}{c}{} \\ {} \\ {} \\ {}\end{tabular} & \begin{tabular}{c}{} \\ {} \\ {} \\ {}\end{tabular} & \begin{tabular}{c}{} \\ {} \\ {} \\ {}\end{tabular} & \begin{tabular}{c}-1.65 \\ -18.36* \\ -10.36* \\ -6.01*\end{tabular} & \begin{tabular}{c}15.88* \\ -0.50 \\ 4.39* \\ 7.36*\end{tabular} & \begin{tabular}{c}17.92* \\ 5.98* \\ 14.85* \\ 14.75*\end{tabular} & \begin{tabular}{c}16.54* \\ 4.36* \\ 11.35* \\ 17.22*\end{tabular} \\
\addlinespace[0.5em]
\bottomrule
\end{tabular}
\vspace{0.1cm}
\begin{tablenotes}[para,flushleft]
\large
\textbf{Note:} This table reports Diebold--Mariano test statistics based on the daily cross-sectional mean of squared-error differences (row minus column), following the forecast-comparison approach of \citet{gu2020empirical}. Positive statistics favor the column specification; negative statistics favor the row specification. All comparisons use the same stock-date observations. Heteroskedasticity- and autocorrelation-consistent (HAC) standard errors use one lag. An asterisk denotes significance at the 5\% level using a two-sided asymptotic standard normal test, without adjustment for multiple comparisons. Input windows are measured in trading days. The benchmark is CatBoost trained on U.S. excess returns.
\end{tablenotes}
\end{threeparttable}
\end{adjustbox}
\end{table}
\vspace*{\fill}
\end{landscape}

\vspace*{\fill}
\begin{table}[h]
  \centering
  \begin{threeparttable}
    \begin{adjustbox}{width=0.82\textwidth, center}
    \scriptsize
    \captionsetup{width=\linewidth}
    \caption{Finance-Native Pre-Trained TSFMs - Spread Portfolio Performance}
    \label{Spread_Portfolio_pre_trained_table}
    \begin{minipage}{\linewidth}
      \renewcommand{\arraystretch}{1.2}
      \begin{tabular}{c|cccc|cccc|cccc}
        \toprule
        \textbf{Model} & \multicolumn{4}{c|}{\textbf{Benchmark}} & \multicolumn{4}{c|}{\textbf{Chronos (Tiny)}} & \multicolumn{4}{c}{\textbf{Chronos (Mini)}}  \\
        \textbf{Window Size} & 5 & 21 & 252 & 512 & 5 & 21 & 252 & 512 & 5 & 21 & 252 & 512 \\
        \midrule
        Low & -31.09 & -35.64 & -34.19 & -35.11 & 14.95 & 4.33 & -8.55 & -12.36 & 16.03 & 3.85 & -13.08 & -15.67 \\ % & 14.08 & 0.17 & -16.68 & -20.16 \\
        \addlinespace[3pt]
        2 &  -0.80 & -2.89 & -3.83 & -3.98 & 12.96 & 7.77 & 1.47 & -1.21 & 12.77 & 8.05 & 0.29 & -1.19 \\ % 12.07 & 6.62 & -0.34 & -2.30 \\
        \addlinespace[3pt]
        3 & 4.33 & 4.56 & 2.80 & 3.33 & 11.67 & 9.12 & 5.44 & 2.70 & 11.05 & 8.64 & 5.22 & 3.30 \\ % 11.48 & 8.47 & 3.57 & 4.06 \\
        \addlinespace[3pt]
        4 & 7.51 & 8.36 & 6.40 & 6.87 & 11.05 & 8.97 & 6.47 & 7.22 & 11.61 & 9.65 & 6.34 & 6.72 \\ % 10.40 & 8.30 & 5.36 & 6.13 \\
        \addlinespace[3pt]
        5 & 9.40 & 10.27 & 9.88 & 9.40 & 11.05 & 10.01 & 8.58 & 9.20 & 9.85 & 9.56 & 8.98 & 9.19 \\ % 10.22 & 9.99 & 8.37 & 8.59 \\
        \addlinespace[3pt]
        6 & 11.57 & 11.44 & 12.00 & 12.43 & 9.93 & 10.32 & 10.78 & 11.44 & 10.02 & 10.89 & 10.08 & 11.41 \\ % 10.13 & 10.74 & 10.79 & 11.11 \\
        \addlinespace[3pt]
        7 & 15.27 & 15.26 & 15.66 & 14.08 & 9.00 & 10.97 & 13.51 & 13.93 & 8.55 & 11.81 & 13.92 & 12.50 \\ % 10.16 & 11.66 & 13.00 & 12.68 \\
        \addlinespace[3pt]
        8 & 18.24 & 18.67 & 18.31 & 19.43 & 10.70 & 12.89 & 16.24 & 16.90 & 9.29 & 12.30 & 16.64 & 16.91 \\ % 10.70 & 13.89 & 16.21 & 16.81 \\
        \addlinespace[3pt]
        9 & 25.06 & 26.11 & 27.71 & 27.79 & 9.09 & 15.43 & 20.32 & 21.64 & 9.27 & 14.81 & 22.67 & 22.32 \\ % 10.40 & 15.09 & 22.26 & 22.65 \\
        \addlinespace[3pt]
        High & 53.61 & 56.94 & 58.36 & 58.85 & 12.66 & 23.27 & 38.82 & 43.63 & 14.62 & 23.53 & 42.03 & 47.59 \\ % 13.44 & 28.15 & 50.54 & 53.52 \\
        \addlinespace[3pt]
        $\tfrac{1}{2}(H-L)$ & 42.35 & 46.29 & 46.27 & 46.98 & -1.14 & 9.47 & 23.69 & 28.00 & -0.70 & 9.84 & 27.55 & 31.63 \\ % -0.32 & 13.99 & 33.61 & 36.84 \\
    
        \midrule \midrule

        \textbf{Model} & \multicolumn{4}{c|}{\textbf{Chronos (Small)}} & \multicolumn{4}{c|}{\textbf{TimesFM (8M)}} & \multicolumn{4}{c}{\textbf{TimesFM (20M)}}  \\
        \textbf{Window Size} & 5 & 21 & 252 & 512 & 5 & 21 & 252 & 512 & 5 & 21 & 252 & 512  \\
        \midrule
        Low & 14.08 & 0.17 & -16.68 & -20.16 & 30.70 & 4.43 & -11.43 & -12.24 & 29.87 & 1.79 & -17.60 & -17.09  \\
        \addlinespace[3pt]
        2 & 12.07 & 6.62 & -0.34 & -2.30 & 20.89 & 8.57 & 0.28 & -0.94 & 19.29 & 5.58 & -2.56 & -1.48 \\
        \addlinespace[3pt]
        3 & 11.48 & 8.47 & 3.57 & 4.06 & 16.75 & 8.21 & 3.86 & 3.78 & 16.38 & 5.87 & 2.83 & 2.66 \\
        \addlinespace[3pt]
        4 & 10.40 & 8.30 & 5.36 & 6.13 & 14.38 & 9.48 & 6.56 & 7.01 & 13.73 & 7.87 & 6.61 & 6.05  \\
        \addlinespace[3pt]
        5 & 10.22 & 9.99 & 8.37 & 8.59 & 12.01 & 10.81 & 8.68 & 9.36 & 11.82 & 9.59 & 9.68 & 8.66 \\
        \addlinespace[3pt]
        6 & 10.13 & 10.74 & 10.79 & 11.11 & 10.09 & 11.62 & 11.83 & 11.77 & 10.78 & 12.15 & 12.23 & 12.34 \\
        \addlinespace[3pt]
        7 & 10.16 & 11.66 & 13.00 & 12.68 & 8.28 & 12.42 & 14.05 & 14.15 & 8.81 & 13.58 & 14.48 & 14.74 \\
        \addlinespace[3pt]
        8 & 10.70 & 13.89 & 16.21 & 16.81 & 4.77 & 12.69 & 18.10 & 17.84 & 6.11 & 15.95 & 19.15 & 19.14 \\
        \addlinespace[3pt]
        9 & 10.40 & 15.09 & 22.26 & 22.65 & 2.02 & 15.76 & 22.80 & 22.45 & 2.62 & 17.96 & 25.40 & 24.93 \\
        \addlinespace[3pt]
        High & 13.44 & 28.15 & 50.54 & 53.52 & -6.83 & 19.10 & 38.34 & 39.91 & -6.34 & 22.75 & 42.87 & 43.14 \\
        \addlinespace[3pt]
        $\tfrac{1}{2}(H-L)$ & -0.32 & 13.99 & 33.61 & 36.84 & -18.77 & 7.33 & 24.88 & 26.07 & -18.10 & 10.48 & 30.24 & 30.12 \\
        \bottomrule
      \end{tabular}
      \vspace{0.1cm}
      \begin{tablenotes}[para,flushleft]
        \footnotesize
        \textbf{Note:} This table presents the annualized average returns of decile spread portfolios formed using forecasts from various predictive models across different window sizes (5, 21, 252, and 512 trading days). The benchmark model is CatBoost, the representative benchmark. The time series foundation models (TSFMs) include Chronos (tiny, mini, and small) and TimesFM (with 8 million and 20 million parameters). Inference is performed using the finance-native pre-trained models. For each model and window size, we report standalone decile returns from Low to High. Here, $H$ and $L$ denote the standalone returns of the highest- and lowest-forecast deciles. The final row reports the unit-gross long--short return, $\frac{1}{2}(H-L)$, with long- and short-leg contributions of $\frac{1}{2}H$ and $-\frac{1}{2}L$, respectively. Portfolios are constructed using equal-weighted (EW) decile sorting based on predicted returns.
      \end{tablenotes}
    \end{minipage}
    \end{adjustbox}
  \end{threeparttable}
\end{table}
\vspace*{\fill}
\clearpage

% \vspace*{\fill}
% \begin{figure}[h]
%     \caption{Cumulative Log Returns of Pre-Trained TSFMs: Long and Short Portfolios}
%     \label{Cumulative_log_return_pre_trained_long_and_short}
%     \centering
%     \includegraphics[
%         width=0.70\textwidth,
%         clip,
%         trim=0 0 0 54, % left bottom right top
%     ]{FM_pretrained.pdf} % Adjust the width and trims
%     \begin{minipage}{0.8\textwidth}
%         \footnotesize
%         \textbf{Note:} This figure displays the cumulative log returns of long and short portfolios, separately constructed using various forecasting models using input windows of 5, 21, 252, and 512 trading days. The benchmark model is CatBoost, the representative benchmark. The time series foundation models (TSFMs) include Chronos (tiny, mini, and small) and TimesFM (with 8 million and 20 million parameters). Inference is performed using the finance-native pre-trained models. Each subplot corresponds to a specific horizon, as indicated by the text labels in the upper-left corners. For each model, the solid line denotes the performance of the long portfolio, while the dashed line represents the corresponding short portfolio. The benchmark model (CatBoost) is highlighted in black with bold lines. Shaded areas indicate U.S. recession periods, as defined by the National Bureau of Economic Research (NBER). The model-based long--short portfolios use equal weighting within each leg and are scaled to unit gross exposure. The S\&P 500 is shown separately as a market reference.
%     \end{minipage}
% \end{figure}
% \vspace*{\fill}
% \clearpage

\subsection{Detailed Scaling Figures} \label{appendix:scaling_figures}
\begin{figure}[h] %%%% ----------------------- fixed
    \caption{Impact of Global Data on Forecasting Performance (Benchmarks)}
    \label{impact_benchmark_scale_forecasting}
    \centering
    \includegraphics[
        width=0.75\textwidth,
        clip,
        trim=0 56 0 63, % left bottom right top
    ]{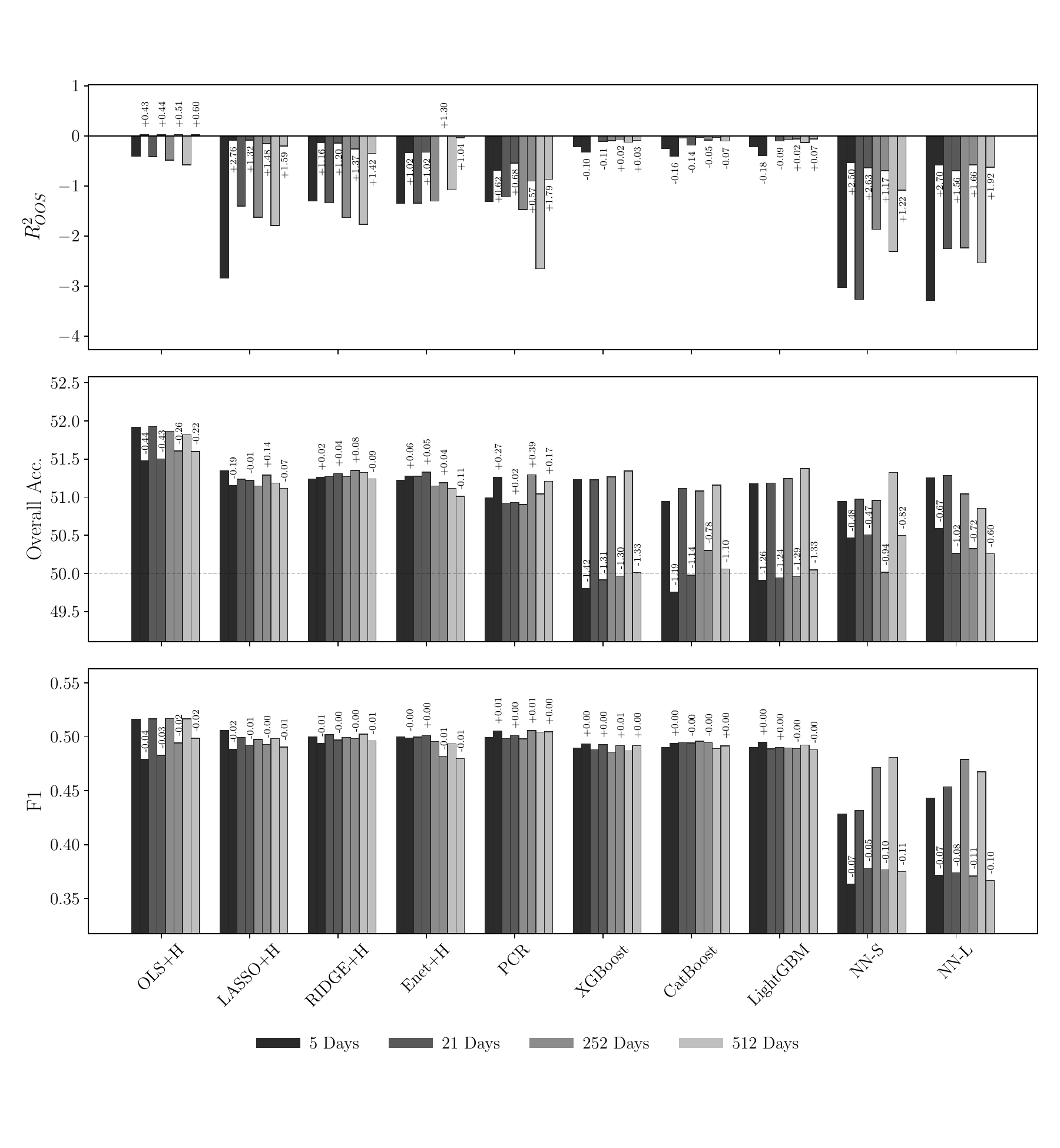}
    \begin{minipage}{0.8\textwidth}
        \footnotesize
        \textbf{Note:} This figure compares forecasting performance metrics ($\RsqOOS$, overall accuracy, and F1) for benchmark models under two regimes: U.S.-only (left bar) and global (right bar) data. Bars of the same color correspond to the same input window length (5, 21, 252, and 512 trading days). Each pair of adjacent bars illustrates the performance change attributable to expanding the training data from U.S.-only to Global. Benchmark models include linear (OLS+H, LASSO+H, RIDGE+H, Enet+H, and PCR), ensemble (XGBoost, CatBoost, and LightGBM), and neural network (NN-S and NN-L) models. `Overall Acc.' denotes overall directional accuracy, and `F1' refers to the macro-averaged F1 score. Accuracy measures and $\RsqOOS$ are reported in percentage terms; macro-$F_1$ is reported on a zero-to-one scale. In the middle panel, the horizontal line indicates 50\% overall accuracy. `H' indicates that the model is estimated using the Huber loss.
    \end{minipage}
\end{figure}

\begin{figure}[p] %%%% ----------------------- fixed
    \caption{Impact of Global Data on Portfolio Performance (Benchmarks)}
    \label{impact_benchmark_scale_portfolio}
    \centering
    \includegraphics[
        width=0.75\textwidth,
        clip,
        trim=0 56 0 63, % left bottom right top
    ]{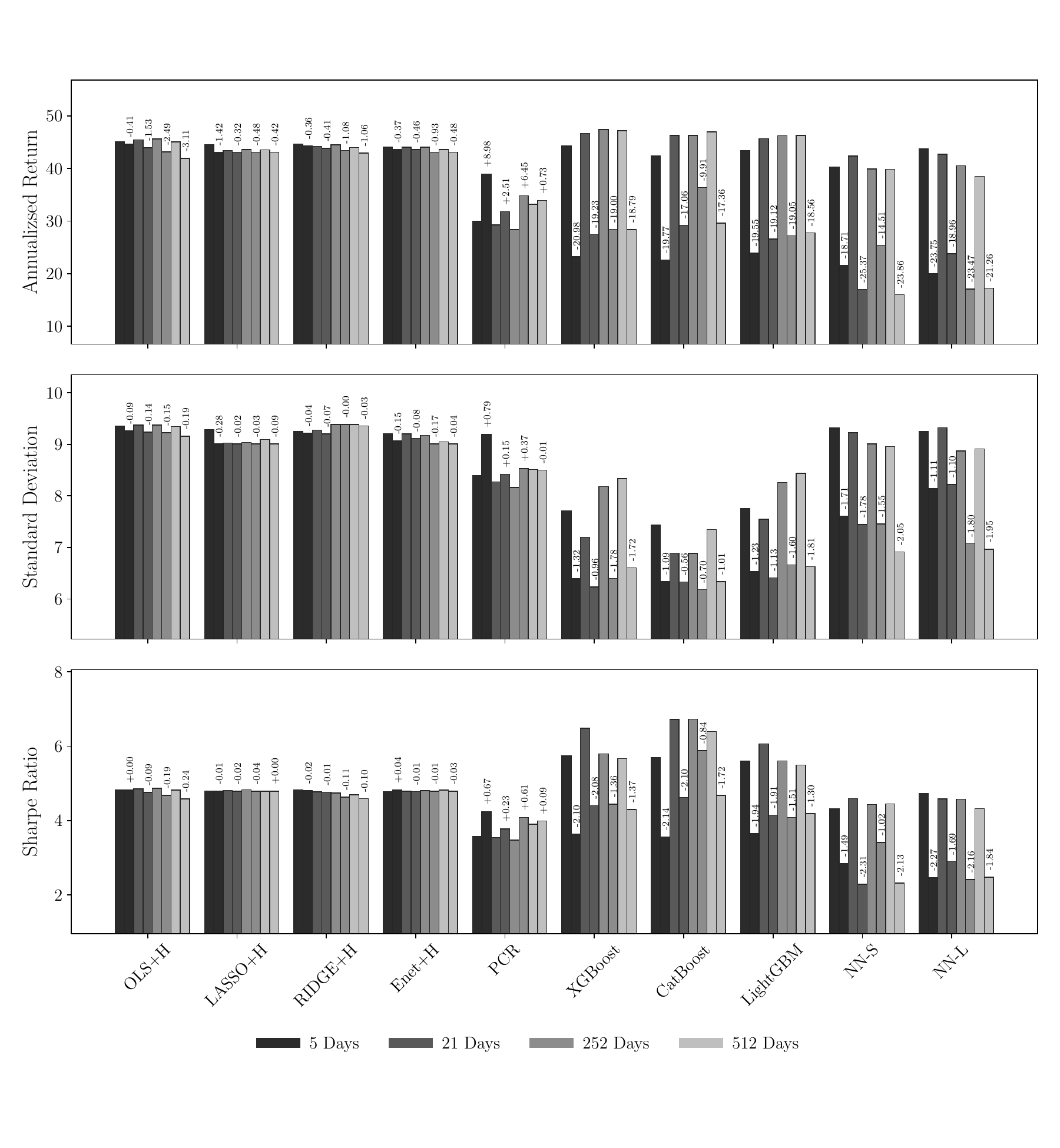}
    \begin{minipage}{0.8\textwidth}
        \footnotesize
        \textbf{Note:} This figure compares portfolio performance metrics (annualized return, standard deviation, and Sharpe ratio) for benchmark models under two regimes: U.S.-only (left bar) and global (right bar). Bars of the same color correspond to the same input window length (5, 21, 252, and 512 trading days). Each pair of adjacent bars illustrates the performance change attributable to expanding the training data from U.S.-only to Global. Benchmark models include linear (OLS+H, LASSO+H, RIDGE+H, Enet+H, and PCR), ensemble (XGBoost, CatBoost, and LightGBM), and neural network (NN-S and NN-L) models. `H' indicates that the model is estimated using the Huber loss.
    \end{minipage}
\end{figure}

This subsection reports the complete visual evidence underlying the representative scaling comparisons in \Cref{FP_scaling_representative} and \Cref{PP_scaling_representative}. The first two figures show how benchmark models respond when their training sample is expanded from U.S. excess returns to global excess returns. The last two figures report the corresponding TSFM comparisons across U.S.-only, global, JKP-augmented, and synthetic-augmented pre-training regimes. The main text summarizes the central patterns using representative numerical changes, while the figures here provide the full model-by-model and window-by-window comparison.\footnote{Regarding the benchmark results in \Cref{impact_benchmark_scale_forecasting} and \Cref{impact_benchmark_scale_portfolio}, one potential concern is that larger neural networks may achieve better performance by capturing more complex nonlinear relationships. To assess this, we extended the network architecture to \(128\) and \(512\) hidden units while keeping all other hyperparameters (activation function, optimizer, learning rate, and regularization) unchanged. For ease of comparison, we also report the results for the smaller architectures with 8 and 32 units (NN-S and NN-L), which are the models used throughout this study. The forecasting, portfolio, and spread portfolio results are presented in \Crefrange{FP_NN_extended}{PP_NN_extended_spread} of the \hyperref[pp:internet-title]{Internet Appendix}. Comparing across model architectures, we find that increasing the number of hidden units beyond 32 yields mixed results. At the 512-day window in the U.S. sample, $\RsqOOS$ improves from $-2.54\%$ for 32 hidden units to $-2.07\%$ and $-2.14\%$ for 128 and 512 units, respectively. This pattern is not uniform across windows or metrics: at 5 days, the larger U.S. networks have more negative $\RsqOOS$ than the 32-unit network. Portfolio performance is broadly similar across network sizes in the U.S. sample, while the global sample shows larger, non-monotonic differences. Although the SR occasionally improves with increased model size, these gains are not monotonic, as performance sometimes declines with higher complexity. The larger networks do not overturn the strong performance of tree-based ensembles, but rankings depend on the metric, training universe, and input window. Overall, these findings reinforce the strong squared-error and gross portfolio performance of the ensemble benchmarks relative to the neural networks considered.}
\vspace*{\fill}
\vspace*{\fill}

\begin{figure}[p] % ----------------- Fixed
    \caption{Impact of Global and Augmented Pre-Training on Forecasting Performance}
    \label{impact_TSFM_scale_forecasting}
    \centering
    \includegraphics[
        width=0.75\textwidth,
        clip,
        trim=0 56 0 63, % left bottom right top
    ]{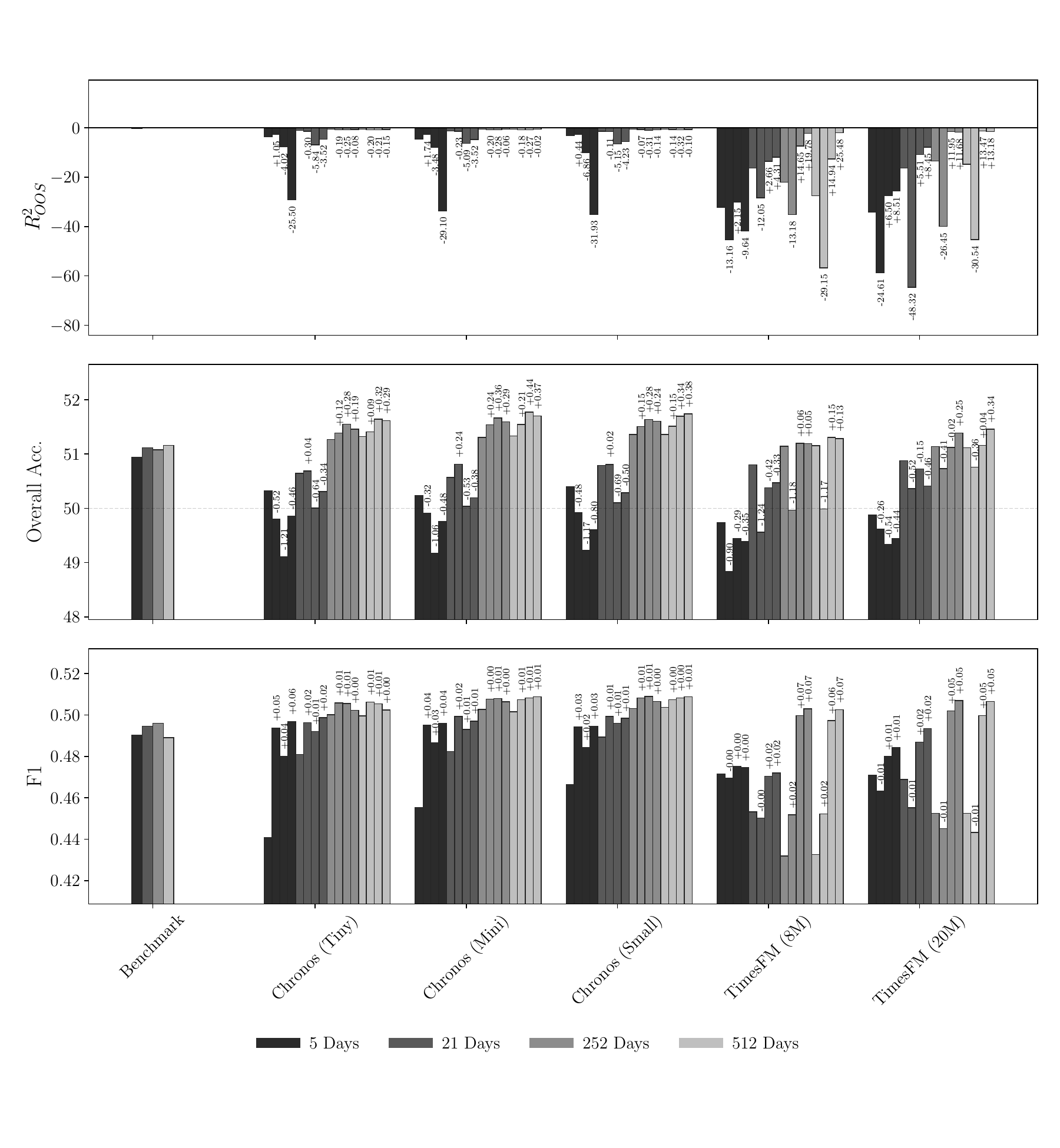}
    \begin{minipage}{0.8\textwidth}
        \footnotesize
        \textbf{Note:} This figure compares forecasting performance metrics ($\RsqOOS$, overall accuracy, and F1) for five models under four pre-training regimes: U.S.-only as the left bar in each set, Global as the second, JKP-augmented as the third, and Synthetic-augmented as the fourth. Bars of the same color correspond to the same input window length (5, 21, 252, and 512 trading days). Each group of adjacent bars depicts the change in performance resulting from expansions of the pre-training data, progressing from U.S.-only to Global, and subsequently to JKP- and synthetic-augmented variants. Models include Chronos (tiny, mini, and small) and TimesFM (with 8 million and 20 million parameters). JKP characteristics used in the augmented data are defined in \citet{jensen2023there}, and the synthetic data are generated following \citet{ansari2024chronos}. The benchmark group presents the results of the CatBoost model, trained on U.S. data. `Overall Acc.' denotes overall directional accuracy, and `F1' refers to the macro-averaged F1 score. Accuracy measures and $\RsqOOS$ are reported in percentage terms; macro-$F_1$ is reported on a zero-to-one scale. In the middle panel, the horizontal line indicates 50\% overall accuracy.
    \end{minipage}
\end{figure}

\begin{figure}[p] % ----------------- Fixed
    \caption{Impact of Global and Augmented Pre-Training on Portfolio Performance}
    \label{impact_TSFM_scale_portfolio}
    \centering
    \includegraphics[
        width=0.75\textwidth,
        clip,
        trim=0 56 0 63, % left bottom right top
    ]{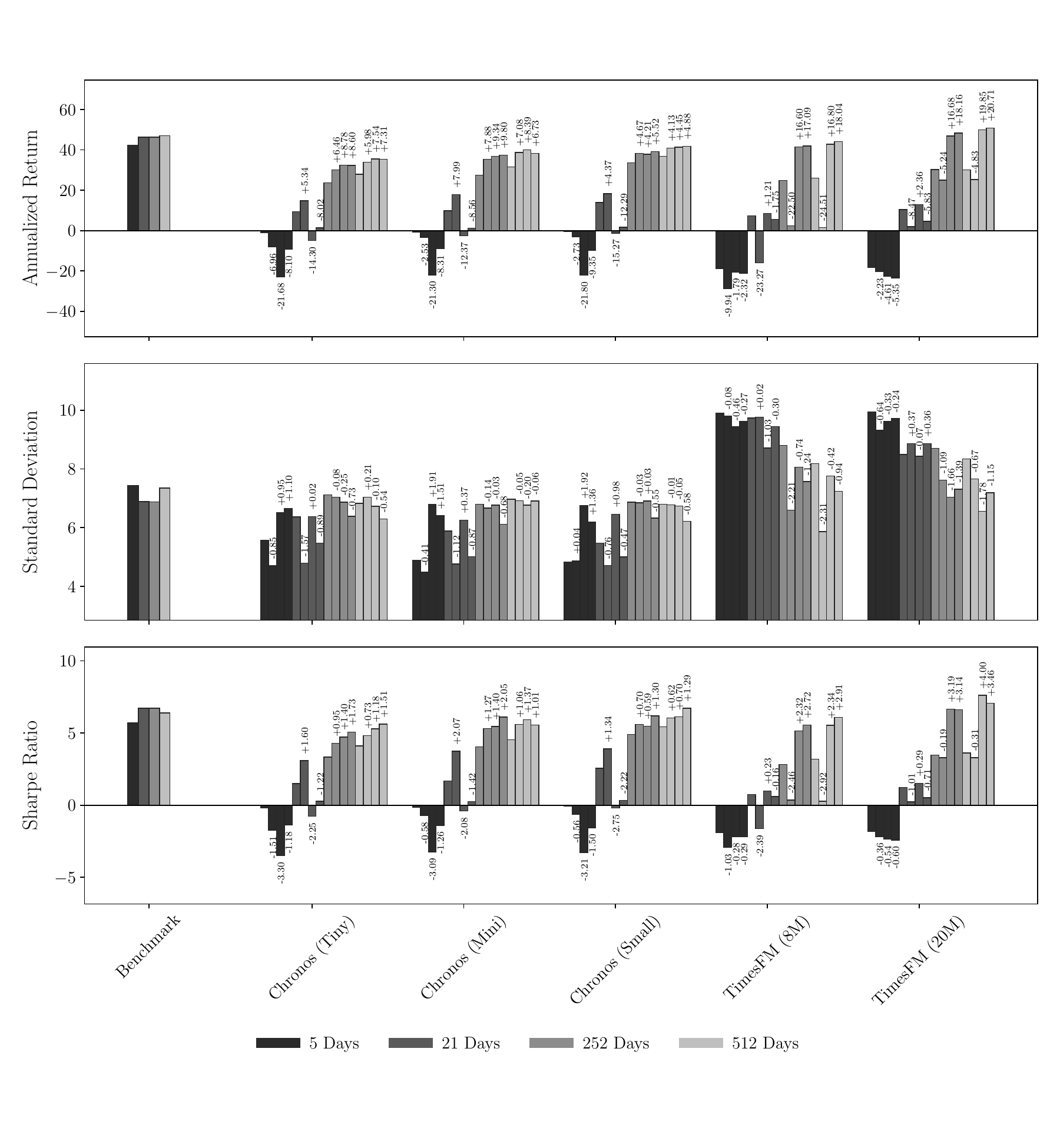}
    \begin{minipage}{0.8\textwidth}
        \footnotesize
        \textbf{Note:} This figure compares portfolio performance metrics (annualized return, standard deviation, and Sharpe ratio) for five models under four pre-training regimes: U.S.-only as the left bar in each set, Global as the second, JKP-augmented as the third, and Synthetic-Augmented as the fourth. Bars of the same color correspond to the same input window length (5, 21, 252, and 512 trading days). Each group of adjacent bars depicts the change in performance resulting from expansions of the pre-training data, progressing from U.S.-only to Global, and subsequently to JKP- and synthetic-augmented variants. Models include Chronos (tiny, mini, and small) and TimesFM (with 8 million and 20 million parameters). JKP characteristics used in the augmented data are defined in \citet{jensen2023there}, and the synthetic data are generated following \citet{ansari2024chronos}. The benchmark group presents the results of the CatBoost model, trained on U.S. data.
    \end{minipage}
\end{figure}

\begin{landscape}
\thispagestyle{landscape}
\begin{table}[p]
\vspace*{\fill} % push it down\
\centering
\caption{Comparison of Daily Out-of-Sample Forecasting Performance of Augmented Pre-Trained TSFMs Using Diebold--Mariano Test}
\label{DM_test_TSFMs_Scaling}
\begin{adjustbox}{max width=1.50\textwidth}
\begin{threeparttable}
\renewcommand{\arraystretch}{1.2}
\setlength{\tabcolsep}{3pt}

 \\
\bottomrule
\addlinespace[0.5em]
\end{tabular}
\vspace{0.1cm}
\begin{tablenotes}[para,flushleft]
\large
\textbf{Note:} This table reports Diebold--Mariano test statistics based on the daily cross-sectional mean of squared-error differences (row minus column), following the forecast-comparison approach of \citet{gu2020empirical}. Positive statistics favor the column specification; negative statistics favor the row specification. All comparisons use the same stock-date observations. Heteroskedasticity- and autocorrelation-consistent (HAC) standard errors use one lag. An asterisk denotes significance at the 5\% level using a two-sided asymptotic standard normal test, without adjustment for multiple comparisons. Input windows are measured in trading days. The benchmark model is CatBoost, the representative benchmark. Time series foundation models (TSFMs) include Chronos (tiny, mini, and small) and TimesFM (with 8 million and 20 million parameters). The models are pre-trained from scratch under two training-data regimes: global data augmented with JKP characteristics, and global data augmented with synthetic data of equivalent size to the JKP characteristics. Synthetic-augmented models are identified by `Syn.'. JKP characteristics are defined in \citet{jensen2023there}, and synthetic data are generated following \citet{ansari2024chronos}.
\end{tablenotes}
\end{threeparttable}
\end{adjustbox}
\end{table}
\end{landscape}

\begin{figure} %% fixed ----------------------------------
    \caption{Yearly SRs of Long--Short Portfolios (Benchmarks)}
    \label{yearly_sharpe_long_short_benchmarks}
    \centering
    \includegraphics[
        width=0.75\textwidth,
        clip,
        trim=0 0 0 5, % left bottom right top
    ]{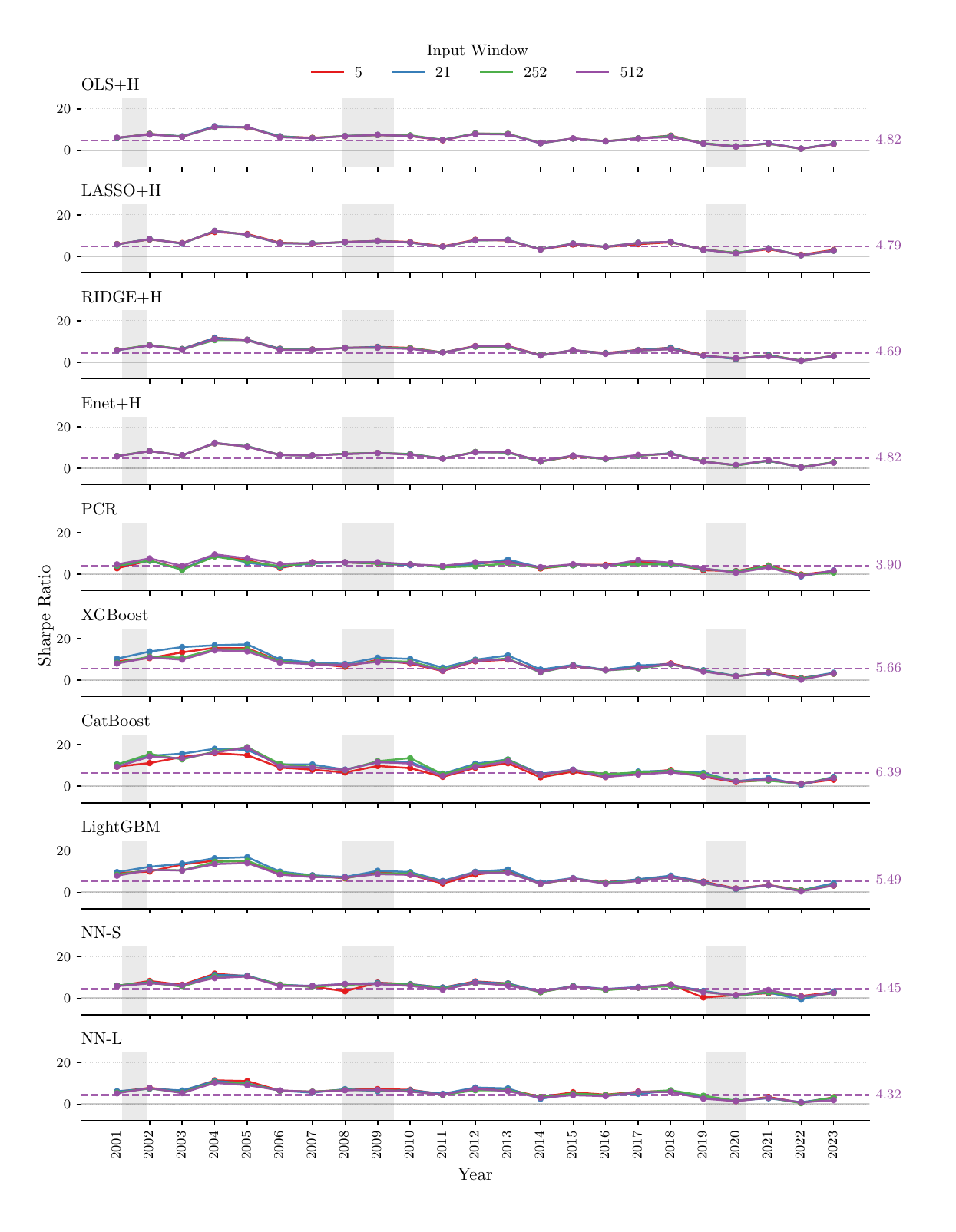} % Adjust the width and trims
    
    \begin{minipage}{0.8\textwidth}
        \footnotesize
        \textbf{Note:} The figure reports annual SRs from 2001 to 2023 for long--short EW portfolios. Results are presented by model across multiple window sizes (5, 21, 252, and 512 trading days). The set of models includes linear methods (OLS+H, LASSO+H, RIDGE+H, Enet+H, and PCR), ensemble models (XGBoost, CatBoost, and LightGBM), and neural networks (NN-S and NN-L), all trained using U.S. excess returns. `H' indicates that the model is estimated using the Huber loss. The dashed horizontal line in each panel denotes the full-sample SR of the corresponding 512-day specification, with its value reported on the right-hand side of the panel. Shaded regions correspond to U.S. recession periods as identified by the National Bureau of Economic Research (NBER).
    \end{minipage}
\end{figure}

\begin{landscape}
\thispagestyle{landscape}
\begin{table}\centering\vfill
\begin{threeparttable}
\begin{adjustbox}{width=1.0\textwidth, center}
\scriptsize
\captionsetup{width=\linewidth}
\caption{Long--Short Portfolio SRs under Position-Based Cost Scenarios (Benchmarks)}
\label{sharpe_ratio_with_transaction_costs_benchmarks}
\begin{minipage}{\linewidth}
\renewcommand{\arraystretch}{1.2}%
\begin{tabular}{c|*{16}{c}}
\toprule
\multirow{2}{*}{\textbf{Model}} 
& \multicolumn{4}{c}{0 bps} 
& \multicolumn{4}{c}{20 bps} 
& \multicolumn{4}{c}{40 bps} 
& \multicolumn{4}{c}{Mixed} \\
\cmidrule(lr){2-5} \cmidrule(lr){6-9} \cmidrule(lr){10-13} \cmidrule(lr){14-17}
& 5 & 21 & 252 & 512 
& 5 & 21 & 252 & 512 
& 5 & 21 & 252 & 512 
& 5 & 21 & 252 & 512  \\
\midrule

\multicolumn{17}{l}{\textbf{U.S.}} \\
OLS+H       & 4.82 & 4.85 & 4.87 & 4.82 & -0.57 & -0.53 & -0.51 & -0.57 & -5.96 & -5.90 & -5.89 & -5.96 & -0.00 & 0.04 & 0.06 & -0.01 \\
LASSO+H  & 4.80 & 4.81 & 4.82 & 4.79 & -0.63 & -0.77 & -0.75 & -0.75 & -6.06 & -6.35 & -6.33 & -6.29 & -0.06 & -0.19 & -0.17 & -0.18 \\
RIDGE+H & 4.83 & 4.77 & 4.74 & 4.69 & -0.62 & -0.67 & -0.63 & -0.68 & -6.07 & -6.10 & -6.00 & -6.05 & -0.05 & -0.09 & -0.07 & -0.12 \\
Enet+H & 4.78 & 4.79 & 4.80 & 4.82 & -0.69 & -0.69 & -0.69 & -0.75 & -6.16 & -6.17 & -6.18 & -6.32 & -0.11 & -0.11 & -0.12 & -0.17 \\
PCR  & 3.57 & 3.54 & 3.48 & 3.90 & -2.43 & -2.55 & -2.70 & -2.02 & -8.44 & -8.65 & -8.87 & -7.95 & -1.83 & -1.95 & -2.10 & -1.44 \\
XGBoost   & 5.74 & 6.48 & 5.80 & 5.66 & -0.79 & -0.52 & -0.36 & -0.39 & -7.33 & -7.52 & -6.52 & -6.44 & -0.15 & 0.13 & 0.20 & 0.17 \\
CatBoost   & 5.70 & 6.72 & 6.72 & 6.39 & -1.08 & -0.60 & -0.60 & -0.47 & -7.86 & -7.91 & -7.92 & -7.32 & -0.41 & 0.09 & 0.09 & 0.16 \\
LightGBM   & 5.60 & 6.06 & 5.60 & 5.49 & -0.90 & -0.62 & -0.50 & -0.49 & -7.40 & -7.30 & -6.60 & -6.46 & -0.27 & 0.00 & 0.05 & 0.06 \\
NN-S   & 4.32 & 4.59 & 4.43 & 4.45 & -1.08 & -0.87 & -1.16 & -1.18 & -6.49 & -6.33 & -6.76 & -6.80 & -0.49 & -0.30 & -0.60 & -0.61 \\
NN-L   & 4.73 & 4.58 & 4.57 & 4.32 & -0.72 & -0.82 & -1.11 & -1.33 & -6.17 & -6.23 & -6.79 & -6.99 & -0.14 & -0.25 & -0.54 & -0.77 \\
\midrule

\multicolumn{17}{l}{\textbf{Global}} \\
OLS+H       & 4.82 & 4.76 & 4.68 & 4.58 & -0.62 & -0.70 & -0.79 & -0.92 & -6.07 & -6.15 & -6.25 & -6.43 & -0.05 & -0.13 & -0.22 & -0.36 \\
LASSO+H  & 4.79 & 4.79 & 4.79 & 4.79 & -0.81 & -0.81 & -0.81 & -0.81 & -6.40 & -6.40 & -6.40 & -6.40 & -0.23 & -0.23 & -0.23 & -0.23 \\
RIDGE+H & 4.81 & 4.76 & 4.63 & 4.59 & -0.66 & -0.72 & -0.74 & -0.80 & -6.13 & -6.19 & -6.11 & -6.18 & -0.09 & -0.14 & -0.19 & -0.24 \\
Enet+H & 4.82 & 4.78 & 4.79 & 4.79 & -0.74 & -0.75 & -0.81 & -0.81 & -6.30 & -6.27 & -6.40 & -6.40 & -0.16 & -0.17 & -0.23 & -0.23 \\
PCR  & 4.24 & 3.77 & 4.08 & 3.99 & -1.25 & -2.21 & -1.83 & -1.94 & -6.73 & -8.19 & -7.74 & -7.87 & -0.67 & -1.62 & -1.25 & -1.36 \\
XGBoost   & 3.64 & 4.40 & 4.44 & 4.30 & -4.24 & -3.68 & -3.43 & -3.33 & -12.12 & -11.77 & -11.31 & -10.96 & -3.22 & -2.66 & -2.46 & -2.40 \\
CatBoost   & 3.56 & 4.62 & 5.88 & 4.67 & -4.39 & -3.34 & -2.27 & -3.28 & -12.33 & -11.30 & -10.42 & -11.23 & -3.36 & -2.35 & -1.28 & -2.28 \\
LightGBM   & 3.66 & 4.15 & 4.09 & 4.18 & -4.06 & -3.72 & -3.48 & -3.42 & -11.78 & -11.58 & -11.05 & -11.02 & -3.06 & -2.72 & -2.54 & -2.49 \\
NN-S   & 2.84 & 2.28 & 3.41 & 2.32 & -3.79 & -4.49 & -3.35 & -4.98 & -10.41 & -11.26 & -10.11 & -12.27 & -2.98 & -3.74 & -2.69 & -4.27 \\
NN-L   & 2.46 & 2.89 & 2.41 & 2.48 & -3.73 & -3.24 & -4.71 & -4.76 & -9.92 & -9.37 & -11.84 & -12.00 & -3.03 & -2.60 & -4.03 & -4.07 \\

\bottomrule
\end{tabular}

\vspace{0.1cm}
\begin{tablenotes}[para,flushleft]\footnotesize
\textbf{Note:} This table reports annualized Sharpe ratios of equal-weighted (EW) long--short decile portfolios formed using daily forecasts from different benchmark models across input window lengths of 5, 21, 252, and 512 trading days. Performance is evaluated under four hypothetical daily position-based cost scenarios: no costs (0 bps), fixed costs of 20 bps and 40 bps, and a mixed-cost structure where small- and large-cap firms face different estimated trading costs following \citet{frazzini2012trading}. In the mixed specification, the estimated costs correspond roughly to 21.27 bps for small-cap stocks and 11.21 bps for large-cap stocks. Models evaluated include linear (OLS+H, LASSO+H, RIDGE+H, Enet+H, and PCR), ensemble (XGBoost, CatBoost, and LightGBM), and neural network (NN-S and NN-L) models. `H' indicates that the model is estimated using the Huber loss. The top panel presents results based on training with U.S. excess return data, while the bottom panel presents results based on training with global excess return data. Costs are charged daily on the absolute weight of every selected position, including retained positions, with total gross exposure normalized to one. The fixed-cost scenarios therefore deduct 20 or 40 bps from the portfolio return each day. These are position-based cost scenarios, not turnover-based estimates.

\end{tablenotes}
\end{minipage}
\end{adjustbox}
\end{threeparttable}
\vfill
\end{table}
\end{landscape}

% This subsection reports the complete international benchmark results for the seven markets summarized in \Cref{international_representative_comparison}.

\begin{landscape}
\thispagestyle{landscape}
\subsection{Complete International Benchmark Results} \label{complete_international_benchmark_results}
\vspace*{\fill} % push it down\
\begin{table}[h]
  \centering
  \begin{threeparttable}
    \begin{adjustbox}{width=1.2\textwidth, center}
    \scriptsize
    \captionsetup{width=\linewidth}
    \caption{Benchmark Models - $\RsqOOS$ (International)}
    \label{international_benchmark_forecasting_table}
    \begin{minipage}{\linewidth}
        \renewcommand{\arraystretch}{1.3}
      \begin{tabular}{c|cccc|cccc|cccc|cccc|cccc}
        \toprule
        \textbf{Model} & \multicolumn{4}{c|}{\textbf{OLS+H}} & \multicolumn{4}{c|}{\textbf{LASSO+H}} & \multicolumn{4}{c|}{\textbf{RIDGE+H}} & \multicolumn{4}{c|}{\textbf{Enet+H}} & \multicolumn{4}{c}{\textbf{PCR}} \\
        \textbf{Window Size} & 5 & 21 & 252 & 512 & 5 & 21 & 252 & 512 & 5 & 21 & 252 & 512 & 5 & 21 & 252 & 512 & 5 & 21 & 252 & 512 \\
        \midrule
        HKG & 0.24 & 0.27 & 0.27 & 0.30 & 0.28 & 0.30 & 0.28 & 0.28 & 0.29 & 0.33 & 0.28 & 0.25 & 0.18 & 0.20 & 0.32 & 0.33 & -0.05 & -0.21 & -0.08 & -0.17 \\
        TWN & -0.81 & -0.79 & -0.89 & -0.94 & -1.60 & -1.61 & -1.78 & -1.87 & -1.87 & -1.82 & -2.14 & -2.29 & -2.23 & -2.24 & -1.28 & -1.37 & -2.73 & -1.89 & -2.98 & -2.74 \\
        KOR & -0.64 & -0.64 & -0.72 & -0.78 & -1.24 & -1.27 & -1.37 & -1.44 & -1.46 & -1.53 & -1.70 & -1.75 & -1.72 & -1.75 & -0.99 & -1.06 & -2.05 & -1.23 & -2.60 & -2.08 \\
        DEU & 1.24 & 1.26 & 1.39 & 1.44 & 2.42 & 2.44 & 2.57 & 2.64 & 2.68 & 2.70 & 2.98 & 3.12 & 2.95 & 2.98 & 2.09 & 2.19 & 3.16 & 2.16 & 2.73 & 2.90 \\
        GBR & -0.22 & -0.22 & -0.24 & -0.26 & -0.62 & -0.61 & -0.71 & -0.78 & -0.70 & -0.71 & -0.81 & -0.92 & -0.94 & -0.94 & -0.45 & -0.53 & -1.31 & -0.99 & -1.35 & -1.41 \\
        IND & -0.20 & -0.19 & -0.19 & -0.19 & -0.16 & -0.17 & -0.24 & -0.25 & -0.29 & -0.27 & -0.31 & -0.35 & -0.46 & -0.48 & -0.06 & -0.09 & -0.90 & -0.50 & -0.62 & -0.64 \\
        AUS & 0.48 & 0.49 & 0.52 & 0.54 & 0.85 & 0.86 & 0.87 & 0.88 & 1.03 & 1.04 & 1.05 & 1.06 & 1.04 & 1.05 & 0.76 & 0.78 & 0.98 & 0.58 & 0.61 & 0.57 \\
        
        \midrule \midrule
        \textbf{Model} & \multicolumn{4}{c|}{\textbf{XGBoost}} & \multicolumn{4}{c|}{\textbf{CatBoost}} & \multicolumn{4}{c|}{\textbf{LightGBM}} & \multicolumn{4}{c|}{\textbf{NN-S}} & \multicolumn{4}{c}{\textbf{NN-L}} \\
        \textbf{Window Size} & 5 & 21 & 252 & 512 & 5 & 21 & 252 & 512 & 5 & 21 & 252 & 512 & 5 & 21 & 252 & 512 & 5 & 21 & 252 & 512 \\
        \midrule

        HKG & 0.59 & 0.88 & 0.98 & 0.95 & 0.65 & 1.07 & 1.11 & 1.08 & 0.54 & 1.03 & 0.99 & 1.10 & 0.73 & 0.57 & 0.12 & 0.14 & 0.53 & 0.65 & 0.78 & 0.91 \\
        TWN & -0.50 & 0.05 & -0.03 & -0.21 & -0.35 & 0.15 & 0.27 & 0.16 & -0.62 & 0.04 & -0.07 & -0.12 & -1.33 & -1.26 & -1.69 & -1.97 & -1.39 & -1.67 & -1.30 & -1.31 \\
        KOR & -0.59 & -0.44 & -0.36 & -0.51 & -0.70 & -0.60 & -0.30 & -0.24 & -0.99 & -0.49 & -0.48 & -0.45 & -0.82 & -1.03 & -1.53 & -1.19 & -1.35 & -1.18 & -1.18 & -0.84 \\
        DEU & 4.59 & 5.84 & 6.22 & 6.11 & 4.77 & 6.06 & 6.56 & 6.30 & 4.50 & 5.98 & 6.32 & 6.31 & 3.00 & 3.02 & 2.93 & 2.41 & 3.39 & 3.15 & 3.53 & 3.41 \\
        GBR & -0.19 & 0.14 & 0.22 & 0.12 & -0.23 & 0.25 & 0.40 & 0.28 & -0.26 & 0.21 & 0.23 & 0.22 & -0.61 & -0.76 & -0.84 & -1.08 & -0.68 & -0.72 & -0.75 & -0.67 \\
        IND & 2.70 & 3.99 & 4.11 & 3.68 & 3.11 & 4.23 & 4.71 & 4.23 & 2.89 & 4.09 & 4.14 & 4.03 & 1.33 & 1.15 & 0.40 & 0.69 & 1.40 & 0.96 & 1.40 & 1.29 \\
        AUS & 2.06 & 2.45 & 2.65 & 2.50 & 1.98 & 2.47 & 2.66 & 2.54 & 2.04 & 2.55 & 2.74 & 2.71 & 1.17 & 1.16 & 1.04 & 0.80 & 1.30 & 1.17 & 1.19 & 1.16 \\
        \bottomrule
      \end{tabular}
      \vspace{0.1cm}
      \begin{tablenotes}[para,flushleft]
        \footnotesize
        \textbf{Note:} This table reports the out-of-sample $R^2$ ($\RsqOOS$) from forecasts of benchmark models trained on global data evaluated across input window lengths of 5, 21, 252, and 512 trading days from 2001 to 2023. $\RsqOOS$ is reported in percentage terms and averaged with equal weight across the 23 evaluation years. Benchmark models include linear (OLS+H, LASSO+H, RIDGE+H, Enet+H, and PCR), ensemble (XGBoost, CatBoost, and LightGBM), and neural network (NN-S and NN-L) models. `H' indicates that the model is estimated using the Huber loss. The set of countries is determined by three criteria: the relative size of the equity market (measured by total market capitalization), the availability of short-selling mechanisms, subject to historical market-specific restrictions, and the availability of reliable data. The ordering of countries is arbitrary.
      \end{tablenotes}
    \end{minipage}
    \end{adjustbox}
  \end{threeparttable}
\end{table}
\vspace*{\fill}
\end{landscape}

\begin{landscape}
\vspace*{\fill}
\thispagestyle{landscape}
\begin{table}[h]
\vspace*{\fill} % push it down\
  \centering
  \begin{threeparttable}
    \begin{adjustbox}{width=1.2\textwidth, center}
    \scriptsize
    \captionsetup{width=\linewidth}
    \caption{Benchmark Models - SR (International)}
    \label{international_benchmark_portfolio_table}
    \begin{minipage}{\linewidth}
        \renewcommand{\arraystretch}{1.3}
      \begin{tabular}{c|cccc|cccc|cccc|cccc|cccc}
        \toprule
        \textbf{Model} & \multicolumn{4}{c|}{\textbf{OLS+H}} & \multicolumn{4}{c|}{\textbf{LASSO+H}} & \multicolumn{4}{c|}{\textbf{RIDGE+H}} & \multicolumn{4}{c|}{\textbf{Enet+H}} & \multicolumn{4}{c}{\textbf{PCR}} \\
        \textbf{Window Size} & 5 & 21 & 252 & 512 & 5 & 21 & 252 & 512 & 5 & 21 & 252 & 512 & 5 & 21 & 252 & 512 & 5 & 21 & 252 & 512 \\
        \midrule

        HKG & 1.71 & 1.52 & 1.39 & 1.30 & 1.02 & 1.02 & 1.02 & 1.02 & 1.62 & 1.36 & 1.28 & 1.28 & 1.20 & 1.24 & 1.02 & 1.02 & 1.20 & 0.50 & 0.99 & 0.87 \\
        TWN & -1.51 & -1.24 & -1.44 & -1.37 & -1.52 & -1.52 & -1.52 & -1.52 & -1.58 & -1.37 & -1.47 & -1.41 & -1.70 & -1.65 & -1.52 & -1.52 & -1.35 & -1.39 & -1.22 & -1.38 \\
        KOR & -0.01 & -0.05 & -0.02 & 0.07 & 0.15 & 0.15 & 0.15 & 0.15 & 0.01 & -0.05 & -0.00 & -0.00 & 0.12 & 0.12 & 0.15 & 0.15 & 0.16 & -0.04 & 0.05 & -0.14 \\
        DEU & 5.93 & 5.48 & 5.17 & 5.03 & 4.78 & 4.78 & 4.78 & 4.78 & 5.91 & 5.50 & 5.30 & 5.32 & 5.21 & 5.29 & 4.78 & 4.78 & 4.76 & 3.03 & 3.64 & 3.70 \\
        GBR & 0.94 & 0.93 & 0.86 & 0.75 & 0.57 & 0.57 & 0.57 & 0.57 & 0.90 & 0.76 & 0.97 & 0.88 & 0.78 & 0.73 & 0.57 & 0.57 & 0.75 & 0.24 & 0.61 & 0.42 \\
        IND & 0.54 & 0.84 & 0.81 & 0.91 & 0.30 & 0.30 & 0.30 & 0.30 & 0.58 & 0.70 & 0.76 & 0.59 & 0.11 & 0.06 & 0.30 & 0.30 & 0.25 & 0.42 & 0.70 & 0.76 \\
        AUS & 6.42 & 5.55 & 5.64 & 5.27 & 4.93 & 4.93 & 4.93 & 4.93 & 6.42 & 5.81 & 5.31 & 5.27 & 5.46 & 5.49 & 4.93 & 4.93 & 5.61 & 3.45 & 3.71 & 3.59 \\
        
        \midrule \midrule
        \textbf{Model} & \multicolumn{4}{c|}{\textbf{XGBoost}} & \multicolumn{4}{c|}{\textbf{CatBoost}} & \multicolumn{4}{c|}{\textbf{LightGBM}} & \multicolumn{4}{c|}{\textbf{NN-S}} & \multicolumn{4}{c}{\textbf{NN-L}} \\
        \textbf{Window Size} & 5 & 21 & 252 & 512 & 5 & 21 & 252 & 512 & 5 & 21 & 252 & 512 & 5 & 21 & 252 & 512 & 5 & 21 & 252 & 512 \\
        \midrule

        HKG & 1.63 & 1.81 & 1.88 & 1.74 & 1.75 & 1.86 & 1.93 & 1.92 & 1.69 & 1.77 & 1.79 & 1.77 & 0.66 & 0.83 & 0.78 & 0.68 & 1.17 & 0.95 & 0.44 & 0.42 \\
        TWN & 0.25 & 0.96 & 1.12 & 0.48 & 0.42 & 1.50 & 1.59 & 1.15 & 0.47 & 1.28 & 0.79 & 0.59 & -0.98 & -0.64 & -1.12 & -0.95 & -1.03 & -1.01 & -0.67 & -0.89 \\
        KOR & 0.06 & 0.19 & 0.41 & 0.16 & 0.19 & 0.30 & 0.28 & 0.29 & 0.12 & 0.10 & 0.17 & 0.15 & 0.10 & -0.04 & -0.08 & -0.08 & 0.36 & 0.14 & -0.01 & 0.01 \\
        DEU & 5.69 & 5.95 & 5.85 & 5.73 & 5.59 & 6.08 & 5.92 & 5.90 & 5.61 & 5.91 & 5.93 & 5.99 & 2.86 & 2.56 & 3.16 & 2.45 & 2.62 & 2.78 & 2.37 & 2.60 \\
        GBR & 1.31 & 1.63 & 1.67 & 1.52 & 1.27 & 1.94 & 2.35 & 1.88 & 1.40 & 1.80 & 1.57 & 1.67 & 0.48 & 0.41 & 0.59 & 0.11 & 0.13 & 0.37 & 0.52 & 0.56 \\
        IND & 5.34 & 6.00 & 5.58 & 5.17 & 5.49 & 6.44 & 6.74 & 6.16 & 5.22 & 6.13 & 5.58 & 5.31 & 1.18 & 1.21 & 1.28 & 1.43 & 1.13 & 1.47 & 1.39 & 1.65 \\
        AUS & 5.88 & 6.03 & 6.15 & 6.10 & 5.89 & 6.16 & 6.02 & 6.16 & 5.97 & 6.02 & 6.14 & 6.16 & 2.87 & 2.29 & 2.79 & 2.14 & 2.81 & 3.29 & 2.44 & 2.62 \\
        \bottomrule
      \end{tabular}
      \vspace{0.1cm}
      \begin{tablenotes}[para,flushleft]
        \footnotesize
        \textbf{Note:} This table reports the out-of-sample Sharpe ratios of long--short portfolios constructed from forecasts of benchmark models trained on global data, evaluated across input window lengths of 5, 21, 252, and 512 trading days from 2001 to 2023. Benchmark models include linear (OLS+H, LASSO+H, RIDGE+H, Enet+H, and PCR), ensemble (XGBoost, CatBoost, and LightGBM), and neural network (NN-S and NN-L) models. `H' indicates that the model is estimated using the Huber loss. Portfolios are formed using decile sorting based on model forecasts, with equal weighting across stocks. The set of countries is determined by three criteria: the relative size of the equity market (measured by total market capitalization), the availability of short-selling mechanisms, subject to historical market-specific restrictions, and the availability of reliable data. The ordering of countries is arbitrary.
      \end{tablenotes}
    \end{minipage}
    \end{adjustbox}
  \end{threeparttable}
\end{table}
\vspace*{\fill}
\end{landscape}

% This subsection reports the complete international TSFM results for the pre-trained models summarized in \Cref{international_representative_comparison}.

\begin{landscape}
\thispagestyle{landscape}
\subsection{Complete International TSFM Results} \label{complete_international_tsfm_results}
\vspace*{\fill} % push it down\
\begin{table}[h]
  \centering
  \begin{threeparttable}
    \begin{adjustbox}{width=1.2\textwidth, center}
    \scriptsize
    \captionsetup{width=\linewidth}
    \caption{Finance-Native Pre-Trained TSFMs - $\RsqOOS$ (International)}
    \label{international_tsfm_forecasting_table}
    \begin{minipage}{\linewidth}
        \renewcommand{\arraystretch}{1.3}
      \begin{tabular}{c|cccc|cccc|cccc|cccc|cccc}
          \toprule
          & \multicolumn{20}{c}{\textbf{Global}} \\
          \cmidrule(lr){2-21}
        \textbf{Model} & \multicolumn{4}{c|}{\textbf{Chronos (Tiny)}} & \multicolumn{4}{c|}{\textbf{Chronos (Mini)}} & \multicolumn{4}{c|}{\textbf{Chronos (Small)}} & \multicolumn{4}{c|}{\textbf{TimesFM (8M)}} & \multicolumn{4}{c}{\textbf{TimesFM (20M)}} \\
        \textbf{Window Size} & 5 & 21 & 252 & 512 & 5 & 21 & 252 & 512 & 5 & 21 & 252 & 512 & 5 & 21 & 252 & 512 & 5 & 21 & 252 & 512 \\
        \midrule

        HKG & -0.78 & -0.40 & -0.75 & -1.14 & -0.65 & -0.67 & -0.84 & -1.13 & -1.11 & -0.43 & -0.48 & -0.88 & -31.64 & -23.15 & -31.64 & -46.53 & -43.24 & -51.53 & -32.18 & -34.42 \\
        TWN & -1.04 & -1.27 & -2.68 & -2.91 & -1.36 & -1.44 & -3.66 & -4.07 & -1.65 & -1.60 & -4.67 & -4.74 & -30.96 & -22.34 & -27.66 & -39.14 & -42.39 & -50.72 & -34.00 & -38.71 \\
        KOR & -1.07 & -0.69 & -0.82 & -1.04 & -1.13 & -0.38 & -1.73 & -2.15 & -1.40 & -1.39 & -1.67 & -2.30 & -30.69 & -22.74 & -21.64 & -31.83 & -40.20 & -46.77 & -26.77 & -37.65 \\
        DEU & -0.66 & 0.96 & 1.50 & 1.08 & -0.91 & 1.28 & 2.52 & 2.24 & -0.73 & 1.82 & 3.17 & 2.91 & -36.70 & -22.84 & -23.12 & -22.39 & -46.81 & -49.51 & -18.79 & -17.75 \\
        GBR & -1.20 & -0.64 & -0.59 & -0.74 & -1.25 & -0.59 & -0.47 & -0.63 & -1.14 & -0.60 & -0.59 & -0.77 & -32.31 & -22.44 & -24.49 & -30.96 & -43.15 & -49.79 & -25.59 & -30.12 \\
        IND & 0.38 & 2.06 & 2.44 & 2.11 & 0.24 & 3.19 & 4.17 & 3.95 & 0.50 & 3.66 & 4.55 & 4.41 & -31.69 & -22.63 & -17.11 & -17.98 & -41.00 & -48.93 & -29.95 & -27.73 \\
        AUS & -0.87 & 0.24 & 0.27 & 0.49 & -1.10 & 0.34 & 0.67 & 0.69 & -0.97 & 0.32 & 0.75 & 0.72 & -35.06 & -22.98 & -18.45 & -13.95 & -46.18 & -52.62 & -26.34 & -28.02 \\
        
        \midrule
          & \multicolumn{20}{c}{\textbf{JKP-Augmented}} \\
        \midrule

        HKG & -4.67 & -2.69 & -0.75 & -0.91 & -3.83 & -2.62 & -0.93 & -1.38 & -5.59 & -2.16 & -0.96 & -1.26 & -22.57 & -10.28 & -3.53 & -7.43 & -19.95 & -6.61 & -0.06 & 0.13 \\
        TWN & -4.81 & -3.05 & -2.39 & -2.09 & -4.89 & -3.21 & -3.66 & -4.09 & -5.64 & -3.40 & -4.38 & -4.63 & -23.15 & -12.74 & -5.82 & -10.33 & -20.53 & -10.50 & -1.83 & -1.54 \\
        KOR & -4.33 & -1.63 & -0.93 & -0.93 & -4.90 & -2.08 & -1.22 & -1.19 & -4.69 & -2.84 & -1.83 & -2.62 & -23.27 & -13.76 & -5.28 & -9.84 & -20.73 & -11.87 & -1.58 & -2.22 \\
        DEU & -5.43 & -1.52 & 1.84 & 1.80 & -5.54 & -1.47 & 2.37 & 2.26 & -5.44 & -0.95 & 3.90 & 3.53 & -23.70 & -6.57 & 3.70 & 3.39 & -22.07 & -1.75 & 7.29 & 7.69 \\
        GBR & -4.92 & -3.36 & -0.52 & -0.69 &-4.85 & -3.64 & -0.56 & -0.75 & -6.20 & -3.85 & -0.69 & -0.74 & -22.95 & -11.69 & -3.94 & -7.66 & -21.04 & -8.68 & -1.09 & -0.97 \\
        IND & -3.70 & -0.84 & 2.79 & 2.42 & -3.90 & -0.22 & 3.52 & 2.67 & -3.74 & -0.56 & 4.01 & 3.60 & -20.86 & -9.49 & 0.12 & -3.49 & -18.36 & -5.61 & 4.29 & 4.65 \\
        AUS & -4.48 & -1.26 & 0.34 & 0.50 & -5.17 & -2.04 & 0.39 & 0.70 & -5.05 & -1.45 & 0.79 & 0.95 & -23.62 & -9.26 & -0.57 & -2.56 & -21.57 & -5.12 & 2.16 & 2.49 \\

        \midrule
          & \multicolumn{20}{c}{\textbf{Synthetic-Augmented}} \\
        \midrule

        HKG & -18.86 & -2.27 & -0.62 & -0.91 & -21.23 & -3.12 & -0.59 & -0.62 & -23.89 & -3.48 & -0.45 & -0.46 & -32.41 & -8.46 & -0.94 & -0.98 & -18.09 & -4.92 & -0.06 & -0.05 \\
        TWN & -24.41 & -7.01 & -2.68 & -2.55 & -25.19 & -6.04 & -2.68 & -2.59 & -23.57 & -5.80 & -2.61 & -2.78 & -33.29 & -10.95 & -3.54 & -3.12 & -19.64 & -7.74 & -2.45 & -2.31 \\
        KOR & -20.53 & -6.04 & -0.77 & -1.00 & -23.28 & -9.44 & -0.99 & -0.85 & -26.36 & -6.08 & -0.83 & -1.07 & -31.94 & -12.03 & -3.93 & -3.97 & -20.12 & -9.25 & -3.09 & -3.39 \\
        DEU & -17.74 & -1.05 & 1.49 & 1.48 & -20.81 & -0.86 & 2.12 & 2.19 & -23.48 & -1.69 & 3.01 & 2.56 & -33.70 & -5.73 & 5.37 & 5.88 & -20.39 & -0.80 & 7.20 & 7.49 \\
        GBR & -21.44 & -2.29 & -0.62 & -0.97 & -25.95 & -3.12 & -0.56 & -0.73 & -29.62 & -3.30 & -0.58 & -0.69 & -32.35 & -9.66 & -1.97 & -2.01 & -19.69 & -6.33 & -1.53 & -1.54 \\
        IND & -18.77 & -4.52 & 1.65 & 1.81 & -19.37 & -4.15 & 2.98 & 2.31 & -19.95 & -3.74 & 3.71 & 3.07 & -29.72 & -7.74 & 2.15 & 2.24 & -16.77 & -4.07 & 3.23 & 3.52 \\
        AUS & -18.73 & -2.05 & 0.15 & 0.25 & -22.34 & -1.99 & 0.31 & 0.75 & -27.64 & -1.69 & 0.45 & 0.80 & -33.38 & -7.86 & 1.23 & 1.45 & -20.10 & -3.64 & 1.93 & 2.07 \\

        \bottomrule
      \end{tabular}
      \vspace{0.1cm}
      \begin{tablenotes}[para,flushleft]
        \footnotesize
        \textbf{Note:} This table reports the out-of-sample $R^2$ ($\RsqOOS$) from forecasts of the time series foundation models (TSFMs) pre-trained on global data (top panel), JKP-augmented data (middle panel), and synthetic-augmented data (bottom panel), evaluated across input window lengths of 5, 21, 252, and 512 trading days from 2001 to 2023. $\RsqOOS$ is reported in percentage terms and averaged with equal weight across the 23 evaluation years. JKP characteristics used in the augmented data are defined in \citet{jensen2023there}, and the synthetic data are generated following \citet{ansari2024chronos}. The TSFMs include Chronos (tiny, mini, and small) and TimesFM (with 8 million and 20 million parameters). Inference is performed using the finance-native pre-trained models. The set of countries is determined by three criteria: the relative size of the equity market (measured by total market capitalization), the availability of short-selling mechanisms, subject to historical market-specific restrictions, and the availability of reliable data. The ordering of countries is arbitrary.
      \end{tablenotes}
    \end{minipage}
    \end{adjustbox}
  \end{threeparttable}
\end{table}
\vspace*{\fill} % push it down\
\end{landscape}

\begin{landscape}
\thispagestyle{landscape}
\begin{table}
\vspace*{\fill} % push it down\
  \centering
  \begin{threeparttable}
    \begin{adjustbox}{width=1.2\textwidth, center}
    \scriptsize
    \captionsetup{width=\linewidth}
    \caption{Finance-Native Pre-Trained TSFMs - SR (International)}
    \label{international_tsfm_portfolio_table}
    \begin{minipage}{\linewidth}
        \renewcommand{\arraystretch}{1.3}
      \begin{tabular}{c|cccc|cccc|cccc|cccc|cccc}
          \toprule
          & \multicolumn{20}{c}{\textbf{Global}} \\
          \cmidrule(lr){2-21}

        \textbf{Model} & \multicolumn{4}{c|}{\textbf{Chronos (Tiny)}} & \multicolumn{4}{c|}{\textbf{Chronos (Mini)}} & \multicolumn{4}{c|}{\textbf{Chronos (Small)}} & \multicolumn{4}{c|}{\textbf{TimesFM (8M)}} & \multicolumn{4}{c}{\textbf{TimesFM (20M)}} \\
        \textbf{Window Size} & 5 & 21 & 252 & 512 & 5 & 21 & 252 & 512 & 5 & 21 & 252 & 512 & 5 & 21 & 252 & 512 & 5 & 21 & 252 & 512 \\
        \midrule

        HKG & 0.23 & 0.73 & 1.10 & 0.79 & 0.19 & 0.73 & 1.11 & 0.91 & 0.22 & 0.81 & 1.41 & 1.37 & -1.40 & -1.15 & -0.20 & 0.10 & -0.78 & 0.12 & 0.80 & 0.82 \\
        TWN & 0.31 & 0.45 & 1.19 & 0.99 & 0.26 & 1.07 & 1.57 & 1.23 & 0.42 & 0.63 & 1.29 & 1.30 & 0.49 & 0.40 & -0.08 & 0.20 & 0.36 & 0.07 & -0.42 & -0.58 \\
        KOR & -0.25 & 0.32 & 0.77 & 0.30 & -0.04 & 0.29 & 0.38 & 0.68 & -0.07 & 0.21 & 0.64 & 0.75 & -0.22 & -0.15 & 0.35 & 0.06 & -0.08 & -0.11 & 0.35 & 0.21 \\
        DEU & -0.27 & 1.81 & 2.53 & 2.75 & 0.13 & 2.02 & 3.11 & 3.24 & 0.29 & 2.19 & 3.42 & 3.40 & -3.53 & -1.14 & 1.02 & 0.97 & -2.70 & 0.69 & 3.11 & 3.48 \\
        GBR & -0.18 & 0.34 & 1.39 & 1.54 & -0.26 & 0.82 & 1.33 & 1.57 & -0.18 & 0.81 & 1.47 & 1.77 & -1.12 & -0.98 & 0.16 & 0.23 & -0.85 & -0.19 & 0.89 & 0.86 \\
        IND & 2.81 & 4.79 & 5.63 & 5.09 & 2.92 & 5.57 & 6.21 & 6.08 & 3.08 & 5.60 & 6.71 & 6.40 & 0.42 & 0.38 & 0.76 & 0.55 & 1.00 & 1.44 & 1.12 & 1.58 \\
        AUS & 0.02 & 1.81 & 2.30 & 2.63 & -0.31 & 2.17 & 2.77 & 2.67 & 0.17 & 2.26 & 2.65 & 2.99 & -4.86 & -2.52 & 0.53 & 0.89 & -3.84 & -0.12 & 2.49 & 2.78 \\
        
        \midrule
          & \multicolumn{20}{c}{\textbf{JKP-Augmented}} \\
        \midrule

        HKG & -0.46 & 0.37 & 1.09 & 0.96 & -0.25 & 0.35 & 1.10 & 0.93 & -0.13 & 0.27 & 1.09 & 1.16 & -0.98 & -0.05 & 0.97 & 1.24 & -0.96 & 0.21 & 1.42 & 1.58 \\
        TWN & 0.52 & 0.78 & 0.86 & 1.28 & 0.48 & 0.70 & 1.57 & 1.45 & 0.53 & 1.14 & 1.77 & 1.88 & 0.32 & 0.25 & 0.01 & 0.16 & 0.17 & 0.07 & 1.13 & 1.16 \\
        KOR & -0.25 & 0.39 & 0.73 & 0.27 & -0.12 & 0.40 & 0.72 & 0.48 & -0.24 & 0.22 & 0.92 & 0.32 & -0.17 & 0.03 & 0.21 & 0.03 & -0.16 & -0.08 & 0.69 & 0.66 \\
        DEU & -2.16 & 0.24 & 2.73 & 2.69 & -2.41 & 0.27 & 3.01 & 3.14 & -2.19 & 0.58 & 3.20 & 3.27 & -2.79 & 1.04 & 4.26 & 4.56 & -2.63 & 2.35 & 5.56 & 5.77 \\
        GBR & -1.03 & -0.03 & 1.37 & 1.62 & -0.90 & 0.10 & 1.32 & 1.61 & -0.96 & -0.04 & 1.41 & 1.55 & -0.87 & -0.13 & 1.59 & 1.46 & -0.98 & 0.18 & 1.89 & 1.81 \\
        IND & 1.27 & 3.48 & 4.71 & 4.27 & 1.33 & 3.29 & 5.48 & 5.11 & 1.74 & 3.47 & 6.11 & 5.77 & 0.57 & 1.76 & 4.76 & 4.73 & 0.72 & 3.14 & 6.31 & 6.39 \\
        AUS & -2.52 & -0.18 & 2.18 & 2.46 & -2.74 & 0.03 & 2.28 & 2.21 & -2.32 & 0.54 & 2.59 & 2.77 & -4.31 & 0.14 & 4.82 & 4.99 & -4.75 & 1.29 & 5.39 & 5.84 \\

        \midrule
          & \multicolumn{20}{c}{\textbf{Synthetic-Augmented}} \\
        \midrule

        HKG & -0.44 & 0.27 & 0.85 & 1.27 & -0.49 & 0.54 & 1.12 & 1.18 & -0.42 & 0.26 & 1.12 & 1.33 & -0.87 & -0.27 & 1.03 & 1.12 & -1.09 & 0.27 & 1.28 & 1.72 \\
        TWN & 0.34 & 0.57 & 1.01 & 1.26 & 0.70 & 0.88 & 1.14 & 1.34 & 0.82 & 0.82 & 1.29 & 1.37 & 0.50 & 0.34 & 0.41 & 0.44 & 0.38 & 0.58 & 0.85 & 0.96 \\
        KOR & -0.11 & -0.01 & 0.73 & 0.37 & -0.13 & 0.00 & 0.83 & 0.33 & -0.44 & -0.22 & 0.65 & 0.53 & -0.08 & 0.03 & 0.11 & 0.14 & -0.14 & -0.22 & 0.23 & 0.32 \\
        DEU & -0.87 & 0.47 & 2.71 & 2.75 & -1.25 & 0.46 & 3.22 & 3.36 & -1.81 & 0.70 & 3.10 & 3.20 & -2.88 & 0.26 & 4.91 & 5.20 & -3.48 & 1.45 & 5.65 & 5.89 \\
        GBR & -0.76 & -0.20 & 0.80 & 1.16 & -0.87 & 0.20 & 1.58 & 1.81 & -0.79 & 0.30 & 1.46 & 1.88 & -0.66 & -0.13 & 1.10 & 1.40 & -0.74 & 0.23 & 1.73 & 1.61 \\
        IND & 2.07 & 2.65 & 4.03 & 3.83 & 1.70 & 3.03 & 5.26 & 4.97 & 1.63 & 3.18 & 5.53 & 5.34 & 0.88 & 1.46 & 4.73 & 4.76 & 0.81 & 2.79 & 5.24 & 4.87 \\
        AUS & -1.61 & 0.45 & 2.05 & 2.16 & -2.08 & 0.13 & 2.11 & 2.45 & -2.41 & 0.45 & 2.35 & 2.55 & -4.30 & -0.73 & 4.84 & 5.28 & -4.56 & 0.33 & 5.31 & 5.57 \\
        
        \bottomrule
      \end{tabular}
      \vspace{0.1cm}
      \begin{tablenotes}[para,flushleft]
        \footnotesize
        \textbf{Note:} This table reports the out-of-sample Sharpe ratios of long--short portfolios constructed from forecasts of the time series foundation models (TSFMs) pre-trained on global data (top panel), JKP-augmented data (middle panel), and synthetic-augmented data (bottom panel), evaluated across input window lengths of 5, 21, 252, and 512 trading days from 2001 to 2023. JKP characteristics used in the augmented data are defined in \citet{jensen2023there}, and the synthetic data are generated following \citet{ansari2024chronos}. The TSFMs include Chronos (tiny, mini, and small) and TimesFM (with 8 million and 20 million parameters). Inference is performed using the finance-native pre-trained models. Portfolios are formed using decile sorting based on model forecasts, with equal weighting across stocks. The set of countries is determined by three criteria: the relative size of the equity market (measured by total market capitalization), the availability of short-selling mechanisms, subject to historical market-specific restrictions, and the availability of reliable data. The ordering of countries is arbitrary.
      \end{tablenotes}
    \end{minipage}
    \end{adjustbox}
  \end{threeparttable}
\end{table}
\end{landscape}

\subsection{Distributional Evidence} \label{dis_evidence_appdedix}
\begin{figure}[htbp]
    \caption{Chronos TSFM Forecasts over the Next 30 Trading Days of Selected U.S. Equity Excess Returns}
  \centering
  \begin{threeparttable}

    \label{chronos_excess_return_30_days_forecast}

    \begin{adjustbox}{width=0.93\textwidth, center}
      \includegraphics[
        width=\linewidth,
        clip,
        trim=10 23 0 20
      ]{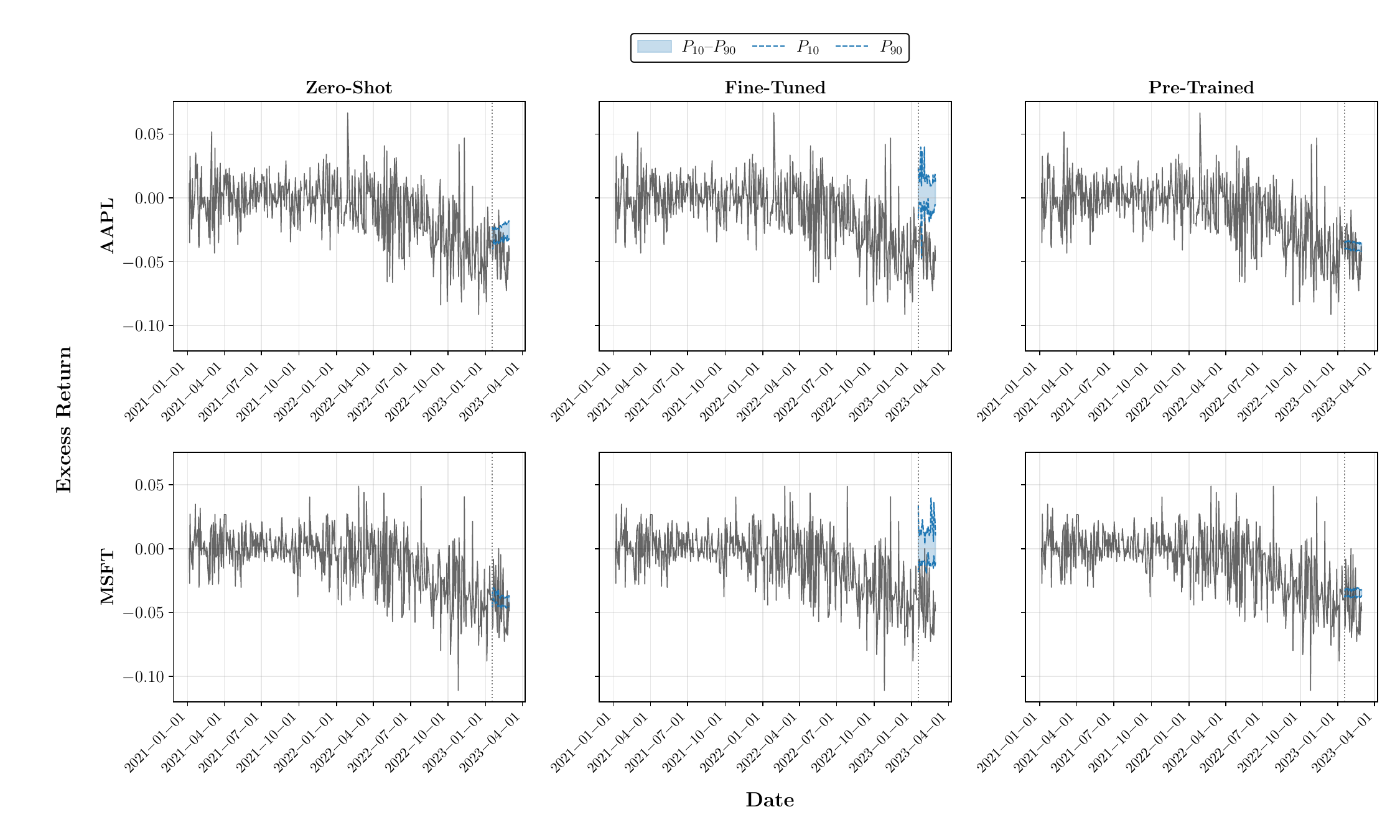}
    \end{adjustbox}

    \begin{tablenotes}[para,flushleft]
      \setlength{\itemsep}{0pt}
      \setlength{\parskip}{3pt}
      \footnotesize
      \justifying
      \textbf{Note:} This figure presents interval forecasts for horizons of 1--30 trading days ($P_{10}$--$P_{90}$) of daily excess returns for selected U.S. stocks (AAPL and MSFT), generated using Chronos TSFMs. The left column displays zero-shot forecasts from the publicly available Chronos model, the middle column shows forecasts based on a fine-tuned, publicly available Chronos model, and the right column reports forecasts from a finance-native Chronos model pre-trained from scratch on global excess returns. All models use the small configuration. The black lines represent realized excess returns, the shaded blue regions denote 80\% predictive intervals, and the dashed blue lines indicate the 10\textsuperscript{th} and 90\textsuperscript{th} percentile boundaries. The vertical dotted line marks the forecast start date. The forecasting horizon spans 30 trading days, using a historical context of 512 observations. Consistent with \Cref{zero_shot_results}, the number of generated samples is fixed at 20.
    \end{tablenotes}

  \end{threeparttable}
\end{figure}

The international results in \Cref{intenational_results} evaluate model performance through aggregate forecasting and portfolio metrics. To complement this evidence, \Cref{chronos_excess_return_30_days_forecast} provides a distributional illustration of individual U.S. equity excess-return forecasts from three Chronos (Small) variants: the off-the-shelf zero-shot model, a model fine-tuned from the public checkpoint, and a model pre-trained from scratch on global excess returns. This comparison links the distributional illustration to the international and scaling results through the shared global pre-training sample.

\Cref{chronos_excess_return_30_days_forecast} is an illustrative distributional comparison rather than a separate test: it shows how the training regime changes the predictive distributions at horizons of 1--30 trading days for a fixed 512-day historical context. The zero-shot and fine-tuned versions inherit representations learned largely outside financial return forecasting, whereas the globally pre-trained model has learned directly from a large panel of equity excess returns across markets. The global Chronos forecasts therefore provide a visual counterpart to the table evidence, illustrating differences in the shape and location of the forecast distribution at the individual-stock level. Together with the aggregate results, this comparison highlights a limitation of transfer from generic TSFM pre-training: the broad cross-domain generalization associated with LLMs does not automatically carry over to financial return forecasting with the models evaluated here. Exposure to massive and heterogeneous time-series corpora does not, by itself, ensure useful financial representations, and conventional fine-tuning does not reliably overcome this limitation. The separate portfolio results provide evidence of more useful long-window ranking signals under finance-native pre-training; the selected distributional examples alone do not establish economic usefulness or probabilistic calibration.

The predictive intervals differ in width and location across stocks and training regimes. These selected examples do not establish whether the nominal 80\% intervals attain their stated coverage; assessing probabilistic calibration would require a systematic out-of-sample coverage analysis across forecast origins and stocks. This is consistent with the broader empirical message of the study: global data scale can help TSFMs learn more useful return representations when combined with domain-aligned training, but it does not eliminate the uncertainty inherent in daily stock returns. Thus, the generalization promise motivating TSFMs requires careful qualification in finance, where neither generic pre-training scale nor subsequent adaptation guarantees reliable forecasting or economic value.

\clearpage

\subsection{Detailed Model Configuration Results}
\label{model_configuration_results}
\vspace*{\fill}
\begin{figure}[H]
    \caption{Impact of Hyperparameter Choice on Forecasting Performance}
    \label{impact_TSFM_HP_forecasting}
    \centering
    \includegraphics[
        width=0.75\textwidth,
        clip,
        trim=0 56 0 63, % left bottom right top
    ]{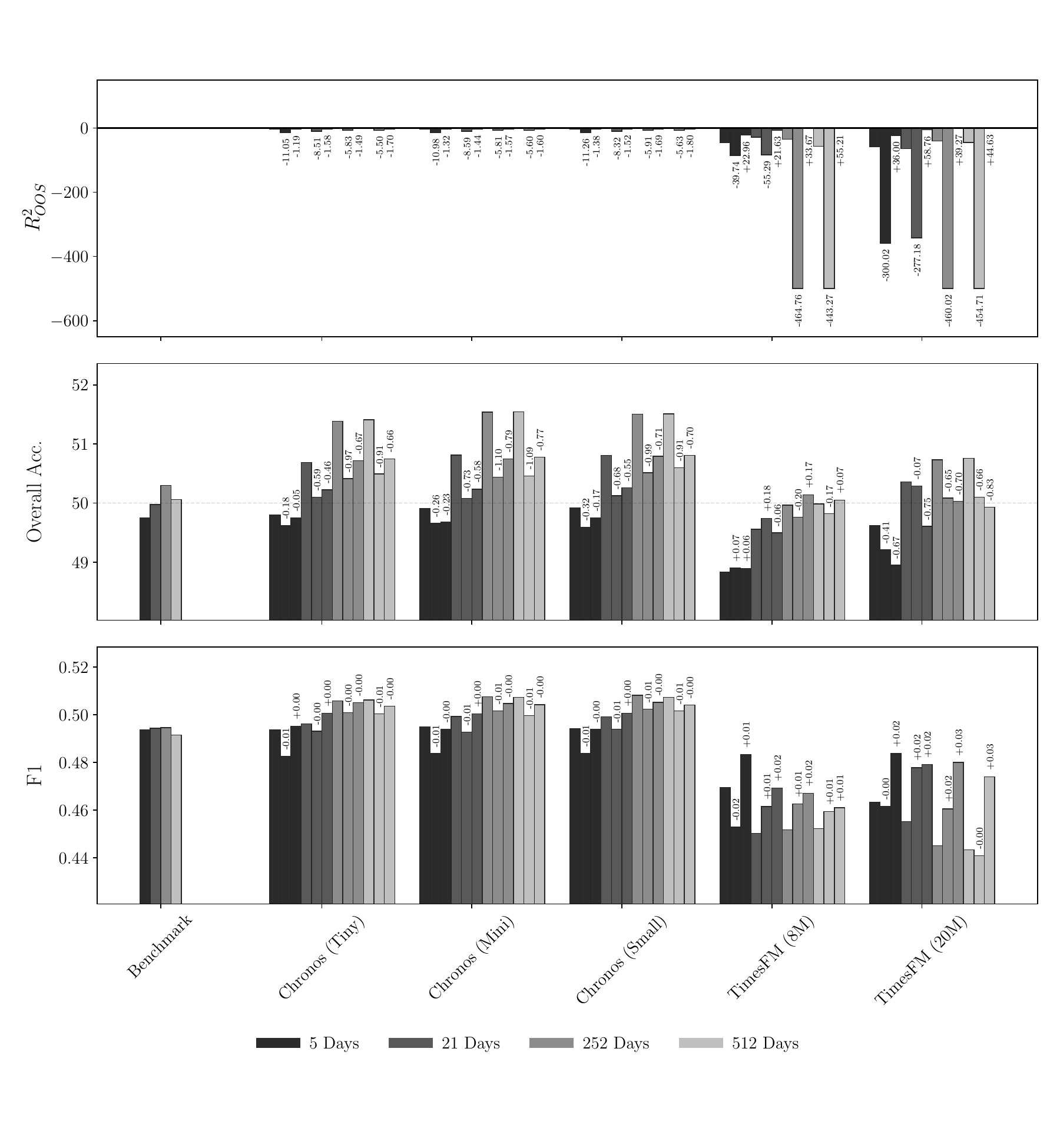} % Adjust the width and trims
    \begin{minipage}{0.8\textwidth}
        \footnotesize
        \textbf{Note:} This figure compares forecasting performance metrics ($\RsqOOS$, overall accuracy, and F1) for five models under three model configurations. For Chronos, the left bar reports results for models pre-trained on global data with the default quantization range. The middle bar presents results obtained by restricting the tokenizer’s quantization range from its default interval of $[-15, 15]$ to $[-2, 2]$, thereby allowing us to evaluate how a tighter binning interval affects model performance. The right bar corresponds to a dynamic strategy in which the quantization bounds are recalibrated annually to the 1\textsuperscript{st} and 99\textsuperscript{th} percentiles of pooled pre-training excess returns normalized by their pooled mean absolute value. For TimesFM, the left bar again reports results for models pre-trained on global data with default hyperparameters. The middle bar presents results obtained by reducing the default input patch length from 32 to 8, while the right bar shows results when increasing the input patch length to 128. Bars of the same color correspond to the same input window length (5, 21, 252, and 512 trading days). Models include Chronos (tiny, mini, and small) and TimesFM (with 8 million and 20 million parameters), all pre-trained from scratch. The benchmark group presents the results of CatBoost trained on global excess returns and evaluated on U.S. stocks using input windows of 5, 21, 252, and 512 trading days. `Overall Acc.' denotes overall directional accuracy, and `F1' refers to the macro-averaged F1 score. Accuracy measures and $\RsqOOS$ are reported in percentage terms; macro-$F_1$ is reported on a zero-to-one scale. In the middle panel, the horizontal line indicates 50\% overall accuracy. To enhance interpretability of the plots, $\RsqOOS$ values were clipped at $-500\%$ in order to mitigate the influence of extreme negative outliers on the visual scale.
    \end{minipage}
\end{figure}
\vspace*{\fill}

\begin{figure}
    \caption{Impact of Hyperparameter Choice on Portfolio Performance}
    \label{impact_TSFM_HP_performance}
    \centering
    \includegraphics[
        width=0.75\textwidth,
        clip,
        trim=0 56 0 63, % left bottom right top
    ]{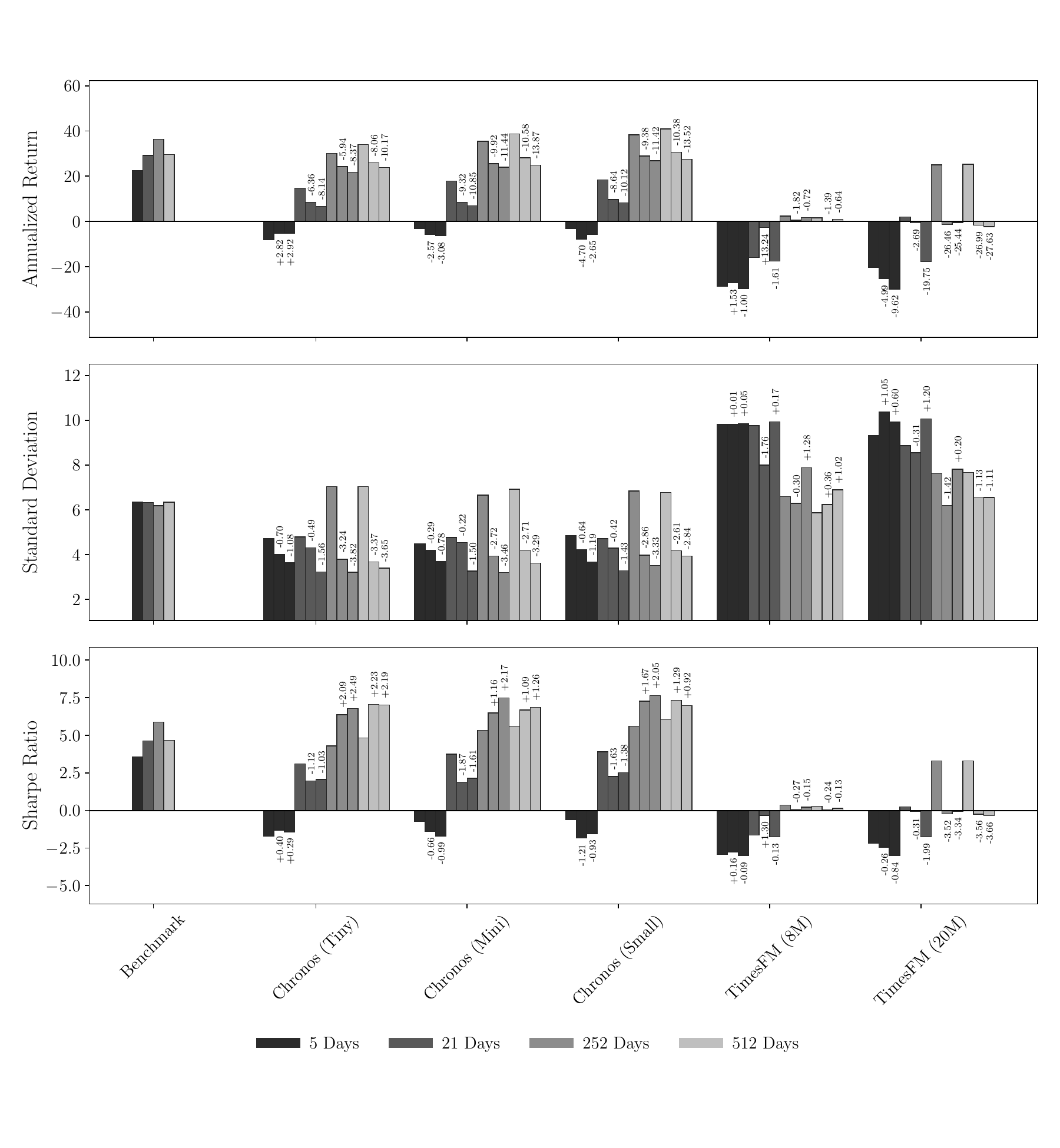} % Adjust the width and trims
    \begin{minipage}{0.8\textwidth}
        \footnotesize
        \textbf{Note:} This figure compares portfolio performance metrics (annualized return, standard deviation, and Sharpe ratio) for five models under three model configurations. For Chronos, the left bar reports results for models pre-trained on global data with the default quantization range. The middle bar presents results obtained by restricting the tokenizer’s quantization range from its default interval of $[-15, 15]$ to $[-2, 2]$, thereby allowing us to evaluate how a tighter binning interval affects model performance. The right bar corresponds to a dynamic strategy in which the quantization bounds are recalibrated annually to the 1\textsuperscript{st} and 99\textsuperscript{th} percentiles of pooled pre-training excess returns normalized by their pooled mean absolute value. For TimesFM, the left bar again reports results for models pre-trained on global data with default hyperparameters. The middle bar presents results obtained by reducing the default input patch length from 32 to 8, while the right bar shows results when increasing the input patch length to 128. Bars of the same color correspond to the same input window length (5, 21, 252, and 512 trading days). Models include Chronos (tiny, mini, and small) and TimesFM (with 8 million and 20 million parameters), all pre-trained from scratch. The benchmark group presents the results of CatBoost trained on global excess returns and evaluated on U.S. stocks using input windows of 5, 21, 252, and 512 trading days.
    \end{minipage}
\end{figure}

\clearpage

\end{bibunit}

% ============================================================
% INTERNET APPENDIX
% ============================================================

\clearpage

\begin{bibunit}

\include{internet_appendix}

% References for Internet Appendix only
\clearpage
\putbib[bib_list]

\end{bibunit}

\end{document}